\documentclass[12pt,a4paper]{article} 
\pdfoutput=1

\usepackage{jheppub}
\usepackage{epstopdf}
\usepackage{graphicx}
\usepackage{epsfig}
\usepackage{dcolumn}  
\usepackage{amssymb} 
\usepackage{amsmath}
\usepackage{amsfonts}
\usepackage{etex}
\usepackage{slashed}  
\usepackage{youngtab}
\usepackage[mathscr]{euscript}
\usepackage{epsfig}
\usepackage[utf8]{inputenc}
\usepackage{xcolor}
\usepackage{mathrsfs}
\usepackage{cases}
\usepackage{bbding}
\usepackage{pifont}
\usepackage{dcolumn}
\usepackage{graphicx}
\usepackage[all]{xy}
\usepackage{titlesec}
\usepackage{ytableau}
\usepackage{collectbox}
\usepackage{centernot}
\usepackage[utf8]{inputenc}
\usepackage{color}
\usepackage{young}
\usepackage[most]{tcolorbox}
\usepackage{dsfont}
\usepackage[mathscr]{euscript}
\usepackage{mathrsfs}

\DeclareMathAlphabet{\mathpzc}{OT1}{pzc}{m}{it}

\usepackage{tikz}
\usepackage{tikz-cd}
\usetikzlibrary{tikzmark,calc,arrows,shapes,decorations.pathmorphing,decorations.pathreplacing,decorations.markings}
\usetikzlibrary{arrows}
\usetikzlibrary{automata}
\usetikzlibrary{positioning}

\newcommand\sqbox[1]{{
	\setbox0=\hbox{\mbox{$\Box$}}
	\setbox1=\hbox{\mbox{\raisebox{0.35ex}{\tiny #1}}}
	\mbox{\raisebox{-0.2ex}{\rlap{\hbox to \wd0{\hss{\box1}\hss}}\box0}}
}}

\titleclass{\subsubsubsection}{straight}[\subsection]

\newcounter{subsubsubsection}[subsubsection]
\renewcommand\thesubsubsubsection{\thesubsubsection.\arabic{subsubsubsection}}

\titleformat{\subsubsubsection}{\normalfont\normalsize\bfseries}{\thesubsubsubsection}{1em}{}
\titlespacing*{\subsubsubsection}{0pt}{3.25ex plus 1ex minus .2ex}{1.5ex plus .2ex}

\makeatletter
\renewcommand\paragraph{\@startsection{paragraph}{5}{\z@}%
  {3.25ex \@plus1ex \@minus.2ex}%
  {-1em}%
  {\normalfont\normalsize\bfseries}}
\renewcommand\subparagraph{\@startsection{subparagraph}{6}{\parindent}%
  {3.25ex \@plus1ex \@minus .2ex}%
  {-1em}%
  {\normalfont\normalsize\bfseries}}
\def\toclevel@subsubsubsection{4}
\def\toclevel@paragraph{5}
\def\toclevel@paragraph{6}
\def\l@subsubsubsection{\@dottedtocline{4}{7em}{4em}}
\def\l@paragraph{\@dottedtocline{5}{10em}{5em}}
\def\l@subparagraph{\@dottedtocline{6}{14em}{6em}}
\makeatother

\newdimen\tableauside\tableauside=1.0ex
\newdimen\tableaurule\tableaurule=0.4pt
\newdimen\tableaustep
\def\phantomhrule#1{\hbox{\vbox to0pt{\hrule height\tableaurule width#1\vss}}}
\def\phantomvrule#1{\vbox{\hbox to0pt{\vrule width\tableaurule height#1\hss}}}
\def\sqr{\vbox{%
		\phantomhrule\tableaustep
		\hbox{\phantomvrule\tableaustep\kern\tableaustep\phantomvrule\tableaustep}%
		\hbox{\vbox{\phantomhrule\tableauside}\kern-\tableaurule}}}
\def\squares#1{\hbox{\count0=#1\noindent\loop\sqr
		\advance\count0 by-1 \ifnum\count0>0\repeat}}
\def\tableau#1{\vcenter{\offinterlineskip
		\tableaustep=\tableauside\advance\tableaustep by-\tableaurule
		\kern\normallineskip\hbox
		{\kern\normallineskip\vbox
			{\gettableau#1 0 }%
			\kern\normallineskip\kern\tableaurule}%
		\kern\normallineskip\kern\tableaurule}}
\def\gettableau#1 {\ifnum#1=0\let\next=\null\else
	{{\tiny\yng(1)}}s{#1}\let\next=\gettableau\fi\next}

\newcommand{\includeCroppedPdf}[2][]{%
    \IfFileExists{./#2-crop.pdf}{}{%
        \immediate\write18{pdfcrop #2 #2-crop.pdf}}%
    \includegraphics[#1]{#2-crop.pdf}}
        
\newcommand\Kappa{\mathrm{K}}

\newcommand\funny{novel}

\graphicspath{{./figures/}}

\def\be{ \begin{eqnarray} }
\def\ee{ \end{eqnarray}}
\def\lm{\limits}

\def\Tr{{\rm Tr}}
\def\p{\partial}

\def\CG {{\cal G}}
\def\CH {{\cal H}}

\def\CN {{\cal N}}

\def\CG {{\cal G}}
\def\CH {{\cal H}}

\def\CQ {{\cal Q}}

\def\IC{\mathbb{C}}

\def\IP{\mathbb{P}}
\def\IR{{\mathbb{R}}}

\def\IZ{{\mathbb{Z}}}

\def\fg{\mathfrak{g}}

\def\fl{\mathfrak{l}}

\def\fC{\mathfrak{C}}

\def\fH{\mathfrak{H}}

\def\fN{\mathfrak{N}}
\def\fR{\mathfrak{R}}
\def\fS{\mathfrak{S}}

\newcommand{\CC}{\mathcal{C}}

\newcommand{\sQ}{{}^{\sharp} Q}
\newcommand{\sW}{{}^{\sharp}W}

\colorlet{mygreen}{green!10}

\def\myh{\psi}
\def\myH{\psi}
\def\mybH{\Psi}
\def\mys{\texttt{s}}
\def\myp{\mathpzc{z}_{+}{}}
\def\myz{\mathpzc{z}_{-}{}}

\title{Shifted Quiver Yangians and Representations\\ from BPS Crystals}

\date{May 2021}

\author{Dmitry Galakhov$^{a,b}$, Wei Li$^c$ and Masahito Yamazaki$^a$} 
\affiliation{$^a$Kavli Institute for the Physics and Mathematics of the Universe (WPI),\\
\hspace*{0.3cm }University of Tokyo, 5-1-5 Kashiwanoha,  Kashiwa, Chiba 277-8583, Japan}
\affiliation{$^b$Institute for Information Transmission Problems,\\ 
\hspace*{0.3cm}Bolshoy Karetny per.\ 19, build.\ 1, Moscow, 127994, Russia}
\affiliation{$^c$Institute of Theoretical Physics, Chinese Academy of Sciences,\\
\hspace*{0.3cm} 55 Zhongguancun East Road, 100190 Beijing, P.R.\ China} 

\emailAdd{dmitrii.galakhov@ipmu.jp, galakhov@itep.ru, weili@mail.itp.ac.cn, masahito.yamazaki@ipmu.jp}

\abstract{We introduce a class of new algebras, the shifted quiver Yangians, 
as the BPS algebras for type IIA string theory on general toric Calabi-Yau three-folds.
We construct representations of the shifted quiver Yangian from general subcrystals of the canonical crystal.
We derive our results via  equivariant localization for supersymmetric quiver quantum mechanics for various framed quivers, where the framings are determined by the shape of the subcrystals.

Our results unify many known BPS state counting problems,
including open BPS counting,  non-compact D4-branes, and wall crossing phenomena,
simply as different representations of the shifted quiver Yangians.
Furthermore, most of our representations seem to be new, and this suggests the existence of a zoo of BPS state counting problems yet to be studied in detail.} 

\begin{document}

\maketitle
\makeatletter
\g@addto@macro\bfseries{\boldmath}
\makeatother

\section{Introduction}

It has been a fascinating problem in supersymmetric gauge theories and string theory 
to identify the \emph{BPS algebras} \cite{Harvey:1996gc} underlying the BPS (Bogomol'nyi--Prasad--Sommerfield) states and their enumerative invariants. While there have been many developments in this direction, it is still in general a difficult question to identify the BPS algebra explicitly.

It has recently been realized \cite{Li:2020rij,Galakhov:2020vyb} that the \emph{BPS quiver Yangian}, defined in~\cite{Li:2020rij}, is precisely the BPS algebra for type IIA string theory compactified on an arbitrary toric Calabi-Yau three-fold.\footnote{The problem of identifying the BPS algebra for toric Calabi-Yau manifolds was posed previously in \cite{Rapcak:2018nsl}. The references \cite{Li:2020rij,Galakhov:2020vyb} answered this question.}
The quiver Yangian incorporates and generalizes discussions on affine Yangians~\cite{SV,Maulik:2012wi,Tsymbaliuk,Tsymbaliuk:2014fvq,Prochazka:2015deb,Gaberdiel:2017dbk} and
$\mathcal{W}$ algebras~\cite{Alday:2009aq,Wyllard:2009hg,Gaiotto:2017euk,Rapcak:2019wzw}, and contains the cohomological Hall algebra (CoHA)~\cite{Kontsevich:2010px,Rapcak:2018nsl,Rapcak:2020ueh}
as a Borel subalgebra. The quiver Yangian has a well-defined representation on the torus fixed-point set of the BPS moduli space; the fixed points are known to be described by configurations of the statistical model of crystal melting~\cite{Szendroi,MR2836398,MR2592501,Ooguri:2008yb,Yamazaki:2010fz},
which are known mathematically to enumerate the generalized Donaldson-Thomas (DT) invariants \cite{DT}.\footnote{
For the special case of $X=\mathbb{C}^3$, this representation reduces to the plane partition representation of the affine Yangian of $\mathfrak{gl}_1$, studied previously in \cite{Tsymbaliuk, Tsymbaliuk:2014fvq, Prochazka:2015deb, Datta:2016cmw, Gaberdiel:2017dbk}, and for toroidal algebras in \cite{Feigin:2013fga, Feigin1204, Bershtein_2018, feigin2012}, see also a more recent work \cite{2021arXiv210405841B} on this subject. For toric Calabi-Yau threefolds without compact $4$-cycles, the quiver Yangians coincide with affine Yangians for Lie superalgebras $\mathfrak{gl}_{m|n}$ \cite{Li:2020rij, Rapcak:2020ueh, Galakhov:2020vyb}. 
For toric Calabi-Yau threefolds with compact $4$-cycles, the quiver Yangians seem to be new algebras and their detailed properties are yet to be explored.}
In fact, the algebra itself can be bootstrapped from its action on these crystal representations~\cite{Li:2020rij}. 
Moreover, such an action can be justified physically by supersymmetric equivariant localization of an $\mathcal{N}=4$ quiver quantum mechanics \cite{Galakhov:2020vyb}, which is the world-volume theory on the D-branes.

\bigskip

The goal of this paper is to extend the considerations of \cite{Li:2020rij, Galakhov:2020vyb} and construct a vast zoo of new representations of the quiver Yangian.
Instead of the particular crystal from \cite{Li:2020rij} (called the canonical crystal in this paper), we consider a general subset of the crystal, and construct a representation of the quiver Yangian acting on its molten crystals.
Such a general discussion requires us to extend the definition of the quiver Yangian to the \emph{shifted quiver Yangian}, which is a new algebra we define in this paper.

The \emph{shapes} of these subcrystals translate to different \emph{framings} of the quivers and  the eigenvalues of the ground states (a.k.a.\ the ground state charge functions) of the representations. 
Aforementioned methods of BPS algebra construction --    $\mathcal{N}=4$ quiver quantum mechanics and bootstrap -- are applicable to these new \emph{framed} quivers and ground state charge functions, respectively. 
We show the resulting BPS algebra is indeed the shifted quiver Yangian.

\bigskip

Our discussion incorporates many of the known phenomena in the 
studies of BPS states in the literature, such as 
the wall crossing phenomena  and the inclusion of various non-compact D-branes (some of which gives rise to the so-called ``open/closed BPS state counting''). All these different-looking BPS counting problems are now unified as different representations of the shifted quiver Yangian.
Interestingly, many of our representations do not seem to have known counterparts in the BPS state counting problem 
in the literature, and our results suggest that there will be many BPS state counting problems which are yet to be identified/studied,
even inside the realm of non-compact toric Calabi-Yau geometries.

\bigskip

The plan of this paper is as follows. 
In Section~\ref{sec:SQY}, we define the shifted quiver Yangian. 
In Section~\ref{sec:subcrystals}, we describe the representations of the shifted quiver Yangian in terms of crystal melting.
In Section~\ref{sec:BPSalgebras}, we show that the shifted quiver Yangian  is indeed the BPS algebra of the $\mathcal{N}=4$ quiver quantum mechanics system specified by a framed quiver.
Sections~\ref{sec:wallcrossing} and \ref{sec:Open} demonstrate our main results with examples of open BPS counting and BPS wall-crossing, respectively.
In Section~\ref{sec:genera_rep} we discuss examples of representations which have not been discussed previously in the literature.
We end with a summary and comments on open problems in Section~\ref{sec:summary}.
We include two appendices containing technical materials.

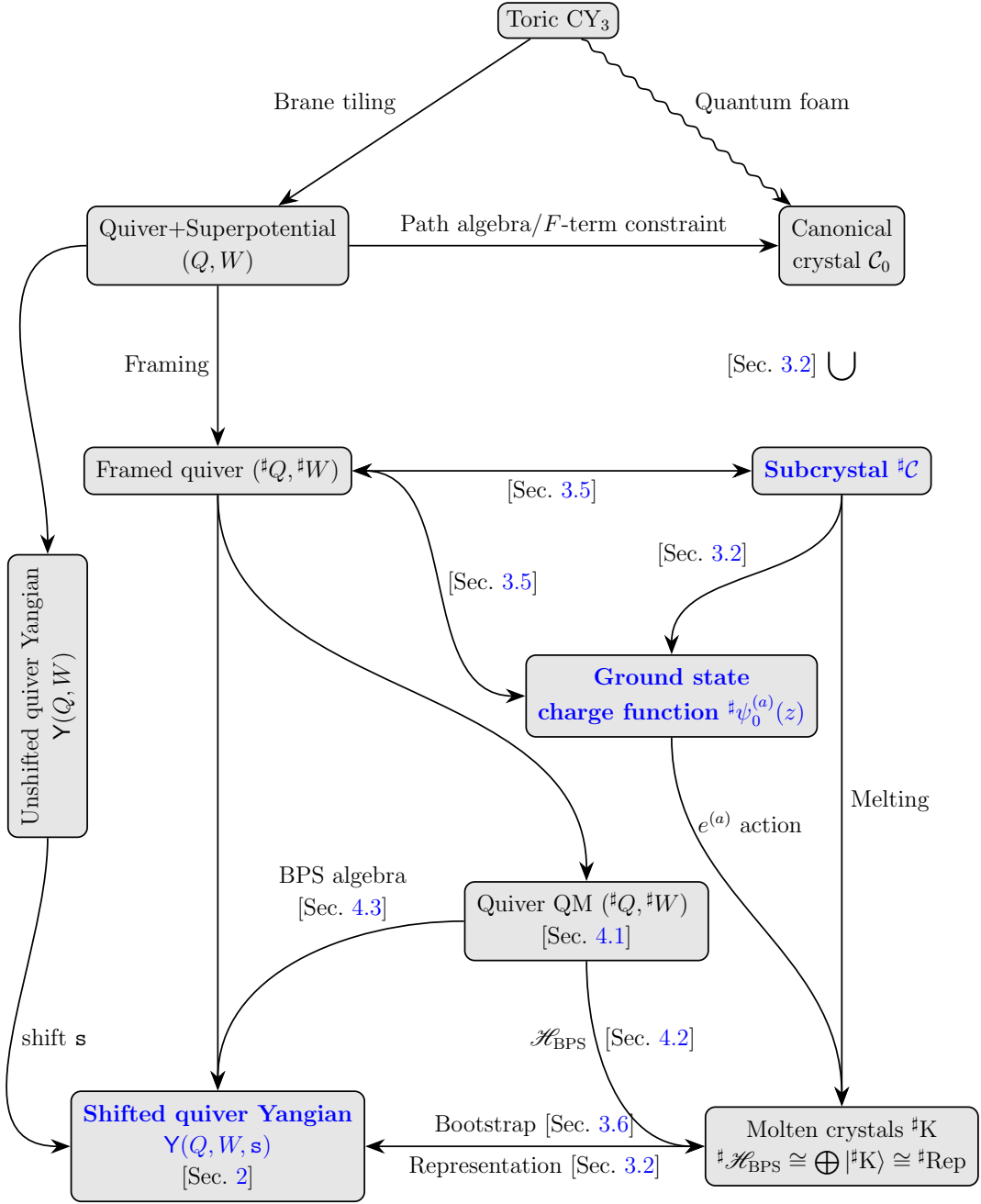
\begin{figure}[htbp]
\resizebox{1.0\textwidth}{!}{
$
\begin{array}{c}
\begin{tikzpicture}[every path/.style={thick, >={Stealth[scale=1.5]}}]
\tikzset{blob/.style={shape          
	= rectangle,
	rounded corners=5pt,
	draw,                  
	fill           = black!10!white,
	align          = center,
	thick}}
\node[blob] (QS) at (-1,0) 
{$\begin{array}{c}
     \mbox{Quiver+Superpotential}  \\
     (Q,W) 
\end{array}$};
\node[blob] (TCY) at (5,4) 
{Toric CY$_3$};
\node[blob] (USQY) at (-4,-8) 
{\rotatebox[origin=c]{90}{$\begin{array}{c}
     \mbox{Unshifted quiver Yangian} \\
      \mathsf{Y}(Q,W)
\end{array}$}};
\node[blob] (SQY) at (-1,-16) 
{$\begin{array}{c}
    \textcolor{blue}{ \mbox{\bf Shifted quiver Yangian}} \\
    \textcolor{blue}{ \mathsf{Y}(Q,W, \mys)}\\
     \mbox{[Sec.~\ref{sec:SQY}]}
\end{array}$};
\node[blob] (CC) at (10,0) 
{$\begin{array}{c}
     \mbox{Canonical}  \\
     \mbox{crystal }\CC_0
\end{array}$};
\node[blob] (SC) at (10,-4) 
{$\begin{array}{c}
     \textcolor{blue}{\mbox{\bf Subcrystal }{}^{\sharp}\CC}
\end{array}$};
\node[blob] (QQM) at (5.5,-12) 
{$\begin{array}{c}
     \mbox{Quiver QM  $(\sQ,\sW)$ }  \\
     \mbox{[Sec.~\ref{ssec:QQM}]}
\end{array}$};
\node[blob] (MC) at (10,-16)
{$\begin{array}{c}
     \mbox{Molten crystals }{}^{\sharp}\Kappa\\ {}^{\sharp}\mathscr{H}_{\rm BPS}\cong \bigoplus|{}^{\sharp}\Kappa\rangle \cong {}^{\sharp}\textrm{Rep}
\end{array}$};
\node[blob] (FQ) at (-1,-4) 
{$\begin{array}{c}
     \mbox{Framed quiver $(\sQ,\sW)$}\\
\end{array}$};
\node[blob] (VCF) at (7,-8) 
{$\begin{array}{c}
     \textcolor{blue}{\mbox{\bf Ground state}}\\ 
     \textcolor{blue}{\mbox{\bf charge function } {^{\sharp}\psi}_0^{(a)}(z)}
\end{array}$};
\path
(TCY) edge[->] node[above left] {Brane tiling} (QS) 
(TCY) edge[->,decorate, decoration={snake, segment length=10pt, amplitude=1pt}] node[above right] {Quantum foam} (CC) 
(SC) edge[draw=none]
node {\rotatebox[origin=c]{90}{\scalebox{2.0}{$\subset$}}}
node[left] {[Sec.~\ref{ssec:rep_subcry}]\, \, }
(CC)
(SC) edge[->]
node[right] {Melting}
(MC);
\draw[->] (QS.east) to[out=0,in=180]
node[pos=0.5,above] {Path algebra/$F$-term constraint} 
(CC.west);
\draw[->] (QS.south) to[out=270,in=90]
node[left] {Framing}
(FQ.north);
\draw[<->] (FQ.east) to[out=0,in=180] 
node[below]{ [Sec.~\ref{ssec:framedQW}]}
(SC.west);
\draw[<->] (FQ.east) to[out=0,in=180] 
node[right] {[Sec.~\ref{ssec:framedQW}]}
(VCF.west);
\draw[->]  (FQ.south) to[out=270,in=90] 
(SQY.north);
\draw[->]  (FQ.south) to[out=270,in=90] (QQM.north);
\draw[->] (
QQM.west) to[out=180,in=90]
node [pos=0.15,above left=-0.2cm] {
$\begin{array}{c}
     \mbox{BPS algebra}  \\
     \mbox{[Sec.~\ref{ssec:BPSalgebra}]}
\end{array}$}
(SQY.north);
\draw[->] (SC.south) to[out=270,in=90] 
node[above left] {[Sec.~\ref{ssec:rep_subcry}]}
(VCF.north);
\draw[->] (QQM.south) to[out=270,in=180] 
node[pos=0.3,left] {$\mathscr{H}_{\textrm{BPS}}$ 
}
node[pos=0.3,right] { [Sec.~\ref{ssec:MotlenCrystalQQM}]}
(MC.west);
\draw[<->] (SQY.east) to[out=0,in=180] 
node[pos=0.5,above] {Bootstrap [Sec.~\ref{ssec:bootstrap}]} 
node[pos=0.5,below] {Representation [Sec.~\ref{ssec:rep_subcry}]} 
(MC.west);
\draw[->] (QS.west) to[out=180,in=90] 
node[right] {} 
(USQY.north);
\draw[->] (USQY.south) to[out=270,in=180] 
node[right] {shift $\mys$} 
(SQY.west);
\draw[->] (VCF.south) to[out=270,in=90] 
node[pos=0.2,right] {$e^{(a)}$ action} 
(MC.north);
\end{tikzpicture}
\end{array}
$
} 
\label{fig.overview}
\caption{Interrelations of various concepts discussed in this paper.
Those colored in blue are new concepts introduced in this paper.}
\end{figure}

The main players of this story, together with the sections they appear in, are summarized  in Figure~\ref{fig.overview}.

\clearpage

\section{Shifted quiver Yangians}
\label{sec:SQY}

In this section, we give the definition of the shifted quiver Yangians and their essential properties. 
While this section is self-contained, our definition here closely follows the definition of the quiver Yangian (unshifted quiver Yangian in the terminology of this paper) 
in \cite{Li:2020rij}, for which readers are referred to for many relevant details.

\subsection{Defining data and generators}
\label{ssec:data}

One of the defining data for the shifted quiver Yangian is a pair $(Q,W)$, where $Q=(Q_0, Q_1)$ is the quiver diagram, with $Q_0$ and $Q_1$ being the set of vertices $\{a\}$ and the set  of arrows $\{I\}$, respectively, and $W$ is the superpotential, which consists of monomial terms corresponding to closed loops $\{L\}$ in the quiver $Q$. 
The pair $(Q,W)$ defines an $\mathcal{N}=4$ supersymmetric quiver quantum mechanics \cite{Denef:2002ru}.
In this paper, we will consider those $(Q,W)$ that give rise to the quantum mechanics theory whose vacuum moduli space describes a toric Calabi-Yau three-fold.\footnote{The combinatorial relations between the quiver data $(Q, W)$
and the toric data have been studied extensively in the literature on brane tilings \cite{Hanany:2005ve,Franco:2005sm,Hanany:2005ss}, see e.g.\ \cite{Kennaway:2007tq,Yamazaki:2008bt} for reviews.}
The definition of the shifted quiver Yangians, however,  works for more general choices of $Q$ and $W$.
\bigskip

We assign a flavor charge (an equivariant parameter) $h_I$ for each arrow $I\in Q_1$, subject to the following constraints.
First, each term in the superpotential $W$ should have total charge zero, i.e.\ the charge assignment represents a symmetry preserving the superpotential $W$.
Correspondingly, for each loop $L$ in the quiver $Q$ associated with a term in $W$, we have a loop constraint:
\begin{equation}\label{eq:loop-constraint}
\textrm{ loop constraint: } \quad \sum_{I\in \textrm{loop }L} h_I=0 \;.
\end{equation}
In addition, there are gauge symmetries that correspond to the $U(1)$ isometries of the toric Calabi-Yau geometry.
As a result, up to gauge redundancies we are also free to impose a constraint
\begin{equation}\label{eq-vertex-constraint-toric}
\textrm{ vertex constraint: } \quad
\sum_{I\in a} \textrm{sign}_a(I) \, h_I=0\,,
\end{equation}
for each vertex $a$, and $\textrm{sign}_a(I)$ is $+1$ if the arrow $I$ flows towards the vertex $a$, $-1$ if the arrow $I$ flows outwards from the vertex $a$, and $0$ otherwise.
Having imposed both types of constraints, we are left with two independent parameters we denote as $\mathsf{h}_1$ and $\mathsf{h}_2$ for toric Calabi-Yau three-folds \cite{Li:2020rij}.
These two parameters correspond to the two parameters for equivariant torus actions preserving the holomorphic Calabi-Yau three-form.

\bigskip

The shifted quiver Yangian consists of a triplet of generators  $(e^{(a)}(z), f^{(a)}(z),  \myh^{(a)}(z))$ for each quiver vertex $a\in Q_0$,
with the mode expansion
\begin{equation}\label{eq-mode-expansion-toric}
e^{(a)}(z)\equiv\sum^{+\infty}_{n=0}\frac{e^{(a)}_n}{z^{n+1}} \,, \qquad \myh^{(a)}(z)\equiv \sum^{+\infty}_{n=-\infty}\frac{\myh^{(a)}_n}{z^{n+1+\mys^{(a)}}}\,, \qquad f^{(a)}(z)\equiv \sum^{+\infty}_{n=0}\frac{f^{(a)}_n}{z^{n+1}} \;.
\end{equation}
Here the set of integers $\mys=\{\mys^{(a)}\}$ in the expansion of $\psi^{(a)}(z)$ for $a\in Q_0$ characterize the ``shift" in the shifted quiver Yangian. Note that the shift affects the mode expansions of $\myh^{(a)}(z)$ only.\footnote{In this section we will only consider the untruncated versions of the algebras. 
The truncation of the algebra will be considered in Section.~\ref{ssec:truncations}. The truncation will in general change the shift of the algebra.}

For toric Calabi-Yau threefolds without compact $4$-cycles, 
\begin{equation}\label{eq:psi_condition_wo4}
\psi^{(a)}_{n<-2}=0 \quad \textrm{and} \quad \psi^{(a)}_{n=-1}=1\,,
\end{equation}
for all $a\in Q_0$.
Namely, for the case without compact $4$-cycles, the mode expansion for $\psi^{(a)}(z)$ in \eqref{eq-mode-expansion-toric} can be written as 
\begin{equation}\label{eq-mode-expansion-toric-wo4}
 \myh^{(a)}(z)\equiv \sum^{+\infty}_{n=-1}\frac{\myh^{(a)}_n}{z^{n+1+\mys^{(a)}}}\,.
\end{equation}

For toric Calabi-Yau threefolds with compact $4$-cycles, since the mode expansion of $\psi^{(a)}$ is from $-\infty$ to $+\infty$, the information of the shift $\mys^{(a)}$ is somewhat lost,\footnote{One way to retain this information even for the case with compact $4$-cycles is to invoke the vacuum state $|\varnothing\rangle$ and demand that $\psi^{(a)}_{n<-2}|\varnothing\rangle=0$ and $ \psi^{(a)}_{n=-1}|\varnothing\rangle=|\varnothing\rangle$, analogous to the condition \eqref{eq:psi_condition_wo4} for the case without  compact $4$-cycles.}
because the expansion of $\psi^{(a)}(z)$ in \eqref{eq-mode-expansion-toric} can also be written as
\begin{equation}\label{eq-mode-expansion-toric-infinite}
\myh^{(a)}(z)\equiv \sum^{+\infty}_{n=-\infty}\frac{\tilde{\myh}^{(a)}_n}{z^{n+1}}\,.
\end{equation}

\subsection{Relations in fields}
\label{ssec:RelationField}

We can now define 
the shifted quiver Yangian $\mathsf{Y}(Q,W,\mys)$
by the following OPE(Operator Product Expansion)-like relations:
\begin{tcolorbox}[ams align]\label{eq-OPE-toric}
\begin{aligned}
\myh^{(a)}(z)\, \myh^{(b)}(w)&=   \myh^{(b)}(w)\, \myh^{(a)}(z)  \;,\\
\myh^{(a)}(z)\, e^{(b)}(w)&\simeq  \varphi^{b\Rightarrow a}(z-w)\, e^{(b)}(w)\, \myh^{(a)}(z)  \;, \\
e^{(a)}(z)\, e^{(b)}(w)&\sim  (-1)^{|a||b|}  \varphi^{b\Rightarrow a}(z-w) \, e^{(b)}(w)\, e^{(a)}(z)  \;, \\
\myh^{(a)}(z)\, f^{(b)}(w)&\simeq   \varphi^{b\Rightarrow a}(z-w)^{-1} \, f^{(b)}(w)\,\myh^{(a)}(z) \;,\\
f^{(a)}(z)\, f^{(b)}(w)&\sim  (-1)^{|a||b|} \varphi^{b\Rightarrow a}(z-w)^{-1}\,  f^{(b)}(w)\, f^{(a)}(z)   \;,\\
\left[e^{(a)}(z),f^{(b)}(w) \right\} &\sim -  \delta^{a,b} \frac{\myh^{(a)}(z)-\myh^{(b)}(w)}{z-w}  \;,
\end{aligned}
\end{tcolorbox}
\noindent where throughout this paper ``$\simeq$" means equality up to $z^n w^{m\geq 0}$ terms,\footnote{For Calabi-Yau threeforlds without compact 4-cycles, ``$\simeq$" means equality up to $z^{n\geq -\mys^{(a)}}w^m$ and $z^n w^{m\geq 0}$ terms.} and ``$\sim$" means equality up to $z^{n\geq 0} w^{m}$ and $z^{n} w^{m\geq 0}$ terms.
The bracket $[e^{(a)}(z),f^{(b)}(w)\}$ denotes 
the anti-commutator $\{ e^{(a)}(z),f^{(b)}(w)\}$ when both $a$ and $b$ are odd, and the commutator $[e^{(a)}(z),f^{(b)}(w)]$ otherwise.
The $\mathbb{Z}_2$-grading of the generators $e_n^{(a)}$ and $f_n^{(a)}$ is given by 
\begin{equation}\label{eq.Z2_grading}
|a|=\begin{cases}
0 & (\exists I\in Q_1 \,\,\, \textrm{such that $I$ starts and ends at $a$} ) \;,\\
1 & (\textrm{otherwise}) \;,
\end{cases}
\end{equation}
while the operators $\myh_n^{(a)}$ are always even.
The function (the bond factor) $\varphi^{a\Rightarrow b}(z)$ in (\ref{eq-OPE-toric}) is 
defined as
\begin{tcolorbox}[ams equation]\label{eq-charge-atob}
\varphi^{a\Rightarrow b} (u)\equiv \frac{\prod_{I\in \{b\rightarrow a\}}(u+h_{I})}{\prod_{I\in \{a\rightarrow b\}}(u-h_{I})} \;,
\end{tcolorbox}
\noindent 
where $\{a\rightarrow b\}$ denotes the set of arrows from vertex $a$ to vertex $b$.
(When there is no arrow between vertices $a$ and $b$ in the quiver, 
we define $\varphi^{a\Rightarrow b} (u) =\varphi^{b\Rightarrow a} (u) =1$.)
The function $\varphi^{a\Rightarrow b}(z)$ satisfies the \emph{reflection property}\footnote{Note that for a chiral quiver with some $|a \rightarrow b|+|b\rightarrow a|$ being odd, this reflection property means that we need to choose an ordering for the vertices $a$ and $b$ when writing down the $e-e$ and $f-f$ relations in \eqref{eq-OPE-toric}.} 
\begin{equation}\label{eq.varphi_sym}
\varphi^{a\Rightarrow b} (u) \, \varphi^{b\Rightarrow a} (-u)=(-1)^{\left|a\to b\right|+\left|b\to a\right|} \;,
\end{equation}
where $|a\rightarrow b|$ denotes the number of arrows from vertex $a$ to vertex $b$.
Finally, we emphasize that the bond factor $\varphi^{a\Rightarrow b}(u)$ (\ref{eq-charge-atob}) should be treated as a ``formal" rational function: 
namely, all the factors in its numerator and denominator need to be kept even when the charges $h_I$ take some special values such that some factors of the numerator and the denominator cancel each other.

\bigskip

When we set  $\mys^{(a)}=0$, we recover the quiver Yangian defined earlier in \cite{Li:2020rij}.
In this paper, in order to distinguish this special case from more general shifted quiver Yangians, we call the algebra $\mathsf{Y}(Q,W, \mys=0)$ the \emph{unshifted quiver Yangian}.\footnote{All the quiver Yangians in  \cite{Li:2020rij} correspond to the canonical crystals, see Section \ref{ssec:canon_cry}. Without truncation (see Section \ref{ssec:truncations}), a quiver Yangian $\mathsf{Y}(Q,W,\mys)$ that corresponds to the canonical crystal should have shift $\mys^{(a)}=\delta_{a,1}$. 
The definition of \cite{Li:2020rij}  included one truncation factor $(z+C)$ for all the $\psi^{(a=1)}(z)$ generators, therefore the algebras in  \cite{Li:2020rij} are unshifted quiver Yangians, i.e.\ $\mys^{(a)}=0$ for all $a\in Q_0$. } 
\bigskip

Finally, we mention another mode expansion for the Cartan generators $\psi^{(a)}(z)$ of the shifted quiver Yangians that is more natural in the study of their representations (to be discussed in Section~\ref{sec:subcrystals}).
Instead of the expansion of $\psi^{(a)}(z)$ in \eqref{eq-mode-expansion-toric}, we have 
\begin{equation}\label{eq:psi-to-H}
\psi^{(a)}(z)\equiv {}^{\sharp}\psi^{(a)}_{0}(z)\cdot\phi^{(a)}(z) \qquad \textrm{with}\quad ^{\sharp}\psi^{(a)}_0(z)\equiv \frac{\prod^{\mys^{(a)}_{-}}_{\beta=1}(z-\myz^{(a)}_{\beta}) }{\prod^{\mys^{(a)}_{+}}_{\alpha=1}(z-\myp^{(a)}_{\alpha})} \;,
\end{equation}
where we have the shift $\mys^{(a)} = \mys^{(a)}_{+}  - \mys^{(a)}_{-}$ and the new Cartan generators $\phi^{(a)}(z)$ have the mode expansion
\begin{equation}
    \phi^{(a)}(z)=  \sum^{+\infty}_{n=-\infty}\frac{\phi^{(a)}_n}{z^{n+1}} \;,
\end{equation}
so that $\psi^{(a)}_n$ are linear combinations of $\phi^{(a)}_n$.

In this description of the shifted quiver Yangian, the ``shift" is  
encoded in the set of rational functions $\{{}^{\sharp}\psi_0^{(a)}(z)\}$ defined in (\ref{eq:psi-to-H}), and the shifts are characterized by $\mys^{(a)}_{+}$ poles $\{\myp^{(a)}_{\alpha}\}$ and $\mys^{(a)}_{-}$ zeros $\{\myz^{(a)}_{\alpha}\}$ for each vertex $a\in Q_0$.
To emphasize this we sometimes use the notation  $\mathsf{Y}(Q,W, {}^{\sharp}\psi)$ for the algebra $\mathsf{Y}(Q,W, \mys)$.

The special cases of $\mathsf{Y}(Q,W, {}^{\sharp}\psi)$ with $\mys^{(a)}_{+}=0$ or $\mys^{(a)}_{-}=0$  were discussed previously in \cite{Kodera:2016faj,Rapcak:2018nsl} for some examples of quivers associated with Lie superalgebras $\mathfrak{gl}_{m|n}$.\footnote{Shifted Yangians associated with finite-dimensional Lie algebras were discussed earlier in 
\cite{MR2199632,MR2456464,MR3248988}.}
Our definition here generalizes the definition to a large class of quivers, which are in general not associated with 
finite-dimensional Lie (super)algebras.

\subsection{Relations in modes}

The quadratic relation (\ref{eq-OPE-toric}) in terms of the fields can then be translated in terms of modes using the expansion (\ref{eq-mode-expansion-toric}):
\begin{tcolorbox}[ams align]\label{eq-OPE-modes-toric}
\begin{aligned}
&\left[\myh^{(a)}_n \, , \, \myh^{(b)}_m\right]=0 \;,\\
&\sum^{|b\rightarrow a|}_{k=0}(-1)^{|b\rightarrow a|-k} \, \sigma^{b\rightarrow a}_{|b\rightarrow a|-k}\,  [\myh^{(a)}_n\, e^{(b)}_m]_k =\sum^{|a\rightarrow b|}_{k=0} \sigma^{a\rightarrow b}_{|a\rightarrow b|-k}\, [ e^{(b)}_m\, \myh^{(a)}_n]^k  \;,\\
& \sum^{|b\rightarrow a|}_{k=0}(-1)^{|b\rightarrow a|-k} \, \sigma^{b\rightarrow a}_{|b\rightarrow a|-k}\,  [e^{(a)}_n\, e^{(b)}_m]_k =(-1)^{|a||b|}\sum^{|a\rightarrow b|}_{k=0} \sigma^{a\rightarrow b}_{|a\rightarrow b|-k}\, [ e^{(b)}_m\, e^{(a)}_n]^k  \;,\\
& \sum^{|a\rightarrow b|}_{k=0} \sigma^{a\rightarrow b}_{|a\rightarrow b|-k}\, [\myh^{(a)}_n\, f^{(b)}_m]_k = \sum^{|b\rightarrow a|}_{k=0}(-1)^{|b\rightarrow a|-k} \, \sigma^{b\rightarrow a}_{|b\rightarrow a|-k}\,  [ f^{(b)}_m\, \myh^{(a)}_n]^k  \;,\\
&\sum^{|a\rightarrow b|}_{k=0} \sigma^{a\rightarrow b}_{|a\rightarrow b|-k}\, [f^{(a)}_n\, f^{(b)}_m]_k =(-1)^{|a||b|} \sum^{|b\rightarrow a|}_{k=0}(-1)^{|b\rightarrow a|-k} \, \sigma^{b\rightarrow a}_{|b\rightarrow a|-k}\,  [ f^{(b)}_m\, f^{(a)}_n]^k  ,\\
&\left[e^{(a)}_n\, , \, f^{(b)}_m \right\}=\delta^{a,b}\,\myh^{(a)}_{n+m-\mys^{(a)}} \;.
\end{aligned}
\end{tcolorbox}
\noindent 
Here the mode numbers for $e^{(a)}$ and $f^{(a)}$ always take the range $m\in\mathbb{Z}_{\geq 0}$, whereas for the Cartan generators $\myh^{(a)}$, the mode number has the range $m\in\mathbb{Z}_{\geq 0}$ for the Calabi-Yau three-fold without compact $4$-cycle but $m\in\mathbb{Z}$ for those with compact $4$-cycles.
We have defined 
\begin{equation}\label{eq-ABn}
\begin{aligned}
\left[A_n\, B_m\right]_k& \equiv \sum^{k}_{j=0} (-1)^j\,\tbinom{k}{j} \,A_{n+k-j}\, B_{m+j}\,, \qquad \\
\left[B_m\,A_n\right]^k &\equiv \sum^{k}_{j=0} (-1)^j\,\tbinom{k}{j} \,B_{m+j}\, A_{n+k-j}\,;
\end{aligned}
\end{equation}
$\sigma^{a\rightarrow b}_k$ denotes the $k^{\textrm{th}}$ elementary symmetric sum of the set $\{h_I\}$ with $I\in \{a\rightarrow b\}$; and finally $|a\rightarrow b|$ stands for  the number of arrows from vertex $a$ to vertex $b$ in the quiver $Q$.
\bigskip

When $\mys$ is finite, the quadratic relations between the Cartan generators $\psi^{(a)}$ and the $e^{(a)}/f^{(a)}$ generators simplify for the first few $\psi^{(a)}_n$ modes.
To derive these ``initial conditions", one simply plugs in the mode expansion \eqref{eq-mode-expansion-toric}
(with $\psi^{(a)}_{-1}=1$)
into the $\psi-e$ and $\psi-f$ relations in  (\ref{eq-OPE-modes-toric}) with $n=-|a\rightarrow b| =-|b\rightarrow a|$, and gets
\begin{equation}\label{eq-psi-e-f-initial-0}
\begin{aligned}
\left[\psi^{(a)}_0\,,\,e^{(b)}_m\right] &=\left(\sigma^{a\rightarrow b}_{1}+\sigma^{b\rightarrow a}_{1}\right)\,e^{(b)}_{m}=\left(\sum_{I\in \{a\rightarrow b\}}h_I+\sum_{I\in \{b\rightarrow a\}}h_I\right) \,e^{(b)}_{m} \;,\\
\left[\psi^{(a)}_0\,,\,f^{(b)}_m\right] &=-\left(\sigma^{a\rightarrow b}_{1}+\sigma^{b\rightarrow a}_{1}\right)\,f^{(b)}_{m}=-\left(\sum_{I\in \{a\rightarrow b\}}h_I+\sum_{I\in \{b\rightarrow a\}}h_I\right) \,f^{(b)}_{m} \;,
\end{aligned}
\end{equation}
where we have used $\psi^{(a)}_{-1}=1$.
Next, for $|a\rightarrow b|= |b\rightarrow a|\geq 2$, consider $n=-|a\rightarrow b|+1$, which gives
\begin{equation}\label{eq-psi-e-f-initial-1}
\begin{aligned}
\left[\psi^{(a)}_1\,,\,e^{(b)}_m\right] &=\left(\sigma^{a\rightarrow b}_{2}-\sigma^{b\rightarrow a}_{2}\right)\,e^{(b)}_{m}
+\left(\sigma^{b\rightarrow a}_{1}\, \psi^{(a)}_0\, e^{(b)}_{m}+\sigma^{a\rightarrow b}_{1}\, e^{(b)}_{m}\, \psi^{(a)}_0\right) \;,\\
\left[\psi^{(a)}_1\,,\,f^{(b)}_m\right] &=-\left(\sigma^{a\rightarrow b}_{2}-\sigma^{b\rightarrow a}_{2}\right)\,f^{(b)}_{m}
-\left(\sigma^{a\rightarrow b}_{1}\, \psi^{(a)}_0\, f^{(b)}_{m}+\sigma^{b\rightarrow a}_{1}\, f^{(b)}_{m}\, \psi^{(a)}_0\right) \;.
\end{aligned}
\end{equation}
The conditions with $\psi^{(a)}_{\ell \geq 2}$, if they exist, 
can be derived similarly.
These ``initial conditions" are useful in the attempt of mapping the quiver Yangian to other types of algebras such as $\mathcal{W}$ algebras.

\subsection{Some properties}

Let us now comment on some properties of the
shifted quiver Yangian $\mathsf{Y}(Q,W,\mys)$.
Our discussion here is brief since it is parallel to the 
case of the unshifted quiver Yangian in \cite[section 4.3]{Li:2020rij}.

\begin{itemize}

\item We have the triangular decomposition
\begin{align}\label{eq_triangular}
\mathsf{Y}_{(Q,W)}= \mathsf{Y}_{(Q,W)}^{+} \oplus \mathsf{B}_{(Q,W)} \oplus \mathsf{Y}_{(Q,W)}^{-}  \;,
\end{align}
where $\mathsf{Y}_{(Q,W)}^{+}, \mathsf{Y}_{(Q,W)}^{-}, \mathsf{B}_{(Q,W)}$ are generated by the $e^{(a)}_n, f^{(a)}_n, \psi^{(a)}_n$'s, respectively.

\item We have a $\mathbb{Z}_2$ automorphism
\begin{align}
e^{(a)}(z) \leftrightarrow f^{(a)}(z)\;,
\quad 
\psi^{(a)}(z) \leftrightarrow \psi^{(a)}(z)^{-1} \;,
\end{align}
which exchanges $\mathsf{Y}_{(Q,W)}^{+}$ and $\mathsf{Y}_{(Q,W)}^{-}$ while preserving $\mathsf{B}_{(Q,W)}$.
\bigskip

\item In addition to the  $\mathbb{Z}_2$ grading introduced in \eqref{eq.Z2_grading},  for each vertex $a\in Q_0$ we can define an associated $\mathbb{Z}$ grading $\mathrm{deg}_a$ by 
\begin{align}
\mathrm{deg}_a (e^{(b)}_n) = \delta_{a,b} \;, \quad
\mathrm{deg}_a (\psi^{(b)}_n)= 0 \;, \quad
\mathrm{deg}_a (f^{(b)}_n) = - \delta_{a,b} \;.
\end{align}
We can also define a filtration on the algebra
by 
\begin{align}\label{level_grading}
\mathrm{deg}_{\rm level} (e^{(b)}_n) = 
\mathrm{deg}_{\rm level} (f^{(b)}_n) = n+\tfrac{1}{2} \;, \quad
\mathrm{deg}_{\rm level} (\psi^{(b)}_n)= n+1 -\mys^{(b)} \;.
\end{align}

\item The spectral shift $z\to z-\varepsilon$ generates an automorphism of the algebra.

\item Recall that the equivariant parameters are the flavor symmetry charge assignments of the 
quiver quantum mechanics. This comes with ambiguities arising from gauge symmetry,
i.e.\ the shift 
\begin{align}\label{eq.h_shift}
h_I \to h'_I=h_I + \varepsilon_a  \, \textrm{sign}_a(I)  \;,
\end{align}
where 
\begin{align}\label{eq.sign_def}
\textrm{sign}_a(I) \equiv \begin{cases}
+1 & \qquad (s(I)= a \;, \quad t(I)\ne a) \;, \\
-1 &\qquad (s(I)\ne a \;, \quad t(I)= a)\;, \\
0 &\qquad (\textrm{otherwise})\;,
\end{cases}
\end{align}
and  $s(I)$ and $t(I)$ denote the source and the target of the arrow $I$, respectively.
This generates an automorphism of the algebra for each quiver vertex $a$.

\item 
The quadratic relations (\ref{eq-OPE-toric}), or equivalently the mode version (\ref{eq-OPE-modes-toric}), can be supplemented by extra relations,
as in the unshifted case \cite[section 4.4]{Li:2020rij}.
These new relations are traditionally called the Serre relations, and we would then obtain
a  {\it reduced shifted quiver Yangian }
\begin{align}\label{eq:reducedShiftedQY}
\underline{\mathsf{Y}}_{(Q,W)} = \mathsf{Y}_{(Q,W)}/ \textrm{(Serre relations)} \;.
\end{align}
For cases without compact $4$-cycles the quiver Yangian is identified to be the affine Yangian of $\mathfrak{gl}_{m|n}$,
for which Serre relations are already known \cite{Bezerra:2019dmp}.
For a general toric Calabi-Yau threefold there seems to be no known form of the Serre relations,
and it would be interesting to study this point further.

\end{itemize}

\section{Molten crystal representations of shifted quiver Yangians}
\label{sec:subcrystals}

In the previous Section, we defined the shifted quiver Yangians and listed their basic properties.
In this section, we will explain how to construct and classify their representations.

\subsection{Review: canonical crystals and unshifted quiver Yangians} \label{ssec:canon_cry}

Let us first review the molten crystal representation of the unshifted quiver Yangian constructed in \cite{Li:2020rij}.
\bigskip

The BPS crystal associated to a quiver-superpotential pair $(Q,W)$ was constructed in \cite{Ooguri:2008yb} as the representation of the path algebra of ``quiver with relations" $A_{(Q,W)}$, where the relations mean the $F$-term relations $\partial W/\partial \Phi_I=0$ for each bifundamental $\Phi_I$ associated to an arrow $I$.
Each atom in the crystal corresponds to an $F$-term equivalence class of paths starting from a framed vertex $\mathfrak{o}$, 
which is chosen from the set of vertices $\{a\}$ in the quiver $Q$ to be the origin of the crystal; and the atoms are connected by arrows $\{I\}$ of the quiver, as follows from the definition.
The resulting crystal can be viewed as a three-dimensional lift of the universal cover of the periodic quiver $\tilde{Q}=(Q_0,Q_1,Q_2)$.\footnote{For a quiver quantum mechanics originating from a toric Calabi-Yau three-fold, the quiver-superpotential pair $(Q,W)$ can be recast into a periodic quiver $\tilde{Q}=(Q_0,Q_1,Q_2)$, namely, a quiver $\tilde{Q}$ on the two-dimensional torus so that the faces $Q_2=\{F\}$ of the quiver represent monomial terms in the superpotential $W$.}

To promote the BPS crystal from the representation of the path algebra $A_{(Q,W)}$ to the representation of the (unshifted) quiver Yangian $\mathsf{Y}(Q,W)$, we need to give the crystal a finer structure, in particular, we can 
relate the coordinates of each atom to the  flavor-symmetry charges of the theory.
The two charge assignments are non-R flavor symmetries, and are precisely given by the equivariant parameters $h_I$ introduced previously in Section~\ref{ssec:data} as part of the data of the $\mathsf{Y}(Q,W)$ algebra; the third coordinate is given by an R-symmetry, under which the superpotential has charge $+2$. 
Once we fix a charge assignment,
the charge of an atom is defined as the total flavor charge of the corresponding operator  \cite{Li:2020rij}:
\begin{align}
h(\Box)\equiv\sum_{I\, \in\,  \textrm{path}[\mathfrak{o}\rightarrow \Box\,]} h_I \,, 
\end{align}
where we used the fact that an atom  can be represented by a path from $\mathfrak{o}$.
We call the resulting crystal associated with the quiver-superpotential pair $(Q,W)$ the \emph{canonical crystal}, denoted by $\mathcal{C}_0(Q,W)$, or simply $\mathcal{C}_0$ when there is no risk of confusion.

\bigskip
Here and in what follows we will use the symbol $\Box$ to denote an atom of a crystal. 
The path $\mathfrak{o}\rightarrow \Box$ in a crystal can be projected first to the periodic quiver $\tilde Q$ and further to $Q$.
In $Q$ the projection of $\mathfrak{o}\rightarrow \Box$ ends on a particular quiver vertex $a$, we say in this case that atom $\Box$ is \emph{colored} by color $a$.
If we will need to stress in our notations that some atom $\Box$ has a particular color $a$, we will denote it as $\sqbox{a}$.
\bigskip

Given a crystal $\CC$, we can consider a configuration $\Kappa$ of a molten crystal.
This is a finite subset of the atoms in the crystal such that the following \emph{melting rule} is satisfied: for any atom $\Box$ such that $I\cdot \Box \in \Kappa$ for some arrow $I$, then $\Box$ is also contained in $\Kappa$ \cite{Ooguri:2008yb}. 
Namely, the molten crystal $K$ is the \emph{complement} of a crystal configuration that is $\CC$ with some atoms ``melted away" near the origin.  
\bigskip

The representation of the unshifted quiver Yangian in \cite{Li:2020rij} is given by
\begin{tcolorbox}[ams align]\label{eq.efpsi-action}
\begin{aligned}
\begin{aligned}
\myH^{(a)}(z)|\Kappa\rangle&= \mybH_{\Kappa}^{(a)}(z)|\Kappa\rangle \;,\\
e^{(a)}(z)|\Kappa\rangle &=\sum_{\sqbox{$a$} \,\in \,\textrm{Add}(\Kappa)} 
 \frac{\pm\sqrt{p^{(a)}\textrm{Res}_{u=h(\sqbox{$a$})}\mybH^{(a)}_{\Kappa}(u)}}{z-h(\sqbox{$a$})}|\Kappa+\sqbox{$a$}\rangle \;,\\
f^{(a)}(z)|\Kappa\rangle &=\sum_{\sqbox{$a$}\, \in\, \textrm{Rem}(\Kappa)}
\frac{\pm\sqrt{q^{(a)}\textrm{Res}_{u=h(\sqbox{$a$})}\mybH^{(a)}_{\Kappa}(u)}}{z-h(\sqbox{$a$})}|\Kappa-\sqbox{$a$}\rangle \;.\\
\end{aligned}
\end{aligned}
\end{tcolorbox}
\medskip
\noindent
Here the signs $q^{(a)}=1$ and $p^{(a)}\equiv \varphi^{a\Rightarrow a}(0)=\pm 1$ in \eqref{eq.efpsi-action} are related to the statistics of the operators $e^{(a)}(z)$ and $f^{(a)}(z)$. 
The sets
$\textrm{Add}(\Kappa)$ and $\textrm{Rem}(\Kappa)$ denote the sets of atoms that can be added to or removed from the crystal configuration $\Kappa$, respectively, so that the result is again an allowed crystal configuration satisfying the melting rule.
The $\pm$ signs in the coefficients of the $e^{(a)}(z)$ and $f^{(a)}(z)$ actions depend on both the initial state $\Kappa$ and the atom $\sqbox{$a$}$, and can be fixed once we determine the statistics of the algebra. 
The ``charge function'' $\mybH^{(a)}_{\Kappa}(u)$
and the ``vacuum charge function'' $\psi^{(a)}_0(z)$ are defined by 
\begin{tcolorbox}[ams align]\label{eq.efpsi-action-2}
\begin{aligned}
 \mybH^{(a)}_{\Kappa}(u)&\equiv \psi^{(a)}_0(u) \prod_{b\in Q_0} \prod_{\sqbox{$b$}\in \Kappa} \varphi^{b\Rightarrow a}(u-h(\sqbox{$b$})) \;,\\
 \quad\qquad \psi^{(a)}_0(z)&\equiv \left(\frac{1}{z} \right)^{\delta_{a,\mathfrak{o}}}\;.
\end{aligned}
\end{tcolorbox}
\noindent 
Note that the vacuum charge function $\psi_0^{(a)}$ is non-trivial only for ``the framed vertex'' $a=\mathfrak{o}$,
which we choose to be one of the vertices of the quiver.

What is crucial in this representation is that the poles of the charge function correspond to the atoms that can be added or removed from the crystal. 
This ensures that the residues in the action of $e^{(a)}$ and $f^{(a)}$ (inside the square root in \eqref{eq.efpsi-action}) are non-vanishing, and hence we can generate all the possible states $|\Kappa \rangle$ starting with the vacuum.
(When these properties do not hold we obtain a reducible representation of the algebra, as we will discuss in Section~\ref{ssec:reducible}.)

\subsection{Representations from subcrystals}\label{ssec:rep_subcry}

We are now ready to discuss more general representations.
This time we consider a representation space spanned by configurations of molten crystal starting from a subcrystal
${}^{\sharp} \mathcal{C}$ of the canonical crystal $\mathcal{C}_0$.

\bigskip

To construct a representation associated with ${}^{\sharp} \mathcal{C}$ we proceed in several steps.
\begin{enumerate}
\item First, we decompose a subcrystal ${}^{\sharp} \mathcal{C}$ into a superposition of multiple canonical crystals, each after an appropriate translation and weighted with plus/minus multiplicities.
Namely, a subcrystal ${}^{\sharp} \mathcal{C}$ can be viewed as a superposition of positive/negative crystals (see Figure~\ref{fig.F1}(a)).

The decomposition procedure itself takes a few steps.
Suppose one carves out a subcrystal ${}^{\sharp} \mathcal{C}$ from the canonical crystal $\mathcal{C}_0$.
We would like to determine the coordinates (with respect to $\mathcal{C}_0$) of the starting atoms of each positive and negative crystals that ${}^{\sharp} \mathcal{C}$ is decomposed into.
\begin{enumerate}
\item\label{starter} 
The starting atoms of the positive crystals near the origin are called \emph{starters}. 
Without loss of generality, we 
move  ${}^{\sharp} \mathcal{C}$ as close as possible to the origin of $\mathcal{C}_{0}$ in the direction orthogonal to the 2D projection, such that as many as possible starters can have depth zero
(i.e.\ lie on the surface of the crystal).
\item The positive crystals headed by the starters would eventually meet and start to overlap inside $\mathcal{C}_0$. 
To avoid over-counting the overlapping region inside ${}^{\sharp} \mathcal{C}$, we need to add the ``negative" crystal started at their meeting point to cancel the redundant region.
When the overlapping region is the overlap of $n\geq 2$ positive crystals, we need to add $n-1$ copies of negative crystals, all starting at the meeting point of these $n$ positive crystals. 
The atom at the meeting point, i.e.\ the atom that heads the negative crystal, is called \emph{pauser}, because it corresponds to a zero in the charge function and thus ``slows down" the speed of the crystal growth.
The order of a pauser is defined as $n-1$, the number of copies of the negative crystals. 
An order-one pauser is called simple pauser, etc.
Note that the positions, colors, and orders of the pausers are fixed once the positions of the starters are given. See Figure~\ref{fig.F1}(i).

\item 
The negative crystals might also meet and start to overlap in ${}^{\sharp} \mathcal{C}$.
We then need to add positive crystals to cancel the overlaps of the negative crystals.
The starting atoms of these positive crystals are again starters. 
Then these secondary positive crystals might intersect, prompting us to add secondary negative crystals, etc.
This procedure follows the inclusion-exclusion principle and continues until we determine all the starters and pausers. See Figure~\ref{fig.F1}(ii).

\item Finally, one can stop the growth of ${}^{\sharp} \mathcal{C}$ along certain directions by adding negative crystals starting at appropriate places.
The atom that heads such a negative crystal is called \emph{stopper}, since the crystal ${}^{\sharp} \mathcal{C}$ stops growing at the position of this atom. See Figure~\ref{fig.F1}(iii).
\end{enumerate}
Using this procedure, we can represent any subcrystal ${}^{\sharp} \mathcal{C}$ by such linear superpositions of (possibly infinite) translations of canonical crystals $\mathcal{C}_0$ .
Finally, note that to define a subcrystal ${}^{\sharp} \mathcal{C}$, we only need to specify the set of its starters and stoppers, because the set of the pausers is determined by the positions of its starters.
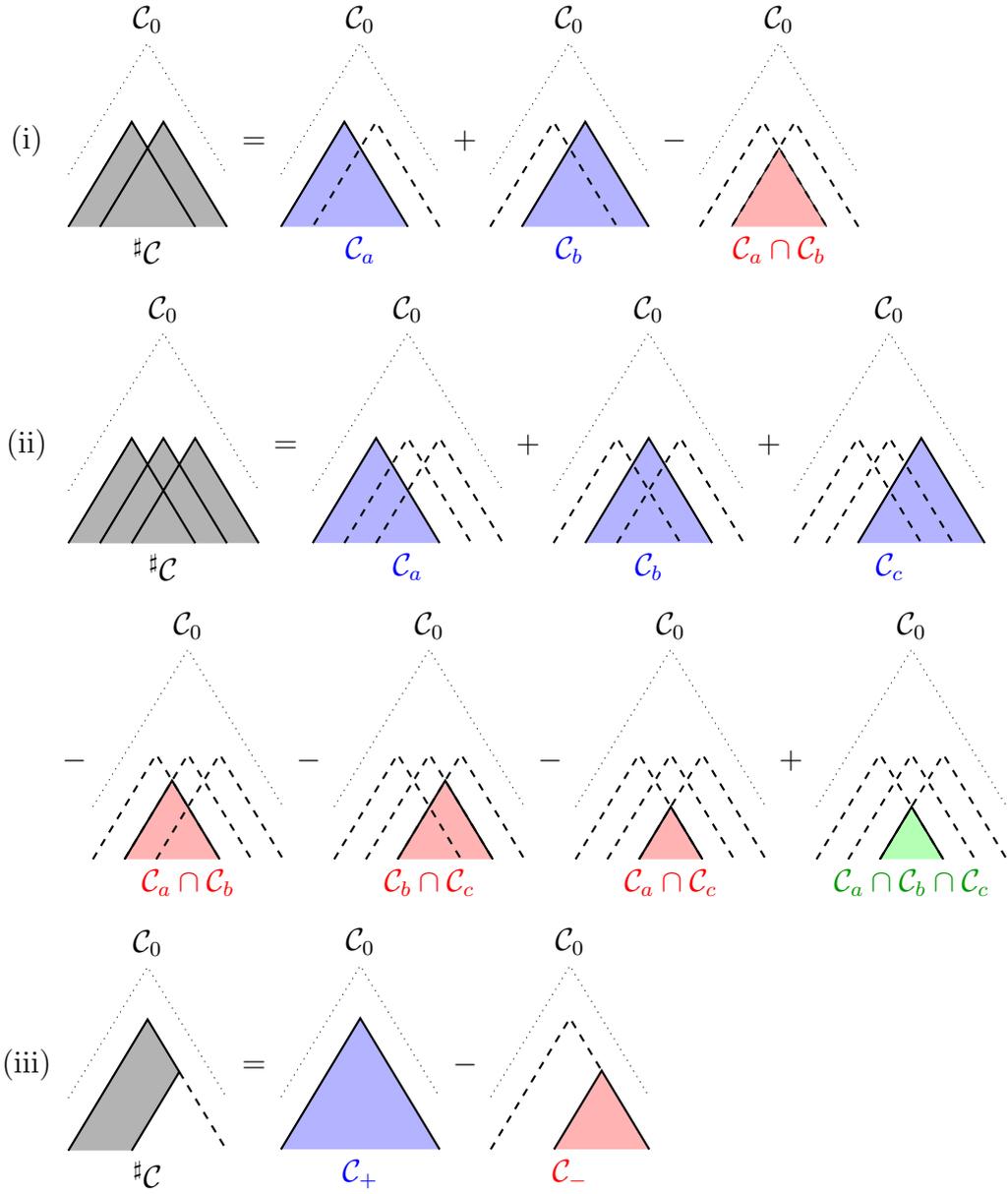
\begin{figure}[ht!]
\begin{center}
$
\begin{array}{cl}
\mbox{(i)} & \begin{array}{c}
\begin{tikzpicture}[xscale=0.6,rotate=-135,scale=0.5]
\draw[dotted] (5,0) -- (0,0) -- (0,5);
\draw[fill=white!40!gray] (6,1) -- (2,1) -- (2,5);
\draw[fill=white!40!gray] (1,6) -- (1,2) -- (5,2);
\draw[thick]  (6,1) -- (2,1) -- (2,5) (1,6) -- (1,2) -- (5,2);
\node[above] at (0,0) {$\CC_0$};
\node[below] at (3.5,3.5) {${}^{\sharp}\CC$};
\end{tikzpicture}
\end{array} = 
\begin{array}{c}
\begin{tikzpicture}[xscale=0.6,rotate=-135,scale=0.5]
\draw[dotted] (5,0) -- (0,0) -- (0,5);
\draw[fill=white!70!blue] (6,1) -- (2,1) -- (2,5);
\draw[thick]  (6,1) -- (2,1) -- (2,5);
\draw[thick,dashed] (1,6) -- (1,2) -- (5,2);
\node[above] at (0,0) {$\CC_0$};
\node[below] at (3.5,3.5) {$\color{blue}\CC_{a}$};
\end{tikzpicture}
\end{array} +
\begin{array}{c}
\begin{tikzpicture}[xscale=0.6,rotate=-135,scale=0.5]
\draw[dotted] (5,0) -- (0,0) -- (0,5);
\draw[fill=white!70!blue] (1,6) -- (1,2) -- (5,2);
\draw[thick,dashed] (6,1) -- (2,1) -- (2,5);
\draw[thick] (1,6) -- (1,2) -- (5,2);
\node[above] at (0,0) {$\CC_0$};
\node[below] at (3.5,3.5) {$\color{blue}\CC_{b}$};
\end{tikzpicture}
\end{array}-
\begin{array}{c}
\begin{tikzpicture}[xscale=0.6,rotate=-135,scale=0.5]
\draw[dotted] (5,0) -- (0,0) -- (0,5);
\draw[fill=white!70!red] (5,2) -- (2,2) -- (2,5);
\draw[thick,dashed]  (6,1) -- (2,1) -- (2,5) (1,6) -- (1,2) -- (5,2);
\node[above] at (0,0) {$\CC_0$};
\node[below] at (3.5,3.5) {$\color{red}\CC_{a}\cap\CC_b$};
\end{tikzpicture}
\end{array}\\
\mbox{(ii)} & \begin{array}{c}
\begin{tikzpicture}[xscale=0.6,rotate=-135,scale=0.5]
\draw[dotted] (6,0) -- (0,0) -- (0,6);
\draw[fill=white!40!gray] (7,1) -- (3,1) -- (3,5);
\draw[fill=white!40!gray] (6,2) -- (2,2) -- (2,6);
\draw[fill=white!40!gray] (5,3) -- (1,3) -- (1,7);
\draw[thick] (7,1) -- (3,1) -- (3,5);
\draw[thick] (6,2) -- (2,2) -- (2,6);
\draw[thick] (5,3) -- (1,3) -- (1,7);
\node[above] at (0,0) {$\CC_0$};
\node[below] at (4,4) {${}^{\sharp}\CC$};
\end{tikzpicture}
\end{array} = 
\begin{array}{c}
\begin{tikzpicture}[xscale=0.6,rotate=-135,scale=0.5]
\draw[dotted] (6,0) -- (0,0) -- (0,6);
\draw[fill=white!70!blue] (7,1) -- (3,1) -- (3,5);
\draw[thick] (7,1) -- (3,1) -- (3,5);
\draw[thick,dashed] (6,2) -- (2,2) -- (2,6);
\draw[thick,dashed] (5,3) -- (1,3) -- (1,7);
\node[above] at (0,0) {$\CC_0$};
\node[below] at (4,4) {$\color{blue}\CC_{a}$};
\end{tikzpicture}
\end{array}+
\begin{array}{c}
\begin{tikzpicture}[xscale=0.6,rotate=-135,scale=0.5]
\draw[dotted] (6,0) -- (0,0) -- (0,6);
\draw[fill=white!70!blue] (6,2) -- (2,2) -- (2,6);
\draw[thick,dashed] (7,1) -- (3,1) -- (3,5);
\draw[thick] (6,2) -- (2,2) -- (2,6);
\draw[thick,dashed] (5,3) -- (1,3) -- (1,7);
\node[above] at (0,0) {$\CC_0$};
\node[below] at (4,4) {$\color{blue}\CC_{b}$};
\end{tikzpicture}
\end{array}+
\begin{array}{c}
\begin{tikzpicture}[xscale=0.6,rotate=-135,scale=0.5]
\draw[dotted] (6,0) -- (0,0) -- (0,6);
\draw[fill=white!70!blue] (5,3) -- (1,3) -- (1,7);
\draw[thick,dashed] (7,1) -- (3,1) -- (3,5);
\draw[thick,dashed] (6,2) -- (2,2) -- (2,6);
\draw[thick] (5,3) -- (1,3) -- (1,7);
\node[above] at (0,0) {$\CC_0$};
\node[below] at (4,4) {$\color{blue}\CC_{c}$};
\end{tikzpicture}
\end{array}\\
 & -\begin{array}{c}
\begin{tikzpicture}[xscale=0.6,rotate=-135,scale=0.5]
\draw[dotted] (6,0) -- (0,0) -- (0,6);
\draw[fill=white!70!red] (6,2) -- (3,2) -- (3,5);
\draw[thick] (6,2) -- (3,2) -- (3,5);
\draw[thick,dashed] (7,1) -- (3,1) -- (3,5);
\draw[thick,dashed] (6,2) -- (2,2) -- (2,6);
\draw[thick,dashed] (5,3) -- (1,3) -- (1,7);
\node[above] at (0,0) {$\CC_0$};
\node[below] at (4,4) {$\color{red}\CC_{a}\cap\CC_{b}$};
\end{tikzpicture}
\end{array}
-\begin{array}{c}
\begin{tikzpicture}[xscale=0.6,rotate=-135,scale=0.5]
\draw[dotted] (6,0) -- (0,0) -- (0,6);
\draw[fill=white!70!red] (2,6) -- (2,3) -- (5,3);
\draw[thick] (2,6) -- (2,3) -- (5,3);
\draw[thick,dashed] (7,1) -- (3,1) -- (3,5);
\draw[thick,dashed] (6,2) -- (2,2) -- (2,6);
\draw[thick,dashed] (5,3) -- (1,3) -- (1,7);
\node[above] at (0,0) {$\CC_0$};
\node[below] at (4,4) {$\color{red}\CC_{b}\cap\CC_{c}$};
\end{tikzpicture}
\end{array}
-\begin{array}{c}
\begin{tikzpicture}[xscale=0.6,rotate=-135,scale=0.5]
\draw[dotted] (6,0) -- (0,0) -- (0,6);
\draw[fill=white!70!red] (3,5) -- (3,3) -- (5,3);
\draw[thick]  (3,5) -- (3,3) -- (5,3);
\draw[thick,dashed] (7,1) -- (3,1) -- (3,5);
\draw[thick,dashed] (6,2) -- (2,2) -- (2,6);
\draw[thick,dashed] (5,3) -- (1,3) -- (1,7);
\node[above] at (0,0) {$\CC_0$};
\node[below] at (4,4) {$\color{red}\CC_{a}\cap\CC_{c}$};
\end{tikzpicture}
\end{array}
+\begin{array}{c}
\begin{tikzpicture}[xscale=0.6,rotate=-135,scale=0.5]
\draw[dotted] (6,0) -- (0,0) -- (0,6);
\draw[fill=white!70!green] (3,5) -- (3,3) -- (5,3);
\draw[thick]  (3,5) -- (3,3) -- (5,3);
\draw[thick,dashed] (7,1) -- (3,1) -- (3,5);
\draw[thick,dashed] (6,2) -- (2,2) -- (2,6);
\draw[thick,dashed] (5,3) -- (1,3) -- (1,7);
\node[above] at (0,0) {$\CC_0$};
\node[below] at (4,4) {$\color{black!40!green}\CC_{a}\cap\CC_{b}\cap\CC_{c}$};
\end{tikzpicture}
\end{array}\\
\mbox{(iii)} & \begin{array}{c}
\begin{tikzpicture}[xscale=0.6,rotate=-135,scale=0.5]
\draw[dotted] (5,0) -- (0,0) -- (0,5);
\draw[fill=white!40!gray] (6,1) -- (1,1) -- (1,3) -- (4,3);
\draw[thick]   (6,1) -- (1,1) -- (1,3) -- (4,3);
\draw[thick,dashed] (1,3) -- (1,6);
\node[above] at (0,0) {$\CC_0$};
\node[below] at (3.5,3.5) {${}^{\sharp}\CC$};
\end{tikzpicture}
\end{array}=
\begin{array}{c}
\begin{tikzpicture}[xscale=0.6,rotate=-135,scale=0.5]
\draw[dotted] (5,0) -- (0,0) -- (0,5);
\draw[fill=white!70!blue] (6,1) -- (1,1) -- (1,6);
\draw[thick]   (6,1) -- (1,1) -- (1,6);
\node[above] at (0,0) {$\CC_0$};
\node[below] at (3.5,3.5) {$\color{blue}\CC_{+}$};
\end{tikzpicture}
\end{array}-
\begin{array}{c}
\begin{tikzpicture}[xscale=0.6,rotate=-135,scale=0.5]
\draw[dotted] (5,0) -- (0,0) -- (0,5);
\draw[fill=white!70!red] (4,3) -- (1,3) -- (1,6);
\draw[thick,dashed]   (6,1) -- (1,1) -- (1,3);
\draw[thick] (4,3) -- (1,3) -- (1,6);
\node[above] at (0,0) {$\CC_0$};
\node[below] at (3.5,3.5) {$\color{red}\CC_{-}$};
\end{tikzpicture}
\end{array}\\
\end{array}
$
\caption{The subcrystal $^{\sharp} \CC$ of the canonical crystal can be represented as a superposition of positive/negative canonical crystals.}
\label{fig.F1}
\end{center}
\end{figure}
\bigskip

\item Next we translate the decomposition of the subcrystal ${}^{\sharp} \mathcal{C}$ into the charge functions of the ground state of the corresponding representation, denoted by ${}^{\sharp}\psi^{(a)}_0(z)$ for $a\in Q_0$.

The subcrystal ${}^{\sharp} \mathcal{C}$ corresponds to the ground state of a representation ${}^{\sharp}\textrm{Rep}$ of the shifted quiver Yangian $\mathsf{Y}(Q,W,\mys)$, where the pair $(Q,W)$ maps one-to-one to the canonical crystal $\mathcal{C}_0$ and the shift $\mys$ is determined by the shape of the subcrystal ${}^{\sharp}\CC$.

Consider the molten crystal configurations from the subcrystal ${}^{\sharp} \mathcal{C}$, i.e.\ the complements of crystal configurations with some atoms melted away from the surface of ${}^{\sharp} \mathcal{C}$.
They together furnish the excited states of the representation ${}^{\sharp}\textrm{Rep}$. 
All these states can be obtained by applying the creation operator $e^{(a)}(z)$ on the ground state iteratively. 
Therefore, the ground state charge functions ${}^{\sharp}\psi^{(a)}_0(z)$ should encode the shape of the subcrystal ${}^{\sharp} \mathcal{C}$.

Let us see how the three types of leading atoms (starters, pausers, and stoppers) are captured by ${}^{\sharp}\psi^{(a)}_0(z)$.
\begin{enumerate}
\item When applying the creation operator $e^{(a)}(z)$ on the ground state, the atoms that one can add are the starters, each heading a positive crystal. 
Therefore, the coordinate function of each starter gives a pole in ${}^{\sharp}\psi_0^{(a)}(z)$:
\begin{equation}
\myp^{(a)}=h(\sqbox{$a$})\,, \quad \textrm{ when $\sqbox{$a$}$ is a starter} \;.
\end{equation}
When there are multiple poles in ${}^{\sharp}\psi_0^{(a)}(z)$, the crystal starts at these starters and grows from them simultaneously.
\item Each order-$n$ pauser gives rise to an order-$n$ zero in ${}^{\sharp}\psi_0^{(a)}(z)$:
\begin{equation}
\myz^{(a)}=h(\sqbox{$a$})\,, \quad \textrm{ when $\sqbox{$a$}$ is a pauser} \;.
\end{equation}

\item Each stopper gives rise to a simple zero in ${}^{\sharp}\psi_0^{(a)}(z)$:
\begin{equation}
\myz^{(a)}=h(\sqbox{$a$})\,, \quad \textrm{ when $\sqbox{$a$}$ is a stopper} \;.
\end{equation}

\end{enumerate}

Putting all these together, we have the ground state charge function:
\begin{tcolorbox}[ams align]\label{eq:psi0_summary}
{}^{\sharp}\psi^{(a)}_0(z)=\frac{\prod^{\mys^{(a)}_{-}}_{\beta=1}(z-\myz^{(a)}_{\beta}) }{\prod^{\mys^{(a)}_{+}}_{\alpha=1}(z-\myp^{(a)}_{\alpha})} \;,
\end{tcolorbox}
\noindent
and $\{\myp^{(a)}_{\alpha}\}$ corresponds to the set of all starters of color $a$, and $\mys^{(a)}_{+}$ is the size of this set, whereas $\{\myz^{(a)}_{\beta}\}$ corresponds to the set of all pausers (with multiplicity given by the order) and stoppers of color $a$, and $\mys^{(a)}_{-}$ is the size of this set.

\item
We can then use the same formulas \eqref{eq.efpsi-action} and \eqref{eq.efpsi-action-2}, with the ground state charge function $\psi^{(a)}_0(z)$ replaced by the ground state charge function ${}^{\sharp}\psi^{(a)}_0(z)$, to define the representation ${}^{\sharp}\textrm{Rep}$ of the shifted quiver Yangian $\mathsf{Y}(Q,W,\mys)$, where  the shift $\mys=\{\mys^{(a)}\}$ with 
\begin{equation}\label{shift}
\mys^{(a)}\equiv\mys^{(a)}_{+}-\mys^{(a)}_{-}\,.
\end{equation}
The proof that this indeed gives a representation of $\mathsf{Y}(Q,W,\mys)$
is essentially the same as in the previous discussion for the unshifted quiver Yangian in \cite{Li:2020rij}.

\end{enumerate}

We have explained the procedure of determining the ground state charge function of a representation, by decomposing the subcrystal in terms of the superposition of positive/negative canonical crystals.
There is actually another way of determining the ground state charge function: by considering the ground state of a non-trivial representation as an excited state of the vacuum representation and then computing the charge function from the definition~\eqref{eq.efpsi-action-2}. 
For a large class of representations, the crystal configuration of the ground state has ``removable" atoms only at infinity, namely, there are sufficient amount of cancellations between the contributions from neighboring atoms that all the ``removing poles" are pushed to  infinity. 
Therefore, although viewed a priori as an excited state in the vacuum representation, the end result of evaluating the charge function via \eqref{eq.efpsi-action-2} gives rise to a charge function of a ground state, i.e.\ they can be annihilated by all the $f^{(a)}(z)$ generators, of a non-vacuum representation.

This method has been used to compute the ground state charge function for non-trivial representations of the affine Yangian of $\mathfrak{gl}_1$, and it works for ordinary representations~\cite{Prochazka:2015deb,Gaberdiel:2017dbk}, the conjugate representations~\cite{Gaberdiel:2018nbs} (the so-called high-wall representations that are relevant for the counting of the PT invariants~\cite{Gaberdiel:2018nbs, Gaiotto:2020dsq}), and even the representations in the twisted sectors~\cite{Datta:2016cmw}.
(Note that since this method only works for those representation whose ground state only has ``removable" atoms at infinity, we cannot use it to study e.g.\ the finite chamber representation in Section~\ref{ssec:finite_chamber} nor those ``novel" representations in Section~\ref{sec:genera_rep}.)

In this paper, we also generalize this method to generic toric Calabi-Yau threefolds. 
Since the computation is rather technical, we leave it to Appendix~\ref{sec:app_manual_charge_f}. 
In this approach, the relevant molten crystal configuration consists of infinitely many atoms, and the resulting charge function might need a regularization,\footnote{For the affine Yangian of $\mathfrak{gl}_1$, the contributions from the infinite number of atoms cancel and there is no need for regularization.} for more details see Appendix~\ref{sec:app_manual_charge_f}. 
But after the appropriate regularization, the resulting charge function agrees with the result~\eqref{eq:psi0_summary} computed via the method introduced in this subsection.

Comparing the two methods, one can see that the positive/negative crystal method introduced in this paper is superior. 
First, the new method is more general, applicable to both the infinite-dimensional  and the finite-dimensional  representations, whereas the old method only works for the infinite-dimensional representations.\footnote{We could remove those ``removing poles'' by hand, and thus obtain the correct ground state charge function even with this method. 
However, this is not very natural and hence we will not do so in this paper. We instead study those ``finite" representations using the positive/negative crystal method in the main text. 
}
Second, even for the infinite representations, with the new method there is no need to consider an infinite number of atoms and hence  no need for regularization  --  one can directly read off the ground state charge function \eqref{eq:psi0_summary} from the position of the starters, pausers, and stoppers. 
Lastly, the positive/negative crystal method allows us to construct and classify previously unknown representations easily, see Section~\ref{sec:genera_rep}.

\subsection{Irreducibility}\label{subsec.convex}

The representations we constructed in the previous subsection, by defining the subcrystal in terms of the positions of its starters, pausers, and stoppers, are in general reducible.
First of all, the subcrystal can have multiple connected components. 
In the rest of this paper we focus on those subcrystals that are simply connected. 
However, even for such a simply-connected subcrystal ${}^{\sharp} \mathcal{C}$, the associated representation ${}^{\sharp}\textrm{Rep}$ can still be reducible.

\begin{figure}[ht!]
\begin{center}
    \begin{tikzpicture}
				\begin{scope}[rotate=-135]
					\begin{scope}[scale=0.5]
					    \draw[white,fill=red,opacity=0.1] (9,0) -- (6.5,2.5) -- (0,2.5) -- (0,0) -- cycle;
					    \draw[white,fill=blue,opacity=0.1]  (6.5,2.5) -- (0,2.5) -- (0,9) -- (0.5,8.5) -- (0.5,3.5) -- (5.5,3.5) -- cycle;
					    \draw[white,fill=green,opacity=0.1]   (5.5,3.5) -- (0.5,3.5) -- (0.5,8.5) -- cycle;
						\foreach \x in {0,...,8}
						\foreach \y in {0,...,8}
						{
							\pgfmathparse{int(\x+\y-7)}
							\let\r\pgfmathresult
							\ifnum \r > 0
							\breakforeach
							\fi
							\pgfmathparse{int(10*(\x+\y))}
						    \draw[->](\x+0.1,\y)  -- (\x+0.9,\y);
							\draw[->](\x,\y+0.1)  -- (\x,\y+0.9);
						}
						\draw[ultra thick] (0,9) -- (0,0) -- (9,0);
						\begin{scope}[shift={(2,5)}]
						    \draw[white,fill=white] (4,0) -- (0,0) -- (0,4) -- cycle;
							\foreach \x in {0,...,4}
							\foreach \y in {0,...,4}
							{
								\pgfmathparse{int(\x+\y-3)}
								\let\r\pgfmathresult
								\ifnum \r > 0
								\breakforeach
								\fi
							\draw[dashed] (\x+0.1,\y)  -- (\x+0.9,\y);
							\draw[dashed] (\x,\y+0.1)  -- (\x,\y+0.9);
							}
							\draw[ultra thick] (0,5) -- (0,0) -- (5,0);
						\end{scope}
						\draw[ultra thick,->,>=stealth',orange] (0,0) -- (0,3) -- (4,3);
						\draw[ultra thick,->,>=stealth',black!40!green] (0,0) -- (1,0) -- (1,1) -- (4,1) -- (4,3);
						\shade[ball color = blue] (0,0) circle (0.2);
						\shade[ball color = violet] (0,3) circle (0.2);
						\shade[ball color = violet] (1,4) circle (0.2);
						\shade[ball color = red] (2,5) circle (0.2);
						\node[above right] at (0,2) {$\color{orange}=0$};
						\node[left] at (3,-3) {\color{blue} starter};
						\draw[->,>=stealth'] (3,-3) -- (0.5,-0.5);
						\node[left] at (9,0) {positive crystal};
						\node[right] at (2,10) {negative crystal};
						\node[right] at (-3,6) {\color{violet} shadow stoppers};
						\node[right] at (-1,8) {\color{red} stopper};
						\draw[->,>=stealth'] (-3,6) -- (-0.5,3.5);
						\draw[->,>=stealth'] (-3,6) to[out=315,in=135] (0.5,4.5);
						\draw[->,>=stealth'] (-1,8) -- (1.5,5.5);
					\end{scope}
				\end{scope}
			\end{tikzpicture}
\caption{The irreducibility of the representation requires the stoppers of the subcrystal to have relative depth-zero. 
In this figure, this condition is violated,
leading to three irreducible components, depicted in three different colors (pink, blue, and green).}
\label{fig.irreducibility}
\end{center}
\end{figure}
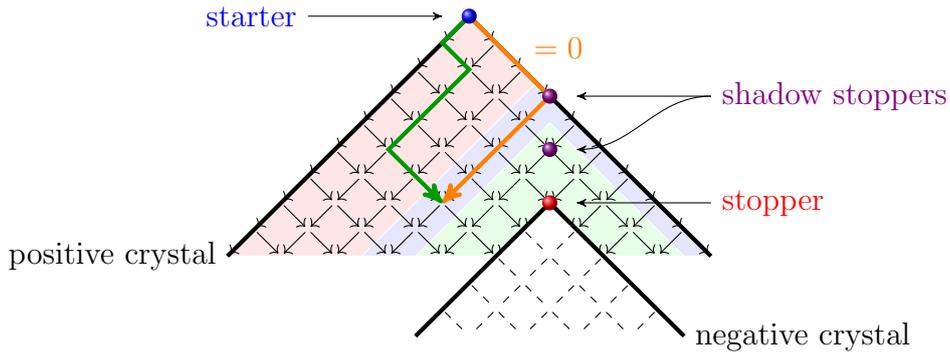

To see this, let us consider the example of Figure~\ref{fig.irreducibility}.
Inside the canonical crystal $\mathcal{C}_{0}$, whose starter is colored blue, we have a negative crystal, headed by a stopper at the location $\Box_2$.
This creates a zero $z=h(\Box_2)$ of the ground state charge function.

Now the important point is that the relations between atoms and the ground state charge function is not one-to-one
(even if we assume generic equivariant parameters).\footnote{We will discuss the case of non-generic equivariant parameters in Section \ref{ssec:truncations}.}
This is because the equivariant parameters satisfy the loop constraint.
Namely, a zero $z=h(\Box_2)$ inevitably means that there are zeros
at all the atoms $\Box_3$ whose two-dimensional projection coincide with $\Box_2$.

We can now see that this can make the representation reducible.
Consider a path shown in orange in Figure~\ref{fig.irreducibility}. The path is $F$-term equivalent to another path shown in green,
which goes through atoms corresponding to zeros of the ground state charge function.
This means that the path gives zero both ways, making the representation reducible.

Let us now determine the irreducible component of a simply-connected subcrystal.
Without loss of generality (since the essential information of the subcrystal ${}^{\sharp}\CC$ is its shape), we can use the translational invariance to move the subcrystal ${}^{\sharp}\CC$ as close to the origin of the canonical crystal $\CC_0$ as possible, such that as many as possible starters have depth-zero.\footnote{If we do not do this, we can simply replace the ``depth" of the stopper below by ``relative depth", i.e. the difference between the depth of the stopper and the smallest depth of the starters.}
When the stopper atom has a depth bigger than zero, those atoms with the same coordinate function $h(\Box_2)$ but with lower depth are called the ``shadows" of the stopper. 
When we grow the crystal by applying the creation operators $e^{(a)}(z)$ on the ground state, we would hit the stopper shadow with depth-zero first, and the crystal growth along that direction would stop there.
Namely it is the depth-zero shadow of the stopper that determines the first irreducible component of the subcrystal.
Peeling off this first irreducible component, we can start the growth of the crystal again and the depth-one shadow of the stopper determines the second irreducible component. 
Repeating this way until we reach the true stopper, we can see that having a stopper of depth-$n$ means that there are (at least) $n+1$ irreducible components of the subcrystal. 
In the example shown in Figure~\ref{fig.irreducibility}, the stopper (colored red) has depth-two, therefore there are two stopper shadows (colored purple) with depth-zero and depth-one. 
Therefore the reducible representation that is given by the subcrystal of Figure~\ref{fig.irreducibility} has three irreducible components (colored orange, blue, and green, respectively).

\subsection{Truncations of shifted quiver Yangians}\label{ssec:truncations}

One assumption in the discussion of the irreducibility above is that
the equivariant parameters are generic  --  if we impose genericity
we expect that the two atoms  $\Box_1, \Box_2 \in \mathcal{C}$ have the same 
values $h(\Box_1), h(\Box_2)$ if and only if the two atoms are located at the same point in the 
two-dimensional projection. 

This changes, however, when the equivariant parameters are not generic. 
In this case
it could happen that the ground state charge function creates more zeros than intended,
making the representation reducible. 

We can further enrich the story by replacing the 
ground state charge function \eqref{eq:psi0_summary} by 
a slightly more general ansatz:
\begin{equation}\label{eq:psi0_summary_mod}
{}^{\sharp}\psi_0^{(a)}(z)=t^{(a)}(z) \frac{\prod^{\mys^{(a)}_{-}}_{\beta=1}(z-\myz^{(a)}_{\beta}) }{\prod^{\mys^{(a)}_{+}}_{\alpha=1}(z-\myp^{(a)}_{\alpha})} \;.
\end{equation}

If we assume that $t^{(a)}(z)$ is a rational function with no poles or zeros at any of the atoms of the subcrystal $^{\sharp}\mathcal{C}$,
then this will not affect the construction of the representation as well as its irreducibility at generic equivariant parameters, 
except that the shift of the shifted Yangian $\mathsf{Y}(Q,W,{}^{\sharp}\psi)$ is affected by the degree of $t^{(a)}(z)$.

Such a factor was already encountered when we discussed the vacuum representations of the unshifted quiver Yangians in \cite{Li:2020rij}. 
In \eqref{eq.efpsi-action-2}, the charge function $\psi_0^{\mathfrak{o}}$ has a zero at 
$z=-C$, where $C$ is the value of the central element and is taken to be generic. 
As long as $C$ does not correspond to the locations of the atoms $h(\Box)$, this does not have any effect on the structure of the crystal.

The factor $t^{(a)}(z)$, however, can affect the irreducibility of the representations when we consider non-generic equivariant parameters. 
This is because the $t^{(a)}(z)$ can create new poles/zeros at the locations of the crystal.

Now, suppose for a shifted quiver Yangian $\mathsf{Y}(Q,W,\mys)$, the same factor 
$t^{(a)}(z)$ is present in the ground state charge function of all its representations ${}^{\sharp}\textrm{Rep}$, and denote the order of  $t^{(a)}(z)$ by $|t^{(a)}|$, then the truncation factor changes the shift $\mys^{(a)}$ of the $\mathsf{Y}(Q,W,\mys)$ algebra by
\begin{equation}
\mys^{(a)} \rightarrow \tilde{\mys}^{(a)} \equiv \mys^{(a)}-|t^{(a)}|\,,
\end{equation}
and realizes a truncation of the algebra $\mathsf{Y}(Q,W, \tilde{\mys})$.

In \cite{Li:2020rij} this issue was studied for the unshifted quiver Yangians, and the corresponding representations lead to truncations of the algebra.
Moreover such truncations are identified with D4-branes wrapping divisors in the toric Calabi-Yau manifold.
The same story applies for our shifted quiver Yangians of this paper.\footnote{In \cite{Li:2020rij}, a truncation factor $t^{(a)}(z)=z+C$ was introduced already during the definition of the quiver Yangians that correspond to canonical crystals, therefore all the quiver Yangians in \cite{Li:2020rij} are unshifted, in the terminology of the current paper.}

\subsection{From subcrystals to framed quivers}
\label{ssec:framedQW}

For the discussion of quiver supersymmetric quantum mechanics
it is useful to reformulate the statements in terms of the quiver.
This is done by introducing the concept of a framed quiver 
$(\sQ, \sW)$, extending the original quiver data $(Q,W)$.

\bigskip

A framing of the quiver is defined by (1) adding to the original quiver $Q$ a new vertex (called \emph{framing node}) and new arrows between the new vertex and the existing vertices, and (2) appending corresponding new monomial terms to the superpotential $W$.
A vertex $a$ connected to the framing node $f$ by some arrow $f\rightarrow a$ is called a \emph{framed vertex} or a \emph{framed node}.

\bigskip

For the case of the canonical crystal $\mathcal{C}_0$, the framed quiver is simple:
we add an extra node $f$, i.e.\ the framing node, and an arrow $\iota$ from
$f$ to a particular vertex $a$, i.e.\ the framed node. 
When we construct the crystal we consider a set of paths 
starting from the framed node, and this ensures that the crystal growth 
starts with an atom of color $a$.
This framing is called the \emph{canonical framing}.
\bigskip

For a general subcrystal ${}^{\sharp} \mathcal{C}$, each pole of the ground state charge function ${}^{\sharp}\psi^{(a)}_0(z)$ is represented in the framed quiver $\sQ$ by an arrow from the framing node $f$ to 
a framed vertex $a$.
For general $\{{}^{\sharp}\psi^{(a)}_0(z)\}$ with multiple poles, this means that when the path starts from the framed nodes there are multiple choices of arrows in the initial step, and this in turn means that the crystal growth starts at multiple locations.
The charge assignments of the framed arrows should be chosen according to the charges of the starter atoms.

The zeros of ${}^{\sharp}\psi^{(a)}_0(z)$ are  represented by arrows from the vertex $a$ back to the framing node $f$.
In addition, for the zeros of ${}^{\sharp}\psi^{(a)}_0(z)$ we need to add appropriate superpotential terms such that the growth of the crystal indeed pauses or stops there.\footnote{Recall that by ``pausing" we mean that the presence of negative crystal (lead by the pauser) ensures that the overlapping of positive crystals does not give rise to an over-counting of the overlapped regions. 
In some sense, a pauser makes the crystal grow slower, but at the right ``speed".}
Namely, in order to realize the negative crystal, we add an arrow from the framed vertex $a$ going back to the framing node $f$ and also a newly-added superpotential term involving it.
These arrows play the role of the Lagrange multipliers enforcing the constraints.
While the addition of a newly-introduced field $q$ potentially enlarges the vacuum moduli space of the theory, the $F$-term constraint, which follows from taking the derivative of the superpotential with respect to other fields, reduces to that of the original theory when we have $q=0$.
The non-trivial statement is that in a suitable chamber of the moduli space (i.e.\ for a suitable choice of the stability parameter), this is ensured automatically.

\bigskip

Physically, the framing node $f$ represents non-compact D-branes filling some non-compact cycles of the toric Calabi-Yau threefold $X$.
For example, for the BPS wall-crossing, the framing node corresponds to a D6-brane filling the entire $X$.
Because the toric Calabi-Yau threefold $X$ itself is non-compact, its BPS moduli space is also non-compact, therefore we need a regularization procedure, where we start with a compact Calabi-Yau manifold and then take the decoupling limit (cf.\ \cite{Jafferis:2008uf}). 
Since the D6-brane is non-compact at the end of this process, the gauge field on the D6-brane becomes non-dynamical and gives rise to a flavor symmetry of the theory.
(As a mathematical counterpart, this extra framing node of the quiver is called ``frozen'' in the cluster algebra literature.)

Similarly, the effect of adding non-compact D4 and D2 branes, which encompasses ``open/closed" BPS counting and truncation of the BPS algebra, respectively, can also be realized by choosing a framing $\sharp$ of $(\sQ,\sW)$ appropriately.

\subsection{Bootstrapping shifted quiver Yangians from subcrystal representations}
\label{ssec:bootstrap} 

The unshifted quiver Yangians $\mathsf{Y}(Q,W)$ can be bootstrapped from their actions on the molten crystal representations from the canonical crystal $\CC_0$, see Section\ 6.4 of \cite{Li:2020rij}.
This property generalizes to the shifted quiver Yangian $\mathsf{Y}(Q,W,{}^{\sharp}\psi)$ or equivalently $\mathsf{Y}(Q,W,\mys)$.
Namely, although in this paper, we first wrote down the algebraic relations of the shifted quiver Yangians in \eqref{eq-OPE-toric} and then gave their actions on the subcrystal representations \eqref{eq.efpsi-action} and \eqref{eq.efpsi-action-2}, had we not known the algebraic relations, we could have also bootstrapped the relations 
\eqref{eq-OPE-toric} from the actions \eqref{eq.efpsi-action} and \eqref{eq.efpsi-action-2}.
\bigskip

The bootstrap procedure goes exactly the same way as the one in Section\ 6.4 of \cite{Li:2020rij}.
Since all the relations involve a product of two operators at two values of spectral parameters: $A_1(z)$ and $A_2(w)$, where $A_{1,2}$ are from the set $\{e^{(a)},\psi^{(a)}, f^{(a)}\}$ for $a\in Q_0$, and we know the action of a single operator on an arbitrary molten crystal state $ |\Kappa\rangle$, applying $A_1(z)$ and $A_2(w)$ on $|\Kappa\rangle$ in two different orders and comparing the two results $A_1(z)A_2(w)|\Kappa\rangle$ and $A_2(z)A_1(w)|\Kappa\rangle$ then gives a relation $(\star)$ involving $A_1(z)$ and $A_2(w)$ on the arbitrary state $ |\Kappa\rangle$.
Since we assume that the set of molten crystal states $\{ |\Kappa\rangle\}$ furnishes a representation of the algebra, and the relation $(\star)$ is true on any $|\Kappa\rangle$, therefore, the relation $(\star)$ holds as an algebraic relation itself, without the need to refer to any state $|\Kappa\rangle$.
This way we can derive all the relations in \eqref{eq-OPE-toric}.
Furthermore, demanding that the vacuum character of the reduced quiver Yangian \eqref{eq:reducedShiftedQY} reproduces the counting of the molten crystal configurations from the canonical crystal strongly constraints the Serre relations \cite{Li:2020rij}. It would be interesting to 
explicitly identify an appropriate set of Serre relations for the quiver Yangian, such that the resulting algebra (i.e.\ the reduced quiver Yangian) acts faithfully on the crystal.

\bigskip

Let us note that the fact that the quiver Yangians can be bootstrapped from their action on the representations
has been discussed previously in the literature.
In fact, inspired by the observation that the affine Yangian of $\mathfrak{gl}_1$ can be bootstrapped from its action on the set of plane partitions, the affine Yangian of $\mathfrak{gl}_2$ and $\mathfrak{gl}_{1|1}$ (in the ``gluing" bases) were constructed with this bootstrap technique, from their actions on a pair of plane partition ``glued" appropriately according to the dual graphs of the toric diagrams of the $\mathbb{C}^2/\mathbb{Z}_2\times\mathbb{C}$ and the resolved conifold geometries, respectively \cite{Gaberdiel:2018nbs, Li:2019nna, Li:2019lgd}.
The ``colored crystal bases" adopted in \cite{Li:2020rij} makes the generalization of this bootstrap technique to arbitrary toric Calabi-Yau threefolds more transparent and streamlined, because they are particularly adapted to the structure of (the torus fix-points set of) the vacuum moduli space.
However, we expect that this bootstrap technique will work in other bases as well.

Finally we mention the similarity of this bootstrap feature of the quiver Yangians to that of vertex operator algebras (VOAs). 
First of all, it is known that for the special case of the toric Calabi-Yau threefolds without compact $4$-cycles, the quiver Yangians are related to some infinite-dimensional $\mathcal{W}$ algebras \cite{ueda2020affine}.\footnote{The affine Yangian of $\mathfrak{gl}_1$ is isomorphic to the universal enveloping algebra of the $\mathcal{W}_{1+\infty}$ algebra \cite{Prochazka:2015deb, Gaberdiel:2017dbk}.
The affine Yangians of $\mathfrak{gl}_2$ and $\mathfrak{gl}_{1|1}$ are expected to be related to the $\mathfrak{gl}_2$ and $\mathfrak{gl}_{2|2}$ matrix extended $\mathcal{W}_{1+\infty}$ algebras, respectively \cite{Gaberdiel:2018nbs, Li:2019nna, Li:2019lgd}; and for more general toric Calabi-Yau threefolds without compact $4$-cycles, we expect that the affine Yangian of $\mathfrak{gl}_{m|n}$ is related to the $\mathfrak{gl}_{m|n}$ extended $\mathcal{W}_{1+\infty}$ algebra studied in \cite{Creutzig_2019, Creutzig_2019_2, eberhardt2019matrixextended, Rapcak:2019wzw}.}
The algebraic relations of a vertex operator algebra can be reconstructed once we know its action on the vacuum module, due to the  locality property of the VOA \cite{Goddard:1989dp}.
This is similar to the bootstrap story of the quiver Yangian.
However, in the colored crystal bases of the quiver Yangians adopted in the current paper, the bootstrap of the quiver Yangians seems much more efficient and transparent than the corresponding problem in VOA. 
It would be interesting to find the $\mathcal{W}$ algebra counterpart of the quiver Yangians for generic toric Calabi-Yau threefolds (namely, those with compact $4$-cycles) by comparing the bootstrap procedures of the quiver Yangians and of the VOAs.

\section{Quiver BPS algebras}
\label{sec:BPSalgebras}

In this section we will see that the representation of the shifted quiver Yangian described in the previous section
can be derived more physically, namely by equivariant localization on the moduli space of the 
quiver quantum mechanics \cite{Denef:2002ru} associated with the framed quiver $(\sQ, \sW)$ introduced in Section \ref{ssec:framedQW}.
The discussion here closely follows a similar discussion of \cite{Galakhov:2020vyb} for the unshifted case.

\subsection{Quiver quantum mechanics}
\label{ssec:QQM}
	
We are identifying the algebraic construction of the BPS quiver Yangian with a physical construction of the BPS algebra in an $\CN=4$ supersymmetric quantum mechanics. 
The defining data of a quantum mechanics Lagrangian consists of (1) a quiver-superpotential pair $(\sQ,\sW)$ and (2) for each vertex $a\in \sQ_0$ the quiver dimension $d_a$ and the stability parameter $\zeta_a$:
\begin{equation}	
d_a\in \IZ_{\geq 0},\quad \zeta_a\in \IR, \quad \textrm{for}\quad a\in \sQ_0 \,.
\end{equation}
The matter content for quiver gauge theories is identified as follows:
\begin{enumerate}
\item A non-framing node $a\in Q_0$ with dimension $d_a$ corresponds to the gauge group $U(d_a)$. 
In particular, the 1D $\CN=4$ vector multiplet contains the following bosonic fields: 
		\begin{equation}
		\mbox{gauge field }A_a, \; \mbox{real scalar }X_a,\;\mbox{complex scalar }\Phi_a \;. 
		\end{equation}
\item An arrow $I\in\{a\to b\}$ for $a,b \in Q_0$ corresponds to a chiral multiplet charged bi-fundamentally with respect to $U(d_b)\times \overline{U(d_a)}$.   
In particular, the 1D $\CN=4$ chiral multiplet contains the following bosonic fields:
		\begin{equation}
		\mbox{complex scalar }q_I \;.
		\end{equation}
In the Lagrangian, to the chiral multiplets one assigns complex mass parameters identified with corresponding equivariant parameters:
		\begin{equation}
		h_I \;.
		\end{equation}
\item A framing node $f$ is associated with an ungauged flavor symmetry $U(d_f)$, and therefore the corresponding gauge degrees of freedom in the node $f$ are frozen. 
One can introduce additional complex masses $m_i$ to the flavor multiplets by giving VEVs to gauge multiplet fields:
\begin{equation}
		\langle A_f\rangle=0,\quad \langle X_f\rangle=0,\quad \langle \Phi_f\rangle={\rm diag}(m_1,m_2,\ldots, m_{d_f}) \;.
\end{equation}
\end{enumerate}
	
	Physically this situation is usually interpreted as a quantum mechanical particle motion in a target space geometry given by the quiver representation moduli space. 
	We will denote this space as
	\begin{equation}
	\fR_{\vec d,\vec\zeta}(\sQ,\sW)\,.
	\end{equation}
\bigskip

	The $\CN=4$ superalgebra is generated by four supercharges, combined in two doublets: 
\begin{equation}
\CQ_{\alpha}\,, \quad \bar{\CQ}_{\dot\alpha}\,, \qquad\textrm{with}\quad \alpha,\dot\alpha=1,2\,, 
\end{equation}
and satisfying the following relations:
\be\label{SUSY}
\begin{aligned}
		\left\{\CQ_{\alpha},\bar\CQ_{\dot\beta} \right\}&=-2\,\sigma^0_{\alpha\dot\beta}\,\CH-2i\, \sigma_{\alpha\dot\beta}^{\mu}\left(B_a^{\mu}\right)_n^m(\CG^{(a)})_m^n \;,\\
		\left\{\CQ_{\alpha},\CQ_{\beta} \right\}&=-8\sum\lm_{a,b\in Q_0}\sum\lm_{I\in \{a\to b\}}\left[\left(\sigma^{0i}\right)_{\alpha}{}^{\gamma}\,\,\epsilon_{\gamma\beta}\left(\bar q_IB_a^i-B_b^i\bar q_I\right)\right]\overline{\partial_{q_I}W} \;,
\end{aligned}
\ee
where $\CH$ and $\CG^{(a)}$ are the Hamiltonian and the Noether charge operator for the $U(d_a)$ gauge transformation, respectively; and the $4$-vector $B^{\mu}_a$ is
\begin{equation}
	B_a^{\mu=0,1,2,3}=\left(A_a,X_a,{\rm Re}\,\Phi_a,{\rm Im}\,\Phi_a\right)\,.
\end{equation}
\smallskip
	
	A bridge between localizations in physics and those in geometry is constructed when one interprets the fermion fields as differential forms on the cotangent bundle to the target space. 
	In this dictionary, the supercharge is translated into a twisted equivariant differential \cite{Galakhov:2020vyb}:
	\be\label{eq:differential}
	\CQ_1=e^{-\fH}\left(d_X+\bar{\p}_{\Phi}+\bar{\p}_{q}+\iota_V+dW\wedge\right)e^{\fH}\;,
	\ee
which is de Rham in scalar fields $X$ and Dolbeault in complex fields $\Phi$ and $q$; and the function $\fH$ is a Morse height function, whose critical locus corresponds to a canonical quiver $D$-term contribution:
	\be\label{D-term}
	\frac{\p\fH}{\p X_a}=\zeta_a{\mathds{1}}_{d_a\times d_a} +\sum\lm_{x\in Q_0}\sum\lm_{I\in\{a\to x\}}q_Iq_I^{\dagger}-\sum\lm_{y\in Q_0}\sum\lm_{J\in\{y\to a\}}q^{\dagger}_Jq_J,\quad \forall a\in Q_0\;.
	\ee
Finally, the vector field $V$ on $\fR$ is:
\begin{equation}
	V=\sum\lm_{a,b\in Q_0}\sum\lm_{I\in\{a\to b\}}\Tr\,\left(\Phi_bq_I-q_I\Phi_a-h_Iq_I\right)\frac{\p}{\p q_I}\;.
\end{equation}

\subsection{Molten crystals as fixed points}
\label{ssec:MotlenCrystalQQM}

The BPS states in this model are defined as physical ground states. 
The corresponding wave-functions are annihilated by $\CH$ and $\CG$, mapping them one-to-one to gauge-invariant harmonic forms on $\fR$. 
Another relation in the superalgebra \eqref{SUSY} dictates that for the system to have a supersymmetric ground state, the superpotential $W$ also has to be an invariant function with respect to both the gauge and the flavor symmetries. 
As a result, the values of the complex masses, i.e.\   the equivariant weights $h_I$, have to satisfy the loop constraint \eqref{eq:loop-constraint}.
	
The wave-functions of the BPS states occupy a subspace of the whole Hilbert space, called the BPS Hilbert space $\mathscr{H}_{\rm BPS}$.

Using the Hodge decomposition theorem, we identify this space with the equivariant cohomologies of the quiver representation moduli space:
\begin{equation}\label{BPSHilb}
		\mathscr{H}_{\rm BPS}\cong H^{\bullet}_{G_{\IC}}\left(\fR_{\vec d,\vec \zeta}\left(Q,W\right),\CQ_1\right)\,,
\end{equation}
\noindent	where the equivariant action is realized by a complexified gauge group:
\begin{equation}
G_{\IC}=\prod\lm_{a\in Q_0}GL(d_a,\IC)\,.
\end{equation}
	
The standard localization techniques provide us a way to calculate this space via a summation over the critical points of $\fH$ and $W$ that are fixed with respect to the equivariant $G_{\IC}$-action. 
The critical locus of $\fH$ gives rise to a canonical quiver $D$-term \eqref{D-term} constraint (i.e.\ a stability constraint)---following the guidelines of the Narasimhan-Seshadri-Hitchin-Kobayashi correspondence theorem \cite{nakajima1994instantons,Donaldson}, one concludes that the $D$-term constraint can be traded for a stability constraint. 
\bigskip
	
In what follows we will consider a specific locus on the moduli space known as a \emph{cyclic chamber}, which can be characterized by the constraint on the stability parameters:
\begin{equation}
	\zeta_a<0\,, \quad \;
	\textrm{for all non-framing nodes}\;a\in Q_0\;.
\end{equation}
In the cyclic chamber, fixed points on the quiver representation moduli space are in one-to-one correspondence with the molten crystals we describe in Section \ref{sec:subcrystals}.
	
A fixed point labeled by a molten crystal $\Kappa$ defines a classical configuration. 
The IR dynamics is described by a canonical phase with a spontaneously broken gauge symmetry. 
To describe this fixed point quiver representation explicitly, we could choose any fixed basis of our preference. 
One basis choice that makes this picture rather transparent is the one 
in which the vectors are formally assigned to atoms of the molten crystal $\Box\in\Kappa$. 
The resulting representation vector space can be split in graded components according to the colors of the atoms:
\begin{equation}
U_a:=\bigoplus \IC\,\sqbox{$a$}\;,\quad {\rm dim} \, U_a=d_a \;,\quad a\in Q_0\;.
\end{equation}
In the IR, the expectation value of $\Phi_a$ is
	\be\label{fixed_point_Phi}
	\begin{aligned}
		\langle\Phi_a\rangle&\in{\rm End}(U_a)\,, \quad \langle\Phi_a\rangle\sqbox{$b$}=\delta_{ab}\,h\left(\sqbox{$a$}\right)\cdot\sqbox{$a$} \,,
	\end{aligned}
	\ee
and that of $q_I$ is	
\be\label{fixed_point_q}
	\begin{aligned}
		\langle q_{I\in \{a\to b\}}\rangle&\in {\rm Hom}(U_a,U_b)\,,\quad \langle q_{I}\rangle\sqbox{$a$}=\left\{\begin{array}{ll}
			\sqbox{$b$}, & {\rm if}\; I\;{\rm connects}\;\sqbox{$a$}\;{\rm to}\;\sqbox{$b$}\;{\rm in}\;\Kappa \;,\\
			0, & {\rm otherwise} \;.
			\end{array}\right.
	\end{aligned}
	\ee
\bigskip
	
The effective field theory near the configuration $\Kappa$ is a theory of \emph{mesons}, where the meson fields are given by gauge-invariant polynomials of the chiral fields $q_I$. 
In our case, the Wilsonian renormalization group action is one-loop exact, due to supersymmetry.
Therefore we can expand all the polynomials in $q_I$ around their expectation values $\langle q_I\rangle$ up to the first-order quantum corrections $\delta q_I$. 
This means that in the IR, the target space of the theory is effectively the space of the linearized gauge-invariant polynomials of $\{q_I\}$:
	\be\label{lin_meson}
	\fR_{\rm IR}(\Kappa):=\left(\bigoplus\lm_{I\in Q_1}\delta q_I\right)/\,\fg_{\IC}\,,
	\ee
	where the linearized action of the gauge algebra $\fg_{\IC}$ reads:
\begin{equation}
	\delta q_{I\in\{a\to b\}}\mapsto \delta q_I+ g_b\cdot\langle q_I\rangle-\langle q_I\rangle\cdot g_a\,,\quad \textrm{with}\quad g_a\in\fg\fl(d_a,\IC)\,.
\end{equation}

	The meson fields $\mu\in\fR_{\rm IR}(\Kappa)$ have the same quantum numbers as the matrix elements of $q_I$ in the chosen basis, 
	so we could label them by indices $I$, $\sqbox{$a$}_{1}$, and $\sqbox{$b$}_2$, where $\sqbox{$a$}_{1}$, and $\sqbox{$b$}_2$
are a pair of atoms of the crystal $\Kappa$ connected by a $q_I$ matrix element (see \eqref{fixed_point_q}). 
	Such a meson field acquires an effective complex mass in the IR   --  an equivariant weight of the corresponding tangent direction to a fixed point in $\fR$:
	\be
	m_{\IC}\left(\mu_{I;\sqbox{$a$}_1,\sqbox{$b$}_2}\right)=h(\sqbox{$b$}_2)-h(\sqbox{$a$}_1)-h_I \;.
	\ee
	
	Naively, one may encounter a situation when a meson's complex mass is zero. 
	In practice, except for the extreme situations that correspond to reducible representations (which we will consider later), the naive zero meson masses are lifted from zero by superpotential higher order corrections. 
	So there are no massless degrees of freedom, and hence the configuration $\Kappa$ is an isolated fixed point and the  effective theory configuration is described by a single state whose wave function we denote as a ket-vector
	\begin{equation}
	|\Kappa\rangle \;.
	\end{equation}
	
	A geometric characteristic of a fixed point $\Kappa$ can be given in terms of an equivariant Euler class. 
	When calculating this expression we have to take into account the lifting corrections due to the superpotential to the equivariant weights. In \cite{Galakhov:2020vyb} a regularized version of the Euler class was proposed. Suppose we consider a meson vector space $\fN$ with a graded basis $\mu_i$:
\begin{equation}
	\fN:\quad\left\{\mu_i,m_{\IC,i} \right\}_{i=1}^{{\rm dim}\;\fN}\,.
\end{equation}
	For such a vector space the regularized Euler class is defined as:
\be
	{\rm Eul}(\fN):=(-1)^{\left\lfloor\sum\lm_{i:\;m_{\IC,i}=0}\frac{1}{2}\right\rfloor}\prod\lm_{i:\;m_{\IC,i}\neq0}m_{\IC,i}\,,
\ee
where $\lfloor*\rfloor$ is the floor function applied to $*$.
	
	A complete subspace of the Hilbert space containing BPS wave functions for all admissible dimension vectors $\vec d$ can be parameterized by a crystal $\CC$:
	\be\label{Hilb}
	\bigoplus\lm_{\vec d}\mathscr{H}_{{\rm BPS};\vec d,{\vec \zeta}_{\rm cyclic}}\cong \bigoplus\lm_{\Kappa\in {\bf melt}(\CC)}\IC|\Kappa\rangle \;.
	\ee
	In the next subsection we will describe a BPS algebra acting on the Hilbert space \eqref{Hilb}.

\subsection{BPS algebras}
\label{ssec:BPSalgebra}

	To construct the BPS algebra, consider two-dimensional vectors $\vec d$ and $\vec d'$ that satisfy 
\begin{equation}
	d_a'=d_a+1
\end{equation}
	for some single node $a\in Q_0$. The BPS algebra generators act as maps between quantum mechanical systems for target spaces parameterized by $\vec d$ and $\vec d'$:
\begin{equation}
	\begin{array}{c}
		\begin{tikzpicture}
			\node[left] at (0,0) {$\mathscr{H}_{\rm BPS}\left[\fR_{\vec d,\vec \zeta}\left(\sQ,\sW\right)\right]$};
			\node[right] at (2,0) {$\mathscr{H}_{\rm BPS}\left[\fR_{\vec d',\vec \zeta}\left(\sQ,\sW\right)\right]$};
			\draw[->] (0,0.07) -- (2,0.07);
			\draw[<-] (0,-0.07) -- (2,-0.07);
			\node[above] at (1,0.07) {${\bf e}^{(a)}$};
			\node[below] at (1,-0.07) {${\bf f}^{(a)}$};
		\end{tikzpicture}
	\end{array}\,.
\end{equation}
	Explicitly, these maps are given by a Fourier-Mukai (FM) transform acting on the cohomologies of the quiver representation moduli spaces, and hence the BPS Hilbert spaces, analogous to the constructions of \cite{Nakajima_book,Braverman:2016wma}. 
	The Fourier-Mukai kernel is given by the structure sheaf of an incidence locus defined as an inclusion of a sub-representation into a bigger representation:
\begin{equation}
	\fR_{\vec d,\vec \zeta}\left(\sQ,\sW\right)\subset\fR_{\vec d',\vec \zeta}\left(\sQ,\sW\right).
\end{equation}
Consider two fixed-point representations corresponding to two molten crystals $\Kappa$ and $\Kappa'$, respectively.
The representation associated with $\Kappa$  can be embedded into the one associated with $\Kappa'$ only if the crystal configuration $\Kappa$ can be embedded into $\Kappa'$, namely if $\Kappa' =\Kappa+\Box$.
	In the neighborhood of a pair of fixed points $\Kappa$ and $\Kappa+\Box$, the incidence locus constraint is also linearized and can be described by a hyperplane:
	\be
	\fS_{\rm IR}(\Kappa,\Kappa+\Box):=\left\{\fR_{\rm IR}(\Kappa)\subset \fR_{\rm IR}(\Kappa+\Box)\right\}\subset \fR_{\rm IR}(\Kappa)\oplus \fR_{\rm IR}(\Kappa+\Box)\;.
	\ee
	
	The resulting matrix elements read for the raising/lowering BPS algebra generators:\footnote{Extra corrections $\sigma_\pm(\Kappa,\Box)$ in these expressions are mere $(\pm 1)$-sign multipliers defined in \cite[equation (2.62)]{Galakhov:2020vyb}. This is a counterpart of the sign choices in \eqref{eq.efpsi-action}.}
	\be\label{eq:matrix_element_ef}
	\begin{aligned}
		{\bf e}^{(a)}|\Kappa\rangle&=\sum\lm_{\sqbox{$a$}\in{\rm Add}(\Kappa)}\sigma_+(\Kappa,\sqbox{$a$})\frac{{\rm Eul}\left(\fR_{\rm IR}(\Kappa)\right)}{{\rm Eul}\left(\fS_{\rm IR}(\Kappa,\Kappa+\sqbox{$a$})\right)}|\Kappa+\sqbox{$a$}\rangle\;,\\ 
		{\bf f}^{(a)}|\Kappa\rangle&=\sum\lm_{\sqbox{$a$}\in{\rm Rem}(\Kappa)}\sigma_-(\Kappa,\sqbox{$a$})\frac{{\rm Eul}\left(\fR_{\rm IR}(\Kappa)\right)}{{\rm Eul}\left(\fS_{\rm IR}(\Kappa-\sqbox{$a$},\Kappa)\right)}|\Kappa-\sqbox{$a$}\rangle\;.
	\end{aligned}
	\ee
	
To compare later with the quiver Yangians that are bootstrapped from the molten crystal representations, it is convenient to introduce a  \emph{spectral parameter} $z$ to the BPS algebra constructed via (\ref{eq:matrix_element_ef}).
First, notice that the gauge invariant combinations of $\Phi_a$ (\ref{fixed_point_Phi}) commute with the supercharge \eqref{eq:differential} and can be used to extend the  BPS algebra of ${\bf e}^{(a)}$ and ${\bf f}^{(a)}$, in particular, we have
\begin{equation}
{\rm Tr}\,(z-\Phi_a)^{-1} |\Kappa\rangle =\sum_{\sqbox{a}\in \Kappa} \frac{1}{z-h(\sqbox{a})} |\Kappa\rangle\,.
\end{equation}	
Then we define the raising and lowering operators with spectral parameter $z$ as	\cite{Galakhov:2020vyb} 
	\be
	{\bf e}^{(a)}(z)\equiv\left[{\rm Tr}(z-\Phi_a)^{-1},{\bf e}^{(a)}\right] \quad \textrm{and} \quad {\bf f}^{(a)}(z)\equiv-\left[{\rm Tr}(z-\Phi_a)^{-1},{\bf f}^{(a)}\right]\,,
	\ee
whose actions on the crystal state are
	\be\label{eq:efpsi-rep-QM}
	\begin{aligned}
		{\bf e}^{(a)}(z) |\Kappa\rangle&=\sum\lm_{\sqbox{$a$}\in{\rm Add}(\Kappa)}\frac{\sigma_+(\Kappa,\sqbox{$a$})}{z-h(\sqbox{$a$})}\times\frac{{\rm Eul}\left(\fR_{\rm IR}(\Kappa)\right)}{{\rm Eul}\left(\fS_{\rm IR}(\Kappa,\Kappa+\sqbox{$a$})\right)}|\Kappa+\sqbox{$a$}\rangle\,,\\
		 {\bf f}^{(a)}(z)|\Kappa\rangle&=\sum\lm_{\sqbox{$a$}\in{\rm Rem}(\Kappa)}\frac{\sigma_-(\Kappa,\sqbox{$a$})}{z-h(\sqbox{$a$})}\times\frac{{\rm Eul}\left(\fR_{\rm IR}(\Kappa)\right)}{{\rm Eul}\left(\fS_{\rm IR}(\Kappa-\sqbox{$a$},\Kappa)\right)}|\Kappa-\sqbox{$a$}\rangle\,.
    \end{aligned}
	\ee
	
In addition, we can define Cartan operators
	\begin{equation}
	    \boldsymbol{\psi}^{(a)}(z)={}^{\sharp}\psi_0^{(a)}(z)\exp\left[\sum\lm_{b\in\sQ_0}\Tr\;\log\;\varphi^{b\Rightarrow a}(z-\Phi_b)\right]\;,
	\end{equation}
where functions $\varphi$ are bond factors associated with the framed quiver $\sQ$ we defined in \eqref{eq-charge-atob},\footnote{While the expression \eqref{eq-charge-atob} was originally meant for the quiver $Q$ in Section \ref{sec:SQY}, here we apply the same formula to the framed quiver $\sQ$.} and the vacuum charge function ${}^{\sharp}\psi_0^{(a)}(z)$ can also be treated in this picture as a bond factor contribution from ``frozen" degrees of freedom corresponding to the framing node $f$:
\begin{equation}
    {}^{\sharp}\psi_0^{(a)}(z)=\varphi^{f\Rightarrow a}(z)\;.
\end{equation}

A crystal basis $|\Kappa\rangle$ is an eigen-basis of operators $\boldsymbol{\psi}^{(a)}(z)$ with eigenvalues
	\begin{equation}
	    \boldsymbol{\psi}^{(a)}(z)|\Kappa\rangle=\Psi_{\Kappa}^{(a)}(z)|\Kappa\rangle \;.
	\end{equation}
	
	\bigskip
	
The generators 
	\begin{equation}
	{\bf e}^{(a)}(z),\; {\bf f}^{(a)}(z),\; \boldsymbol{\psi}^{(a)}(z)
	\end{equation}
	define the BPS algebra acting on the BPS Hilbert space as a crystal representation. 
\bigskip
	
We should comment that although the expressions for the representations of the shifted quiver Yangian \eqref{eq:efpsi-rep-QM} look different from the bootstrapped expressions \eqref{eq.efpsi-action}, the two representations are isomorphic.
The isomorphism is given by a simple re-scaling of the norms of the states in the representation.
In the bootstrapped representation, the generators ${\bf e}^{(a)}(z)$ and ${\bf f}^{(a)}(z)$ are conjugate with respect to the following norm:
	\begin{equation}
	    \langle \Kappa,\Kappa'\rangle_{\mbox{\tiny bootstrap}}=\delta_{\Kappa,\Kappa'}\;,
	\end{equation}
	whereas the norm associated with the quantum mechanics reads:
	\begin{equation}
	    \langle \Kappa,\Kappa'\rangle_{\mbox{\tiny QM}}={\rm Eul}\left(\fR_{\rm IR}(\Kappa)\right)\;\delta_{\Kappa,\Kappa'} \;.
	\end{equation}
\smallskip
	
We have computed the relations in the BPS algebra acting on different vectors of all the crystal representations listed in the paper. 
	A conclusion supported by diverse numerical evaluations is the following:
	\begin{tcolorbox}[ams align]\label{correspondence}
		\begin{aligned}
			\mbox{BPS Algebra }(\sQ,\sW)\cong \mathsf{Y} (Q,W,\mys)\,,
		\end{aligned}
	\end{tcolorbox}
	\noindent as long as the representations are \emph{irreducible}. 
The shifts $\mys$ in \eqref{correspondence} are defined in a canonical way \eqref{eq:psi0_summary}.

The obstacle to extending this claim to all representations including \emph{reducible} ones is not insurmountable. 
The problem occurs when we are trying to move across a boundary between irreducible components inside a reducible representation. 
Even though the crystals at this boundary are related by embedding:
say, $\Kappa \subset \Kappa+\sqbox{$a$}$, 
the corresponding matrix element is however null:
\begin{equation}
	\langle\Kappa+\sqbox{$a$}\,|\,e^{(a)}(z)\,|\,\Kappa\rangle=0 \:.
\end{equation}
	On the quantum mechanics side, in this situation we hit an IR singularity, which we will discuss in the next subsection. 
	The corresponding matrix element for BPS algebra generator ${\bf e}^{(a)}(z)$ is not well-defined. 
	However if one re-defines ill-defined matrix elements of the BPS algebra to be zero as in \eqref{eq.efpsi-action} we conjecture that the equivalence \eqref{correspondence} can be extended to \emph{all} representations.

\subsection{Reducible representations and IR singularities}
\label{ssec:reducible}

The IR singularities we mentioned in the previous section are standard singularities for low-energy effective actions. 
The masses of the effective particles are functions of initial and renormalization group parameters and are generically unconstrained. 
The parameter space may contain specific loci where the masses of some particles become zero. 
An occasionally massless particle contributes as a resonance pole to the scattering matrices at low energies, and thus it can create a  condensate and produce a singularity in the effective action. 
In this subsection we would like to present our arguments that the reducibility of a BPS algebra representation is closely related to the appearance of massless particles in the low energy spectrum, spoiling isolated crystal vacua.

\subsubsection{\texorpdfstring{Example: $\mathbb{C}^3$}{Example: C(3)}}
Let us first illustrate this phenomenon using the simplest example of a $\IC^3$ framed quiver, with a standard superpotential:
\begin{equation}
\begin{aligned}
\sQ&=\begin{array}{c}
		\begin{tikzpicture}
			\begin{scope}[rotate=-90]
			\draw[thick,->] ([shift=(120:0.75)]0,0.75) arc (120:420:0.75);
			\draw[thick,->] ([shift=(60:0.75)]0,0.75) arc (60:90:0.75);
			\draw[thick,->] ([shift=(90:0.75)]0,0.75) arc (90:120:0.75);
			\draw[fill=white] (0,0) circle (0.2);
			\draw[thick,->] (0.1,-1.35) -- (0.1,-0.75);
			\draw[thick] (0.1,-0.75)  -- (0.1,-0.173205);
			\draw[thick] (-0.1,-1.35) --  (-0.1,-0.75);
			\draw[thick,<-] (-0.1,-0.75) -- (-0.1,-0.173205);
			\begin{scope}[shift={(0,-1.5)}]
				\draw[thick,fill=red] (-0.15,-0.15) -- (-0.15,0.15) -- (0.15,0.15) -- (0.15,-0.15) -- cycle;
			\end{scope}
			\node[right] at (0,1.5) {$B_{1,2,3}$};
			\node[below] at (0.15,-0.75) {$R$};
			\node[above] at (-0.15,-0.75) {$S$};
			\end{scope}
		\end{tikzpicture}
	\end{array},\quad 
	\\
\sW&=\Tr\,B_1[B_2,B_3]\,.
\end{aligned}
\end{equation}
Compared with the canonically framed quiver $(Q,W)_0$ for $\mathbb{C}^3$, the framed quiver $(\sQ,\sW)$ has an additional arrow going from the gauged node back to the framing node $\square$, and note that the superpotential $W$ is unchanged by this additional arrow.
Namely, there are two arrows $R$ and $S$ connecting the framing node $\square$  with the gauge node. 
The superpotential is independent of both $R$ and $S$, therefore the equivariant weights of these fields are unconstrained. 
	The ground state charge function in this case reads:
\begin{equation}
	{}^{\sharp}\psi_0(z)=\frac{z+h_S}{z-h_R} \;.
\end{equation}
The crystal ${}^{\sharp}\CC$ corresponding to $(\sQ,\sW)$ is a $\IC^3$ canonical crystal $\CC_0$ filling an octant in the 3D space. 
\bigskip
	
	The reducibility of this representation depends on the value of the complex parameter $h_R+h_S$. 
Consider two molten crystal configurations: an empty one $|\varnothing\rangle$ and one containing a single atom $|\square\rangle$. 
A simple calculation \eqref{eq:matrix_element_ef} gives the matrix element of the raising operator $\bf e$: 
\begin{equation}
	\langle\square\,|\,{\bf e}\,|\,\varnothing\rangle^2=h_R+h_S \;.
\end{equation}

When $h_R+h_S$ is a generic complex number, not belonging to the 2D crystal lattice, the representation is irreducible. 
However, if $h_R+h_S=0$, the representation becomes reducible, and moreover, $|\varnothing \rangle$ is a one-dimensional irreducible component. 

On the other hand, in the configuration associated with $|\square \rangle$ we have the following expectation values of fields:
\begin{equation}
	\langle B_i\rangle=0 \;, \quad  \langle R\rangle=1 \;,\quad \langle S\rangle=0 \;,\quad \langle\Phi\rangle=h_R\;,
\end{equation}
so that the quantum correction $\delta R$ for the condensed field $R$ is massless, fulfilling the condensation constraint:
\begin{equation}
    m_{\IC}(\delta R)=\langle\Phi\rangle-h_R=0\;.
\end{equation}
	A quantum correction $\delta S$ to the chiral field $S$ parameterizes an effective meson field, with an effective chiral mass given by the following expression:
	\begin{equation}
	m_{\IC}\left(\delta S\right)=-\langle\Phi\rangle-h_S=-(h_R+h_S)=	-\langle\square\,|\,{\bf e}\,|\,\varnothing\rangle^2 \;.
	\end{equation}
Therefore we see that the situation when the representation becomes reducible and the vectors $|\varnothing\rangle$ and $|\square\rangle$ belong to two disjoint irreducible components corresponds to an IR singularity due to an effective massless meson particle.

\subsubsection{General cases}	
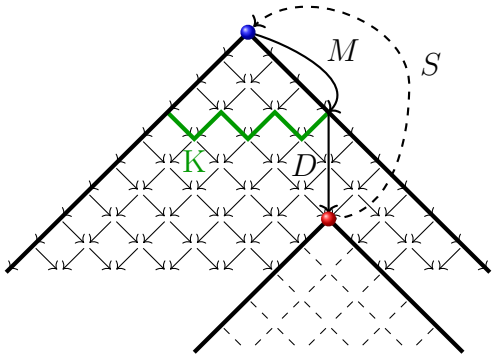
\begin{figure}[ht!]
		\begin{center}
			\begin{tikzpicture}
				\begin{scope}[rotate=-135]
					\begin{scope}[scale=0.5]
						\foreach \x in {0,...,8}
						\foreach \y in {0,...,8}
						{
							\pgfmathparse{int(\x+\y-7)}
							\let\r\pgfmathresult
							\ifnum \r > 0
							\breakforeach
							\fi
							\pgfmathparse{int(10*(\x+\y))}
						    \draw[->](\x+0.1,\y)  -- (\x+0.9,\y);
							\draw[->](\x,\y+0.1)  -- (\x,\y+0.9);
						}
						\draw[ultra thick, black!40!green] (3,0) -- (3,1) -- (2,1) -- (2,2) -- (1,2) -- (1,3) -- (0,3); 
						\draw[ultra thick] (0,9) -- (0,0) -- (9,0);
						\node[below, black!40!green] at (3,1) {$\Kappa$};
						\begin{scope}[shift={(2,5)}]
						    \draw[white,fill=white] (4,0) -- (0,0) -- (0,4) -- cycle;
							\foreach \x in {0,...,4}
							\foreach \y in {0,...,4}
							{
								\pgfmathparse{int(\x+\y-3)}
								\let\r\pgfmathresult
								\ifnum \r > 0
								\breakforeach
								\fi
							\draw[dashed] (\x+0.1,\y)  -- (\x+0.9,\y);
							\draw[dashed] (\x,\y+0.1)  -- (\x,\y+0.9);
							}
							\draw[ultra thick] (0,5) -- (0,0) -- (5,0);
						\end{scope}
						\draw[->,thick] (0,0) to[out=120,in=180] (0,3);
						\draw[thick,->] (0,3) -- (1.9,4.9);
						\draw[thick, dashed,->] (2,5) to[out=135,in=45] (-2,4) to[out=225,in=180] (-0.2,0); 
						\shade[ball color = blue] (0,0) circle (0.2);
						\shade[ball color = red] (2,5) circle (0.2);
						\node[above right] at (-0.5,2) {$M$};
						\node[left] at (1,4) {$D$};
						\node[above right] at (-2,4) {$S$};
					\end{scope}
				\end{scope}
			\end{tikzpicture}
			\caption{Schematized crystal-version depiction for a reducible representation.}\label{fig:depth1}
		\end{center}
	\end{figure}

	The simple logic presented earlier can be extended to the general situation when a negative crystal is placed deeper than the boundaries of the positive crystals (see Figure~\ref{fig:depth1}). In particular, we will argue that this situation leads to an effectively new massless particle in the IR.
	
	Consider a path connecting the leading atom of a positive crystal (the atom with the red dot in Figure~\ref{fig:depth1}) to that of a negative crystal (the blue dot in Figure~\ref{fig:depth1}). 
Decompose this path into two segments: (1) the path going along the positive crystal boundary denoted as $M$ and (2) the path going inside the crystal denoted as $D$. 
Denote the chiral field associated to the arrow going from the gauge node back to the framing node, and corresponding to the head atom of the negative crystal, as $S$. 
As discussed in Section \ref{ssec:framedQW}, the correction to the superpotential due to the arrow $S$ takes the following form:
\begin{equation}
	\delta W=\Tr\,S\cdot D \cdot M\;.
\end{equation}
	This model can describe both the scenarios when the negative crystal is placed at the  boundary of the positive crystal and the one when it is deep in the interior of the positive crystal. 
In the former case map $D$ is just given by the unit in the quiver path algebra.
	
	We start to grow a crystal $\Kappa$ as usual until it reaches an atom located at $M$. 
For such a crystal, the field $M$ condenses and acquires an expectation value
	\begin{equation}
	\langle M\rangle\neq 0 \;,
	\end{equation}
	therefore its complex mass $m_{\IC}(M)=0$. 
After the spontaneous symmetry breaking in the IR, the quantum degrees of freedom $\delta M$ acquire a mass through the Higgs mechanism. 
Equivalently we may say it is deleted from the IR meson space \eqref{lin_meson} by the $\fg_{\IC}$-action. 
Notice that $m_{\IC}(D)=0$ since it does not displace the atom location projected to the 2D $\{\mathsf{h}_1,\mathsf{h}_2\}$-space. From the loop constraint \eqref{eq:loop-constraint} on the superpotential we derive:
	\begin{equation}
	m_{\IC}(S)=-m_{\IC}(D)-m_{\IC}(M)=0\,.
	\end{equation}
	The superpotential contribution corrects this zero mass by:
\begin{equation}\label{deltam}
	|\delta m_{\IC}(S)|^2\sim\left|\frac{\p\delta W}{\p S}\right|^2\sim\left|\langle D\rangle\right|^2\cdot \left|\langle M\rangle\right|^2\sim \left|\langle D\rangle\right|^2\,,
\end{equation}
where in the last correspondence in~\eqref{deltam} we have used the fact that $M$ has a non-zero VEV. 

Now consider the two scenarios separately.
In the first scenario where the negative crystal is located at the boundary of the positive crystal and hence $D$ is just a unit in the quiver path algebra, one can substitute $D$ by just a complex number, so its expectation value is non-zero, and from~\eqref{deltam} the field $S$ has a non-zero effective mass in the IR. 
In the second scenario where the negative crystal is placed deeper inside the positive crystal, according to the melting rule discussed in Section \ref{ssec:canon_cry} we can always pick a molten crystal configuration $\Kappa$ such that it does not contain the atom $D\cdot M$. 
In this case $\langle D\rangle=0$,  and the field $S$ is effectively massless in the IR.

\section{Wall-crossing}
\label{sec:wallcrossing}

The BPS states are known to be affected by the wall-crossing phenomena \cite{Denef:2007vg,Kontsevich:2008fj,Aganagic:2009kf,Gaiotto:2008cd}:
the spectra of the BPS states have a piece-wise dependence on the value of K\"ahler moduli, such that the K\"ahler moduli space is divided into chambers separated by the marginal stability walls. 
A BPS spectrum in a given chamber is well-defined and jumps across the walls. 
This aspect brings certain difficulties to the construction of the BPS algebra outside the cyclic chamber. 

Although there exist various computational tools in the literature that allow one to describe the spectrum and the corresponding invariants for a generic point of the K\"ahler moduli space, a generic structure of the BPS spectrum in an arbitrary chamber remains obscure. 
Generically, outside the cyclic chamber we have not been able to assign some tractable enumerative combinatorics theory to the fixed points in a way similar to how we enumerate the fixed points by the molten crystals in the cyclic chamber. 
This phenomenon occurs even for the simplest toric Calabi-Yau threefolds -- although the fixed point combinatorics is still traceable outside the cyclic chamber, we lose the 3D crystal structure there, hindering the construction of the BPS algebra based on the crystal melting picture.

Therefore in the following we will focus on those cases where the BPS states are still enumerated by 3D crystals even after wall-crossings.

\subsection{Example: wall-crossing in conifold}

The first class of the wall-crossing examples is spanned by quiver mutations, also known  to the physics community as Seiberg dualities \cite{Benini:2014mia}.

A mutation identifies the representation moduli spaces for two quivers with different values of FI parameters and dimensions, in a manner that a cyclic chamber of one quiver may be mapped to a non-cyclic chamber of the other one.

The simplest example of the mutation is captured by the wall-crossing phenomenon in the conifold $\mathcal{O}(-1)\times \mathcal{O}(-1)\rightarrow \mathbb{P}^1$, whose toric diagram and its dual graph are 
\begin{equation}\label{fig-toric-conifold}
\begin{tikzpicture} 
\filldraw [red] (0,0) circle (2pt); 
\filldraw [red] (0,1) circle (2pt); 
\filldraw [red] (1,1) circle (2pt); 
\filldraw [red] (1,0) circle (2pt); 
\node at (-.5,-.5) {(0,0)}; 
\node at (-.5,1.5) {(0,1)}; 
\node at (1.5,1.5) {(1,1)}; 
\node at (1.5,-0.5) {(1,0)}; 
\draw (0,0) -- (0,1); 
\draw (0,1) -- (1,1); 
\draw (0,0) -- (1,0); 
\draw (1,0) -- (1,1); 
\end{tikzpicture}
\qquad \qquad \qquad
\begin{tikzpicture}[scale=0.6] 
\draw[->] (0,0) -- (-1,0); 
\draw[->] (0,0) -- (0,-1); 
\draw (0,0) -- (1,1); 
\draw[->] (1,1) -- (2,1); 
\draw[->] (1,1) -- (1,2); 
\node at (-1.5,0) {3}; 
\node at (2.5,1) {$\hat{3}$}; 
\node at (0,-1.5) {1}; 
\node at (1,2.5) {$\hat{1}$}; 
\end{tikzpicture}
\end{equation}

\subsubsection{\texorpdfstring{Canonical crystal and affine Yangian of $\mathfrak{gl}_{1|1}$}{Canonical crystal and affine Yangian of gl(1|1)}}

The canonically framed quiver-superpotential pair $(Q_0,W_0)$ is	
\begin{equation}\label{eq:conifold_QW0}
\begin{aligned}
Q_0&=\begin{array}{c}
	\begin{tikzpicture}
		\draw[thick] ([shift=(70:2)]1,-1.73205) arc (70:60:2);
		\draw[thick,->] ([shift=(80:2)]1,-1.73205) arc (80:70:2);
		\draw[thick,->] ([shift=(120:2)]1,-1.73205) arc (120:80:2);
		\begin{scope}[yscale=-1]
			\draw[thick] ([shift=(120:2)]1,-1.73205) arc (120:110:2);
			\draw[thick,<-] ([shift=(110:2)]1,-1.73205) arc (110:100:2);
			\draw[thick,<-] ([shift=(100:2)]1,-1.73205) arc (100:60:2);
		\end{scope}
		\draw[fill=white] (0,0) circle (0.2);
		\draw[fill=gray] (2,0) circle (0.2);
		\draw[thick,->] (-1.5,0) -- (-0.2,0);
		\begin{scope}[shift={(-1.5,0)}]
			\draw[fill=red] (-0.15,-0.15) -- (-0.15,0.15) -- (0.15,0.15) -- (0.15,-0.15) -- cycle;
		\end{scope}
		\node[above] at (-0.75,0) {$\iota$};
		\node[above] at (1,0.267949) {$a_1$, $a_2$};
		\node[below] at (1,-0.267949) {$b_1$, $b_2$};
		\node[below] at (-1.5,-0.2) {$\infty$};
		\node[below] at (0,-0.2) {$1$};
		\node[below] at (2,-0.2) {$2$};
	\end{tikzpicture}
\end{array}  \;, \\ 
W_0&=\Tr\;\left[b_2a_2b_1a_1-b_2a_1b_1a_2\right]\,.
\end{aligned}
\end{equation}
The corresponding periodic quiver on a torus, which can be redrawn as an infinite square lattice on a plane universally covering the torus, is
	\be
	\begin{array}{c}
		\begin{tikzpicture}
			\draw[fill=white] (0,0) circle (0.2) (1.5,1.5) circle (0.2) (1.5,-1.5) circle (0.2) (-1.5,1.5) circle (0.2) (-1.5,-1.5) circle (0.2);
			\draw[fill=white!20!gray] (-1.5,0) circle (0.2) (1.5,0) circle (0.2) (0,1.5) circle (0.2) (0,-1.5) circle (0.2);
			\node at (0,0) {$1$};
			\node at (1.5,1.5) {$1$};
			\node at (-1.5,1.5) {$1$};
			\node at (1.5,-1.5) {$1$};
			\node at (-1.5,-1.5) {$1$};
			\node at (-1.5,0) {$2$};
			\node at (1.5,0) {$2$};
			\node at (0,1.5) {$2$};
			\node at (0,-1.5) {$2$};
			\draw[->] (-0.2,0) -- (-1.3,0);
			\node[above] at (-0.75,0) {$a_1$};
			\draw[->] (0.2,0) -- (1.3,0);
			\node[below] at (0.75,0) {$a_2$};
			\draw[->] (1.3,1.5) -- (0.2,1.5);
			\node[above] at (0.75,1.5) {$a_1$};
			\draw[->] (1.3,-1.5) -- (0.2,-1.5);
			\node[below] at (0.75,-1.5) {$a_1$};
			\draw[->] (-1.3,1.5) -- (-0.2,1.5);
			\node[above] at (-0.75,1.5) {$a_2$};
			\draw[->] (-1.3,-1.5) -- (-0.2,-1.5);
			\node[below] at (-0.75,-1.5) {$a_2$};
			\draw[->] (-1.5,0.2) -- (-1.5,1.3);
			\node[left] at (-1.5,0.75) {$b_2$};
			\draw[->] (1.5,0.2) -- (1.5,1.3);
			\node[right] at (1.5,0.75) {$b_2$};
			\draw[->] (0,-1.3) -- (0,-0.2);
			\node[left] at (0,-0.75) {$b_2$};
			\draw[->] (0,1.3) -- (0,0.2);
			\node[right] at (0,0.75) {$b_1$};
			\draw[->] (-1.5,-0.2) -- (-1.5,-1.3);
			\node[left] at (-1.5,-0.75) {$b_1$};
			\draw[->] (1.5,-0.2) -- (1.5,-1.3);
			\node[right] at (1.5,-0.75) {$b_1$};
		\end{tikzpicture}
	\end{array}
	\ee

\begin{figure}[ht!]
    \centering
    \begin{tikzpicture}
    \node at (0,0) {\includegraphics[scale=0.2]{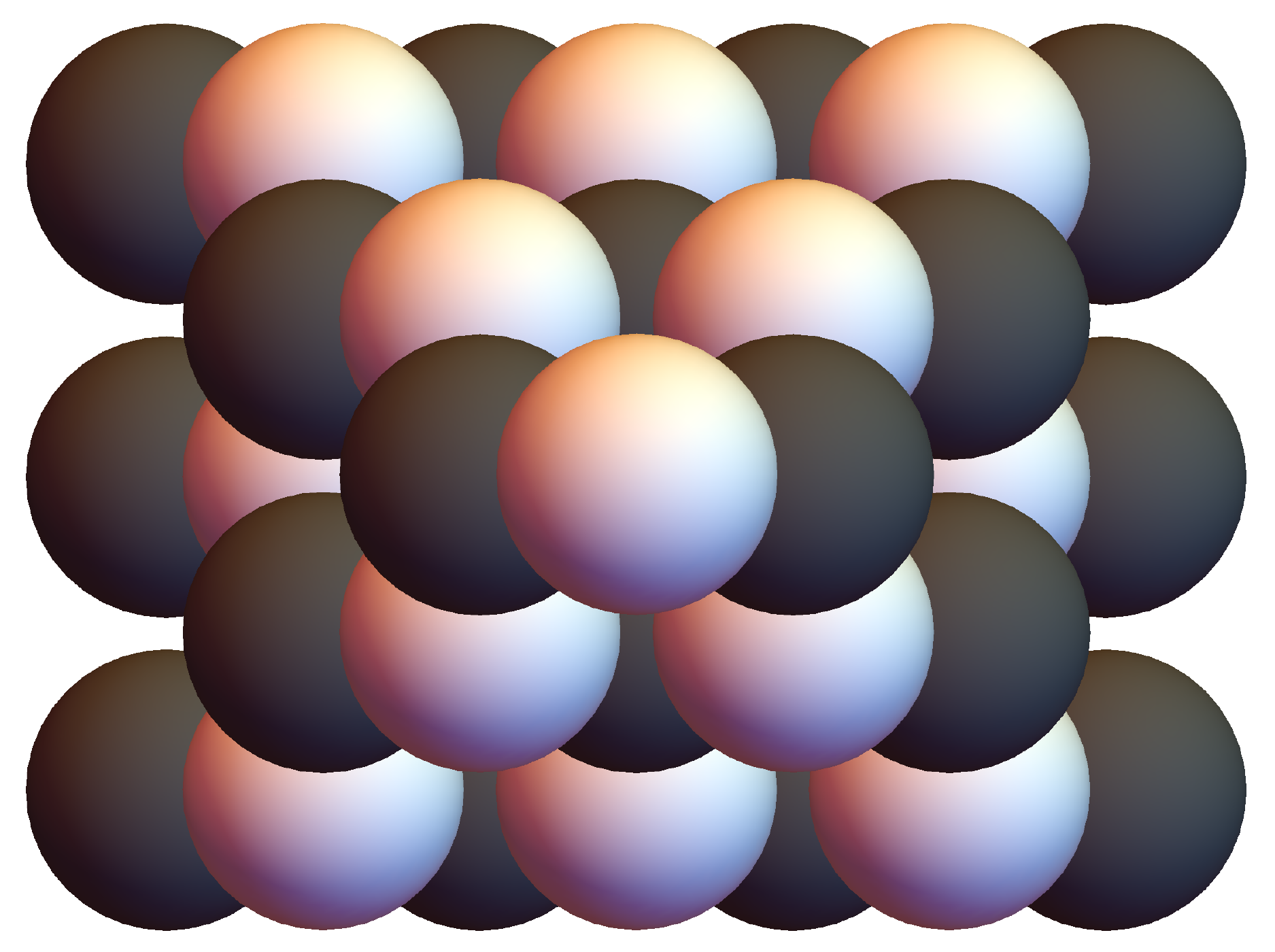}};
    \draw[ultra thick] (0,0) -- (1.5,1.5) (0,0) -- (-1.5,1.5) (0,0) -- (-1.5,-1.5) (0,0) -- (1.5,-1.5);
    \draw[ultra thick,dotted] (1.5,1.5) -- (2,2) (-1.5,1.5) -- (-2,2) (-1.5,-1.5) -- (-2,-2) (1.5,-1.5) -- (2,-2);  
    \end{tikzpicture}
    \caption{The canonical crystal for the resolved conifold (called pyramid partition). It corresponds to the so-called non-commutative DT chamber \cite{Szendroi}, i.e. $m=1$.}
    \label{fig:pyramid}
\end{figure}

The canonical crystal in the case of the conifold is a lift to 3D of the square lattice in a physical 2D plane of weights parameterized by $\mathsf{h}_1$ and $\mathsf{h}_2$. 
The third direction is parameterized by the depth parameter $\bf d$. 
White atoms corresponding to quiver vertex $1$ are filling even levels in $\bf d$, while black atoms corresponding to quiver vertex $2$ are filling odd levels in $\bf d$. 
So we can assign to lattice edges $a_1$, $a_2$, $b_1$, $b_2$ the following 3D lifted vectors (see Figure~\ref{fig:pyramid}): 
	\be\label{eq-basisvector-conifold}
	\vec a_1=(-1,0,1) \;,\; \vec b_1=(0,-1,1)\;,\;
		\vec a_2=(1,0,1)\;,\; \vec b_2=(0,1,1)\;.
	\ee
The loop constraint \eqref{eq:loop-constraint} translates to
\begin{equation}
\mu(a_1)+\mu(a_2)+\mu(b_1)+\mu(b_2)=0\;, 
\end{equation}
and the vertex constraint~\eqref{eq-vertex-constraint-toric} translates to 
\begin{equation}
\mu(a_1)+\mu(a_2)=\mu(b_1)+\mu(b_2)\,.
\end{equation}
After imposing both the loop and vertex constraints, we have
\be\label{m1_conif_weights}
\begin{aligned}
		&\mu(a_1)=-\mathsf{h}_1\,,\quad \mu(a_2)=\mathsf{h}_1\,,\quad \mu(b_1)=-\mathsf{h}_2\,,\quad \mu(b_2)=\mathsf{h}_2\,.\\ 
	\end{aligned}
	\ee

One can then immediately read off the bond factors from the periodic quiver (\ref{tilling-conifold}) by the definition (\ref{eq-charge-atob})
\begin{equation}\label{eq-charge-function-conifold}
 \varphi^{a\Rightarrow a}(u)=1\,,\qquad \varphi^{a\Rightarrow a+1}(u)
 =\frac{(u+\mathsf{h}_{a+1})(u-\mathsf{h}_{a+1})}{(u+\mathsf{h}_{a})(u-\mathsf{h}_{a})}  \;,
\end{equation}
where the indices are understood as mod $2$. 
Accordingly, the resulting algebra is
\begin{equation}\label{eq-algebra-shifted-conifold}
\begin{aligned}
&\textrm{OPE:}\quad
\begin{cases}
\begin{aligned}
\psi^{(a)}(z)\, \psi^{(b)}(w)&\sim \psi^{(b)}(w)\, \psi^{(a)}(z) \;,\\
\psi^{(a)}(z)\, e^{(a)}(w)   &\sim   e^{(a)}(w)\, \psi^{(a)}(z)  \;,\\ 
e^{(a)}(z)\, e^{(a)}(w) & \sim  - e^{(a)}(w)\, e^{(a)}(z)  \;,\\
\psi^{(a)}(z)\, f^{(a)}(w) &  \sim  f^{(a)}(w)\, \psi^{(a)}(z)  \;,\\
 f^{(a)}(z)\, f^{(a)}(w) &  \sim  - f^{(a)}(w)\, f^{(a)}(z) \;, \\
 \psi^{(a+1)}(z)\, e^{(a)}(w)   &\sim \tfrac{(\Delta+\mathsf{h}_{a+1})(\Delta-\mathsf{h}_{a+1})}{(\Delta+\mathsf{h}_{a})(\Delta-\mathsf{h}_{a})}  \, e^{(a)}(w)\, \psi^{(a+1)}(z)  \;,\\ 
e^{(a+1)}(z)\, e^{(a)}(w) &\sim - \tfrac{(\Delta+\mathsf{h}_{a+1})(\Delta-\mathsf{h}_{a+1})}{(\Delta+\mathsf{h}_{a})(\Delta-\mathsf{h}_{a})}  \, e^{(a)}(w)\, e^{(a+1)}(z)  \;,\\
\psi^{(a+1)}(z)\, f^{(a)}(w) &  \sim \tfrac{(\Delta+\mathsf{h}_{a})(\Delta-\mathsf{h}_{a})}{(\Delta+\mathsf{h}_{a+1})(\Delta-\mathsf{h}_{a+1})}   \, f^{(a)}(w)\, \psi^{(a+1)}(z)  \;,\\
 f^{(a+1)}(z)\, f^{(a)}(w) &  \sim - \tfrac{(\Delta+\mathsf{h}_{a})(\Delta-\mathsf{h}_{a})}{(\Delta+\mathsf{h}_{a+1})(\Delta-\mathsf{h}_{a+1})} \,f^{(a)}(w)\, f^{(a+1)}(z) \;,\\
\{e^{(a)}(z)\,, f^{(b)}(w)\}   &=  - \delta^{a,b}\, \frac{\psi^{(a)}(z) - \psi^{(a)}(w)}{z-w}  \;,
\end{aligned}
\end{cases}
\end{aligned}
\end{equation}
where throughout this paper $\Delta$ is defined as 
\begin{equation} 
\Delta\equiv z-w\,.
\end{equation} 
The initial conditions are
\begin{equation}\label{eq-algebra-shifted-conifold-re}
\begin{aligned}
&\textcolor{black}{\textrm{Initial:}}\quad\begin{cases}
\begin{aligned}
&\begin{aligned}
&[\psi^{(a)}_0,e^{(a)}_m] =[\psi^{(a)}_1,e^{(a)}_m] =[\psi^{(a)}_0,f^{(a)}_m] =[\psi^{(a)}_1,f^{(a)}_m] =0\,,\\
& [\psi^{(a+1)}_0,e^{(a)}_m] =0 \;,\\
& [\psi^{(a+1)}_0,f^{(a)}_m] =0 \;, \\
&[\psi^{(a+1)}_1,e^{(a)}_m] = (-1)^a\, (\mathsf{h}_2^2-\mathsf{h}_1^2)\,e^{(a)}_m \;,\\
&[\psi^{(a+1)}_1,f^{(a)}_m] = -  (-1)^a\,(\mathsf{h}_2^2-\mathsf{h}_1^2)\,f^{(a)}_m \;.\\
\end{aligned}
\end{aligned}
\end{cases}
\end{aligned}
\end{equation}
We can check that $\psi^{(0)}_0+\psi^{(1)}_0$ is a central term.
To complete the definition, we supplement these relations with the Serre relation
\begin{equation}
\textrm{Serre}:\quad\begin{cases}\begin{aligned}
&\textrm{Sym}_{z_1,z_2}\, \left\{ e^{(a)}(z_1)\,, \left[  e^{(a+1)}(w_1)\,, \left\{  e^{(a)}(z_2)\,,  e^{(a+1)}(w_2)\right\}\right]\right\} \sim 0  \;,\\
&\textrm{Sym}_{z_1,z_2}\, \left\{ f^{(a)}(z_1)\,, \left[  f^{(a+1)}(w_1)\,, \left\{  f^{(a)}(z_2)\,,  f^{(a+1)}(w_2)\right\} \right]\right\} \sim 0  \;,\\
\end{aligned}
\end{cases}
\end{equation}

\subsubsection{Chamber structure}
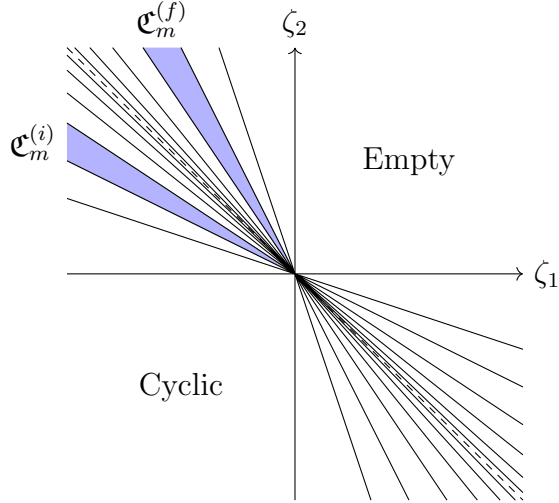
\begin{figure}[ht!]
	\begin{center}
		\begin{tikzpicture}
			\draw[->] (0,-3) -- (0,3);
			\draw[->] (-3,0) -- (3,0);
			\node[right] at (3,0) {$\zeta_{1}$};
			\node[above] at (0,3) {$\zeta_{2}$};
			\draw[dashed] (-3,3) -- (3,-3);
			\node at (-1.5,-1.5) {Cyclic};
			\node at (1.5,1.5) {Empty};
			\draw[fill=blue,opacity=0.3] (-3,1.5) -- (0,0) -- (-3,2);
			\draw[fill=blue,opacity=0.3] (-1.5,3) -- (0,0) -- (-2,3);
			\draw (-3,1) -- (3,-1) (-1,3) -- (1,-3) (-3,1.5) -- (3,-1.5) (-3,2) -- (3,-2) (-1.5,3) -- (1.5,-3)  (-2,3) -- (2,-3) (-3,2.4) -- (3,-2.4) (-2.4,3) -- (2.4,-3) (-3,2.7) -- (3,-2.7) (-2.7,3) -- (2.7,-3) (-3,2.9) -- (3,-2.9) (-2.9,3) -- (2.9,-3);
			\node[left] at (-3,1.75) {$\fC_m^{(i)}$};
			\node[above] at (-1.75,3) {$\fC_m^{(f)}$};
		\end{tikzpicture}
	\caption{The stability chambers of the conifold. } \label{fig:conifold}
	\end{center}
\end{figure}

We study the chamber structure of moduli space associated with the conifold following \cite{MR2836398}.
The marginal stability walls and the chambers in the K\"ahler moduli space are depicted in Figure~\ref{fig:conifold}. 
The entire positive quadrant is occupied by an ``empty" chamber in which there are no supersymmetric BPS states in the spectrum, except for the trivial vacuum where all the quiver dimensions are zero. 
The negative quadrant is filled by the cyclic chamber $\fC_{\rm cyc}$. 

In the two remaining quadrants, there are an infinite number of chambers, labeled by integers and accumulating towards the line 
\begin{equation}
\zeta_1+\zeta_2=0\,.
\end{equation} 
These chambers can be divided into two classes: infinite chambers $\fC^{(i)}_m$ and finite ones $\fC^{(f)}_m$, both labeled by positive integer $m\in\mathbb{N}$.\footnote{This identification will become clear in what follows.}
They are given by their boundaries:
\begin{equation}\label{eq:conif_chamb_FI}
\begin{array}{lllll}
	\fC^{(i)}_m:\quad &\; &(m-1)\, \zeta_{1}+m\, \zeta_{2}>0\,,\quad &\; &m\, \zeta_{1}+(m+1)\, \zeta_{2}<0\,,\\
	\fC^{(f)}_m:\quad &\; &m\, \zeta_{1}+(m-1)\, \zeta_{2}<0\,,\quad &\; &(m+1)\,\zeta_{1}+m\, \zeta_{2}>0\,.
\end{array}
\end{equation}

Let us define linear transformations of dimensions:
\begin{equation}\label{eq:conif_chamb_dim}
\begin{split}
	& \vec d_m^{(i)}=\left(m d_{1}+(1-m)d_{2}\,, \, (1+m)d_{1}-m d_{2}\right)\,,\\
	& \vec d_m^{(f)}=\left(-(m+1)d_{1}+m d_{2}\,, \, -m d_{1}+(m-1)d_{2}\right)\,.\\
\end{split}
\end{equation}
Then we have the following equivalence between quiver representation moduli spaces induced by mutation sequences:
\begin{equation}
\begin{array}{lll}
	\fR\left(\vec d_m^{(i)},\vec \zeta\in \fC_m^{(i)},Q,W\right)&\cong & \fR\left(\vec d,\vec \zeta\in \fC_{\rm cyc},Q_m^{(i)},W_m^{(i)}\right)\;,\\
	\fR\left(\vec d_m^{(f)},\vec \zeta\in \fC_m^{(f)},Q,W\right)&\cong & \fR\left(\vec d,\vec \zeta\in \fC_{\rm cyc},Q_m^{(f)},W_m^{(f)}\right).
\end{array}
\end{equation}
Below we will discuss the quiver mutation on  $\left(Q_m^{(i)},W_m^{(i)}\right)$ and $\left(Q_m^{(f)},W_m^{(f)}\right)$. 
In particular, a single mutation process transforms a quiver through a single marginal stability wall $m\leftrightarrow m+1$, as we explain 
 in Appendix \ref{sec:App_conifold} (see also \cite{Chuang:2008aw}).

\subsubsection{Infinite chambers}
\label{ssec:infinite_chamber_conifold}

A sequence of $m-1$ steps of quiver mutations (for a single step see Appendix~\ref{sec:App_conifold}) on the canonically framed quiver~\eqref{eq:conifold_QW0} for the conifold geometry gives the quivers $Q_m^{(i)}$ and the superpotential $W_m^{(i)}$ in the infinite chamber $\fC^{(i)}_m$ \cite{Chuang:2008aw}:
\be\label{eq:quiv_inf_conif}
	\begin{aligned}
Q_m^{(i)}&=\begin{array}{c}
			\begin{tikzpicture}
				\draw[thick,->] ([shift=(120:3)]1.5,-2.59808) arc (120:80:3);
				\draw[thick,->] ([shift=(80:3)]1.5,-2.59808) arc (80:70:3);
				\draw[thick] ([shift=(70:3)]1.5,-2.59808) arc (70:60:3);
				\draw[thick,->] ([shift=(300:3)]1.5,2.59808) arc (300:260:3);
				\draw[thick,->] ([shift=(260:3)]1.5,2.59808) arc (260:250:3);
				\draw[thick] ([shift=(250:3)]1.5,2.59808) arc (250:240:3);
				\draw[thick,->] (-4,0) -- (-2.5,0);
				\draw[thick,->] (-2.5,0) -- (-2,0);
				\draw[thick,->] (-2,0) -- (-1.5,0);
				\draw[thick] (-1.5,0) -- (0,0);
				\draw[thick] (3,0) to[out=225,in=0] (1.5,-1) (-2.5,-1) to[out=180,in=315] (-4,0);
				\draw[thick,->] (1.5,-1) -- (0,-1);
				\draw[thick,->] (0,-1) -- (-0.5,-1);
				\draw[thick,->] (-0.5,-1) -- (-1,-1);
				\draw[thick] (-1,-1) -- (-2.5,-1);
				\draw[fill=white] (0,0) circle (0.2);
				\draw[fill=gray] (3,0) circle (0.2);
				\node[above] at (0,0.2) {$1$};
				\node[above] at (3,0.2) {$2$};
				\begin{scope}[shift={(-4,0)}]
					\draw[fill=red] (-0.15,-0.15) -- (-0.15,0.15) -- (0.15,0.15) -- (0.15,-0.15) -- cycle;
				\end{scope}
				\node[above] at (1.5,0.401924) {$a_1$, $a_2$};
				\node[above] at (1.5,-0.401924) {$b_1$, $b_2$};
				\node[above] at (-2,0) {$r_1,\ldots, r_m$};
				\node[above] at (-0.5,-1) {$s_1,\ldots,s_{m-1}$};
			\end{tikzpicture}
		\end{array},\\ 
W_m^{(i)}&=\Tr\left[b_2a_2b_1a_1-b_2a_1b_1a_2+\sum\lm_{i=1}^{m-1}s_i\left(a_2r_i-a_1r_{i+1}\right)\right].
	\end{aligned}
	\ee
Below we will also rederive this quiver-superpotential pair from the shape of the corresponding subcrystal, by first decomposing the subcrystal into superpositions of positive/negative crystals, using the procedure of Section~\ref{ssec:rep_subcry} and~\ref{ssec:framedQW}.
The 	fields associated with the arrows in $Q_m^{(i)}$ have the following complex masses, i.e.\ equivariant weights:
	\be\label{m_conif_weights}
	\begin{aligned}
		&\mu(a_1)=-\mathsf{h}_1\,,\quad \mu(a_2)=\mathsf{h}_1\,,\quad \mu(b_1)=-\mathsf{h}_2\,,\quad \mu(b_2)=\mathsf{h}_2\,,\\ 
		&\mu(r_k)=2(k-1)\mathsf{h}_1\,,\quad \mu(s_k)=-(2k-1)\mathsf{h}_1\,.
	\end{aligned}
	\ee
\bigskip

\begin{figure}[ht!]
		\begin{center}
			\begin{tikzpicture}
				\node at (0,0) {\includegraphics[scale=0.3]{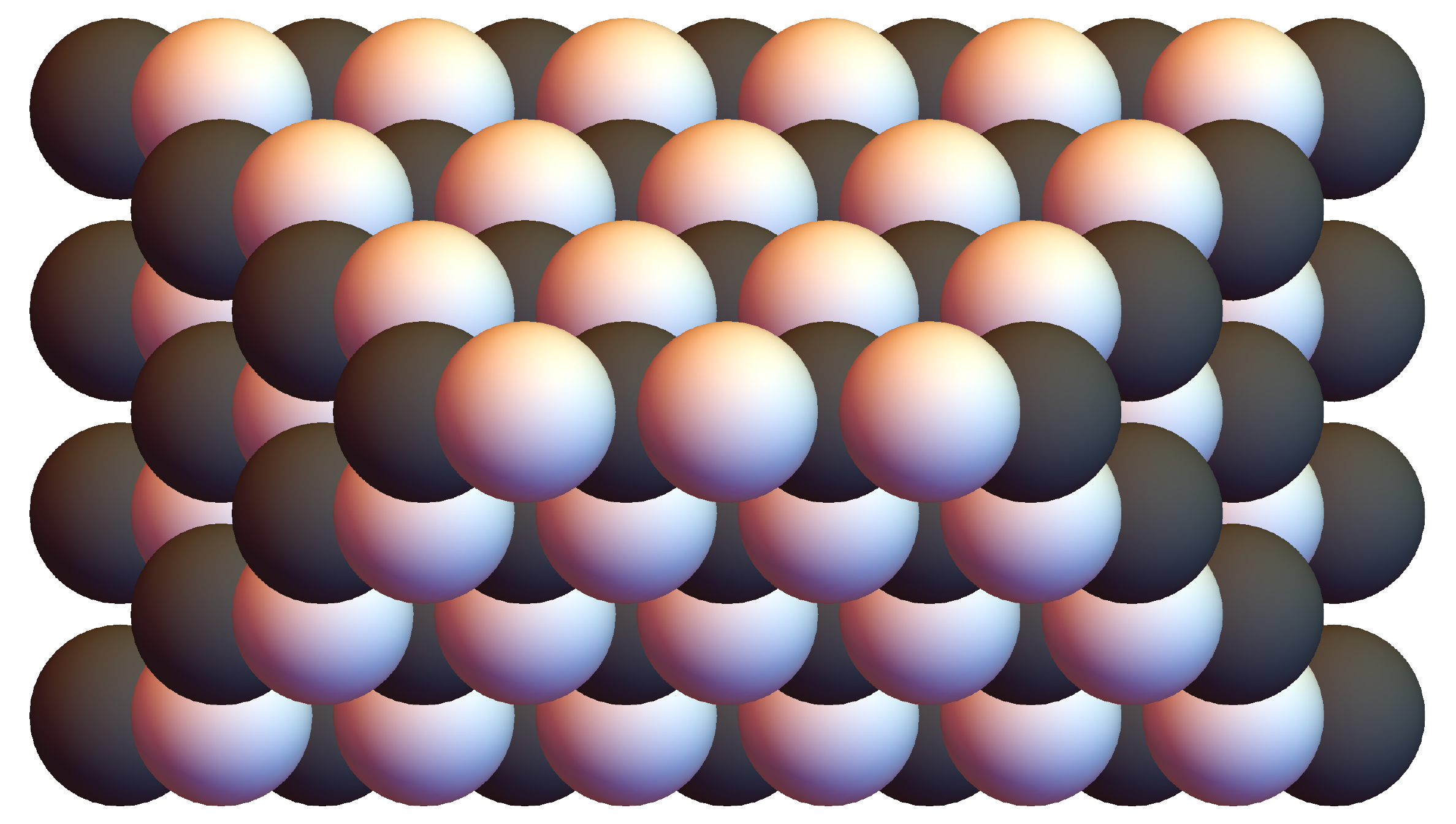}};
				\node at (-5.5,0) {(a)};
				\node at (-5.5,-3) {(b)};
				\begin{scope}[shift={(-3,-3)}]
					\draw[thick] (0,0) -- (3.5,0) (4.5,0) -- (6,0) (-1,1) -- (0,0) -- (-1,-1) (7,1) -- (6,0) -- (7,-1);
					\draw[thick,dashed] (3.5,0) -- (4.5,0);
					\draw[fill=white] (0,0) circle (0.2);
					\draw[fill=gray] (1,0) circle (0.2);
					\draw[fill=white] (2,0) circle (0.2);
					\draw[fill=gray] (3,0) circle (0.2);
					\draw[fill=gray] (5,0) circle (0.2);
					\draw[fill=white] (6,0) circle (0.2);
					\node[below] at (0,-0.2) {$\textsf{x}_1$};
					\node[below] at (1,-0.2) {$\textsf{y}_1$};
					\node[below] at (2,-0.2) {$\textsf{x}_2$};
					\node[below] at (3,-0.2) {$\textsf{y}_2$};
					\node[below] at (5,-0.2) {$\textsf{y}_{m-1}$};
					\node[below] at (6,-0.2) {$\textsf{x}_m$};
				\end{scope}
			\end{tikzpicture}
			\caption{(a) The crystal for the infinite chamber  $\CC_m^{(i)}$ of the resolved conifold,  
			with $m=3$.
			(b) The locations of the starters at $\textsf{x}_k$ with $k=1,2,\dots,m$ and the pausers at $\textsf{y}_k$ with $k=1,2,\dots,m-1$.}\label{fig:inf_quiv_cry}
		\end{center}
	\end{figure}
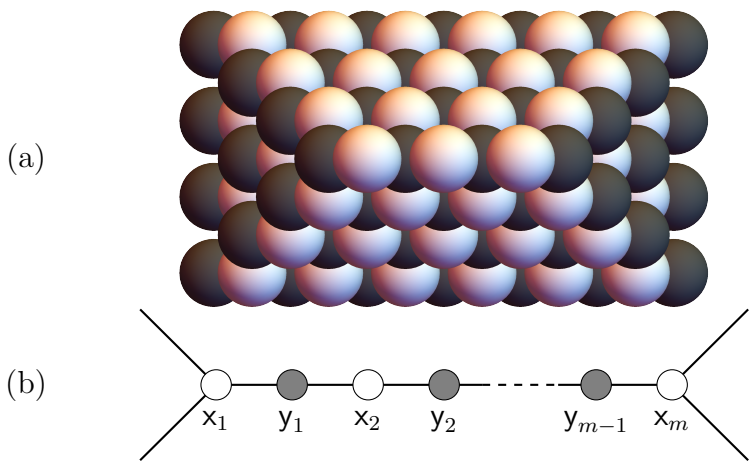

The crystal $\CC_m^{(i)}$ corresponding to the quiver~\eqref{eq:quiv_inf_conif} is a prism-like \emph{infinite} crystal corresponding to a stack of pyramids such that the resulting crystal develops an edge (see Figure~\ref{fig:inf_quiv_cry}(a)) \cite{MR2836398,Szendroi,Chuang:2008aw,2007arXiv0709.3079Y}.
The edge contains exactly $m$ (white) atoms and the case $m=1$ corresponds to the canonical crystal shown in Figure~\ref{fig:pyramid}. 
A generating function for the number $N(w,b)$ of molten crystals containing $w$ white atoms and $b$ black atoms reads \cite{2007arXiv0709.3079Y}:
	\be
	\begin{aligned}
		&Z_m(q_1,q_2)=\sum\lm_{w,b}N_m(w,b)q_1^wq_2^b
		=\left(\prod\lm_{n=1}^{\infty}\left(1-q_1^nq_2^n\right)^{-2n}\right)\\ &\qquad \qquad\qquad \times\left(\prod\lm_{n=1}^{\infty}\left(1+q_1^{m+n-1}q_2^{m+n}\right)^n\right)\left(\prod\lm_{n=m}^{\infty}\left(1+q_1^{-m+n+1}q_2^{-m+n}\right)^n\right).
	\end{aligned}
	\ee

Let us first specify the coordinate system for the atoms in the crystal $\CC^{(i)}_m$. 
We choose the white atom at the left end of the edge as the origin; and the direction to its right is chosen as the $x_1$ direction, the one to its north as the $x_2$ direction, and finally the direction pointing deeper into the crystal as the $x_3$ direction.
An atom at the coordinate $(x_1,x_2,x_3)$ has the equivariant weight (or the coordinate function)
\begin{equation}\label{eq:cfconifoldm}
h(\square)=\mathsf{h}_1 x_1(\square)+\mathsf{h_2} \, x_2(\square)\,,
\end{equation}
which is independent of the $x_3$ direction.
Along the $x_3$ direction, there is a double layer structure, namely, with white and black layers alternating. 
The atoms in the $n^{\textrm{th}}$ white layer and the $n^{\textrm{th}}$ black layer both have $x_3=n-1$.

In the crystal $\CC_m^{(i)}$, the $m$ white atoms (for node $1$)  along the edge (see Figure~\ref{fig:inf_quiv_cry}(b)) are at depth ${\bf d}=0$, with coordinates
\begin{equation}\label{eq:starterxk}
(x_1, x_2,x_3)=(2k-2,0,0)=:\textsf{x}_k \,, \quad \textrm{for}  \quad k=1,\dots, m\,,
\end{equation}
whose coordinate functions~\eqref{eq:cfconifoldm} give the equivariant weights of $m$ arrows $r_k$ going from the framing node to node $1$:
\begin{equation}\label{eq:inf_chamber_s}
\textrm{starter of color $1$:}\qquad 	\chi_k\equiv \mu(r_k)=2(k-1)\, \mathsf{h}_1\;,\quad k=1,\dots, m\,.
\end{equation}
The $(m-1)$ black atoms (for node $2$) sitting between the $m$ white atoms are also at depth ${\bf d}=0$, whose projections to the edge of the prism are shown in  Figure~\ref{fig:inf_quiv_cry}(b).
They have coordinates
\begin{equation}\label{eq:pauseryk}
 (x_1, x_2,x_3)=(2k-1,0,1)=: \textrm{y}_k \,, \quad \textrm{for}  \quad  k=1,\dots, m-1\,,
\end{equation}
whose coordinate functions~\eqref{eq:cfconifoldm} give the equivariant  weights of the $m-1$ Lagrange multiplier fields $s_k$, which are denoted by the $m-1$ arrows going from node $2$ to the framing node:
\begin{equation}\label{eq:inf_chamber_p}
\textrm{pausers of color $2$:}\qquad 	\upsilon_k\equiv -\mu(s_k)=(2k-1)\mathsf{h}_1\;,\quad k=1,\dots, m-1\,.
\end{equation}
Finally, the Lagrange multiplier fields $s_k$ impose relations on the quiver path algebra that ensure that two paths starting from nearest white atoms at $\textsf{x}_k$ and $\textsf{x}_{k+1}$ end on a common black atom at $\textsf{y}_k$. 
	
\bigskip

From the viewpoint of this paper, the crystal $\CC_m^{(i)}$ (shown in Figure~\ref{fig:inf_quiv_cry}(a)) should be considered as a subcrystal of the canonical crystal $\CC_1^{(i)}$ (shown in Figure~\ref{fig:pyramid}).
To obtain the subcrystal $\CC_m^{(i)}$ from the canonical crystal $\CC_1^{(i)}$, one can either (1) remove $m-1$ ``layers" of atoms from $\CC_1^{(i)}$ along certain direction 
or (2) superpose certain number of positive and negative canonical crystals $\CC_1^{(i)}$, positioned according to the starters and pausers, respectively.\footnote{There is no stopper involved because we are not considering truncation for now.}

Let us look at the second approach (explained in Section~\ref{ssec:rep_subcry}) now and leave the first one to Appendix~\ref{sec:app:WC_con}.	
To illustrate the procedure, it is enough to use 
the subcrystal $\CC_3^{(i)}$ as an example. 
In the procedure of Section~\ref{ssec:rep_subcry}, one can see that $\CC_3^{(i)}$ contains three positive (canonical) crystals at level-1: 
$\CC_a$, $\CC_b$, $\CC_c$, where $\CC_{a,b,c}=\CC_{\textsf{x}_1,\textsf{x}_2,\textsf{x}_3}$ and $\CC_{\textsf{x}}$ is the canonical crystal whose leading atom is at the position $\textsf{x}$.
Their intersections gives the three negative (canonical) crystals at level-1: $\CC_{ab}$, $\CC_{bc}$, $\CC_{ac}$.
Then the intersection of theses three negative crystals at level-1 gives one positive crystal at level-2: $\CC_{abc}$.	
	\be
	\begin{array}{cc|cc}
		\CC_a: & \begin{array}{c}
			\includegraphics[trim=0 50 0 0, clip, scale=0.2]{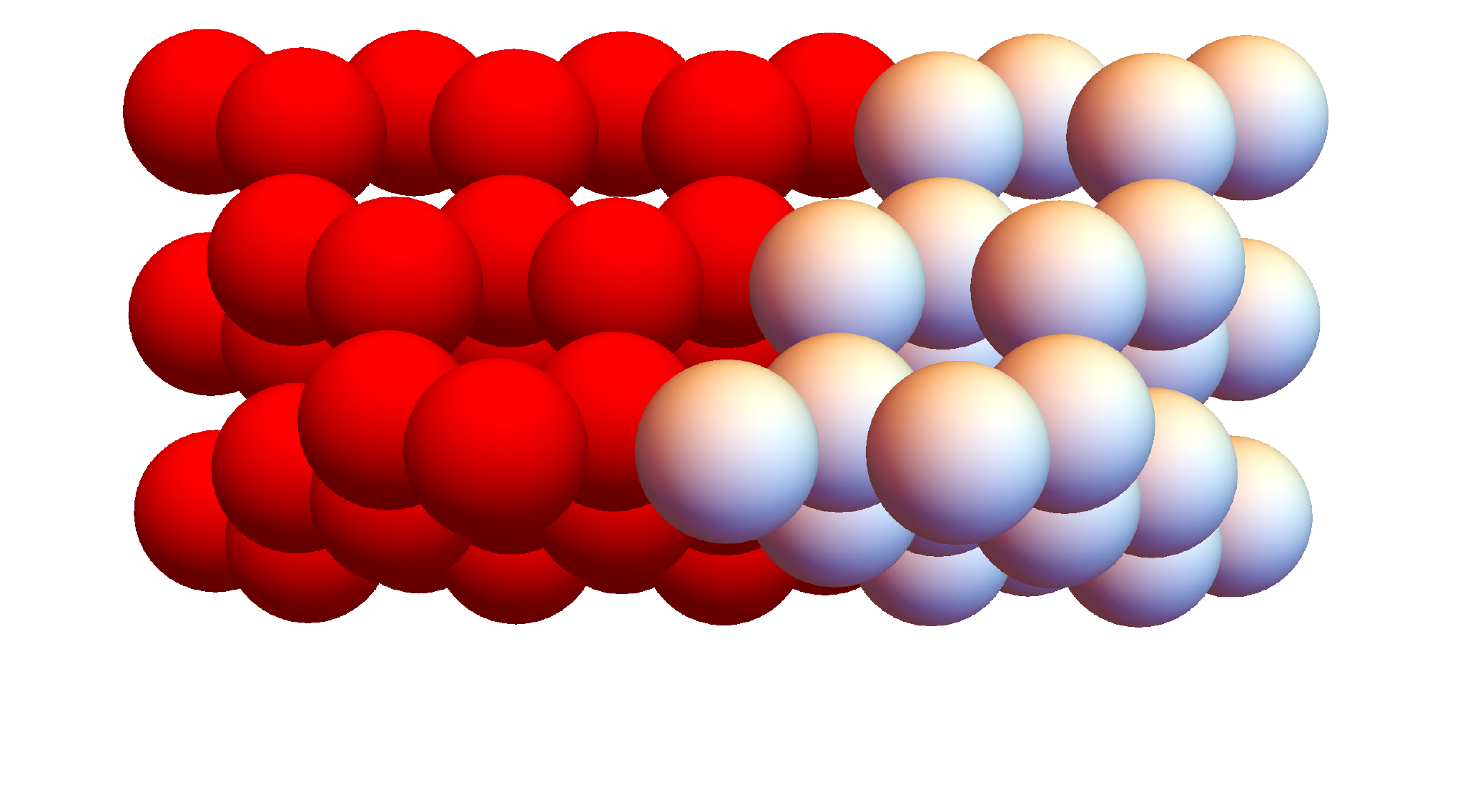}
		\end{array}&
	\CC_b: & \begin{array}{c}
		\includegraphics[trim=0 50 0 0, clip, scale=0.2]{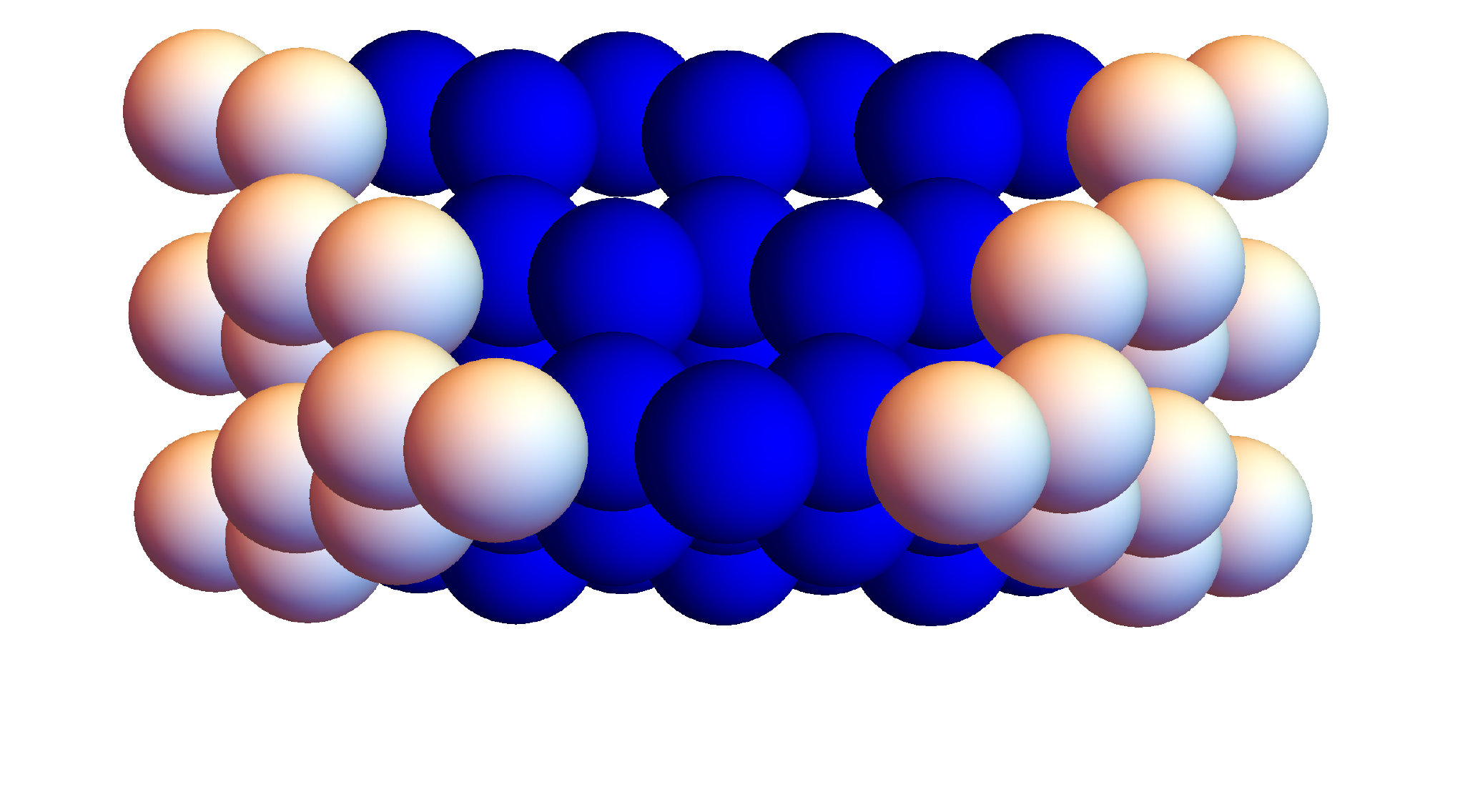}
	\end{array}\\
	\hline 
	\CC_c: & \begin{array}{c}
		\includegraphics[trim=0 50 0 0, clip, scale=0.2]{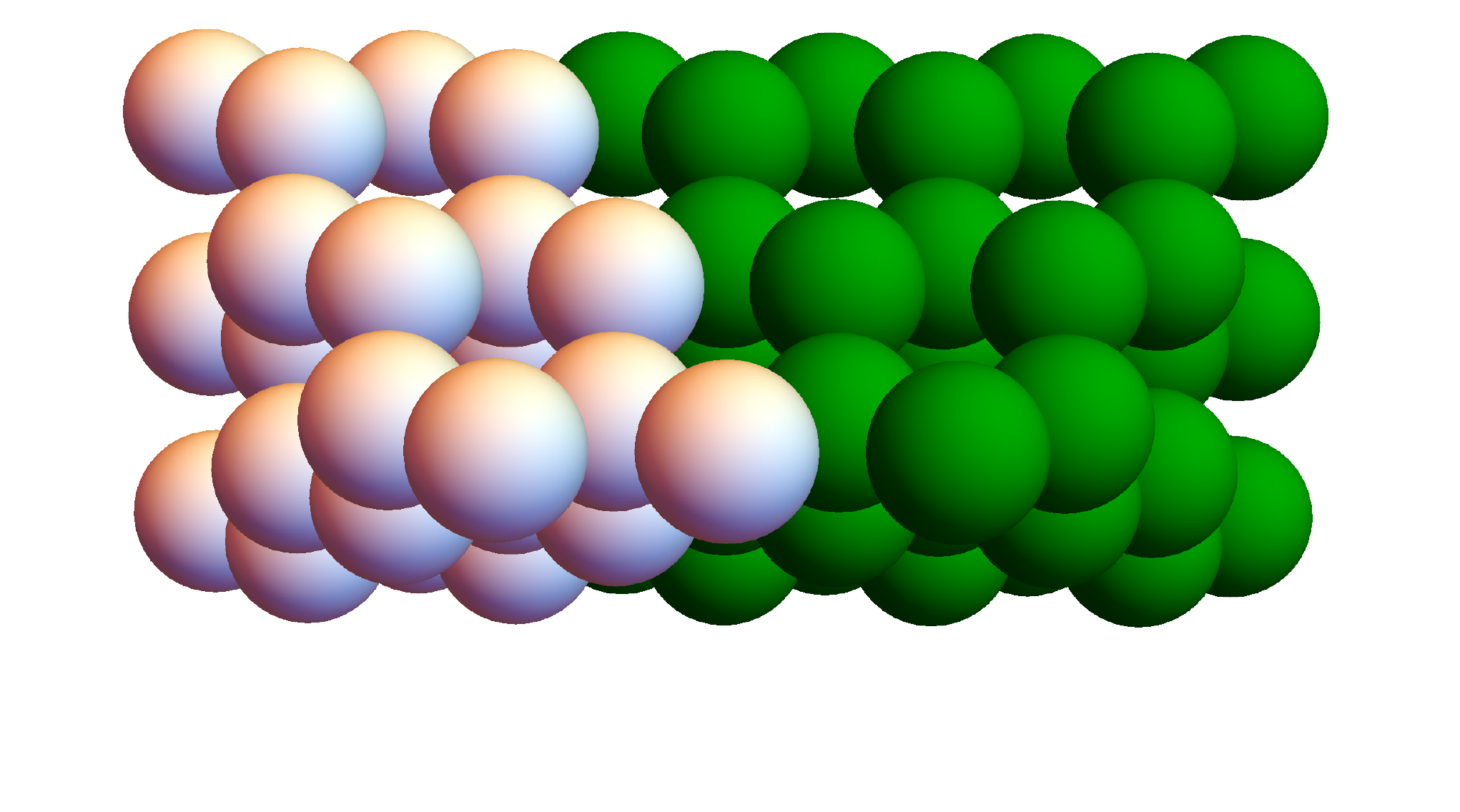}
	\end{array} &
	\CC_{ab} & \begin{array}{c}
		\includegraphics[trim=0 50 0 0, clip, scale=0.2]{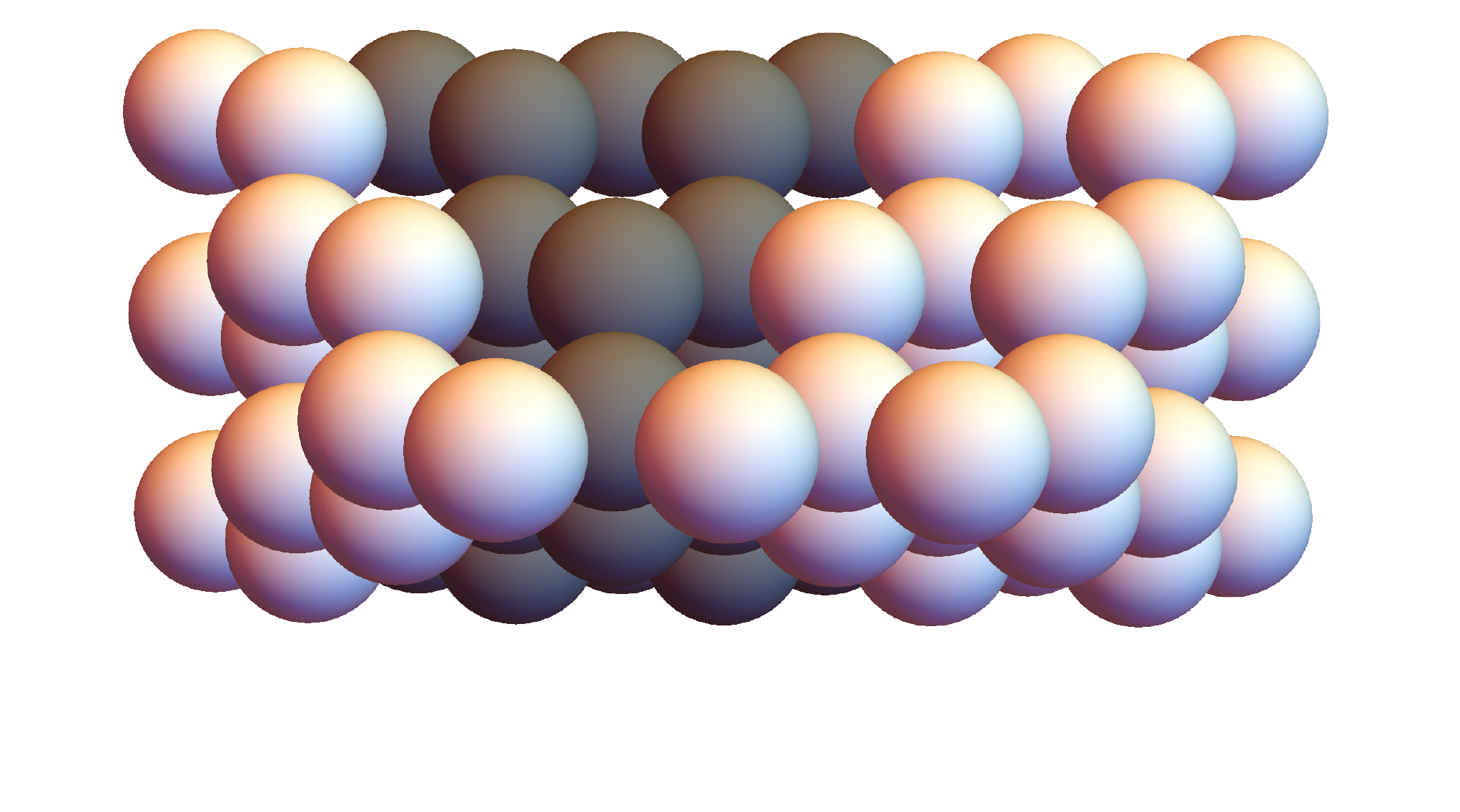}
	\end{array}\\
	\hline
	\CC_{bc}: & \begin{array}{c}
		\includegraphics[trim=0 50 0 0, clip, scale=0.2]{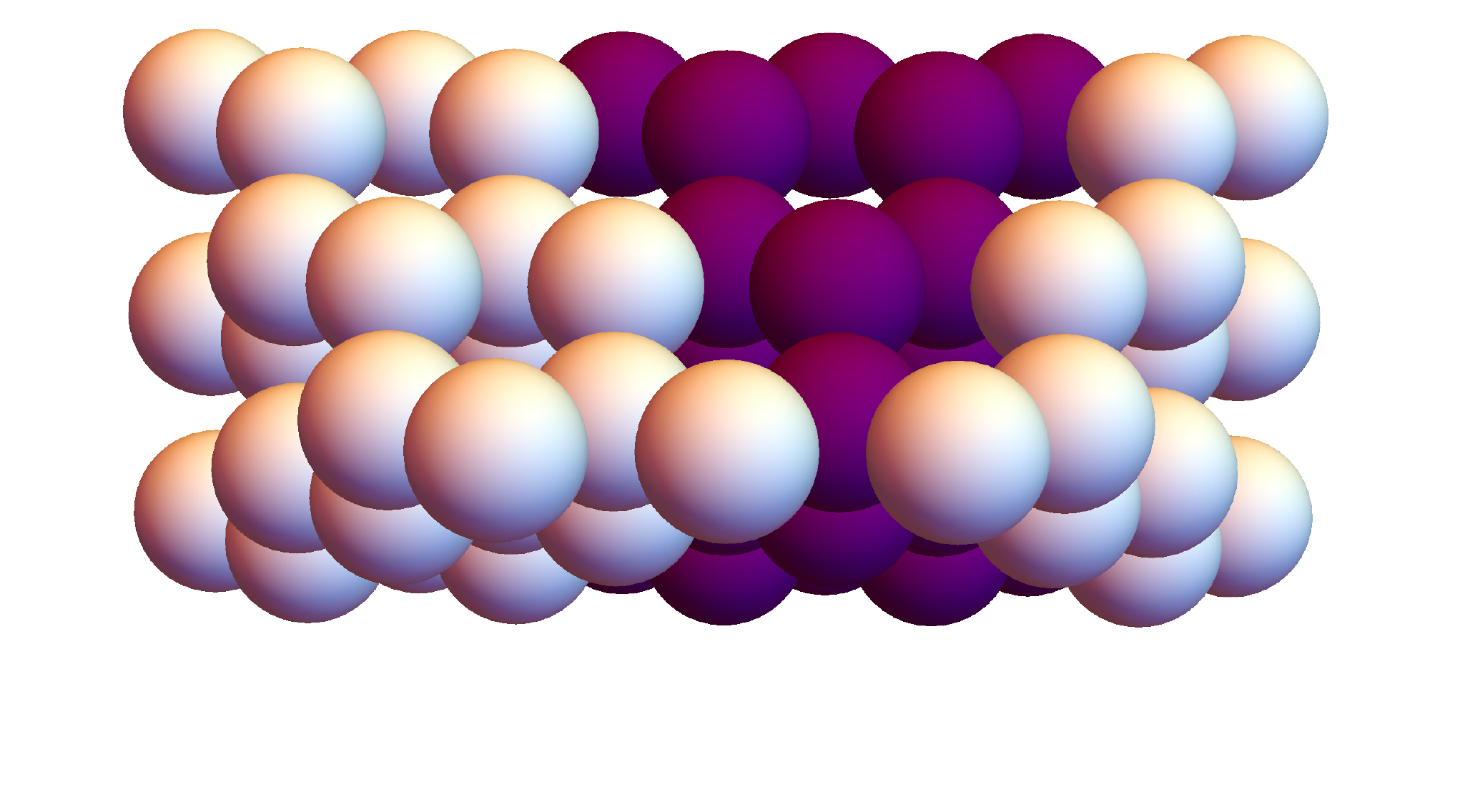}
	\end{array}&
	\CC_{ac}: & \begin{array}{c}
		\includegraphics[trim=0 50 0 0, clip, scale=0.2]{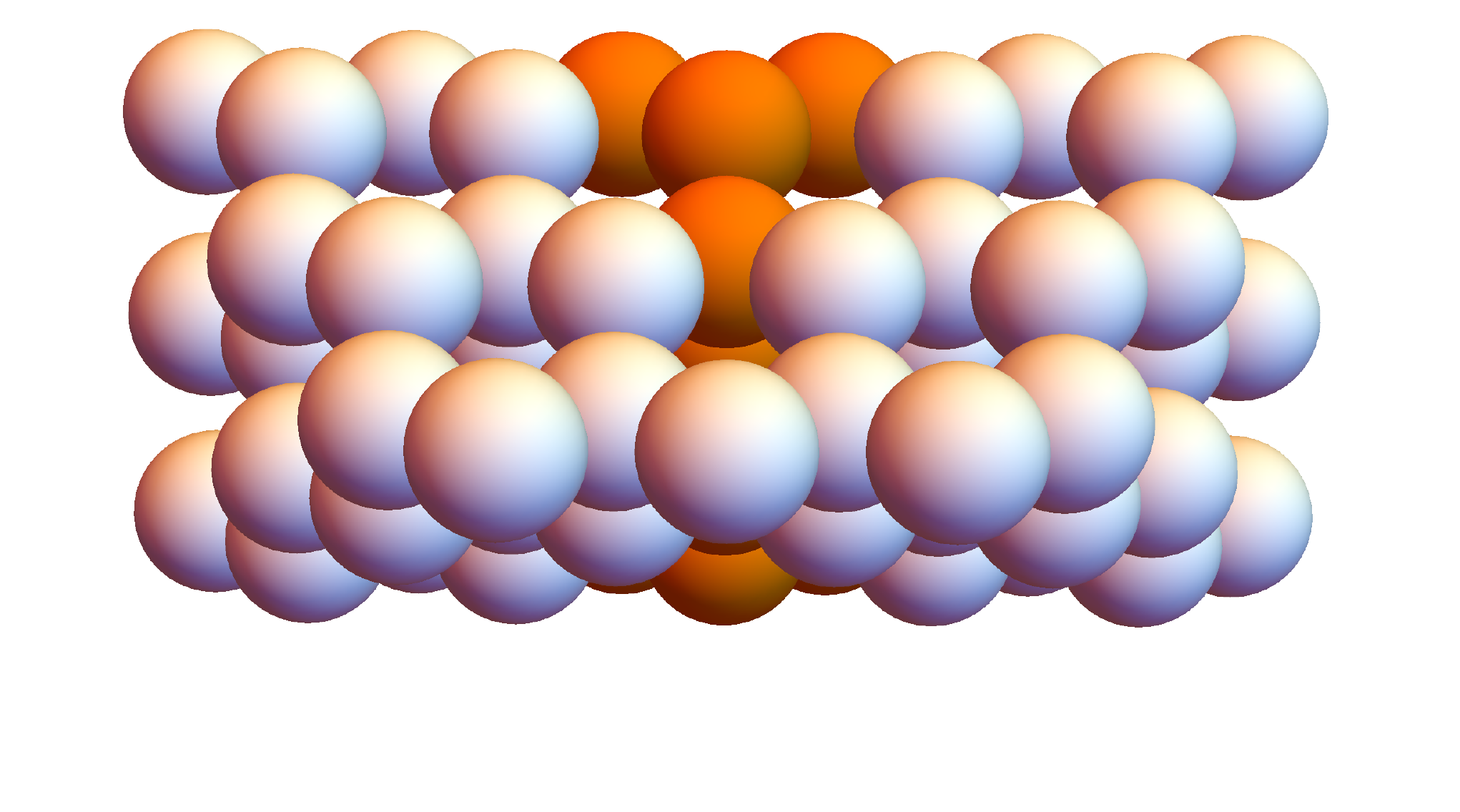}
	\end{array}
	\end{array}
	\ee

In principle this procedure, based on the inclusion-exclusion principle, can be infinite. 
In the current case, one can see that the decomposition procedure terminates at $\CC_{abc}$ because the negative crystal $\CC_{ac}$ and the positive crystal $\CC_{abc}$ coincide and hence cancel each other. 
As a result, the subcrystal $\CC_3^{(i)}$ has a decomposition 
	\be
	\CC_3^{(i)}=\CC_a+\CC_b+\CC_c-\CC_{ab} -\CC_{bc} \;.
	\ee
Therefore,  the positive crystal contributions are only from starter atoms in the prism edge located in points $\chi_k$ in Figure~\ref{fig:inf_quiv_cry}(b), and the negative contributions come only from pausers located in points $\upsilon_k$ in Figure~\ref{fig:inf_quiv_cry}(b). 
For the general crystal we have:
	\be
	\CC_m^{(i)}=\sum\lm_{k=1}^{m}\CC_{\textsf{x}_k}-\sum\lm_{k=1}^{m-1}\CC_{\textsf{y}_k}\;,
	\ee
where  $\CC_{x}$ is a canonical crystal whose leading atom is at the position $\textsf{x}$, $\{\textsf{x}_k\}$ is the set of the coordinates of the $m$ starters , and $\{\textsf{y}_k\}$ is the one for the $m-1$ pausers.
\bigskip

Using~\eqref{eq:psi0_summary} we translate the coordinate function  of the starters \eqref{eq:inf_chamber_s} and pausers \eqref{eq:inf_chamber_p} into the ground state charge function:
	\be\label{eq:GSCF_conifoldm}
	\begin{aligned}
		{}^{\sharp}\psi^{(1)}_{0}=\prod\lm_{k=1}^m\frac{1}{z-\chi_k} \quad \textrm{and} \quad
		{}^{\sharp}\psi^{(2)}_{0}=\prod\lm_{k=1}^{m-1}\left(z-\upsilon_k\right)\,.
	\end{aligned}
	\ee
Therefore, the infinite subcrystal  $\CC_m^{(i)}$, for chamber $m$, gives rise to a representation of the shifted quiver Yangian of  $\mathfrak{gl}_{1|1}$, with shifts given by
\begin{equation}\label{shift_inf}
    \mys = (m, -m+1)\,.
\end{equation}
The net shift $\mys^{(1)} + \mys^{(2)} $ is 1.
One can use the procedure of Section~\ref{ssec:framedQW} to check that the framed quiver and superpotential corresponding to the ground state charge function is indeed  the one obtained by the sequence of quiver mutations~\eqref{eq:quiv_inf_conif}.
\bigskip

Finally, note that there is more than one way to construct the subcrystal $\CC_m^{(i)}$ from the canonical $\CC_1^{(i)}$.
In Appendix~\ref{sec:app:WC_con}, we show how to obtain the crystal $\CC_m^{(i)}$ by removing $m-1$ ``layers" of atoms from the canonical crystal $\CC_1^{(i)}$ along a certain direction; and correspondingly derive the ground state charge function by considering the ground state of the representation given by $\CC_m^{(i)}$ as an excited state in the vacuum representation given by the canonical crystal $\CC_1^{(i)}$.
The resulting ground state charge functions~\eqref{eq-Psi12-gl11-m2} agree with~\eqref{eq:GSCF_conifoldm}, derived via the positive/negative crystal decomposition.

\subsubsection{Finite chambers}
\label{ssec:finite_chamber}

In contrast to the infinite chamber, the crystal in a finite chamber is of finite size \cite{MR2836398}. 
The quiver $Q_m^{(i)}$ and superpotentials $W_m^{(i)}$ in the finite chamber $\CC_m^{(f)}$ are:
\be\label{QW_conif_fin}
	\begin{aligned}
Q_m^{(f)}&=\begin{array}{c}
			\begin{tikzpicture}
				\draw[thick,->] ([shift=(120:3)]1.5,-2.59808) arc (120:80:3);
				\draw[thick,->] ([shift=(80:3)]1.5,-2.59808) arc (80:70:3);
				\draw[thick] ([shift=(70:3)]1.5,-2.59808) arc (70:60:3);
				\draw[thick,->] ([shift=(300:3)]1.5,2.59808) arc (300:260:3);
				\draw[thick,->] ([shift=(260:3)]1.5,2.59808) arc (260:250:3);
				\draw[thick] ([shift=(250:3)]1.5,2.59808) arc (250:240:3);
				\draw[thick,->] (-4,0) -- (-2.5,0);
				\draw[thick,->] (-2.5,0) -- (-2,0);
				\draw[thick,->] (-2,0) -- (-1.5,0);
				\draw[thick] (-1.5,0) -- (0,0);
				\draw[thick] (3,0) to[out=225,in=0] (1.5,-1) (-2.5,-1) to[out=180,in=315] (-4,0);
				\draw[thick,->] (1.5,-1) -- (0,-1);
				\draw[thick,->] (0,-1) -- (-0.5,-1);
				\draw[thick,->] (-0.5,-1) -- (-1,-1);
				\draw[thick] (-1,-1) -- (-2.5,-1);
				\draw[fill=white] (0,0) circle (0.2);
				\draw[fill=gray] (3,0) circle (0.2);
				\node[above] at (0,0.2) {$1$};
				\node[above] at (3,0.2) {$2$};
				\begin{scope}[shift={(-4,0)}]
					\draw[fill=red] (-0.15,-0.15) -- (-0.15,0.15) -- (0.15,0.15) -- (0.15,-0.15) -- cycle;
				\end{scope}
				\node[above] at (1.5,0.401924) {$a_1$, $a_2$};
				\node[above] at (1.5,-0.401924) {$b_1$, $b_2$};
				\node[above] at (-2,0) {$r_1,\ldots, r_m$};
				\node[above] at (-0.5,-1) {$s_0,\ldots,s_{m}$};
			\end{tikzpicture}
		\end{array},\\ 
W_m^{(f)}&=\Tr\left[b_2a_2b_1a_1-b_2a_1b_1a_2+\underline{s_0a_1r_1}+\sum\lm_{i=1}^{m-1}s_{i}(a_2r_i-a_1r_{i+1})+\underline{s_{m}a_2r_m}\right]\,.
	\end{aligned}
	\ee
Compared to the infinite chamber case in~\eqref{eq:quiv_inf_conif} for the same $m$, the quiver $Q_m^{(f)}$ for the finite chamber has two additional fields $s_0$ and $s_m$, whose equivariant weights are
\begin{equation}
    \mu(s_0)=\mathsf{h}_1 \quad \textrm{and} \quad \mu(s_m)=-(2m-1)\mathsf{h}_1\,,
\end{equation}
and the equivariant weights of all the other fields take the same value as those for the $\left(Q_m^{(i)}, W_m^{(i)}\right)$ in the infinite chamber, given in \eqref{m_conif_weights}. 

In accordance with the two additional fields $s_0$ and $s_m$ in the quiver $Q_m^{(f)}$, the superpotential $W_m^{(f)}$ \eqref{QW_conif_fin} has two additional terms, which are underlined.
The two new fields $s_0$ and $s_m$ impose two extra relations on the quiver path algebra:
\begin{equation}
	a_1r_1=0\;,\quad a_2 r_m=0\,,
\end{equation}
which dictate that the crystal should stop its growth at the two corresponding stopper atoms.

\bigskip

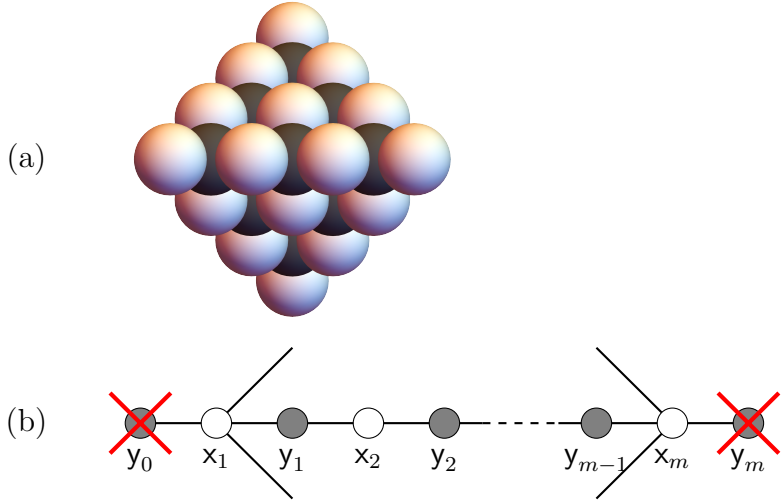
\begin{figure}[ht!]
		\begin{center}
		\begin{tikzpicture}
			\node at (-3.5,0) {(a)};
			\node at (-3.5,-3.5) {(b)};
			\node at (0,0) {\includegraphics[scale=0.18]{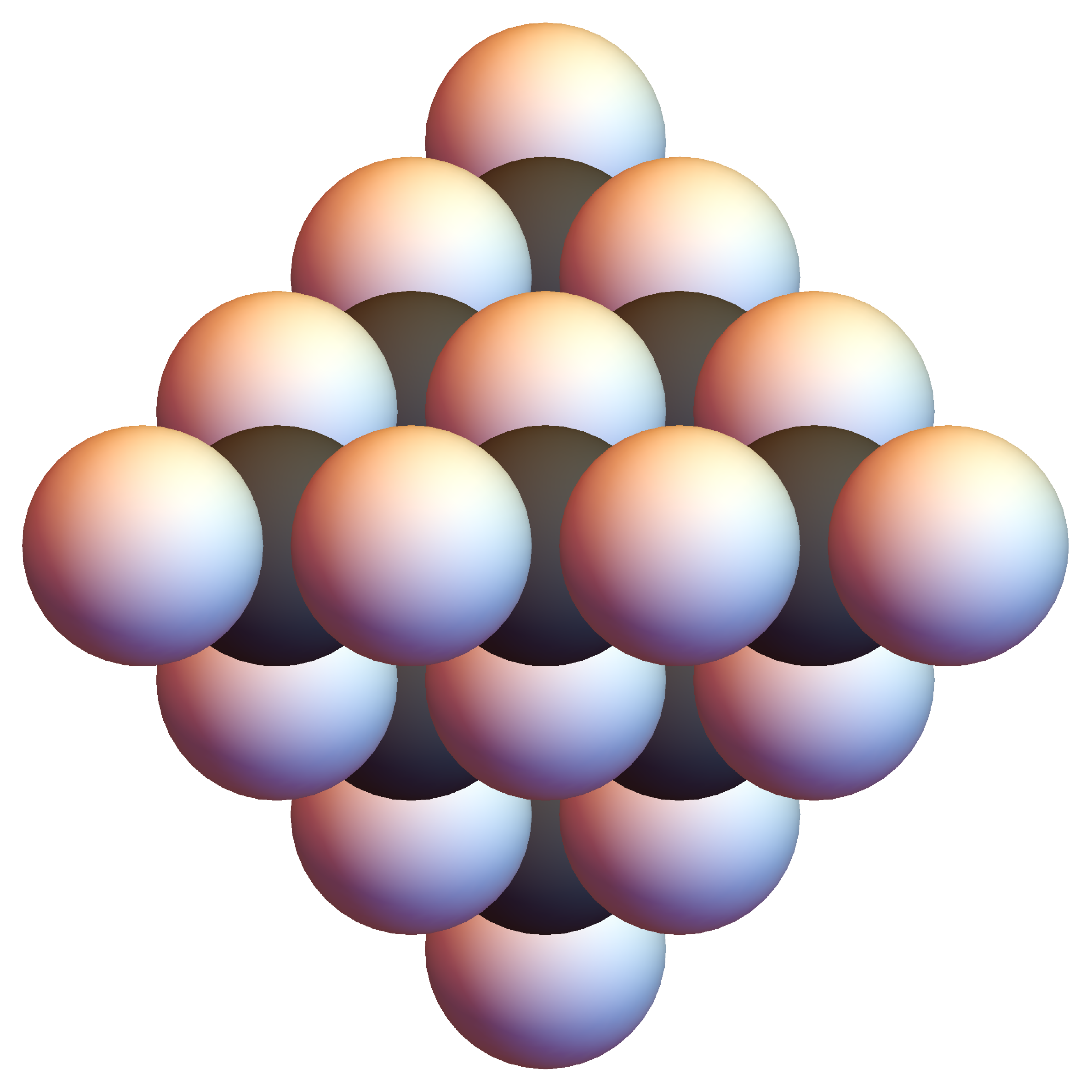}};
			\begin{scope}[shift={(-1,-3.5)}]
			\draw[thick] (-1,0) -- (3.5,0) (4.5,0) -- (7,0) (1,1) -- (0,0) -- (1,-1) (5,1) -- (6,0) -- (5,-1);
			\draw[thick,dashed] (3.5,0) -- (4.5,0);
			\draw[fill=gray] (-1,0) circle (0.2);
			\draw[fill=white] (0,0) circle (0.2);
			\draw[fill=gray] (1,0) circle (0.2);
			\draw[fill=white] (2,0) circle (0.2);
			\draw[fill=gray] (3,0) circle (0.2);
			\draw[fill=gray] (5,0) circle (0.2);
			\draw[fill=white] (6,0) circle (0.2);
			\draw[fill=gray] (7,0) circle (0.2);
			\begin{scope}[shift={(-1,0)}]
				\draw[ultra thick, red] (-0.4,-0.4) -- (0.4,0.4) (0.4,-0.4) -- (-0.4,0.4);
			\end{scope}
			\begin{scope}[shift={(7,0)}]
				\draw[ultra thick, red] (-0.4,-0.4) -- (0.4,0.4) (0.4,-0.4) -- (-0.4,0.4);
			\end{scope}
			\node[below] at (-1,-0.2) {$\textsf{y}_0$};
			\node[below] at (0,-0.2) {$\textsf{x}_1$};
			\node[below] at (1,-0.2) {$\textsf{y}_1$};
			\node[below] at (2,-0.2) {$\textsf{x}_2$};
			\node[below] at (3,-0.2) {$\textsf{y}_2$};
			\node[below] at (5,-0.2) {$\textsf{y}_{m-1}$};
			\node[below] at (6,-0.2) {$\textsf{x}_m$};
			\node[below] at (7,-0.2) {$\textsf{y}_{m}$};
			\end{scope}
		\end{tikzpicture}
	\caption{(a) The crystal for the finite chamber  $\CC_m^{(f)}$ of the resolved conifold,  
			with $m=4$.
			(b) The locations of the starters at $\textsf{x}_k$ with $k=1,2,\dots,m$, the pausers at $\textsf{y}_k$ with $k=1,2,\dots,m-1$, and the stoppers at $\textsf{y}_0$ and $\textsf{y}_{m}$.}
\label{fig:fin_cry_conif}
	\end{center}
	\end{figure}

This means that the corresponding crystal $\CC_m^{(f)}$ can be obtained by cutting off two infinite pyramids from the two ends of an infinite crystal $\CC_m^{(f)}$ (with the same $m$). 
The resulting crystal $\CC_m^{(f)}$ has the shape of a tetrahedron (see Figure~\ref{fig:fin_cry_conif}(a)) and is also called a \emph{finite} type pyramid partition in the literature \cite{MR2836398,Chuang:2008aw}.
Compared to the crystal $\CC_m^{(i)}$ in the infinite chamber for the same $m$, in addition to the $m$ starters  with coordinates at $\textsf{x}_k$ with $k=1,2,\dots,m$ (see \eqref{eq:starterxk}) and the pausers at $\textsf{y}_k$ with $k=1,2,\dots,m-1$ (see \eqref{eq:pauseryk}),  the finite crystal $\CC_m^{(f)}$ has two stoppers at $\textsf{y}_0$ and $\textsf{y}_{m}$.	
The positions of theses starters, stoppers, and pausers for the crystal $\CC_m^{(f)}$  give the equivariant weights of the corresponding arrows via \eqref{eq:cfconifoldm} as in the previous case (see Figure~\ref{fig:fin_cry_conif}(b)):
	\be\label{eq:finite_chamber_sps}
	\begin{aligned}
\textrm{starts of color $1$:}\qquad 	&\chi_k=\mu(r_k)=2(k-1)\mathsf{h}_1\;,\quad k=1,\dots,m\,,\\
\textrm{pausers of color $2$:}\qquad 	&\upsilon_k=-\mu(s_k)=(2k-1)\mathsf{h}_1\;,\quad k=1,\dots,m-1\;,\\
\textrm{stoppers of color $2$:}\qquad 	&\upsilon_k=-\mu(s_k)=(2k-1)\mathsf{h}_1\;,\quad k=0, m \,.
	\end{aligned}
	\ee	
Finally, the partition function for $\CC_m^{(f)}$ reads:
	\be
	Z(q_1,q_2)=\prod\lm_{n=1}^m(1+q_1^{m-n}q_2^{m-n+1})^n \,,
	\ee
from which one can check that  the total number of atoms in $\CC_m^{(f)}$ is:
\begin{equation}
	{\rm max}\;\#=\frac{m(m+1)(2m+1)}{6}\,.
\end{equation}
\bigskip

From \eqref{eq:psi0_summary},  the ground state charge function of the finite representation from the crystal $\CC_m^{(f)}$ is 
	\be\label{eq:finite_chamber_GSCF}
	\begin{aligned}
		{}^{\sharp}\psi^{(1)}_0=\prod\lm_{k=1}^m\frac{1}{z-\chi_k}\quad \textrm{and} \quad
		{}^{\sharp}\psi^{(2)}_0=\prod\lm_{k=0}^{m}\left(z-\upsilon_k\right)\,,
	\end{aligned}
	\ee
where the poles come from the  starters located at $\textsf{x}_k$ in \eqref{eq:finite_chamber_sps} and the zeros from pausers and stoppers located at $\textsf{y}_k$ in \eqref{eq:finite_chamber_sps}.
This is to be compared with the result \eqref{eq:GSCF_conifoldm} for the infinite chamber with the same $m$.
Again, one can use the procedure of Section~\ref{ssec:framedQW} to derive the framed quiver and superpotential pair associated with the ground state charge function \eqref{eq:finite_chamber_GSCF} and check that it gives $\left(Q_m^{(i)}, W_m^{(i)}\right)$ in \eqref{QW_conif_fin}.

Finally, using \eqref{shift}, we see that the finite representation from the crystal $\CC_m^{(f)}$ is a representation of the shifted quiver Yangian of $\mathfrak{gl}_{1|1}$, with shifts given by
\begin{equation}
\mys = (m, -m-1)\,.
\end{equation}
The net shift is $-1$, in contrast to the case in the infinite chamber, whose net shift is $1$ from \eqref{shift_inf}.

\subsection{\texorpdfstring{Example: wall-crossing in $K_{\IP^2}$}{Example: Wall-Crossing in K(P(2))}}
	
Let us now give an example of a toric Calabi-Yau threefold with compact $4$-cycle.

\subsubsection{Canonical crystal and unshifted quiver Yangian}
\label{ssec:canonical_P2}
	Let us consider the geometry $K_{\mathbb{P}^2}$, the canonical bundle over $\mathbb{P}^2$.
	The geometry coincides with $\mathbb{C}^3/\mathbb{Z}_3$, 
	where the action of $\mathbb{Z}_3$ is
	$(z_1, z_2, z_3) \to (\omega z_1, \omega z_2, \omega z_3)$ with $\omega^3=1$.
	The toric diagram and its dual graph are 
	\begin{equation}\label{fig-toric-P2}
		\begin{array}{c}
		\begin{tikzpicture} 
			\filldraw [red] (0,0) circle (2pt); 
			\filldraw [red] (0,-1) circle (2pt); 
			\filldraw [red] (1,1) circle (2pt); 
			\filldraw [red] (-1,0) circle (2pt); 
			\node at (0,0.8) {(0,0)}; 
			\node at (-1.7,0) {(-1,0)}; 
			\node at (1.5,1.5) {(1,1)}; 
			\node at (0,-1.5) {(0,-1)}; 
			\draw (0,0) -- (1,1); 
			\draw (0,0) -- (-1,0); 
			\draw (0,0) -- (0,-1); 
			\draw (-1,0) -- (1,1); 
			\draw (-1,0) -- (0,-1); 
			\draw (0,-1) -- (1,1); 
		\end{tikzpicture}
		\end{array}		\qquad \qquad \qquad
	\begin{array}{c}
		\begin{tikzpicture}[scale=0.6] 
			\draw (0,0) -- (1,0); 
			\draw (0,0) -- (0,1); 
			\draw (1,0) -- (0,1); 
			\draw[->] (0,0) -- (-1,-1); 
			\draw[->] (1,0) -- (3,-1); 
			\draw[->] (0,1) -- (-1,3); 
		\end{tikzpicture}
	\end{array}
	\end{equation}
\bigskip
	
	The canonically framed quiver  is the McKay quiver \cite{reid1997mckay,Ueda:2006jn} for the $\mathbb{Z}_3$-action together with the framing node and one arrow from the framing node to the node $1$:
	\begin{equation}\label{quiver-P2}
Q_0=		\begin{array}{c}
			\begin{tikzpicture}
				\begin{scope}[shift={(0.866025,0)}]
					\begin{scope}[shift={(-0.866025,0)}]
						\draw[thick,->] (0,-1.5) -- (0,-0.5);
						\draw[thick,->] (0,-0.5) -- (0,0);
						\draw[thick,->] (0,0) -- (0,0.5);
						\draw[thick] (0,0.5) -- (0,1.5);
					\end{scope}
					\begin{scope}[rotate=-120]
						\begin{scope}[shift={(-0.866025,0)}]
							\draw[thick,->] (0,-1.5) -- (0,-0.5);
							\draw[thick,->] (0,-0.5) -- (0,0);
							\draw[thick,->] (0,0) -- (0,0.5);
							\draw[thick] (0,0.5) -- (0,1.5);
						\end{scope}
					\end{scope}
					\begin{scope}[rotate=-240]
						\begin{scope}[shift={(-0.866025,0)}]
							\draw[thick,->] (0,-1.5) -- (0,-0.5);
							\draw[thick,->] (0,-0.5) -- (0,0);
							\draw[thick,->] (0,0) -- (0,0.5);
							\draw[thick] (0,0.5) -- (0,1.5);
						\end{scope}
					\end{scope}
				\end{scope}
				\draw[thick,postaction={decorate},decoration={markings, 
		mark= at position 0.5 with {\arrow{>}}}] 
		(-2,-1.5) -- (0,-1.5);
		\begin{scope}[shift={(-2,-1.5)}]
		    \draw[fill=red] (-0.15,-0.15) -- (-0.15,0.15) -- (0.15,0.15) -- (0.15,-0.15) -- cycle;
		\end{scope}
				\draw[fill=white] (0,-1.5) circle (0.2);
				\draw[fill=white] (0,1.5) circle (0.2);
				\draw[fill=white] (2.59808,0) circle (0.2);
				\node[below] at (0,-1.7) {$1$};
				\node[above] at (0,1.7) {$2$};
				\node[right] at (2.79808,0) {$3$};
				\node[left] at (0,0) {$\left(X_i^{(1)},\alpha_i^{(1)}\right)$};
				\node[above right] at (1.29904,0.75) {$\left(X_i^{(2)},\alpha_i^{(2)}\right)$};
				\node[below right] at (1.29904,-0.75) {$\left(X_i^{(3)},\alpha_i^{(3)}\right)$};
			\end{tikzpicture}
		\end{array}
	\end{equation}
The corresponding superpotential is
	\begin{align}\label{W_P2}
		W_0=\sum_{i,j,k=1}^3 \varepsilon^{ijk}\textrm{Tr}(X^{(1)}_i X^{(2)}_j X^{(3)}_k) \;,
	\end{align}
where $\varepsilon^{ijk}$ is the totally antisymmetric tensor .
	
	The loop constraint (\ref{eq:loop-constraint}) from the superpotential is 
	\begin{align}
		\alpha^{(1)}_i+\alpha^{(2)}_j+\alpha^{(3)}_k=0 \qquad \textrm{for} \quad  \{i,j,k\} \in \{ 1,2,3\} \;,
	\end{align}
	and the vertex constraint (\ref{eq-vertex-constraint-toric}) for this case is
	\begin{equation}
		\sum^3_{i=1} \alpha^{(a)}_i = \sum^3_{i=1} \alpha^{(a+1)}_i \qquad \textrm{for} \quad a=1,2,3\,,
	\end{equation}
	which reduces the number of parameters to two, given by the triple $(\mathsf{h}_1, \mathsf{h}_2, \mathsf{h}_3)$:
	\begin{align}\label{eq.P2_final}
		\alpha^{(1)}_i=\alpha^{(2)}_i=\alpha^{(3)}_i=\mathsf{h}_i \quad (i=1,2,3) \;, 
		\qquad \mathsf{h}_1+\mathsf{h}_2+\mathsf{h}_3=0 \;.
	\end{align}
Finally, the periodic quiver corresponding to the canonically framed quiver  and superpotential pair \eqref{quiver-P2} and \eqref{W_P2} is
	\begin{equation}\label{periodicquiver_P2}
		\mbox{\scalebox{0.7}{$
		\begin{tikzpicture}[scale=0.8]
			\filldraw[mygreen] (-3,3)--  (0, 6) -- (0,-3) -- (-3,-6) -- cycle; 
			\node[state]  [regular polygon, regular polygon sides=4, draw=blue!50, very thick, fill=blue!10] (C) at (0,0)  {$3$};
			\node[state]  [regular polygon, regular polygon sides=4, draw=blue!50, very thick, fill=blue!10] (E) at (3,0)  {$1$};
			\node[state]  [regular polygon, regular polygon sides=4, draw=blue!50, very thick, fill=blue!10] (W) at (-3,0)  {$2$};
			\node[state]  [regular polygon, regular polygon sides=4, draw=blue!50, very thick, fill=blue!10] (N) at (0,3)  {$2$};
			\node[state]  [regular polygon, regular polygon sides=4, draw=blue!50, very thick, fill=blue!10] (S) at (0,-3)  {$1$};
			\node[state]  [regular polygon, regular polygon sides=4, draw=blue!50, very thick, fill=blue!10] (a31) at (3,3)  {$3$};
			\node[state]  [regular polygon, regular polygon sides=4, draw=blue!50, very thick, fill=blue!10] (a32) at (3,-3)  {$2$};
			\node[state]  [regular polygon, regular polygon sides=4, draw=blue!50, very thick, fill=blue!10] (a34) at (-3,-3)  {$3$};
			\node[state]  [regular polygon, regular polygon sides=4, draw=blue!50, very thick, fill=blue!10] (a33) at (-3,3)  {$1$};
			\path[->] 
			(C) edge   [thick, red]   node [above] {$\mathsf{h}_2$} (E)
			(W) edge   [thick, red]   node [above] {$\mathsf{h}_2$} (C)
			(a31) edge   [thick, red]   node [right] {$\mathsf{h}_1$} (E)
			(E) edge   [thick, red]   node [right] {$\mathsf{h}_1$} (a32)
			(a33) edge   [thick, red]   node [left] {$\mathsf{h}_1$} (W)
			(W) edge   [thick, red]   node [left] {$\mathsf{h}_1$} (a34)
			(N) edge   [thick, red]   node [above] {$\mathsf{h}_2$} (a31)
			(a33) edge   [thick, red]   node [above] {$\mathsf{h}_2$} (N)
			(a34) edge   [thick, red]   node [above] {$\mathsf{h}_2$} (S)
			(S) edge   [thick, red]   node [above] {$\mathsf{h}_2$} (a32)
			(N) edge   [thick, red]   node [right] {$\mathsf{h}_1$} (C)
			(C) edge   [thick, red]   node [right] {$\mathsf{h}_1$} (S)
			(C) edge   [thick, red]   node [right] {$\mathsf{h}_3$} (a33)
			(a32) edge   [thick, red]   node [right] {$\mathsf{h}_3$} (C)
			(E) edge   [thick, red]   node [right] {$\mathsf{h}_3$} (N)
			(S) edge   [thick, red]   node [right] {$\mathsf{h}_3$} (W)
			;
		\end{tikzpicture}$}}
	\end{equation}
	where we have shown the fundamental regions of the torus as shaded regions.
	Since there is no self-loop in the quiver diagram (\ref{quiver-P2}), all vertices are fermionic:
	\begin{equation}
		|a|=1\,, \qquad \qquad a=1,2,3\,.
	\end{equation}

\bigskip	
	
For the $\mathbb{C}^3/\mathbb{Z}_3$ geometry, the canonical crystal built from the periodic quiver \eqref{periodicquiver_P2} has the same shape as the one for the $\mathbb{C}^3$ geometry, and the $\mathbb{Z}_3$ orbifolding only changes the coloring scheme.
Namely, each molten crystal configuration from the canonical crystal of $\mathbb{C}^3/\mathbb{Z}_3$  is a plane partition, and the coloring scheme is the following.
The 3D octant can be sliced into layers of atoms, with the layers perpendicular to the vector $(1,1,1)$.
Now instead of having a uniform color for all atoms as in  $\mathbb{C}^3$, the color of a box (i.e.\ atom) in the plane partition is defined as   
\begin{equation}\label{eq:coloring_P2}
	{\bf color}(\Box)=\left({\bf slice}(\Box)\;{\rm mod}\;3\right)+1\,,
\end{equation}
where $\bf{slice}(\Box)$ counts which slice the $\Box$ is in, given by
\begin{equation}
 {\bf slice}(\Box) =x_1+x_2+x_3\quad \textrm{for} \quad \Box \textrm{ at coordinate $(x_1,x_2,x_3)$}\,.
\end{equation}
In this convention the box at the origin of the plane partition has coordinate $(0,0,0)$ and has ${\bf color}(\Box)=1$.  

\bigskip
	
From the period quiver \eqref{periodicquiver_P2}, we can read off the bond factors to be	
\begin{equation}\label{eq-charge-function-C3Z3-re}
\begin{aligned}
 &\varphi^{a\Rightarrow a}(u)= 1\,, \\
 &\varphi^{a\Rightarrow a+1}(u)= \frac{1}{\prod_{i=1,2,3} \left(u-\mathsf{h}_i\right)} \equiv \varphi_{-}(u)\,, \\
 & \varphi^{a\Rightarrow a-1}(u)= \prod_{i=1,2,3} \left(u+\mathsf{h}_i\right) \equiv \varphi_{+}(u)\;,
\end{aligned}
\end{equation}
which gives the unshifted quiver Yangian
\begin{equation}\label{eq-OPE-C3Z3}
\begin{aligned}
&\textrm{OPE:}\quad\begin{cases}\begin{aligned}
\psi^{(a)}(z)\, \psi^{(b)}(w)&= \psi^{(b)}(w)\, \psi^{(a)}(z)\;,\\
 \psi^{(a)}(z)\, e^{(a)}(w)   &\simeq   e^{(a)}(w)\, \psi^{(a)}(z) \;,\\ 
e^{(a)}(z)\, e^{(a)}(w) & \sim - e^{(a)}(w)\, e^{(a)}(z) \;,\\
\psi^{(a)}(z)\, f^{(a)}(w) &  \simeq f^{(a)}(w)\, \psi^{(a)}(z) \;,\\
 f^{(a)}(z)\, f^{(a)}(w) &  \sim  - f^{(a)}(w)\, f^{(a)}(z) \\
 \psi^{(a\pm 1)}(z)\, e^{(a)}(w)   &\simeq \varphi^{a\Rightarrow a\pm 1}(\Delta) \, e^{(a)}(w)\, \psi^{(a\pm 1)}(z) \;,\\ 
e^{(a+1)}(z)\, e^{(a)}(w) & \sim -  \varphi^{a\Rightarrow a+1}(\Delta)\, e^{(a)}(w)\, e^{(a+1)}(z)\;, \\
\psi^{(a\pm 1)}(z)\, f^{(a)}(w) &  \simeq  \varphi^{a\Rightarrow a\pm 1}(\Delta)^{-1}\, f^{(a)}(w)\, \psi^{(a\pm 1)}(z) \;,\\
 f^{(a+1)}(z)\, f^{(a)}(w) &  \sim - \varphi^{a\Rightarrow a+1}(\Delta)^{-1}\,f^{(a)}(w)\, f^{(a+1)}(z)\;,\\
\{e^{(a)}(z)\,, f^{(b)}(w)\}  &\sim  - \delta^{a,b}\, \frac{\psi^{(a)}(z) - \psi^{(b)}(w)}{z-w} \;,
\end{aligned}
\end{cases}
\end{aligned}
\end{equation}
\noindent where $a=1,2,3 \in \mathbb{Z}_3$.

\subsubsection{\texorpdfstring{Wall-crossing in $K_{\IP^2}$}{Wall-Crossing in K(P(2))}}
	
As described in~\cite{Aganagic:2010qr}, turning on a strong $B$-filed results in wall-crossing. 
The new BPS states analogous to the case of wall-crossing in the conifold geometry can be described by the molten crystal configurations from certain subcrystals of the canonical crystal.

The subcrystal $\CC_{m}$ that corresponds to the chamber $m$ is given by removing the first $m$ layers of atoms from the canonical crystal, resulting in a truncated triangular pyramid:
	\begin{equation}
		\begin{array}{c}
			\begin{tikzpicture}
				\draw[thick] (0,0) -- (-1,-2) (0,0) -- (0.5,-2) (0,0) -- (1,-1.5);
				\draw[dashed] (-0.5,-1) -- (0.25,-1) -- (0.466667,-0.7) -- cycle;
				\begin{scope}[shift={(4,0)}]
					\draw[thick] (-0.5,-1) -- (-1,-2) (0.25,-1) -- (0.5,-2) (0.466667,-0.7) -- (1,-1.5);
					\draw (-0.5,-1) -- (0.25,-1) -- (0.466667,-0.7) -- cycle;
					\begin{scope}[shift={(0,-0.25)}]
					\draw[blue] (-0.5,-0.75) to [out=90,in=170] (-0.3125, -0.5) to[out=0,in=270] (-0.125, -0.25) to[out=270,in=180] (0.0625, -0.5) to[out=0,in=90] (0.25,-0.75);
					\node[above,blue] at (-0.125, -0.25) {$m$ atoms};
				\end{scope}
				\end{scope}
			\draw[<->] (1.5,-1) -- (2.5,-1);
			\end{tikzpicture}
		\end{array}
	\end{equation}
The top section plane is the right triangle with $m$ atoms in an edge row. 
	
	Consider two atom layers on the top of the truncated triangular pyramid $\CC_m$:
	\begin{equation}\label{diagram_P2}
	\begin{array}{c}
	\includegraphics[scale=0.3,trim = 300 100 0 0]{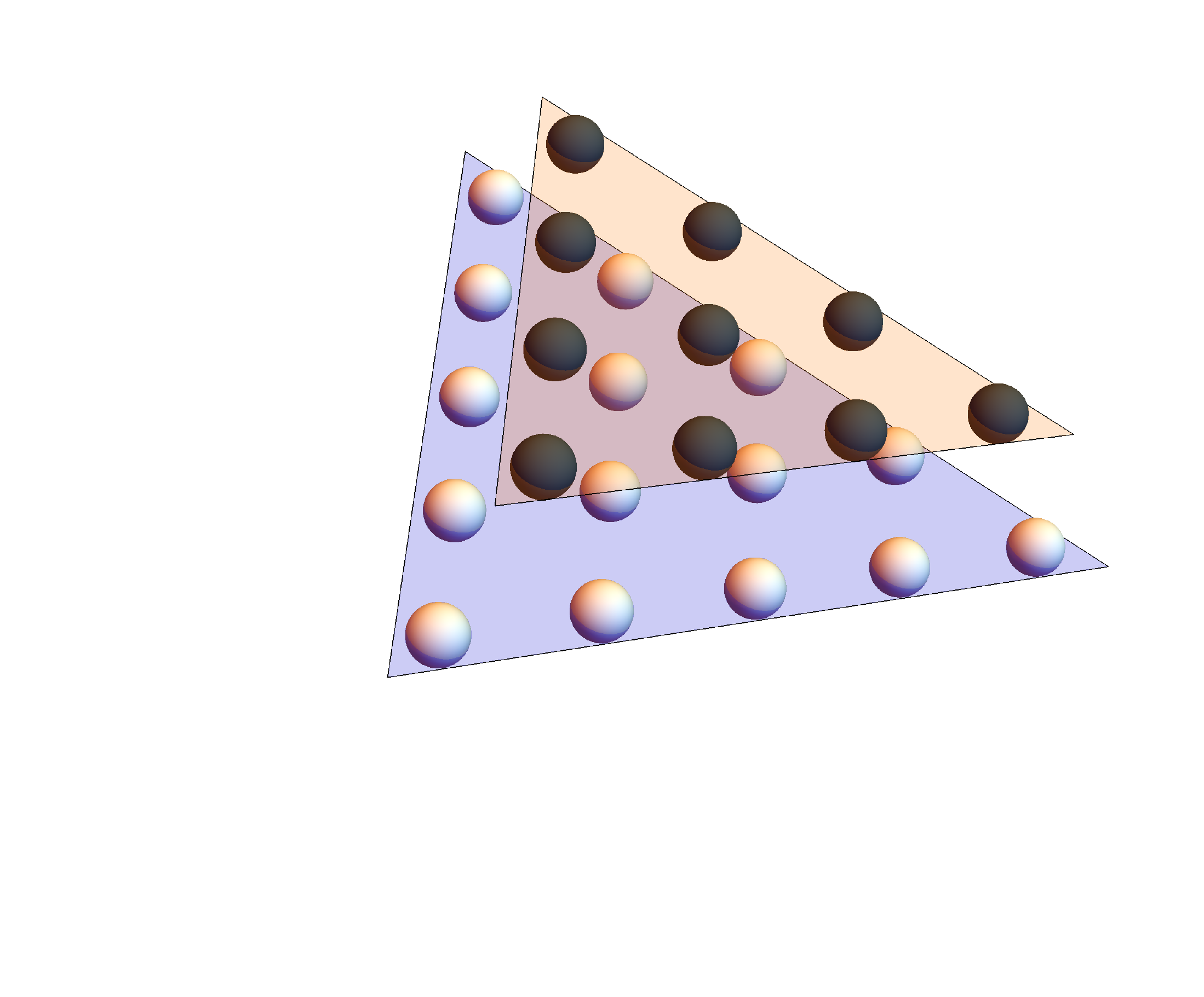}
	\end{array}\quad \begin{array}{c}
	\begin{tikzpicture}[scale=0.6]
		\foreach \x in {0,...,3}
		\foreach \y in {0,...,3}
		{
			\pgfmathparse{int(\x+\y-2)}
			\let\r\pgfmathresult
			\ifnum \r > 0
			\breakforeach
			\fi
			\draw (1.73205 * \x + 0.866025 * \y, 1.5 * \y) -- (-0.866025 + 1.73205 * \x + 0.866025 * \y, -0.5 + 1.5 * \y);
			\draw (1.73205 * \x + 0.866025 * \y, 1.5 * \y) -- (0.866025 + 1.73205 * \x + 0.866025 * \y, -0.5 + 1.5 * \y);
			\draw (1.73205 * \x + 0.866025 * \y, 1.5 * \y) -- (1.73205 * \x + 0.866025 * \y, 1. + 1.5 * \y);
			\draw[fill=black] (1.73205 * \x + 0.866025 * \y, 1.5 * \y) circle (0.3);
		}
		\foreach \x in {0,...,1}
		\foreach \y in {0,...,1}
		{
			\pgfmathparse{int(\x+\y)}
			\let\r\pgfmathresult
			\ifnum \r > 0
			\breakforeach
			\fi
			\draw[fill=red] (1.73205 + 1.73205 * \x + 0.866025 * \y, 1. + 1.5 * \y) circle (0.3);
		}
		\foreach \x in {1,...,3}
		{
			\draw[fill=blue] (-0.866025 + 1.73205 * \x, -0.5) circle (0.3);
		}
		\foreach \x in {1,...,3}
		{
			\draw[fill=blue] (-0.866025 + 0.866025 * \x, -0.5 + 1.5 * \x) circle (0.3);
		}
		\foreach \x in {1,...,3}
		{
			\draw[fill=blue] (2.59808 + 0.866025 * \x, 5.5 - 1.5 * \x) circle (0.3);
		}
		\draw[fill=black!40!green] (-0.866025, -0.5) circle (0.3) (6.06218, -0.5) circle (0.3) (2.59808, 5.5) circle (0.3); 
	\end{tikzpicture}
	\end{array}\quad
	\begin{array}{c}
		\begin{tikzpicture}
			\draw (0,0) circle (0.15) (-0.866025, -0.5) circle (0.15) (0.866025, -0.5) circle (0.15) (0., 1.) circle (0.15);
			\draw[->] (-0.129904, -0.075) -- (-0.736122, -0.425);
			\draw[->] (0.129904, -0.075) -- (0.736122, -0.425);
			\draw[->] (0., 0.15) -- (0., 0.85);
			\node[above left] at (-0.433013, -0.25) {$\mathsf{h}_1$};
			\node[below] at (0.433013, -0.25) {$\mathsf{h}_2$};
			\node[right] at (0,0.5) {$\mathsf{h}_3$};
		\end{tikzpicture}
	\end{array}
	\end{equation}
	We mark atoms using the following color code: 
	\begin{itemize}
		\item[$\color{black} \blacksquare$] Atoms at layer 1 (depth 0) with coordinates:
		\begin{equation}
		\rho_{\alpha,\beta}=(\mathsf{h}_3-\mathsf{h}_1)\alpha+(\mathsf{h}_3-\mathsf{h}_2)\beta \;, \quad \alpha,\beta=0,\ldots,m-1;\;\alpha+\beta\leq m-1 \;. 
		\end{equation}
		\item[$\color{red} \blacksquare$]  Atoms at layer 1 (depth 0) with coordinates:
		\begin{equation}
		\sigma_{\alpha,\beta}=(\mathsf{h}_3-\mathsf{h}_1)\alpha+(\mathsf{h}_3-\mathsf{h}_2)\beta+\mathsf{h}_1 \;,\quad \alpha,\beta=1,\ldots,m-1;\;\alpha+\beta\leq m-1\;.
		\end{equation}
		\item[$\color{blue} \blacksquare$] Atoms at layer 2 (depth 1) at the boundary edges of the triangular cross-section.
		\item[$\color{black!40!green} \blacksquare$]  Atoms at layer 2 (depth 1) at the tips of the triangular cross-section.
	\end{itemize}

We apply our standard methods to describe corresponding quivers and ground state charge functions associated with $\CC_m$. 
\begin{enumerate}
\item
The positive crystals are located at all the positions of atoms at depth $0$. 
They correspond to quiver arrows from the framing node to the node of the quiver, denoted by $R_{\alpha,\beta}$, and the corresponding masses are:
\begin{equation}
\mu\left(R_{\alpha,\beta}\right)=\rho_{\alpha,\beta}\;,\quad \quad \alpha,\beta=0,\ldots, m;\;\alpha+\beta\leq m-1\;.
\end{equation}
\item
In addition to the positive crystals, we have to add some negative crystals to cancel multiple contributions. 
These crystals are located at different positions of atoms of layer 2. 
They are associated to corresponding quiver arrows from a quiver node to the framing node. 
We denote these fields as $S_{\alpha, \beta, i}$, and the corresponding masses read:
\begin{equation}
\mu\left(S_{\alpha,\beta,i}\right)=-\sigma_{\alpha,\beta}\;,\quad \alpha,\beta=0,\ldots, m; \;\alpha+\beta\leq m\;.
\end{equation}
The resulting quiver has the following form:
\begin{equation}
Q_m=\begin{array}{c}
	\begin{tikzpicture}
		\begin{scope}[shift={(0.866025,0)}]
			\begin{scope}[shift={(-0.866025,0)}]
				\draw[thick,->] (0,-1.5) -- (0,-0.5);
				\draw[thick,->] (0,-0.5) -- (0,0);
				\draw[thick,->] (0,0) -- (0,0.5);
				\draw[thick] (0,0.5) -- (0,1.5);
			\end{scope}
			\begin{scope}[rotate=-120]
				\begin{scope}[shift={(-0.866025,0)}]
					\draw[thick,->] (0,-1.5) -- (0,-0.5);
					\draw[thick,->] (0,-0.5) -- (0,0);
					\draw[thick,->] (0,0) -- (0,0.5);
					\draw[thick] (0,0.5) -- (0,1.5);
				\end{scope}
			\end{scope}
			\begin{scope}[rotate=-240]
				\begin{scope}[shift={(-0.866025,0)}]
					\draw[thick,->] (0,-1.5) -- (0,-0.5);
					\draw[thick,->] (0,-0.5) -- (0,0);
					\draw[thick,->] (0,0) -- (0,0.5);
					\draw[thick] (0,0.5) -- (0,1.5);
				\end{scope}
			\end{scope}
		\end{scope}
		\begin{scope}[xscale=-1]
			\begin{scope}[shift={(0.866025,0)}]
				\begin{scope}[rotate=-120]
					\begin{scope}[shift={(-0.866025,0)}]
						\draw[thick,->] (0,-1.5) -- (0,-0.5);
						\draw[thick,->] (0,-0.5) -- (0,0);
						\draw[thick,->] (0,0) -- (0,0.5);
						\draw[thick] (0,0.5) -- (0,1.5);
					\end{scope}
				\end{scope}
				\begin{scope}[rotate=-240]
					\begin{scope}[shift={(-0.866025,0)}]
						\draw[thick,->] (0,-1.5) -- (0,-0.5);
						\draw[thick,->] (0,-0.5) -- (0,0);
						\draw[thick,->] (0,0) -- (0,0.5);
						\draw[thick] (0,0.5) -- (0,1.5);
					\end{scope}
				\end{scope}
			\end{scope}
		\end{scope}
		\draw[fill=white] (0,-1.5) circle (0.2);
		\draw[fill=white] (0,1.5) circle (0.2);
		\draw[fill=white] (2.59808,0) circle (0.2);
		\begin{scope}[shift={(-2.59808,0)}]
			\draw[fill=red] (-0.15,-0.15) -- (-0.15,0.15) -- (0.15,0.15) -- (0.15,-0.15) -- cycle;
		\end{scope}
		\node[below] at (0,-1.7) {$1$};
		\node[above] at (0,1.7) {$2$};
		\node[right] at (2.79808,0) {$3$};
		\node[left] at (0,0) {$X_i^{(1)}$};
		\node[above left] at (-1.29904,0.75) {$S_{\alpha,\beta,i}$};
		\node[above right] at (1.29904,0.75) {$X_i^{(2)}$};
		\node[below right] at (1.29904,-0.75) {$X_i^{(3)}$};
		\node[below left] at (-1.29904,-0.75) {$R_{\alpha,\beta}$};
	\end{tikzpicture}
\end{array}
\end{equation}
The cardinality of the index $i$ in fields $S_{\alpha,\beta,i}$ reflects the multiplicity of negative crystal contributions. This multiplicity is 2 for atoms lying on the face of the atom layer 2 marked with the red color in diagram \eqref{diagram_P2}, 1 for boundary atoms marked with the blue color in \eqref{diagram_P2} and 0 for vertex atoms marked with the green color:
\begin{equation}
	i=1,\ldots,2-\delta_{\alpha,m}-\delta_{\beta,m}-\delta_{\alpha+\beta,m} \;.
\end{equation}
\item
The fields  $S_{\alpha,\beta,i}$ are Lagrange multipliers enforcing additional relations to the quiver path algebra. The number of such relations and the cardinality   are given by the number of nearest neighbors to the corresponding atom at layer 1. 
The corresponding superpotential reads:
\begin{equation}
\begin{split}
	&W_m=\sum_{i,j,k=1}^3 \varepsilon^{ijk}\textrm{Tr}(X^{(1)}_i X^{(2)}_j X^{(3)}_k)\\
	&+\sum\lm_{\alpha=1}^{m-1}\Tr\left(S_{\alpha,0,1}\left(X_1^{(1)}R_{\alpha,0}-X_2^{(1)}R_{\alpha-1,0}\right)\right)\\
	&+\sum\lm_{\alpha=1}^{m-1}\Tr\left(S_{0,\alpha,1}\left(X_1^{(1)}R_{0,\alpha}-X_3^{(1)}R_{0,\alpha-1}\right)\right)\\
	&+\sum\lm_{\alpha=1}^{m-1}\Tr\left(S_{\alpha,m-\alpha,1}\left(X_2^{(1)}R_{\alpha-1,m-\alpha}-X_3^{(1)}R_{\alpha,m-\alpha-1}\right)\right)\\
	&+\scalebox{0.8}{$\displaystyle\sum\lm_{\substack{\alpha,\beta=1\\ \alpha+\beta\leq m-1}}^{m-1}\Tr\left(S_{\alpha,\beta,1}\left(X_1^{(1)}R_{\alpha,\beta}-X_2^{(1)}R_{\alpha-1,\beta}\right)+S_{\alpha,\beta,2}\left(X_1^{(1)}R_{\alpha,\beta}-X_3^{(1)}R_{\alpha,\beta-1}\right)\right)$}\,.
\end{split}
\end{equation}
\end{enumerate}

Summarizing, we derive the following ground state charge functions for the chamber $\CC_m$:
\begin{equation}
    \begin{split}
    &{}^{\sharp}\psi^{(1)}_0(z)=\frac{1}{\prod\lm_{\alpha=0}^{m-1}\prod\lm_{\beta=0}^{m-1-\alpha}\left(z-\rho_{\alpha,\beta}\right)}\;,\\
    &{}^{\sharp}\psi^{(2)}_0(z)=\prod\lm_{\alpha=1}^{m-1}\prod\lm_{\beta=1}^{m-1-\alpha}\left(z-\sigma_{\alpha,\beta}\right)^2\times\prod\lm_{\gamma=1}^{m-1}\left(z-\sigma_{\gamma,0}\right)\left(z-\sigma_{0,\gamma}\right)\left(z-\sigma_{\gamma,m-\gamma}\right)\;,\\
    &{}^{\sharp}\psi^{(3)}_0(z)=1\;,
    \end{split}
\end{equation}
which correspond to the shifts:
\begin{equation}
\mys^{(1)}= \frac{m(m+1)}{2} \;, \quad \mys^{(2)}=-\left(m^2-1\right)\;, \quad  \mys^{(3)} =0 \;.
\end{equation}

\bigskip

Finally, we mention that although we have focused on the infinite chambers $\CC^{(i)}_m$ for the $K_{\IP^2}$ geometry, it is also easy to study the finite chamber $\CC^{(f)}_m$.
Recall that for the conifold case, the finite chamber $\CC^{(f)}_m$ can be obtained by placing two stoppers at the end of the ridge of the $\CC^{(i)}_m$ pyramid (with the same $m$).
Similarly for the $K_{\IP^2}$ geometry, a subcrystal $\CC^{(f)}_m$  for the finite chamber can be obtained by placing three stoppers along the three edges $x_i$ with $i=1,2,3$ and on the same slice.
It is straightforward to write down the corresponding framed quiver and superpotential, the ground state charge function, and the shift for the corresponding finite representation. 

\section{Open BPS states}
\label{sec:Open}

One of the simplest types of representations of the shifted quiver Yangian $\mathsf{Y}(Q,W,\mys)$ counts the open BPS states, i.e.\ the open DT invariants (see e.g.\ \cite{Okounkov:2003sp,Nagao:2009ky,Nagao:2009rq,Yamazaki:2010fz,Sulkowski:2010eg} for molten crystal representations for open DT invariants). In this section, we explain how to use our crystal/quiver construction to characterize them. 

	A growing crystal approaches asymptotically the corresponding $(p, q)$-web of the toric Calabi-Yau geometry \cite{Ooguri:2008yb}. 
This statement can be made precise in terms of amoebae in a dimer model for crystal growth \cite{Kenyon:2003uj,Ooguri:2009ri}. 
Representations whose characters are generating functions for the corresponding open DT invariants have a non-trivial asymptotic behavior at the toric diagram legs given by 2D crystals. 
In what follows we will concentrate on the case of $\IC^3$ where this phenomenon occurs first.\footnote{Being constructed from multiple positive and negative contributions, crystals for open BPS states are rigid. This implies that the only free parameter is a crystal ``center of mass" allowing one to translate the crystal projection in the $(\mathsf{h}_1,\mathsf{h}_2)$-plane as a whole. Poles in the vacuum charge functions -- weights of arrows flowing from the framing node -- are originally unconstrained. Constraints on the positions of the poles are defined through the superpotential, each zero -- a weight for an arrow flowing towards the framing node -- delivers a single constraint.
Therefore a relation between the number of variables and constraints defines a relation between the net numbers of positive and negative shifts:
$$
(\mbox{net }\#\mbox{ of poles})-(\mbox{net }\#\mbox{ of zeroes})=\sum\lm_a \mys_+^{(a)}-\sum\lm_a \mys_-^{(a)}=1.
$$}

\subsection{\texorpdfstring{Example: open BPS states in $\mathbb{C}^3$}{Example: Open BPS States in C(3)}}
\label{ssec:Open_C3}

Let us start by considering the simplest example, $\mathbb{C}^3$.	
The toric diagram and its dual graph of $\mathbb{C}^3$ are
\begin{equation}\label{fig-toric-C3}
\begin{tikzpicture} 
\filldraw [red] (0,0) circle (2pt); 
\filldraw [red] (0,1) circle (2pt); 
\filldraw [red] (1,0) circle (2pt); 
\node at (-0.5,-0.5) {(0,0)}; 
\node at (-0.5,1.5) {(0,1)}; 
\node at (1.5,-0.5) {(1,0)}; 
\draw (0,0) -- (0,1); 
\draw (0,0) -- (1,0); 
\draw (1,0) -- (0,1); 
\end{tikzpicture}
\qquad \qquad \qquad
\begin{tikzpicture} 
\draw[->] (0,0) -- (-1,0); 
\draw[->] (0,0) -- (0,-1); 
\draw[->] (0,0) -- (1,1); 
\node at (-1.5,0) {3}; 
\node at (0,-1.5) {1}; 
\node at (1.5,1.5) {2}; 
\end{tikzpicture}
\quad \;.
\end{equation}

\subsubsection{Review: canonical crystal \texorpdfstring{$\mathcal{C}_0$}{C0} and affine Yangian of \texorpdfstring{$\mathfrak{gl}_1$}{gl1}}

The canonically framed quiver-superpotential pair $(Q_0,W_0)$ is	
\begin{equation}\label{eq:QW0C3}
\begin{aligned}
Q_0&=\begin{array}{c}
		\begin{tikzpicture}
			\begin{scope}[rotate=-90]
			\draw[thick,->] ([shift=(120:0.75)]0,0.75) arc (120:420:0.75);
			\draw[thick,->] ([shift=(60:0.75)]0,0.75) arc (60:90:0.75);
			\draw[thick,->] ([shift=(90:0.75)]0,0.75) arc (90:120:0.75);
			\draw[fill=white] (0,0) circle (0.2);
			\draw[thick,->] (0.0,-1.35) -- (0.0,-0.75);
			\draw[thick] (0.0,-0.75)  -- (0.0,-0.173205);
			\begin{scope}[shift={(0,-1.5)}]
				\draw[thick,fill=red] (-0.15,-0.15) -- (-0.15,0.15) -- (0.15,0.15) -- (0.15,-0.15) -- cycle;
			\end{scope}
			\node[right] at (0,1.5) {$B_{1,2,3}$};
			\node[below] at (0.15,-0.75) {$R$};
			\end{scope}
		\end{tikzpicture}
	\end{array},\quad 
	\\
W_0&=\Tr\,B_1[B_2,B_3]\,.
\end{aligned}
\end{equation}
The periodic quiver corresponding to $(Q_0, W_0)$ in~\eqref{eq:QW0C3} is
\begin{equation}\label{fig-periodic-quiver-C3}
\begin{aligned}
&
\begin{tikzpicture}[scale=1]
\filldraw[mygreen] (-1,-1.73205)--(1,-1.73205)  -- (2,0)--(0,0)--cycle; 
\node[state]  [regular polygon, regular polygon sides=4, draw=blue!50, very thick, fill=blue!10] (a1) at (0,0)  {$1$};
\node[state]  [regular polygon, regular polygon sides=4, regular polygon, regular polygon sides=4, draw=blue!50, very thick, fill=blue!10] (a21) at (-1,-1.73205)  {$1$};
\node[state]  [regular polygon, regular polygon sides=4, draw=blue!50, very thick, fill=blue!10] (a22) at (2,0)  {$1$};
\node[state]  [regular polygon, regular polygon sides=4, draw=blue!50, very thick, fill=blue!10] (a312) at (1,-1.73205)  {$1$};
\path[->] 
(a1) edge   [thick, red]   node [left] {$\mathsf{h}_1$} (a21)
(a22) edge   [thick, red]   node [right] {$\mathsf{h}_1$} (a312)
(a1) edge   [thick, red]   node [above] {$\mathsf{h}_2$} (a22)
(a21) edge   [thick, red]   node [below] {$\mathsf{h}_2$} (a312)
(a312) edge   [thick, red]   node [right] {$\mathsf{h}_3$} (a1)
; 
\end{tikzpicture}
\end{aligned}
\end{equation}
where the fundamental region of the torus is shown as a shaded region, and the parameter $\mathsf{h}_i$ is the charge associated to the field $B_i$ with $i=1,2,3$.
The loop constraint~\eqref{eq:loop-constraint} gives
\begin{equation}\label{eq-loop-constraint-C3}
\mathsf{h}_1+\mathsf{h}_2+\mathsf{h}_3=0 \,,
\end{equation}
which is equivalent to the vertex constraint~\eqref{eq-vertex-constraint-toric} in this case.

From the periodic quiver~\eqref{fig-periodic-quiver-C3} one can construct the canonical crystal $\mathcal{C}_0$, which has the shape of a single 3D octant. 
Each state in this representation is labeled by a plane partition inside the octant. 
The three edges of this 3D octant, and of each plane partition in it, can be labeled by the positive directions of $x$-, $y$- and $z$-axes, whose projections down to 2D correspond precisely to the three legs in the dual graph of the toric diagram~\eqref{fig-toric-C3}.
\bigskip

The canonical crystal $\mathcal{C}_0$ gives the vacuum representation of the unshifted quiver Yangian.
For $\mathbb{C}^3$, there is only one bond factor
\eqref{eq-charge-atob}, which can be read off from the periodic quiver~\eqref{fig-periodic-quiver-C3} to be
\begin{equation}\label{eq-algebraM-C3}
\varphi^{1\Rightarrow 1}(u)=\varphi_3(u)=\frac{(u+\mathsf{h}_1)(u+\mathsf{h}_2)(u+\mathsf{h}_3)}{(u-\mathsf{h}_1)(u-\mathsf{h}_2)(u-\mathsf{h}_3)} \;,
\end{equation}
subject to the loop constraint~\eqref{eq-loop-constraint-C3}.
Plugging this into the general formulae for the OPE relations~\eqref{eq-OPE-toric} and the initial conditions~\eqref{eq-psi-e-f-initial-0} and~\eqref{eq-psi-e-f-initial-1}, and supplementing them with Serre relations, we have the full list of algebra relations of the affine Yangian of $\mathfrak{gl}_1$:
\begin{equation}
\begin{aligned}
&\textrm{OPE:}\quad\begin{cases}\begin{aligned}
&\psi(z)\psi(w)\sim \psi(w)\psi(z) \;,\\
&\begin{aligned}
\psi(z)\, e(w) &  \sim  \varphi_3(\Delta)\, e(w)\, \psi(z)  \;,\\ 
\psi(z)\, f(w) & \sim \varphi_3^{-1}(\Delta)\, f(w)\, \psi(z)  \;,
\end{aligned}
\qquad
\begin{aligned}
e(z)\, e(w) & \sim    \varphi_3(\Delta)\, e(w)\, e(z)  \;,\\
 f(z)\, f(w) &  \sim    \varphi_3^{-1}(\Delta)\, f(w)\, f(z)  \;,
\end{aligned}\\
&[e(z)\,, f(w)]   \sim  - \, \frac{\psi(z) - \psi(w)}{z-w}  \;,
\end{aligned}
\end{cases}\label{box-OPE-gl1-our}
\\
&\textrm{Initial:}\quad\begin{cases}
\begin{aligned}
&\begin{aligned}
&[\psi_0,e_m] = 0 \;,\\
&[\psi_0,f_m] = 0 \;,
\end{aligned}\qquad \qquad
\begin{aligned}
&[\psi_1,e_m] = 0 \;,\\
&[\psi_1,f_m] = 0 \;,
\end{aligned}\qquad \qquad
\begin{aligned}
&[\psi_2,e_m] = 2 \, \sigma_3\, e_m \;,\\
&[\psi_2,f_m] = - 2  \,\sigma_3\, f_m \;,
\end{aligned}
\end{aligned}
\end{cases}
\\
&\textrm{Serre}:\quad\begin{cases}\begin{aligned}
&\textrm{Sym}_{z_1,z_2,z_3}\, (z_2-z_3)\,\left[ e(z_1)\,, \left[ e(z_2)\,, e(z_3)\right] \right]\sim 0  \;,\\
&\textrm{Sym}_{z_1,z_2,z_3}\, (z_2-z_3)\left[ f(z_1)\,, \left[ f(z_2)\,,f(z_3)\right]\right] \sim 0  \;,
\end{aligned}
\end{cases}
\end{aligned}
\end{equation}
\noindent where $\sigma_3\equiv \mathsf{h}_1 \mathsf{h}_2 \mathsf{h}_3$. 
It is straightforward to write down the relations in terms of modes, following (\ref{eq-OPE-modes-toric}).

\subsubsection{\texorpdfstring{Subcrystal $\mathcal{C}_{\vec{Y}}$}{Subcrystal C(Y)}}

\begin{figure}[ht!]
	\begin{center}
		\begin{tikzpicture}
			\node at (0,0) {\includegraphics*[scale=0.3]{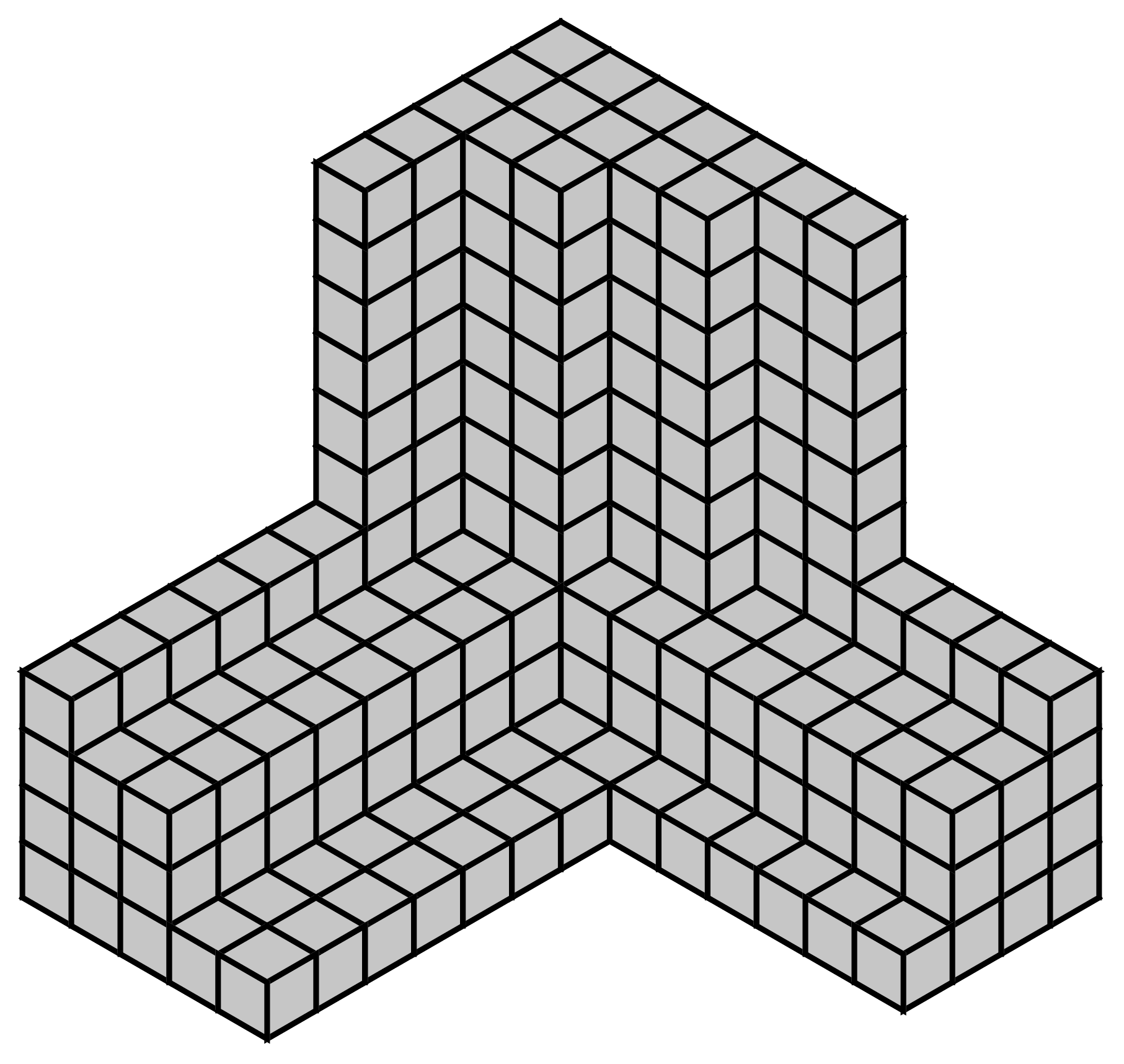}};
			\node at (7,0) {\includegraphics*[scale=0.3]{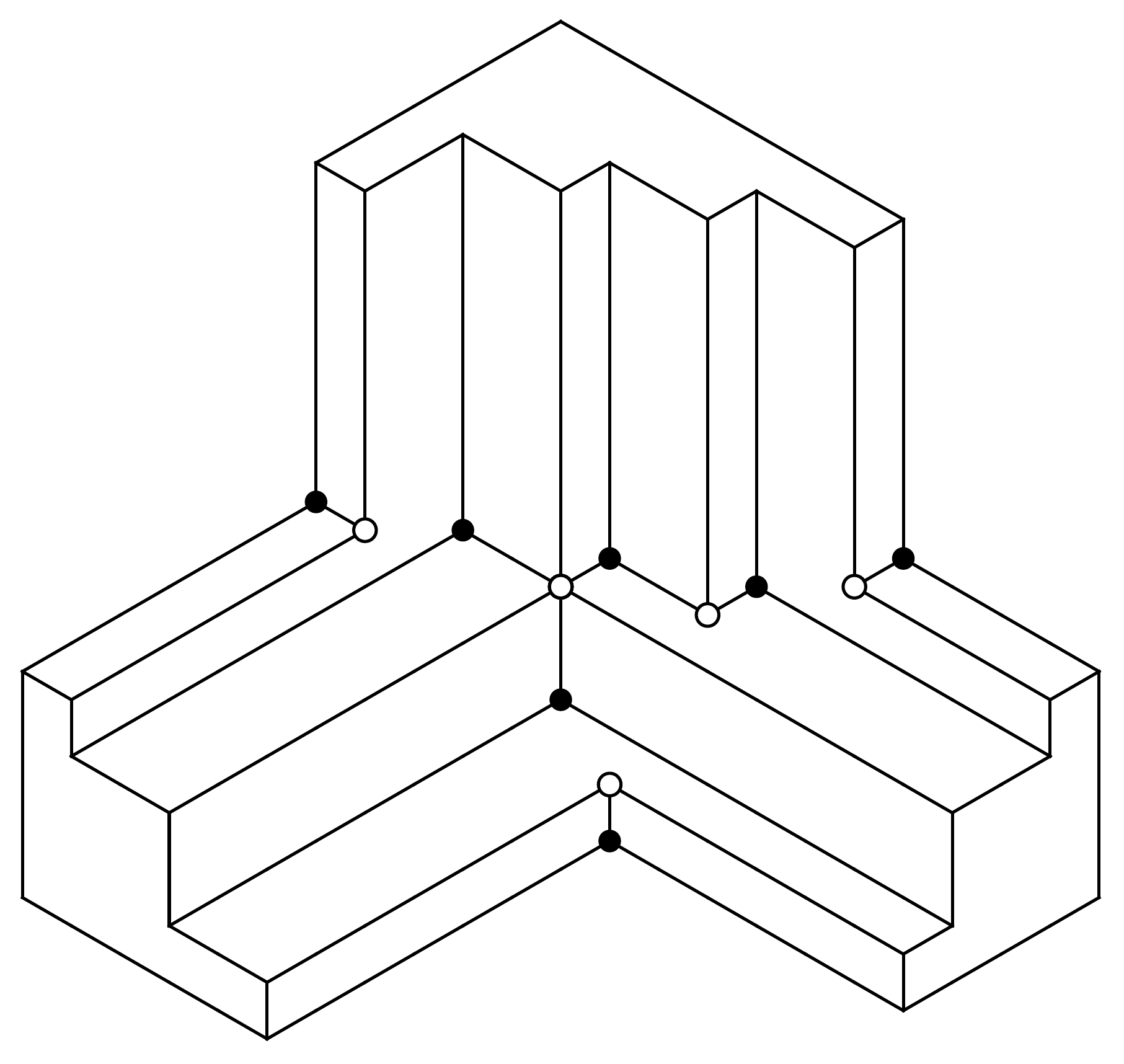}};
			\node at (0,-3.5) {(a)};
			\node at (7,-3.5) {(b)};
			\draw (3.5,-3.5) -- (3.5,3.5);
		\end{tikzpicture}
\caption{For plane partition representation with asymptotics $Y_1=\{4,3,3,1,1\}$, $Y_2=\{4,3,3,1\}$, $Y_3=\{7,5,3,1,1\}$, the ground state configuration (a) and the graph $\mathscr{D}_{\vec Y}$ (b).}
\label{fig:Lowest_tpriplet}
\end{center}
\end{figure}
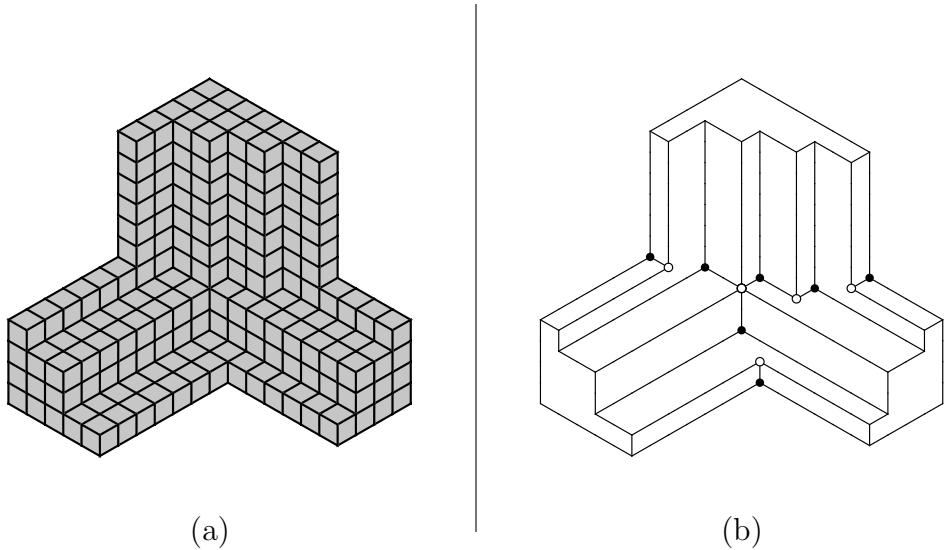

For the canonical crystal $\mathcal{C}_0$, all the plane partitions have trivial asymptotics along the three directions. 
A general representation corresponding to the open BPS states
is labeled by a triplet of Young diagrams 
\begin{equation}\vec{Y}=(Y_1,Y_2,Y_3)
\end{equation} along the three directions. 
The shape of the sub-crystal ${}^{\sharp} \mathcal{C} =\mathcal{C}_{\vec{Y}}$ can be obtained in the following steps.
First, a triplet of Young diagrams $\vec{Y}$ uniquely determines the \emph{minimal configuration} of the plane partition with $\vec{Y}$ as asymptotics, e.g.\ see Figure~\ref{fig:Lowest_tpriplet}(a) for the minimal plane partition with $Y_1=\{4,3,3,1,1\}$, $Y_2=\{4,3,3,1\}$, $Y_3=\{7,5,3,1,1\}$ as asymptotics.
Since it is minimal, it would be killed by the annihilation operator $f(z)$, namely, this minimal plane partition is the ground state of the representation labeled by $\vec{Y}$.
Second, the subcrystal $\mathcal{C}_{\vec{Y}}$ is the complement of this ground state plane partition in the canonical crystal $\mathcal{C}_0$, which is the whole octant in this case.
The excited states in this representation are obtained by applying creation operator $e(z)$, i.e.\ adding more boxes, on the ground state plane partition. (Equivalently, this can also be viewed as melting away atoms from $\mathcal{C}_{\vec{Y}}$.)
By construction, all the plane partitions in the representation $\mathcal{C}_{\vec{Y}}$ have the same asymptotics $\vec{Y}$.

\subsubsection{\texorpdfstring{Ground state charge function $\psi_{\vec{Y}}(z)$}{Ground State Charge Function psi(Y)(z)}}

Let us now translate the shape of the subcrystal $\mathcal{C}_{\vec{Y}}$ into the corresponding ground state charge function $\psi_{\vec{Y}}(z)$.

To achieve this, we first apply the method of Section 3.2 to reconstruct $\mathcal{C}_{\vec{Y}}$ as a union of positive and negative canonical crystals $\mathcal{C}_0$.
Namely, for a given $\mathcal{C}_{\vec{Y}}$, we first determine the minimal plane partition configuration with $\vec{Y}$ as asymptotics. 
Then starting from this minimal plane partition, the atoms one can place at its  convex and concave corners are the origins of the positive and negative crystals, respectively.
Finally, by the reasoning of Section 3.2, the coordinate functions of these convex and concave atoms then corresponds to the poles and zeros of the ground state charge function $\psi_{\vec{Y}}(z)$.

Since the coordinate function of an atom is independent of its depth, one can also carry out the procedure above in the 2D projection. 
Namely, the three asymptotic Young diagrams $\vec{Y}=(Y_1,Y_2,Y_3)$ and the corresponding minimal plane partition can be uniquely characterized by an oriented graph drawn on the 2D hexagon lattice, denoted as $\mathscr{D}_{\vec Y}$.
The advantage of using the 2D graph $\mathscr{D}_{\vec Y}$ is that the coordinate functions of the atoms at the convex and concave corners of the minimal plane partition correspond directly to various intersection points of the external oriented lines (which are fixed by $\vec{Y}$) and some internal lines that are added to complete the graph.
We now explain the procedure of determining the graph $\mathscr{D}_{\vec Y}$ and the associated ground state charge function $\psi_{\vec{Y}}(z)$ in the following steps.
\begin{enumerate}
\item The coordinate system of the 2D crystal lattice can be specified by the length and orientations of the links in the lattice, which we choose as follows
\be
	\begin{array}{c}
		\begin{tikzpicture}
			\path (-0.866025,0) edge[->] node[below left] {$\vec \ell_2$} (0,-0.5) (0,0.5) edge[->] node[above left] {$\vec \ell_1$} (-0.866025,0) (0,-0.5) edge[->] node[right] {$\vec \ell_3$} (0,0.5);
		\end{tikzpicture}
	\end{array}\,
\ee
where $\vec{\ell}_{1,2,3}$ are three 2D vectors that obey $\vec{\ell}_1+\vec{\ell}_2+\vec{\ell}_3=0$.
The position of a vertex  $\mathbf{v}$ in the 2D lattice can be decomposed into
\begin{equation}\label{eq:vecxC3}
\vec{x}(\textbf{v})=\sum n_i\, \vec{\ell}_{i} \quad \textrm{with}\quad n_i>0\,.
\end{equation}
Each vertex in $\mathscr{D}_{\vec Y}$ corresponds to a tower of atoms in the crystal, with depth from $0$ to $\infty$.
But they all have the same coordinate function, namely, for a vertex at the position~\eqref{eq:vecxC3}, all the atoms that project to it share the coordinate function
\begin{equation}
h(\square)=\sum n_i\, \mathsf{h}_{i} \quad \textrm{with}\quad n_i>0\,.
\end{equation} 
One can define the coordinate of a vertex in the 2D lattice  to be 
\begin{equation}\label{eq:vertexdistance}
x(\textbf{v})\equiv\vec{x}(\textbf{v})\cdot (\mathsf{h}_1,\mathsf{h}_2,\mathsf{h}_3)=\sum n_i\, \mathsf{h}_{i} \,,
\end{equation}
for $\vec{x}(\textbf{v})=\sum n_i\, \vec{\ell}_{i}$, with $n_i>0\,$, 
coinciding with the coordinate function of the atoms that project to it. 

\item For a given $\vec{Y}$, we start with three semi-infinite long rows along the three positive directions $x_i$ with $i=1,2,3$, each with a cross section being the corresponding Young diagram $Y_i$.

The 2D projection of these three rows give rise to three ``beams" of external oriented lines parallel to $\ell_i$ respectively. 
The orientation and position of each line in the $i^{\textrm{th}}$ beam are determined by the Young diagram $Y_i$.
First, within each beam, the directions of the external lines alternate, starting and ending with an outgoing arrow:
\begin{equation}\label{fig:2Dgraph}
		\begin{array}{c}
			\begin{tikzpicture}[scale=0.8]
				\begin{scope}[shift={(0,0.866025)}]
					\draw[thick] (-1.5,0) -- (1.5,0);
					\draw[->] (-1,0) -- (-1,2);
					\draw[dashed] (-0.5,0) -- (-0.5,2);
					\draw[dashed] (0,0) -- (0,2);
					\draw[dashed] (0.5,0) -- (0.5,2);
					\draw[->] (1,0) -- (1,2);
					\node[above] at (0,2.2) {$Y_3$};
					\draw[dashed] (-1.2,1) -- (-1.2,2.1) -- (1.2,2.1) -- (1.2,1) -- cycle;
				\end{scope}
				\begin{scope}[rotate=120]
					\begin{scope}[shift={(0,0.866025)}]
						\draw[thick] (-1.5,0) -- (1.5,0);
						\draw[->] (-1,0) -- (-1,2);
						\draw[dashed] (-0.5,0) -- (-0.5,2);
						\draw[dashed] (0,0) -- (0,2);
						\draw[dashed] (0.5,0) -- (0.5,2);
						\draw[->] (1,0) -- (1,2);
						\node[below left] at (0,2) {$Y_1$};
					\end{scope}
				\end{scope}
				\begin{scope}[rotate=240]
					\begin{scope}[shift={(0,0.866025)}]
						\draw[thick] (-1.5,0) -- (1.5,0);
						\draw[->] (-1,0) -- (-1,2);
						\draw[dashed] (-0.5,0) -- (-0.5,2);
						\draw[dashed] (0,0) -- (0,2);
						\draw[dashed] (0.5,0) -- (0.5,2);
						\draw[->] (1,0) -- (1,2);
						\node[below right] at (0,2) {$Y_2$};
					\end{scope}
				\end{scope}
				\draw (1.2,2.36603) --  (5,3) (1.2,2.36603) --  (5,1);
				\begin{scope}[shift={(5,2)},scale=1.4]
					\draw (0,-1) -- (0,1) -- (4.5,1) -- (4.5,-1) -- cycle;
					\draw[->,thick] (0.5,-0.7) -- (0.5,0.7);
					\draw[<-,thick] (1,-0.7) -- (1,0.7);
					\draw[->,thick] (1.5,-0.7) -- (1.5,0.7);
					\draw[<-,thick] (2,-0.7) -- (2,0.7);
					\draw[->,thick] (2.5,-0.7) -- (2.5,0.7);
					\draw[dashed] (3,-0.7) -- (3,0.7);
					\draw[<-,thick] (3.5,-0.7) -- (3.5,0.7);
					\draw[->,thick] (4,-0.7) -- (4,0.7);
					\begin{scope}[shift={(0.5,0)}]
						\node[above] at (0.25,0) {$u_1$};
						\draw[<->] (0,0) -- (0.5,0);
					\end{scope}
					\begin{scope}[shift={(1,0)}]
						\node[above] at (0.25,0) {$v_1$};
						\draw[<->] (0,0) -- (0.5,0);
					\end{scope}
					\begin{scope}[shift={(1.5,0)}]
						\node[above] at (0.25,0) {$u_2$};
						\draw[<->] (0,0) -- (0.5,0);
					\end{scope}
					\begin{scope}[shift={(2,0)}]
						\node[above] at (0.25,0) {$v_2$};
						\draw[<->] (0,0) -- (0.5,0);
					\end{scope}
					\begin{scope}[shift={(3.5,0)}]
						\node[above] at (0.25,0) {$v_p$};
						\draw[<->] (0,0) -- (0.5,0);
					\end{scope}
				\end{scope}
				\begin{scope}[shift={(6,-2)}]
					\draw[thick] (0,0) -- (0,2) -- (1,2) -- (1,0.5) -- (2.5,0.5) -- (2.5,0) -- cycle;
					\draw[dashed] (0,-0.25) -- (3,-0.25);
					\draw[thick] (0,-0.5) -- (3.5,-0.5) -- (3.5,-1) -- (0,-1) -- cycle; 
					\node[above] at (0.5,2) {$u_1$};
					\node[left] at (1,1.25) {$v_1$};
					\node[above] at (1.75,0.5) {$u_2$};
					\node[right] at (2.5,0.25) {$v_2$};
					\node[right] at (3.5,-0.75) {$v_p$};
				\end{scope}
			\end{tikzpicture}
		\end{array}
\end{equation} 
Second, the distances between adjacent arrows are given by the hook lengths of the boundary cells of $Y_i$, namely, the lengths of the edges of  $Y_i$, see~\eqref{fig:2Dgraph}.
Finally, the distance between the first arrow of the $i^{\textrm{th}} $ beam and $\ell_i$ is given by the height of $Y_i$.

\item The three semi-infinite rows would intersect, and determine the minimal plane partition configuration fixed by $\vec{Y}$.
For a given $\vec{Y}$, one can draw the minimal plane partition directly in 3D in order to locate all the convex and concave corners.

Equivalently, in the 2D graph $\mathscr{D}_{\vec Y}$, one can let the external lines intersect, and the intersection points (i.e.\ the vertices in the internal part of $\mathscr{D}_{\vec Y}$) should correspond one-to-one to the corners of the minimal plane partition. 
To match all the corners, one needs to add some internal lines, in order that (1) for each vertex the arrows are either all incoming or all outgoing, and (2) all vertices in $\mathscr{D}_{\vec Y}$ are either valence-$3$ or valence-$6$.\footnote{This rule of adding internal lines is to match the vertices to the corners of the minimal plane partition.}

The resulting corners, or equivalently, the vertices in $\mathscr{D}_{\vec Y}$ fall into the following three types:
\begin{itemize}
\item Concave corner $\Longleftrightarrow$ outgoing valence-$3$ vertex (denoted as black vertex in $\mathscr{D}_{\vec Y}$).

In the decomposition of the subcrystal $\mathcal{C}_{\vec{Y}}$ into superpositions of positive/negative $\mathcal{C}_0$ crystals, a concave corner is a place where one can add an atom $\square$, which sits at the origin of a positive crystal.
Namely, each concave corner corresponds to a starter. 
We denote the set of starters for a given $\vec{Y}$ as $s(\vec{Y})$. 

Projecting to 2D, one can immediately see that a concave corner in the minimal plane partition corresponds to a vertex with three outgoing arrows in the following directions:
\begin{equation}
		\begin{array}{c}
			\begin{tikzpicture}[scale=0.6]
				\begin{scope}
					\draw[thick, ->] (0,0) -- (0,1);
				\end{scope}
				\begin{scope}[rotate=120]
					\draw[thick, ->] (0,0) -- (0,1);
				\end{scope}
				\begin{scope}[rotate=240]
					\draw[thick, ->] (0,0) -- (0,1);
				\end{scope}
				\draw[fill=black] (0,0) circle (0.15);
			\end{tikzpicture}
		\end{array}\,,
\end{equation}
and such a vertex is colored black in $\mathscr{D}_{\vec Y}$.
We denote the set of black vertices as $\mathbf{s}(\vec{Y})$, which maps one-to-one to the set of starters $s(\vec{Y})$.

For the example in Figure~\ref{fig:Lowest_tpriplet}, there are seven starters, at the positions
\begin{equation}\label{eq:starter_cor_C3}
(0,7,4),\; (1,5,3),\; (2,3,3),\; (3,1,3),\; (3,3,1),\; (4,5,0),\; (5,0,4)\;,
\end{equation}
with coordinate function $h(\square)=x(\mathbf{v})$ values:
\begin{equation}\label{eq:starter_C3}
\begin{split}
7 \mathsf{h}_2+4 \mathsf{h}_3\;,\; 4 \mathsf{h}_2+2 \mathsf{h}_3\;,\; \mathsf{h}_2+\mathsf{h}_3\;,\; 2 \mathsf{h}_1+2 \mathsf{h}_3\;,\\ 2 \mathsf{h}_1+2 \mathsf{h}_2\;,\; 4 \mathsf{h}_1+5 \mathsf{h}_2\;,\; 5 \mathsf{h}_1+4 \mathsf{h}_3\;.
\end{split}
\end{equation}

\item Half-convex corner $\Longleftrightarrow$ incoming valence-$3$ vertex (denoted as white vertex in $\mathscr{D}_{\vec Y}$).

A half-convex corner in the minimal plane partition is the intersection point of two positive crystals, and therefore corresponds to the origin of a negative crystal. 
This is a simple pauser, i.e.\ it gives rise to a simple zero in the ground state charge function. 
We denote the set of simple pausers for a given $\vec{Y}$ as $p_1(\vec{Y})$.

Projecting down to 2D, a half-convex corner maps to a vertex with three incoming arrows, and there are $6$ possible configurations for the arrows:
\begin{equation}
		\begin{array}{c}
			\begin{tikzpicture}[scale=0.6]
				\begin{scope}
					\draw[thick, <-] (0,0.15) -- (0,1);
				\end{scope}
				\begin{scope}[rotate=60]
					\draw[thick, <-] (0,0.15) -- (0,1);
				\end{scope}
				\begin{scope}[rotate=120]
					\draw[thick, <-] (0,0.15) -- (0,1);
				\end{scope}
				\begin{scope}[rotate=180]
					\draw[thick, <-,white] (0,0.15) -- (0,1);
				\end{scope}
				\begin{scope}[rotate=240]
					\draw[thick, <-,white] (0,0.15) -- (0,1);
				\end{scope}
				\begin{scope}[rotate=300]
					\draw[thick, <-,white] (0,0.15) -- (0,1);
				\end{scope}
				\draw[thick, fill=white] (0,0) circle (0.15);
			\end{tikzpicture}
		\end{array}\,,\quad
		\begin{array}{c}
			\begin{tikzpicture}[scale=0.6]
				\begin{scope}[rotate=60]
					\begin{scope}
						\draw[thick, <-] (0,0.15) -- (0,1);
					\end{scope}
					\begin{scope}[rotate=60]
						\draw[thick, <-] (0,0.15) -- (0,1);
					\end{scope}
					\begin{scope}[rotate=120]
						\draw[thick, <-] (0,0.15) -- (0,1);
					\end{scope}
					\begin{scope}[rotate=180]
						\draw[thick, <-,white] (0,0.15) -- (0,1);
					\end{scope}
					\begin{scope}[rotate=240]
						\draw[thick, <-,white] (0,0.15) -- (0,1);
					\end{scope}
					\begin{scope}[rotate=300]
						\draw[thick, <-,white] (0,0.15) -- (0,1);
					\end{scope}
					\draw[thick, fill=white] (0,0) circle (0.15);
				\end{scope}
			\end{tikzpicture}
		\end{array}\,,\quad
		\begin{array}{c}
			\begin{tikzpicture}[scale=0.6]
				\begin{scope}[rotate=120]
					\begin{scope}
						\draw[thick, <-] (0,0.15) -- (0,1);
					\end{scope}
					\begin{scope}[rotate=60]
						\draw[thick, <-] (0,0.15) -- (0,1);
					\end{scope}
					\begin{scope}[rotate=120]
						\draw[thick, <-] (0,0.15) -- (0,1);
					\end{scope}
					\begin{scope}[rotate=180]
						\draw[thick, <-,white] (0,0.15) -- (0,1);
					\end{scope}
					\begin{scope}[rotate=240]
						\draw[thick, <-,white] (0,0.15) -- (0,1);
					\end{scope}
					\begin{scope}[rotate=300]
						\draw[thick, <-,white] (0,0.15) -- (0,1);
					\end{scope}
					\draw[thick, fill=white] (0,0) circle (0.15);
				\end{scope}
			\end{tikzpicture}
		\end{array}\,,\quad
		\begin{array}{c}
			\begin{tikzpicture}[scale=0.6]
				\begin{scope}[rotate=180]
					\begin{scope}
						\draw[thick, <-] (0,0.15) -- (0,1);
					\end{scope}
					\begin{scope}[rotate=60]
						\draw[thick, <-] (0,0.15) -- (0,1);
					\end{scope}
					\begin{scope}[rotate=120]
						\draw[thick, <-] (0,0.15) -- (0,1);
					\end{scope}
					\begin{scope}[rotate=180]
						\draw[thick, <-,white] (0,0.15) -- (0,1);
					\end{scope}
					\begin{scope}[rotate=240]
						\draw[thick, <-,white] (0,0.15) -- (0,1);
					\end{scope}
					\begin{scope}[rotate=300]
						\draw[thick, <-,white] (0,0.15) -- (0,1);
					\end{scope}
					\draw[thick, fill=white] (0,0) circle (0.15);
				\end{scope}
			\end{tikzpicture}
		\end{array}\,,\quad
		\begin{array}{c}
			\begin{tikzpicture}[scale=0.6]
				\begin{scope}[rotate=240]
					\begin{scope}
						\draw[thick, <-] (0,0.15) -- (0,1);
					\end{scope}
					\begin{scope}[rotate=60]
						\draw[thick, <-] (0,0.15) -- (0,1);
					\end{scope}
					\begin{scope}[rotate=120]
						\draw[thick, <-] (0,0.15) -- (0,1);
					\end{scope}
					\begin{scope}[rotate=180]
						\draw[thick, <-,white] (0,0.15) -- (0,1);
					\end{scope}
					\begin{scope}[rotate=240]
						\draw[thick, <-,white] (0,0.15) -- (0,1);
					\end{scope}
					\begin{scope}[rotate=300]
						\draw[thick, <-,white] (0,0.15) -- (0,1);
					\end{scope}
					\draw[thick, fill=white] (0,0) circle (0.15);
				\end{scope}
			\end{tikzpicture}
		\end{array}\,,\quad
		\begin{array}{c}
			\begin{tikzpicture}[scale=0.6]
				\begin{scope}[rotate=300]
					\begin{scope}
						\draw[thick, <-] (0,0.15) -- (0,1);
					\end{scope}
					\begin{scope}[rotate=60]
						\draw[thick, <-] (0,0.15) -- (0,1);
					\end{scope}
					\begin{scope}[rotate=120]
						\draw[thick, <-] (0,0.15) -- (0,1);
					\end{scope}
					\begin{scope}[rotate=180]
						\draw[thick, <-,white] (0,0.15) -- (0,1);
					\end{scope}
					\begin{scope}[rotate=240]
						\draw[thick, <-,white] (0,0.15) -- (0,1);
					\end{scope}
					\begin{scope}[rotate=300]
						\draw[thick, <-,white] (0,0.15) -- (0,1);
					\end{scope}
					\draw[thick, fill=white] (0,0) circle (0.15);
				\end{scope}
			\end{tikzpicture}
		\end{array}\,.
\end{equation}
We color such vertices as white and denote the set of them for a given $\vec{Y}$ as $\mathbf{p}_1(\vec{Y})$, which maps one-to-one to the set of simple pausers $p_1(\vec{Y})$.

For the example in Figure~\ref{fig:Lowest_tpriplet}, there are four simple pausers, at the positions
\begin{equation}\label{eq:pauser1_cor_C3}
(1,7,4),\;(2,5,3),\;(4,5,1),\;(5,1,4) \;,
\end{equation}
with coordinate function $h(\square)=x(\mathbf{v})$ values:
\begin{equation}\label{eq:pauser1_C3}
6 \mathsf{h}_2+3 \mathsf{h}_3,\;3 \mathsf{h}_2+\mathsf{h}_3,\;3 \mathsf{h}_1+4 \mathsf{h}_2,\;4 \mathsf{h}_1+3 \mathsf{h}_3 \;.
\end{equation}

\item Full-convex corner $\Longleftrightarrow$ incoming valence-$6$ vertex (denoted as white vertex in $\mathscr{D}_{\vec Y}$).

In the minimal plane partition, it is also possible to have full-convex corners, which correspond to the (simultaneous) intersection points of three positive crystals. 
An atom added at such a corner is the origin of two copies of negative crystals overlapping each other, hence is a double pauser, contributing a double zero to the ground state charge function.
We denote the set of double pauser for a given $\vec{Y}$ by $p_2(\vec{Y})$.

Projecting to 2D, a full-convex corner, i.e.\ a double pauser, corresponds to a vertex with $6$ incoming arrows:
\begin{equation}
		\begin{array}{c}
			\begin{tikzpicture}[scale=0.6]
				\begin{scope}
					\draw[thick, <-] (0,0.15) -- (0,1);
				\end{scope}
				\begin{scope}[rotate=60]
					\draw[thick, <-] (0,0.15) -- (0,1);
				\end{scope}
				\begin{scope}[rotate=120]
					\draw[thick, <-] (0,0.15) -- (0,1);
				\end{scope}
				\begin{scope}[rotate=180]
					\draw[thick, <-] (0,0.15) -- (0,1);
				\end{scope}
				\begin{scope}[rotate=240]
					\draw[thick, <-] (0,0.15) -- (0,1);
				\end{scope}
				\begin{scope}[rotate=300]
					\draw[thick, <-] (0,0.15) -- (0,1);
				\end{scope}
				\draw[thick, fill=white] (0,0) circle (0.15);
			\end{tikzpicture}
		\end{array}
\end{equation}
We again color them as white and denote the set of them by $\mathbf{p}_2(\vec{Y})$, which maps one-to-one to the set of double pausers $p_2(\vec{Y})$.

For the example in Figure~\ref{fig:Lowest_tpriplet}, there is one double pauser, at the position \begin{equation}\label{eq:pauser2_cor_C3}
    (3,3,3)\,,
\end{equation} and 
with coordinate function
\begin{equation}\label{eq:pauser2_C3}
h(\square)=x(\mathbf{v})=0\,.
\end{equation}
\end{itemize}

\item Collecting the contribution from $s(\vec{Y})$, $p_1(\vec{Y})$, and $p_2(\vec{Y})$ together, we can write down the ground state charge function immediately:
\be\label{psi_0_C3_open}
	\psi_{\vec{Y}}(z)=\frac{\prod\lm_{\square \in p_1(\vec{Y})}(z-h(\square))\cdot \prod\lm_{\square\in p_2(\vec{Y})}(z-h(\square))^2}{\prod\lm_{\square\in s(\vec{Y})}(z-h(\square))} \,,
\ee
with $h(\square)$ in $s(\vec{Y})$, $p_1(\vec{Y})$, and $p_2(\vec{Y})$ given in \eqref{eq:starter_C3}, \eqref{eq:pauser1_C3}, and \eqref{eq:pauser2_C3}, respectively.
Since the number of poles is the same as the number of zeros in \eqref{psi_0_C3_open}, both $7$, the ground state charge function describes a representation of the unshifted affine Yangian of $\mathfrak{gl}_1$, as expected. 

\end{enumerate}

\subsubsection{Framed quiver and superpotential}

The quiver and superpotential pair that correspond to the canonical crystal $\mathcal{C}_0$, i.e.\ the one with trivial $\vec{Y}$, are given in~\eqref{eq:QW0C3}.
Let us now determine the framed quiver and superpotential pair $(Q_{\vec{Y}},W_{\vec{Y}})$ for the subcrystal $\mathcal{C}_{\vec{Y}}$.

As explained in Section~\ref{ssec:framedQW}, the framed quiver $Q_{\vec{Y}}$ can be obtained by adding arrows to the canonically framed quiver $Q_0$: an arrow from the framing node to the gauge node for each stopper and an arrow from the gauge node back to the framing node for each pauser.
Accordingly, the superpotential needs to be modified by additional terms that enforce the loop constraints for the new loops in $Q_{\vec{Y}}$. 
In summary, we have
\begin{equation}
\begin{aligned}
	Q_{\vec Y}&=\begin{array}{c}
		\begin{tikzpicture}[rotate=-90]
			\draw[thick,->] ([shift=(120:0.75)]0,0.75) arc (120:420:0.75);
			\draw[thick,->] ([shift=(60:0.75)]0,0.75) arc (60:90:0.75);
			\draw[thick,->] ([shift=(90:0.75)]0,0.75) arc (90:120:0.75);
			\draw[fill=white] (0,0) circle (0.2);
			\draw[thick,->] (0.1,-1.35) -- (0.1,-0.85);
			\draw[thick,->] (0.1,-0.85) -- (0.1,-0.75);
			\draw[thick,->] (0.1,-0.75) -- (0.1,-0.65);
			\draw[thick] (0.1,-0.65) -- (0.1,-0.173205);
			\draw[thick] (-0.1,-1.35) -- (-0.1,-0.85);
			\draw[thick,<-] (-0.1,-0.85) -- (-0.1,-0.75);
			\draw[thick,<-] (-0.1,-0.75) -- (-0.1,-0.65);
			\draw[thick,<-] (-0.1,-0.65) -- (-0.1,-0.173205);
			\begin{scope}[shift={(0,-1.5)}]
				\draw[thick,fill=red] (-0.15,-0.15) -- (-0.15,0.15) -- (0.15,0.15) -- (0.15,-0.15) -- cycle;
			\end{scope}
			\node[right] at (0,1.5) {$B_{1,2,3}$};
			\node[below] at (0.15,-0.75) {$R_b$};
			\node[above] at (-0.15,-0.75) {$S_w$, $\tilde S_{\tilde{w},\alpha}$};
		\end{tikzpicture}
	\end{array}, 
	\\
 W_{\vec{Y}}&=B_1[B_2,B_3] +\sum\lm_{z\in p_{1,2}(\vec{Y})}\delta W_z\,,
\end{aligned}
\end{equation}
where $R_{b}$ denotes the chiral field that corresponds to the arrow $b$ from the framing node to the gauge node; and similarly the chiral field $S_w$ and the doublet of chiral fields $\tilde{S}_{\tilde{w}, \alpha}$ with $\alpha=1,2$ correspond to the arrow $w$ and $\tilde{w}$ from the gauge node back to the framing node, respectively. 
Using the dictionary among the arrows in the framing quiver, the origins of the positive/negative crystals, and the poles/zeros in the ground state charge function, we have the following correspondence:
\begin{equation}
\begin{aligned}
&R\,\,\,\longleftrightarrow\,\,\, \textrm{arrow from f.n. to g.n.} \,\,\, \longleftrightarrow \,\,\, \textrm{starter} \,\,\, \longleftrightarrow \,\,\, \textrm{pole in $\psi_{\vec{Y}}(z)$}\;,\\
&S\,\,\,\longleftrightarrow\,\,\, \textrm{arrow from g.n. to f.n.} \,\,\, \longleftrightarrow \,\,\, \textrm{simple pauser} \,\,\, \longleftrightarrow \,\,\, \textrm{simple zero in $\psi_{\vec{Y}}(z)$}\;,\\
&\tilde{S}\,\,\,\longleftrightarrow\,\,\, \textrm{arrow from g.n. to f.n.} \,\,\, \longleftrightarrow \,\,\, \textrm{double pauser} \,\,\, \longleftrightarrow \,\,\, \textrm{double zero in $\psi_{\vec{Y}}(z)$}\;,
\end{aligned}
\end{equation}
where g.n.\ stands for the gauge node and f.n.\ for the framing node.
The masses of these chiral fields are given by $(\pm 1)\times$ the coordinate functions of the corresponding starters/pausers, with $\pm$ for the starter and the pauser, respectively: 
\begin{equation}
\begin{split}
R_b&\quad\textrm{with}\quad \mu(R_b)=h(\square_b)\;,\quad \square_b \in{s}({\vec Y})\;,\\
S_w&\quad\textrm{with}\quad \mu(S_w)=-h(\square_w)\;,\quad \square_w \in{p}_1({\vec Y})\;,\\
\tilde{S}_{\tilde{w},\alpha}&\quad\textrm{with}\quad \mu(\tilde S_{\tilde{w},\alpha})=-h(\square_{\tilde{w}})\;,\quad \square_{\tilde{w}}\in{ p}_2(\vec Y),\;\alpha=1,2\;.
	\end{split}
\end{equation} 
	
Recall from Section~\ref{ssec:framedQW} that for each arrow $z$ from the framed vertex back to the framing node, which corresponds to a zero in $\psi_{\vec{Y}}(z)$,  we need to add a term $\delta W_z$ into the superpotential.
The expression for the correction term $\delta W_z$ depends on whether the arrow corresponds to a simple zero or a double zero in $\psi_{\vec{Y}}(z)$.	\begin{enumerate}
\item For $w\in{\bf p}_1({\vec Y})$ there are two possible types of situations:
\begin{equation}
		\begin{array}{c}
			\begin{tikzpicture}[scale=0.6]
				\begin{scope}
					\draw[thick, <-] (0,0.15) -- (0,2);
				\end{scope}
				\begin{scope}[rotate=60]
					\draw[thick, <-] (0,0.15) -- (0,2);
					\draw[fill=black] (0,2) circle (0.15);
					\node[below left] at (0,1) {$j\vec h_2$};
				\end{scope}
				\begin{scope}[rotate=300]
					\draw[thick, <-] (0,0.15) -- (0,2);
					\draw[fill=black] (0,2) circle (0.15);
					\node[below right] at (0,1) {$i\vec h_1$};
				\end{scope}
				\draw[thick, fill=white] (0,0) circle (0.15);
			\end{tikzpicture}
		\end{array},\quad \delta W_w=S_{w}\left(B_1^iR_{b_1}-B_2^jR_{b_2}\right) \;,
\end{equation}
\begin{equation}
		\begin{array}{c}
			\begin{tikzpicture}[scale=0.6]
				\begin{scope}
					\draw[thick, <-] (0,0.15) -- (0,2);
				\end{scope}
				\begin{scope}[rotate=60]
					\draw[thick, <-] (0,0.15) -- (0,2);
					\draw[fill=black] (0,2) circle (0.15);
					\node[below left] at (0,1) {$j\vec h_2$};
				\end{scope}
				\begin{scope}[rotate=120]
					\draw[thick, <-] (0,0.15) -- (0,2);
				\end{scope}
				\draw[thick, fill=white] (0,0) circle (0.15);
				\begin{scope}[shift={(2,0)}]
					\begin{scope}
						\draw[thick, ->] (0,0.15) -- (0,2);
					\end{scope}
					\begin{scope}[rotate=120]
						\draw[thick, ->] (0,0.15) -- (0,2);
					\end{scope}
					\draw[fill=black] (0,0) circle (0.15);
					\node[right] at (0,0.8) {$k\vec h_3$};
				\end{scope}
				\begin{scope}[rotate=-60]
					\draw[<-,dashed] (0,0.15) -- (0,2.3094);
					\node[above] at (0,1) {$i\vec h_1$};
				\end{scope}
			\end{tikzpicture}
		\end{array},\quad \delta W_w=S_w\left(B_2^jR_{b_1}-B_3^kB_1^iR_{b_2}\right).
\end{equation}
		In both cases we have to impose a Lagrange constraint through the superpotential that crystals growing from two nearest black vertices interfere properly.
We could reshuffle a sequence of $B_1$ and $B_3$ generators in the second term of this superpotential correction. All these arrangements are equivalent.
		\item For $w\in{\bf p}_2({\vec Y})$ we have two interference constraints and, therefore, two Lagrange fields:
		\begin{equation}
		\begin{array}{c}
			\begin{tikzpicture}[scale=0.6]
				\begin{scope}
					\draw[thick, <-] (0,0.15) -- (0,2);
				\end{scope}
				\begin{scope}[rotate=60]
					\draw[thick, <-] (0,0.15) -- (0,2);
					\draw[fill=black] (0,2) circle (0.15);
					\node[below left] at (0,1) {$j\vec h_2$};
				\end{scope}
				\begin{scope}[rotate=120]
					\draw[thick, <-] (0,0.15) -- (0,2);
				\end{scope}
				\begin{scope}[rotate=180]
					\draw[thick, <-] (0,0.15) -- (0,2);
					\draw[fill=black] (0,2) circle (0.15);
					\node[right] at (0,1.3) {$k\vec h_3$};
				\end{scope}
				\begin{scope}[rotate=240]
					\draw[thick, <-] (0,0.15) -- (0,2);
				\end{scope}
				\begin{scope}[rotate=300]
					\draw[thick, <-] (0,0.15) -- (0,2);
					\draw[fill=black] (0,2) circle (0.15);
					\node[below right] at (0,1) {$i\vec h_1$};
				\end{scope}
				\draw[thick, fill=white] (0,0) circle (0.15);
			\end{tikzpicture}
		\end{array},\quad \delta W_w=\tilde S_{w,1}\left(B_1^iR_{b_1}-B_2^jR_{b_2}\right)+\tilde S_{w,2}\left(B_1^iR_{b_1}-B_3^kR_{b_3}\right).
		\end{equation}
		Two interference constraints come from two pairings of the black vertices. 
Clearly, three black vertices could be linked in two pairs in three different ways producing three different superpotential corrections. 
All these three superpotential options are equivalent.
	\end{enumerate}

\subsubsection{\texorpdfstring{Representation of affine Yangian of $\mathfrak{gl}_1$}{Representation of Affine Yangian of gl(1)}}
The representation associated with the canonical crystal $\mathcal{C}_{0}$, spanned by all plane partitions with trivial asymptotics, is the vacuum representation of the affine Yangian of $\mathfrak{gl}_1$.

For non-trivial asymptotics $\vec{Y}$, the representation $\textrm{Rep}_{\vec{Y}}$ associated with the subcrystal $\mathcal{C}_{\vec{Y}}$ is spanned by all plane partitions with asymptotics $\vec{Y}$ and is a non-vacuum representation of the (unshifted) affine Yangian of $\mathfrak{gl}_1$.
The representation can be specified by its ground state charge function~\eqref{psi_0_C3_open}.
The fact that there is no shift involved can be seen from the fact that the number of its poles is equal to the number of its zeros, making the shift $\mys=0$.

Finally, note that since the affine Yangian of $\mathfrak{gl}_1$ is isomorphic to the universal enveloping algebra of $\mathcal{W}_{1+\infty}$ algebra, the representation $\textrm{Rep}_{\vec{Y}}$ is also a representation of the $\mathcal{W}_{1+\infty}$ algebra.
The plane partition representations discussed in this subsection, with all possible asymptotics $\textrm{Rep}_{\vec{Y}}$, give a very transparent and geometric way of constructing representations for the $\mathcal{W}_{1+\infty}$ algebra.
Furthermore, the truncations of these representations become the representations of the rational $\mathcal{W}_{N,k}$ algebras.

\subsection{Example: open BPS states in \texorpdfstring{$(\mathbb{C}^2/\mathbb{Z}_2)\times \mathbb{C}$}{C2/Z2xC}}
\label{ssec:Open_Z2}

\subsubsection{Canonical crystal and affine Yangian of \texorpdfstring{$\mathfrak{gl}_2$}{gl2}}
\label{sssec:canonical_gl2}
For $\mathbb{C}^2/\mathbb{Z}_2$$\times$$\mathbb{C}$, the toric diagram and its dual graph are
\begin{equation}\label{fig-toric-Z2-1}
\begin{tikzpicture} 
\filldraw [red] (0,0) circle (2pt); 
\filldraw [red] (0,1) circle (2pt); 
\filldraw [red] (0,2) circle (2pt); 
\filldraw [red] (1,0) circle (2pt); 
\node at (-.5,-.5) {(0,0)}; 
\node at (-.5,1) {(0,1)}; 
\node at (-.5,2.5) {(0,2)}; 
\node at (1.5,-0.5) {(1,0)}; 
\draw (0,0) -- (0,2); 
\draw (0,0) -- (1,0); 
\draw (1,0) -- (0,2); 
\end{tikzpicture}
\qquad \qquad \qquad
\begin{tikzpicture} 
\draw[->] (0,0) -- (-1,0); 
\draw[->] (0,0) -- (0,-1); 
\draw (0,0) -- (1,1); 
\draw[->] (1,1) -- (-1,1); 
\draw[->] (1,1) -- (3,2); 
\node at (-1.5,0) {3}; 
\node at (-1.5,1) {$\hat{3}$}; 
\node at (0,-1.5) {1}; 
\node at (3.5,2.5) {$\hat{1}$}; 
\end{tikzpicture}
\end{equation}
Its associated quiver diagram is the $A_2$-quiver
\begin{equation}\label{figure-quiver-Z2}
Q=
\begin{array}{c}
\begin{tikzpicture}
\draw[thick,postaction={decorate},decoration={markings,
		mark= at position 0.7 with {\arrow{>}},mark= at position 0.6 with {\arrow{>}}}] (0,0) to[out=30,in=150] node[above] {$(A_1,\alpha_{1}), \,\, (B_2,\beta_{2})$} (5,0);
\draw[thick,postaction={decorate},decoration={markings,
		mark= at position 0.7 with {\arrow{>}},mark= at position 0.6 with {\arrow{>}}}] (5,0) to[out=210,in=330] node[below] {$(B_1,\beta_1), \,\, (A_2,\alpha_{2})$} (0,0);
\draw[thick,postaction={decorate},decoration={markings,
		mark= at position 0.7 with {\arrow{>}}}] (0,0) to[out=180,in=225]  (-0.7,0.7) to [out=45,in=90] (0,0);
\begin{scope}[shift={(5,0)}]
\begin{scope}[xscale=-1]
    \draw[thick,postaction={decorate},decoration={markings,
		mark= at position 0.7 with {\arrow{>}}}] (0,0) to[out=180,in=225]  (-0.7,0.7) to [out=45,in=90] (0,0);
\end{scope}
\end{scope}
\draw[thick,postaction={decorate},decoration={markings,
		mark= at position 0.7 with {\arrow{>}}}] (-2,-0.5) to (-1,-0.5) to[out=0,in=225] (0,0);
\node[above left] at (-0.7,0.7) {$(C_1, \gamma_1)$};
\node[above right] at (5.7,0.7) {$(C_2,\gamma_2)$};
\draw[fill=white] (0,0) circle (0.2);
\draw[fill=gray] (5,0) circle (0.2);
\begin{scope}[shift={(-2,-0.5)}]
    \draw[fill=red] (-0.15,-0.15) -- (-0.15,0.15) -- (0.15,0.15) -- (0.15,-0.15) -- cycle;
\end{scope}
\node[below] at (0,-0.2) {$1$};
\node[below] at (5,-0.2) {$2$};
\end{tikzpicture}
\end{array}
\end{equation}
with super-potential 
\begin{equation}
W=\textrm{Tr}[-C_1\, A_1\, B_1+C_1 \, B_2\, A_2 -C_2\, A_2\, B_2+C_2 \, B_1\, A_1]  \;.
\end{equation}
Both vertices are bosonic:
\begin{equation}\label{eq-gl2-boson-1}
|a|=0 \,,\qquad a=1,2\,,
\end{equation}
since there is a self-loop for each of them in the quiver (\ref{figure-quiver-Z2}).

The loop constraint (\ref{eq:loop-constraint}) and the vertex constraint (\ref{eq-vertex-constraint-toric}) together give
\begin{equation}\label{eq-constraint-gl2}
\begin{aligned}
&\alpha_1=\alpha_2\equiv \mathsf{h}_1  \;,\qquad \beta_1=\beta_2\equiv \mathsf{h}_2 \;, \qquad \gamma_1=\gamma_2=\gamma\equiv \mathsf{h}_3 \;,\\
&\textrm{and}\qquad \mathsf{h}_1+\mathsf{h}_2+\mathsf{h}_3=0 \;.
\end{aligned}
\end{equation}

The two-dimensional projection of the crystal, the periodic quiver, 
is given by:
\begin{equation}\label{tilling-conifold}
\begin{tikzpicture}[scale=0.57]
\filldraw[mygreen] (0,3)-- (-3,0)--(0,-3)-- (3,0) -- cycle; 
\node[state]  [regular polygon, regular polygon sides=4, draw=blue!50, very thick, fill=blue!10] (a1) at (0,0)  {$1$};
\node[state]  [regular polygon, regular polygon sides=4, draw=blue!50, very thick, fill=blue!10] (a21) at (3,0)  {$2$};
\node[state]  [regular polygon, regular polygon sides=4, draw=blue!50, very thick, fill=blue!10] (a22) at (-3,0)  {$2$};
\node[state]  [regular polygon, regular polygon sides=4, draw=blue!50, very thick, fill=blue!10] (a41) at (0,3)  {$2$};
\node[state]  [regular polygon, regular polygon sides=4, draw=blue!50, very thick, fill=blue!10] (a42) at (0,-3)  {$2$};
\node[state]  [regular polygon, regular polygon sides=4, draw=blue!50, very thick, fill=blue!10] (a31) at (3,3)  {$1$};
\node[state]  [regular polygon, regular polygon sides=4, draw=blue!50, very thick, fill=blue!10] (a32) at (3,-3)  {$1$};
\node[state]  [regular polygon, regular polygon sides=4, draw=blue!50, very thick, fill=blue!10] (a34) at (-3,-3)  {$1$};
\node[state]  [regular polygon, regular polygon sides=4, draw=blue!50, very thick, fill=blue!10] (a33) at (-3,3)  {$1$};
\path[->] 
(a1) edge   [thick, red]   node [above] {$\mathsf{h}_2$} (a21)
(a22) edge   [thick, red]   node [above] {$\mathsf{h}_2$} (a1)
(a31) edge   [thick, red]   node [right] {$\mathsf{h}_1$} (a21)
(a21) edge   [thick, red]   node [right] {$\mathsf{h}_1$} (a32)
(a33) edge   [thick, red]   node [left] {$\mathsf{h}_1$} (a22)
(a22) edge   [thick, red]   node [left] {$\mathsf{h}_1$} (a34)
(a41) edge   [thick, red]   node [above] {$\mathsf{h}_2$} (a31)
(a33) edge   [thick, red]   node [above] {$\mathsf{h}_2$} (a41)
(a34) edge   [thick, red]   node [above] {$\mathsf{h}_2$} (a42)
(a42) edge   [thick, red]   node [above] {$\mathsf{h}_2$} (a32)
(a41) edge   [thick, red]   node [right] {$\mathsf{h}_1$} (a1)
(a1) edge   [thick, red]   node [right] {$\mathsf{h}_1$} (a42)
(a1) edge   [thick, red]   node [right] {$\mathsf{h}_3$} (a33)
(a32) edge   [thick, red]   node [right] {$\mathsf{h}_3$} (a1)
(a21) edge   [thick, red]   node [right] {$\mathsf{h}_3$} (a41)
(a42) edge   [thick, red]   node [right] {$\mathsf{h}_3$} (a22)
;
\end{tikzpicture}
\end{equation}

The canonical crystal for the \texorpdfstring{$(\mathbb{C}^2/\mathbb{Z}_n)\times \mathbb{C}$}{C2/ZnxC} geometry has the same shape but different coloring as the canonical crystal for the $\mathbb{C}^3$ geometry.
Namely, each molten crystal configuration from the canonical crystal of  \texorpdfstring{$(\mathbb{C}^2/\mathbb{Z}_n)\times \mathbb{C}$}{C2/ZnxC} still has the shape of a plane partition. 
The orbifolding determines the coloring scheme to be 
\begin{equation}\label{eq:color_def_Zn}
 {\bf color}(\Box) =(x_1+x_2) \mod n +1 \quad  \textrm{for} \quad \Box \textrm{ at coordinate $(x_1,x_2,x_3)$}\,,
\end{equation}
where the $\Box$ at the origin has the coordinate $(0,0,0)$ and $ {\bf color}(\Box) =1$.
Note that the color of a $\Box$ is independent of its $x_3$ coordinate, following from the orbifold action.
In this subsection, we focus on $n=2$, without loss of generality.

\bigskip

The bond factor can be read off from the periodic quiver \eqref{tilling-conifold} to be
\begin{equation}\label{eq:bond_Z2}
\begin{aligned}
&\varphi^{a\Rightarrow a}(u)=\frac{u+\mathsf{h}_3}{u-\mathsf{h}_3} \qquad \textrm{and}\qquad\varphi^{a+1\Rightarrow a}(u)=\frac{(u+\mathsf{h}_1)(u+\mathsf{h}_2)}{(u-\mathsf{h}_1)(u-\mathsf{h}_2)} \;.
\end{aligned}
\end{equation}
Plugging the bond factor \eqref{eq:bond_Z2} and the statistics factor \eqref{eq-gl2-boson-1} into the general formula for the algebraic relation \eqref{eq-OPE-toric} we obtain the affine Yangian of $\mathfrak{gl}_2$. 

\subsubsection{Subcrystals for open BPS states}

We have explained that the canonical crystal for the  \texorpdfstring{$(\mathbb{C}^2/\mathbb{Z}_n)\times \mathbb{C}$}{C2/ZnxC} geometry with all $n\in \mathbb{N}$ has the same shape as the one for $\mathbb{C}^3$.  
Therefore, same as for the open BPS states in the $\mathbb{C}^3$ geometry considered in Section~\ref{ssec:Open_C3}, the open BPS states for the \texorpdfstring{$(\mathbb{C}^2/\mathbb{Z}_n)\times \mathbb{C}$}{C2/ZnxC} geometry are also given by the subcrystal fixed by the three asymptotic Young diagrams $\vec{Y}=(Y_1,Y_2,Y_3)$ along the $x_{1,2,3}$ directions of the plane partition. 
Namely, the coloring scheme just inherits the one from the canonical crystal. 

\begin{figure}[ht!]
	\begin{center}
		\begin{tikzpicture}
			\node at (0,0) {\includegraphics*[scale=0.3]{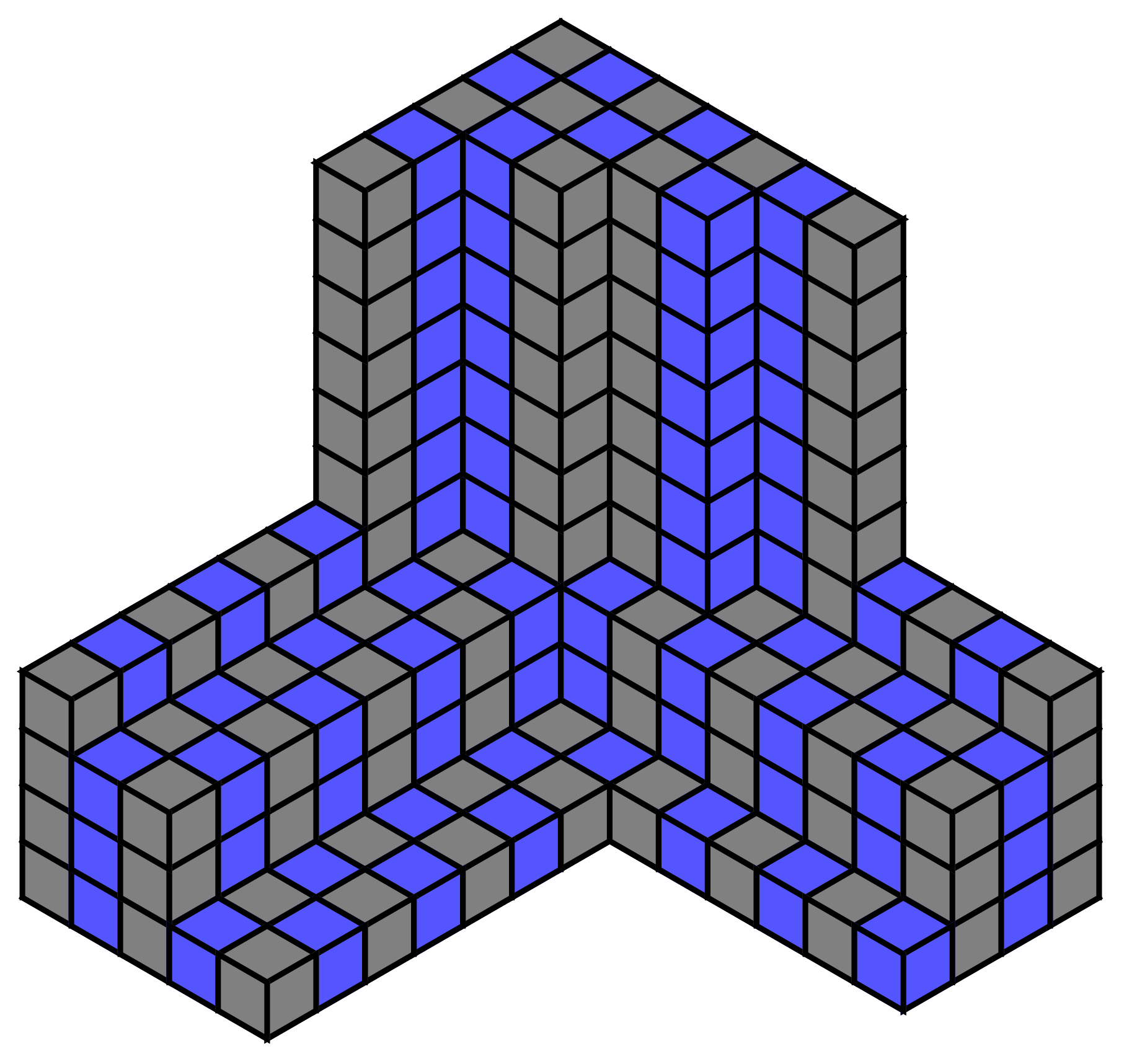}};
		\end{tikzpicture}
\caption{The subcrystal for the open BPS states in \texorpdfstring{$(\mathbb{C}^2/\mathbb{Z}_n)\times \mathbb{C}$}{C2/ZnxC} geometry, with asymptotics $Y_1=\{4,3,3,1,1\}$, $Y_2=\{4,3,3,1\}$, $Y_3=\{7,5,3,1,1\}$}.
\label{fig:Lowest_tpriplet_colored}
\end{center}
\end{figure}

Below, for concreteness, let us focus on $n=2$. 
Figure~\ref{fig:Lowest_tpriplet_colored} gives an example of the subcrystal for the open BPS states in the \texorpdfstring{$(\mathbb{C}^2/\mathbb{Z}_n)\times \mathbb{C}$}{C2/ZnxC} geometry, with the same asymptotics as the $\mathbb{C}^3$ example in Figure~\ref{fig:Lowest_tpriplet}, i.e.\ $Y_1=\{4,3,3,1,1\}$, $Y_2=\{4,3,3,1\}$, $Y_3=\{7,5,3,1,1\}$.
We will now determine the corresponding ground state charge function, the framed quiver and superpotential, and the shift.

The coordinates of the starters and pausers in the subcrystal Figure~\ref{fig:Lowest_tpriplet_colored} are already given in \eqref{eq:starter_cor_C3}, \eqref{eq:pauser1_cor_C3}, and \eqref{eq:pauser2_cor_C3}, inherited from Figure~\ref{fig:Lowest_tpriplet_colored}. We just need to assign colors to them according to \eqref{eq:color_def_Zn} with $n=2$.
Namely, the information of these starters and pausers  that enter the ground state charge function are their coordinate function $h(\square)$ and their color (defined in \eqref{eq:color_def_Zn} with $n=2$), namely
\begin{equation}
    (h(\square),c)=\left(x(\mathbf{v}),1+\vec x({\bf v})\cdot(1,1,0)\;\mbox{\bf mod}\;2\right)\,.
\end{equation} 

\begin{enumerate}
\item

In Figure~\ref{fig:Lowest_tpriplet_colored}, there are seven starters, at the positions
\begin{equation}\label{eq:starter_Z2}
(0,7,4),\; (1,5,3),\; (2,3,3),\; (3,1,3),\; (3,3,1),\; (4,5,0),\; (5,0,4) \;.
\end{equation}
The $(h(\square),c)$ values of the $7$ starters in \eqref{eq:starter_Z2} are
\begin{equation}
\begin{split}
(7 \mathsf{h}_2+4 \mathsf{h}_3,{\color{blue} 2}),\; (4 \mathsf{h}_2+2 \mathsf{h}_3,{\color{gray} 1}),\; (\mathsf{h}_2+\mathsf{h}_3,{\color{blue} 2}),\; (2 \mathsf{h}_1+2 \mathsf{h}_3,{\color{gray} 1}),\\ (2 \mathsf{h}_1+2 \mathsf{h}_2,{\color{gray} 1}),\; (4 \mathsf{h}_1+5 \mathsf{h}_2,{\color{blue} 2}),\; (5 \mathsf{h}_1+4 \mathsf{h}_3,{\color{blue} 2}) \;.
\end{split}
\end{equation}
\item
Similarly, in Figure~\ref{fig:Lowest_tpriplet_colored}, there are four simple pausers, at the positions
\begin{equation}
(1,7,4),\;(2,5,3),\;(4,5,1),\;(5,1,4) \;,
\end{equation}
with coordinate function and color $(h(\square),c)$ values:
\begin{equation}
(6 \mathsf{h}_2+3 \mathsf{h}_3,{\color{gray} 1}),\;(3 \mathsf{h}_2+\mathsf{h}_3,{\color{blue} 2}),\;(3 \mathsf{h}_1+4 \mathsf{h}_2,{\color{blue} 2}),\;(4 \mathsf{h}_1+3 \mathsf{h}_3,{\color{gray} 1})\;.
\end{equation}
\item
Finally, in Figure~\ref{fig:Lowest_tpriplet_colored}, there is one double pauser, at the position $(3,3,3)$, and 
with coordinate function and color $(h(\square),c)$ values
\begin{equation}
(0,{\color{gray} 1})\,.
\end{equation}
\end{enumerate}

From the positions of these starters and pausers, one can immediately write down the ground state charge functions $\psi^{(a)}_{\vec{Y}}(z)\equiv {}^{\sharp}\psi^{(a)}_{0}(z)$ with $a=1,2$:
\begin{equation}\label{eq:GSCF_Z2}
\begin{split}
&\psi_{\vec{Y}}^{{\color{gray}(1)}}(z)=\frac{\left(z-(6 \mathsf{h}_2+3 \mathsf{h}_3)\right)\left(z-(4 \mathsf{h}_1+3 \mathsf{h}_3)\right) z^2}{\left(z- (4 \mathsf{h}_2+2 \mathsf{h}_3)\right)\left(z- (2 \mathsf{h}_1+2 \mathsf{h}_3)\right)\left(z- (2 \mathsf{h}_1+2 \mathsf{h}_2)\right)}\;,\\
&\psi_{\vec{Y}}^{{\color{blue}(2)}}(z)=\scalebox{0.96}{$\dfrac{\left(z-(3 \mathsf{h}_2+\mathsf{h}_3)\right)\left(z-(3 \mathsf{h}_1+4 \mathsf{h}_2)\right)}{\left(z-(7 \mathsf{h}_2+4 \mathsf{h}_3)\right)\left(z- (\mathsf{h}_2+\mathsf{h}_3)\right)\left(z- (4 \mathsf{h}_1+5 \mathsf{h}_2) \right)\left(z-(5 \mathsf{h}_1+4 \mathsf{h}_3)\right)}$} \;.
\end{split}
\end{equation}
Comparing these ground state charge functions \eqref{eq:GSCF_Z2} with the one for the $\mathbb{C}^3$ geometry with the same asymptotics, given in \eqref{psi_0_C3_open}, we see that $\psi_{\vec{Y}}(z)=\psi_{\vec{Y}}^{{\color{gray}(1)}}(z)\psi_{\vec{Y}}^{{\color{blue}(2)}}(z)$, namely, the latter splits into two factors, one for each color.
This is a general feature: for the $(\mathbb{C}^2/\mathbb{Z}_n)\times \mathbb{C}$ and $\IP_2$ geometries, whose canonical crystals have the same shape as the one for $\mathbb{C}^3$ but different color schemes, the product of the ground state charge functions $\psi^{(a)}_{\vec{Y}}(z)$ for all colors $a$ reproduces the ground state charge function $\psi_{\vec{Y}}(z)$  for $\mathbb{C}^3$ with the same asymptotics $\vec{Y}$.

Counting the number of poles and zeros in the ground state charge function in \eqref{eq:GSCF_Z2}, we see that the subcrystal in Figure~\ref{fig:Lowest_tpriplet_colored} gives rise to a representation of the shifted Yangian of $\mathfrak{gl}_2$, with shift:
\begin{equation}
    \mys=(-1,2)\,.
\end{equation}

From the ground state charge function in \eqref{eq:GSCF_Z2}, we can also derive the corresponding framed quiver and superpotential for the asymptotics $\vec{Y}$ in Figure~\ref{fig:Lowest_tpriplet_colored}:
\begin{equation}
\begin{aligned}
    \sQ&=\;\begin{array}{c}
    \begin{tikzpicture}
    \draw[thick,postaction={decorate},decoration={markings, 
		mark= at position 0.55 with {\arrow{>}},mark= at position 0.45 with {\arrow{>}}}] (0,-1) to[bend left=20] (0,1);
	\draw[thick,postaction={decorate},decoration={markings, 
		mark= at position 0.55 with {\arrow{>}},mark= at position 0.45 with {\arrow{>}}}] (0,1) to[bend left=20] (0,-1);
	\draw[thick,postaction={decorate},decoration={markings, 
		mark= at position 0.5 with {\arrow{>}},mark= at position 0.4 with {\arrow{>}},mark= at position 0.6 with {\arrow{>}}}] (-2,0) to[out=270,in=180] (0,-1);
	\draw[thick,postaction={decorate},decoration={markings, 
		mark= at position 0.35 with {\arrow{<}},mark= at position 0.45 with {\arrow{<}},mark= at position 0.55 with {\arrow{<}},mark= at position 0.65 with {\arrow{<}}}] (-2,0) to[out=300,in=150] (0,-1);
	\draw[thick,postaction={decorate},decoration={markings, 
		mark= at position 0.55 with {\arrow{<}},mark= at position 0.45 with {\arrow{<}}}] (-2,0) to[out=90,in=180] (0,1);
	\draw[thick,postaction={decorate},decoration={markings, 
		mark= at position 0.35 with {\arrow{>}},mark= at position 0.45 with {\arrow{>}},mark= at position 0.55 with {\arrow{>}},mark= at position 0.65 with {\arrow{>}}}] (-2,0) to[out=60,in=210] (0,1);
	\draw[thick,postaction={decorate},decoration={markings,
		mark= at position 0.7 with {\arrow{>}}}] (0,1) to[out=90,in=135] (0.5,1.5) to[out=315,in=0] (0,1);
	\begin{scope}[yscale=-1]
	\draw[thick,postaction={decorate},decoration={markings, 
		mark= at position 0.7 with {\arrow{>}}}] (0,1) to[out=90,in=135] (0.5,1.5) to[out=315,in=0] (0,1);
	\end{scope}
	\draw[fill=gray] (0,-1) circle (0.2);
    \draw[fill=blue] (0,1) circle (0.2);
    \begin{scope}[shift={(-2,0)}]
    \draw[fill=red] (-0.15,-0.15) -- (-0.15,0.15) -- (0.15,0.15) -- (0.15,-0.15) -- cycle;
    \end{scope}
    \node[below left] at (-0.2,-1.2) {$1$};
    \node[above left] at (-0.2,1.2) {$2$};
    \end{tikzpicture}
    \end{array}
    \end{aligned}
\end{equation}

\subsection{Example: open BPS states in \texorpdfstring{$\IP_2$}{P2}}

Let us now study the open BPS states in $K_{\IP_2}$. 
As explained in Section~\ref{ssec:canonical_P2}, the canonical crystal has the same shape as the one for the $\mathbb{C}^3$ geometry, but a coloring scheme given by \eqref{ssec:canonical_P2}.
Therefore the method of studying the open BPS states in $K_{\IP_2}$ is the same as the one used for the case of $(\mathbb{C}^2/\mathbb{Z}_2)\times \mathbb{C}$ in Section~\ref{ssec:Open_Z2}.
\bigskip

	\begin{figure}[ht!]
	\begin{center}
		\begin{tikzpicture}
			\node at (0,0) {\includegraphics*[scale=0.3]{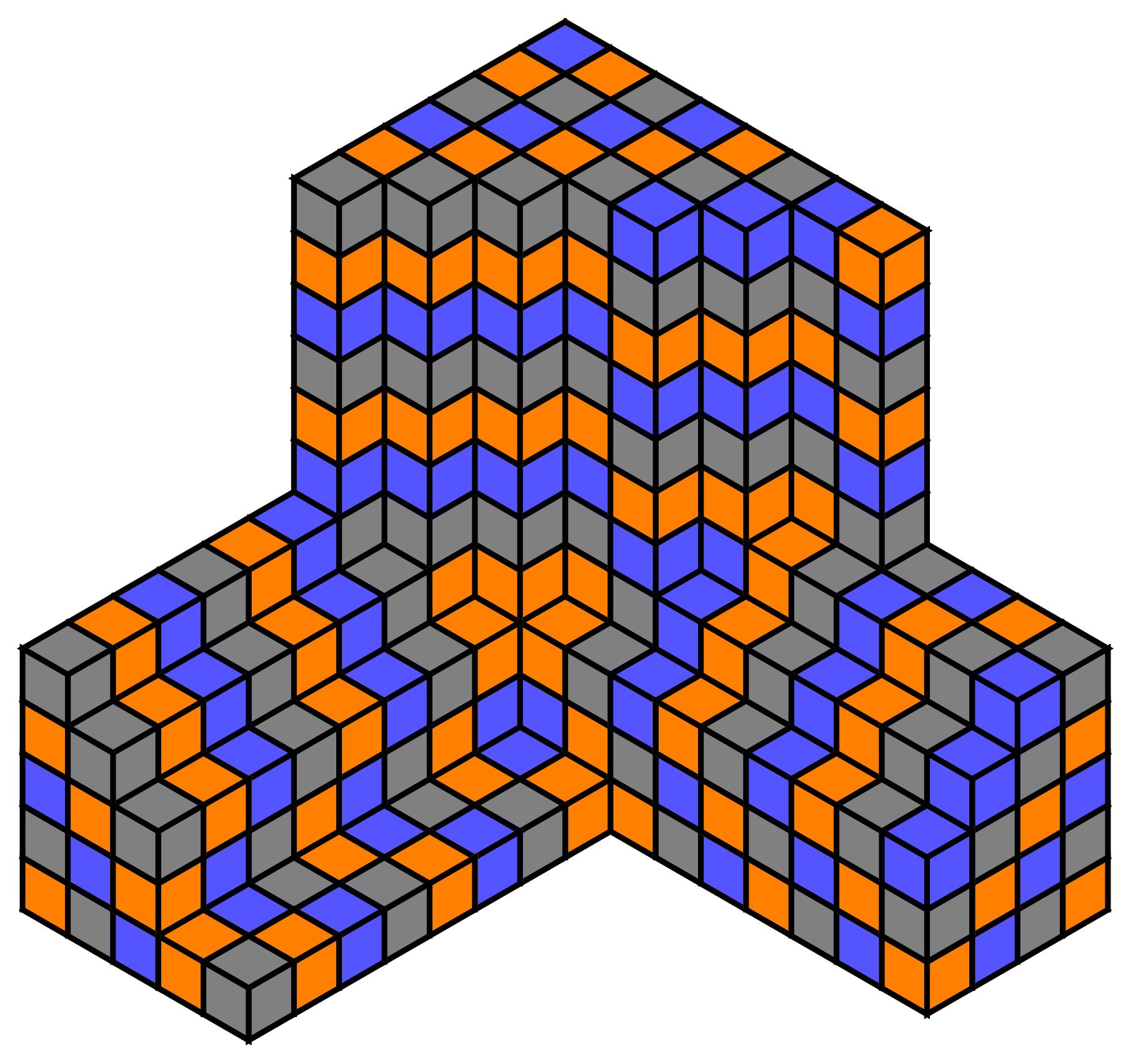}};
			\node at (7,0) {\includegraphics*[scale=0.3]{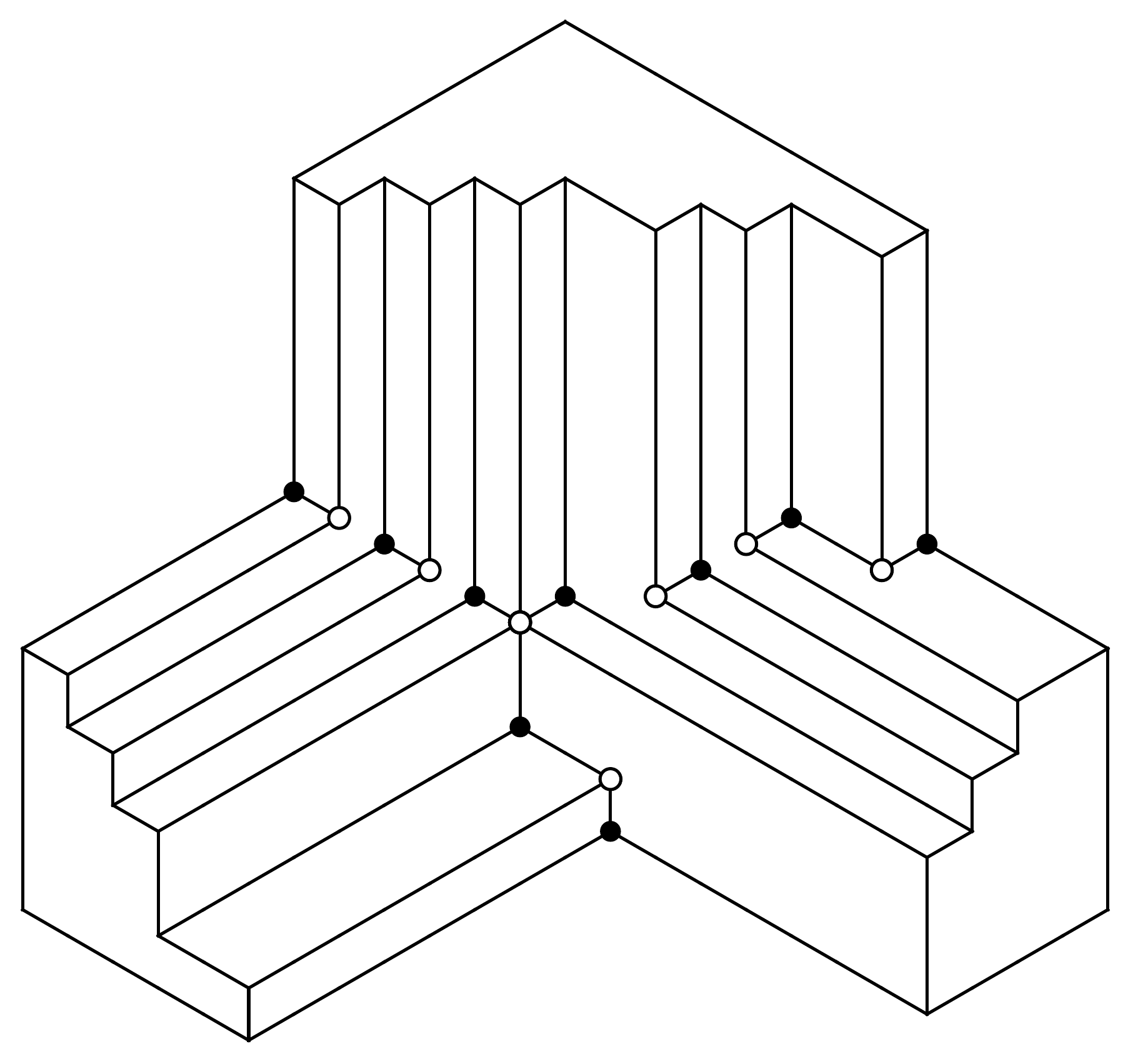}};
			\node at (0,-3.5) {(a)};
			\node at (7,-3.5) {(b)};
			\draw (3.5,-3.5) -- (3.5,3.5);
		\end{tikzpicture}
\caption{A $K_{\IP_2}$ subcrystal given by the asymptotics $Y_1=\{5, 4, 3, 1, 1\}$, $Y_2=\{4, 4, 4, 3, 2\}$, $Y_3=\{8, 6, 5, 3, 2, 1\}$: (a) the ground state configuration and (b) the graph $\mathscr{D}_{\vec Y}$.}.
\label{fig:Lowest_tpriplet_colored_P2}
\end{center}
\end{figure}
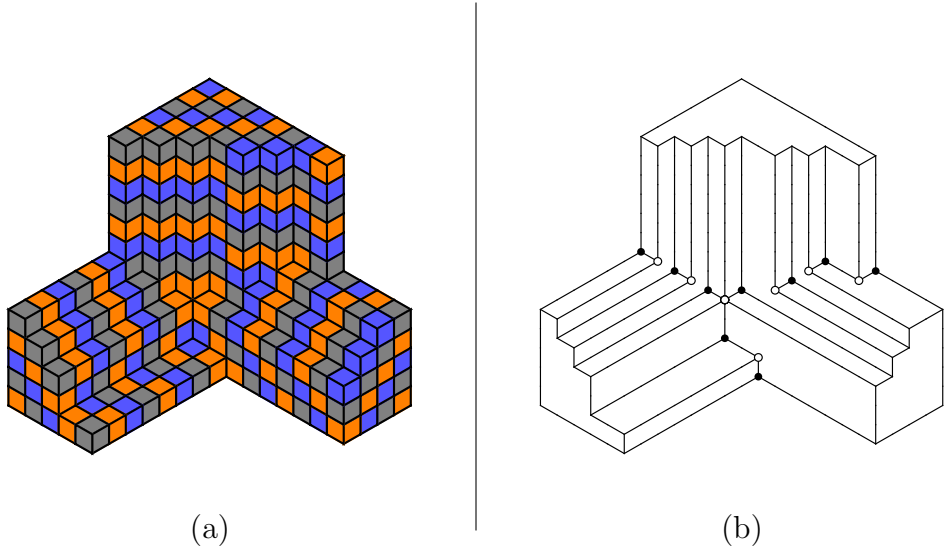
The open BPS states in $K_{\IP_2}$ are again labeled by the three asymptotics Young diagrams $\vec{Y}=(Y_1,Y_2,Y_2)$.
Consider the subcrystal given by the asymptotics $Y_1=\{5, 4, 3, 1, 1\}$, $Y_2=\{4, 4, 4, 3, 2\}$, $Y_3=\{8, 6, 5, 3, 2, 1\}$, shown in Figure~\ref{fig:Lowest_tpriplet_colored_P2}. 
The starters and pausers enter the ground state charge function as the coordinate charge function and the color:
\begin{equation}
(h(\square),c)=\left(x(\mathbf{v}),1+\vec x({\bf v})\cdot(1,1,1)\;{\bf mod}\;3\right) \;.
\end{equation}

Let us list the starters and pausers, together with their $(h(\Box),c)$ values, of the subcrystal in Figure~\ref{fig:Lowest_tpriplet_colored_P2}.
\begin{enumerate}
    \item 
There are nine starters, at the positions
\begin{equation}
\begin{split}
(0,8,5),\;(1,6,5),\;(2,5,4),\;(3,3,3),\;(4,2,3),\\
(4,3,1),\;(4,5,0),\;(5,1,4),\;(6,0,5)\;,
\end{split}
\end{equation}
with  $(h(\square),c)$ values:
\begin{equation}
\begin{split}
(8 \mathsf{h}_2+5 \mathsf{h}_3,{\color{blue}2} ),\;(5 \mathsf{h}_2+4 \mathsf{h}_3,{\color{gray}1} ),\;(3 \mathsf{h}_2+2 \mathsf{h}_3,{\color{orange}3} ),\;(0,{\color{gray}1} ),\;
(2 \mathsf{h}_1+\mathsf{h}_3,{\color{gray}1} ),\\
(3 \mathsf{h}_1+2 \mathsf{h}_2,{\color{orange}3} ),\;(4 \mathsf{h}_1+5 \mathsf{h}_2,{\color{gray}1} ),\;(4 \mathsf{h}_1+3 \mathsf{h}_3,{\color{blue}2} ),\;(6 \mathsf{h}_1+5 \mathsf{h}_3,{\color{orange}3} )\;.
\end{split}
\end{equation}
\item
 There are six simple pausers, at the positions
\begin{equation}
(1,8,5),\;(2,6,5),\;(3,5,4),\;(4,5,1),\;(5,2,4),\;(6,1,5)\;,
\end{equation}
with coordinate-color function $(h(\square),c)$ values:
\begin{equation}
\begin{split}
(7 \mathsf{h}_2+4 \mathsf{h}_3,{\color{orange}3} ),\;(4 \mathsf{h}_2+3 \mathsf{h}_3,{\color{blue}2} ),\;(2 \mathsf{h}_2+\mathsf{h}_3,{\color{gray}1} ),\\
(3 \mathsf{h}_1+4 \mathsf{h}_2,{\color{blue}2} ),\;(3 \mathsf{h}_1+2 \mathsf{h}_3,{\color{orange}3} ),\;(5 \mathsf{h}_1+4 \mathsf{h}_3,{\color{gray}1} )\;.
\end{split}
\end{equation}
\item
There is one double pauser, at the position $(4,3,3)$, and 
with coordinate function $(h(\square),c)$ given by
\begin{equation}
   (\mathsf{h}_1,{\color{blue}2} ) \;.
\end{equation}
\end{enumerate}
Plugging these into the definition of the ground state charge function ${}^{\sharp}\psi^{(a)}_0(z)$ in \eqref{eq:psi0_summary}, we get
\begin{equation}\label{eq:GSCF_P2}
\begin{split}
&\psi_{\vec{Y}}^{{\color{gray}(1)}}(z)=\frac{\left(z-(2 \mathsf{h}_2+ \mathsf{h}_3)\right)\left(z-(5 \mathsf{h}_1+4 \mathsf{h}_3)\right) }{z\left(z- (5 \mathsf{h}_2+4 \mathsf{h}_3)\right)\left(z- (2 \mathsf{h}_1+ \mathsf{h}_3)\right)\left(z- (4 \mathsf{h}_1+5 \mathsf{h}_2)\right)}\,,\\
&\psi_{\vec{Y}}^{{\color{blue}(2)}}(z)=\frac{\left(z-(4 \mathsf{h}_2+3\mathsf{h}_3)\right)\left(z-(3 \mathsf{h}_1+2 \mathsf{h}_3)\right)\left(z- \mathsf{h}_1\right)^2}{\left(z-(8 \mathsf{h}_2+5 \mathsf{h}_3)\right)\left(z- (4\mathsf{h}_1+3\mathsf{h}_3)\right)}\,,\\
&\psi_{\vec{Y}}^{{\color{orange}(3)}}(z)=\frac{\left(z-(7 \mathsf{h}_2+4\mathsf{h}_3)\right)\left(z-(3 \mathsf{h}_1+2 \mathsf{h}_3)\right)}{\left(z-(3 \mathsf{h}_2+2 \mathsf{h}_3)\right)\left(z- (3\mathsf{h}_1+2\mathsf{h}_2)\right)\left(z- (6 \mathsf{h}_1+5 \mathsf{h}_3) \right)}\,.
\end{split}
\end{equation}
Counting the number of the poles and zeros in $\psi^{(a)}_{\vec{Y}}(z)$, we see that the subcrystal given in Figure~\ref{fig:Lowest_tpriplet_colored_P2} gives rise to a representation of the shifted quiver Yangian \eqref{eq-OPE-C3Z3} for the $K_{\IP_2}$ geometry, with the shift \begin{equation}
\mys=(2,-2,1)\,.
\end{equation}
The net shift is $1$, as expected.
\bigskip

From the ground state charge function \eqref{eq:GSCF_Z2} we can derive the  corresponding framed quiver and superpotential for the representation given by Figure~\ref{fig:Lowest_tpriplet_colored_P2}:
\begin{equation}
\begin{aligned}
    \sQ&=\;\begin{array}{c}
    \begin{tikzpicture}
    \draw[thick,postaction={decorate},decoration={markings, 
		mark= at position 0.5 with {\arrow{>}},mark= at position 0.3 with {\arrow{>}},mark= at position 0.4 with {\arrow{>}}}] (-1,0) -- (4,0);
	\draw[thick,postaction={decorate},decoration={markings, 
		mark= at position 0.6 with {\arrow{>}},mark= at position 0.5 with {\arrow{>}},mark= at position 0.4 with {\arrow{>}}}] (4,0) -- (2,-1);
	\draw[thick,postaction={decorate},decoration={markings, 
		mark= at position 0.6 with {\arrow{>}},mark= at position 0.5 with {\arrow{>}},mark= at position 0.4 with {\arrow{>}}}] (2,-1) -- (-1,0);
	\draw[white, fill=white] (1.5,3) to[bend right=15] (2,-1) to[bend left=5] (1.5,3);
	\draw[white, fill=white] (1.5,3) to[bend left=15] (2,-1) to[bend right=5] (1.5,3);
	\draw[thick,postaction={decorate},decoration={markings, 
		mark= at position 0.5 with {\arrow{>}},mark= at position 0.4 with {\arrow{>}},mark= at position 0.6 with {\arrow{>}}}] (1.5,3) to[bend right=10] (2,-1);
	\draw[thick,postaction={decorate},decoration={markings, 
		mark= at position 0.55 with {\arrow{<}},mark= at position 0.45 with {\arrow{<}}}] (1.5,3) to[bend left=10] (2,-1);
	\draw[thick,postaction={decorate},decoration={markings, 
		mark= at position 0.65 with {\arrow{>}},mark= at position 0.55 with {\arrow{>}},mark= at position 0.45 with {\arrow{>}},mark= at position 0.35 with {\arrow{>}}}] (1.5,3) to[bend right=10] (-1,0);
	\draw[thick,postaction={decorate},decoration={markings, 
		mark= at position 0.55 with {\arrow{<}},mark= at position 0.45 with {\arrow{<}}}] (1.5,3) to[bend left=10] (-1,0);
	\draw[thick,postaction={decorate},decoration={markings, 
		mark= at position 0.45 with {\arrow{>}},mark= at position 0.55 with {\arrow{>}}}] (1.5,3) to[bend right=10] (4,0);
	\draw[thick,postaction={decorate},decoration={markings, 
		mark= at position 0.65 with {\arrow{<}},mark= at position 0.55 with {\arrow{<}},mark= at position 0.45 with {\arrow{<}},mark= at position 0.35 with {\arrow{<}}}] (1.5,3) to[bend left=10] (4,0);
	\draw[fill=gray] (-1,0) circle (0.2);
	\draw[fill=blue] (4,0) circle (0.2);
	\draw[fill=orange] (2,-1) circle (0.2);
	\node[left] at (-1.2,0) {$1$};
	\node[right] at (4.2,0) {$2$};
	\node[below] at (2,-1.2) {$3$};
	\begin{scope}[shift={(1.5,3)}]
	    \draw[fill=red] (-0.15,-0.15) -- (-0.15,0.15) -- (0.15,0.15) -- (0.15,-0.15) -- cycle;
	\end{scope}
    \end{tikzpicture}
    \end{array}.
    \end{aligned}
\end{equation}


\subsection{Example: wall-crossing for open BPS states in conifold}

We can combine all the ingredients discussed in this and the previous section.
Namely we can consider open BPS states in general chambers, for a toric Calabi-Yau manifold more general than $\mathbb{C}^3$.
For geometries without compact 4-cycles, the combinatorics for the crystals for such open BPS state countings can be found in \cite{Nagao:2009rq}.
In general, to specify open BPS state counting problems one starts with a crystal for the closed BPS states and then specifies the shape of the crystal by choosing the asymptotic shapes of the crystal, as in the case of the topological vertex \cite{Aganagic:2003db}.
\bigskip

To illustrate this, let us consider as an example the open BPS invariants in an infinite chamber $\CC^{(i)}_m$ for the conifold geometry. 
Without the $D2$-brane, the representation for $\CC^{(i)}_m$ was studied in 
Section~\ref{ssec:infinite_chamber_conifold}, and the shape of the subcrystal, with $m=3$, was given in Figure~\ref{fig:inf_quiv_cry}.

\begin{figure}[ht!]
    \begin{center}
    \begin{tikzpicture}[scale=0.7]
    \node{\rotatebox[origin=c]{180}{\includegraphics[scale=0.28]{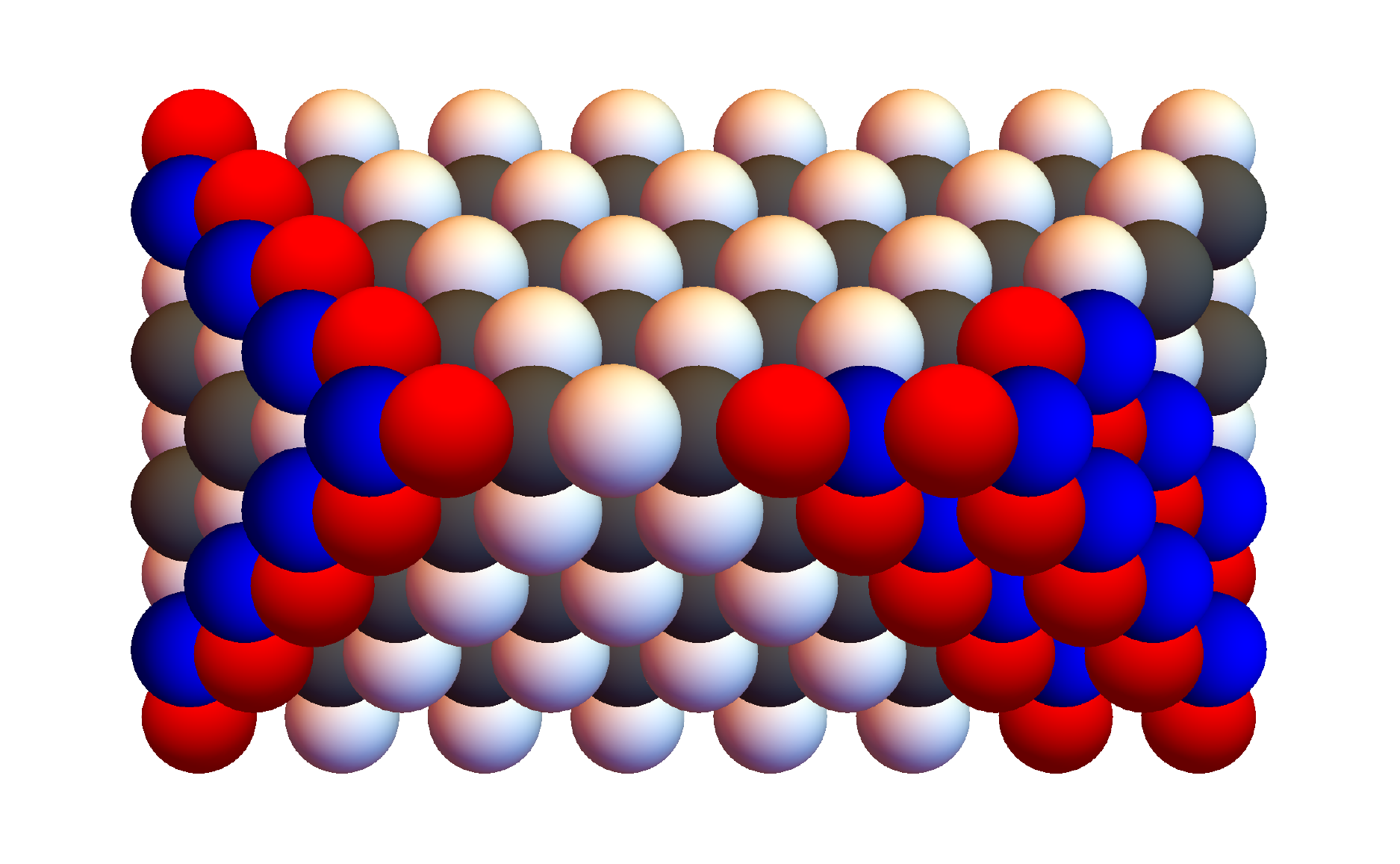}}};
    \node at (9,0) {\rotatebox[origin=c]{180}{\includegraphics[scale=0.28]{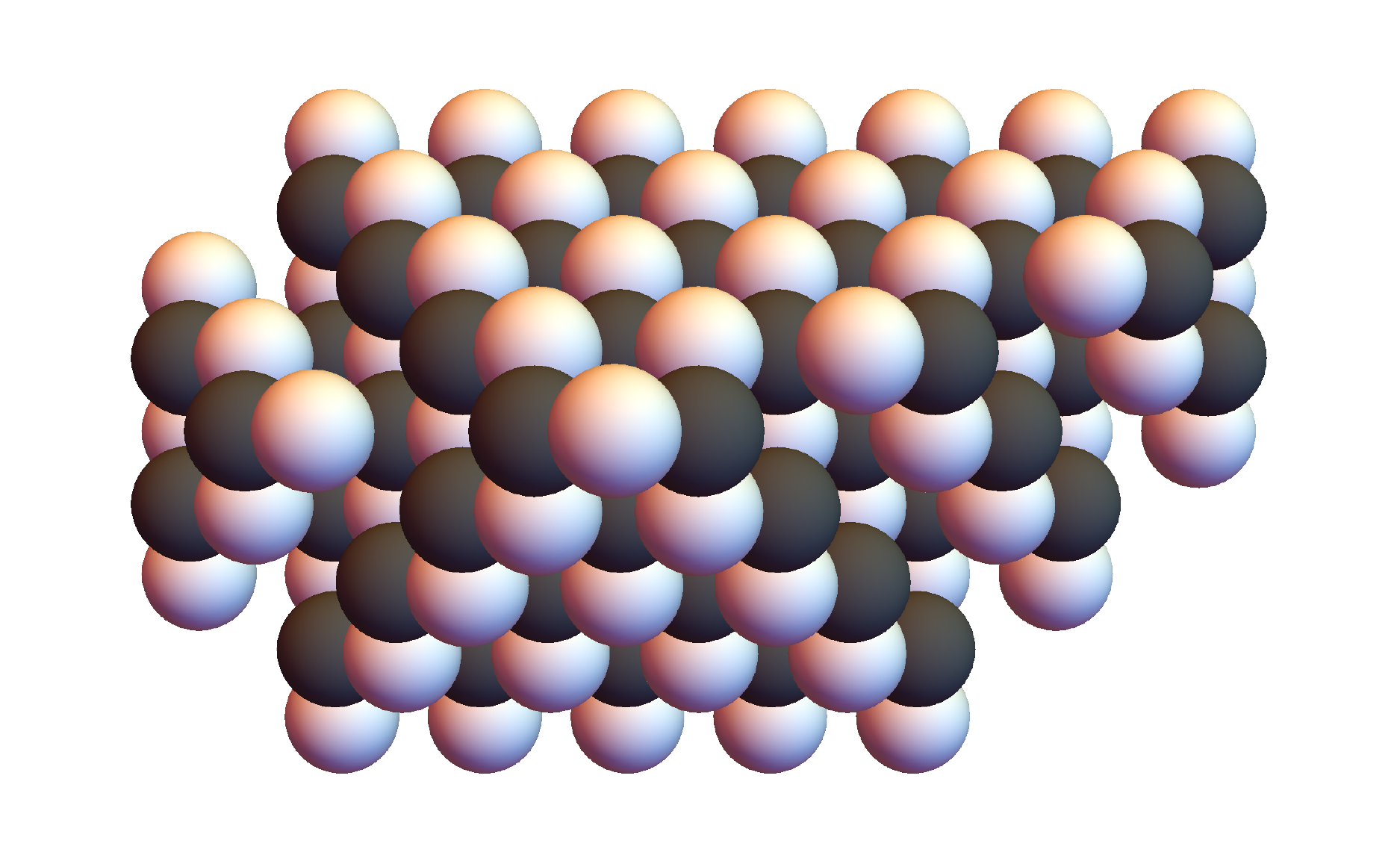}}};
    \draw[ultra thick] (1.4,0) -- (3.5,2.5) (1.4,0) -- (3.5,-2.5) (-1.4,0) -- (-3.5,2.5) (-1.4,0) -- (-3.5,-2.5) (1.4,0) -- (-1.4,0);
    \node[left] at (-3.5,2.5) {$\begin{array}{c}
         \begin{tikzpicture}[scale=0.3,yscale=-1]
         \draw[thick] (0,0) -- (2,0) -- (2,-1) -- (0,-1) -- cycle (1,0) -- (1,-2) -- (0,-2) -- (0,-1);
         \end{tikzpicture}
    \end{array}$};
    \node[left] at (-3.5,-2.5) {$\varnothing$};
    \node[right] at (3.5,2.5) {$\begin{array}{c}
         \begin{tikzpicture}[scale=0.3]
         \draw[thick] (0,0) -- (1,0) -- (1,-1) -- (0,-1) -- cycle;
         \end{tikzpicture}
    \end{array}$};
    \node[right] at (3.5,-2.5) {$\begin{array}{c}
         \begin{tikzpicture}[scale=0.3]
         \draw[thick] (0,0) -- (1,0) -- (1,-1) -- (0,-1) -- cycle;
         \end{tikzpicture}
    \end{array}$};
    \end{tikzpicture}
    \caption{Open BPS state with $\vec Y=\left(\{2,1\},\emptyset,\{1\},\{1\}\right)$ in the infinite chamber $\CC^{(i)}_{m=4}$ of the conifold geometry.
(a) The atoms that lie on the infinite rows with the cross-sections given by $\vec{Y}$ are colored, node $1$ $\rightarrow$ red and node $2$. (b) The red and blue atoms are removed, giving the subcrystal for the open BPS states with $\vec{Y}$ in $\CC^{(i)}_{m=4}$.}
    \label{fig:conifold_open}
    \end{center}
\end{figure}
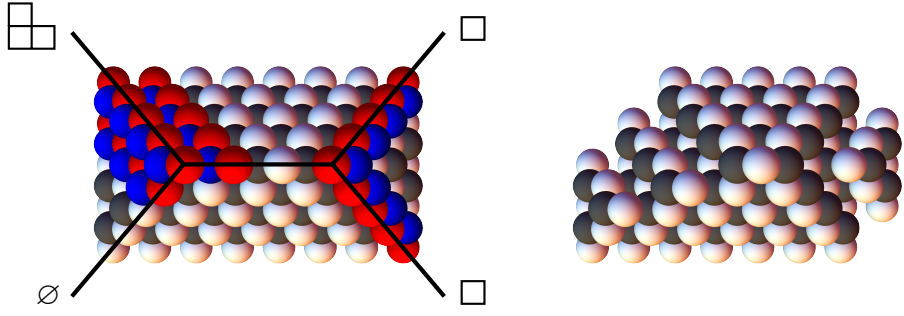

Let us consider $\CC^{(i)}_m$ for $m=4$ with asymptotic $\vec Y=\left(\{2,1\},\emptyset,\{1\},\{1\}\right)$.
Figure~\ref{fig:conifold_open} (a) shows the shape of the subcrystal $\CC^{(i)}_{m=4}$, with the atoms that lie on the infinite rows with the cross-sections given by $\vec{Y}$ colored, node $1$ $\rightarrow$ red and node $2$ $\rightarrow$ blue.
Removing these atoms, we obtain the subcrystal that corresponds to the open BPS states defined by $\vec Y$ in the chamber $\CC^{(i)}_{m=4}$, shown in Figure~\ref{fig:conifold_open} (b).
\bigskip

The corresponding framed quiver and superpotential are a modification of those for the trivial $\vec{Y}$, given in \eqref{eq:quiv_inf_conif}:
\be\label{eq:quiv_inf_conif_Open}
	\begin{aligned}
\sQ&=\begin{array}{c}
			\begin{tikzpicture}
				\draw[thick,->] ([shift=(120:3)]1.5,-2.59808) arc (120:80:3);
				\draw[thick,->] ([shift=(80:3)]1.5,-2.59808) arc (80:70:3);
				\draw[thick] ([shift=(70:3)]1.5,-2.59808) arc (70:60:3);
				\draw[thick,->] ([shift=(300:3)]1.5,2.59808) arc (300:260:3);
				\draw[thick,->] ([shift=(260:3)]1.5,2.59808) arc (260:250:3);
				\draw[thick] ([shift=(250:3)]1.5,2.59808) arc (250:240:3);
				\draw[thick,->] (-4,0) -- (-2.5,0);
				\draw[thick,->] (-2.5,0) -- (-2,0);
				\draw[thick,->] (-2,0) -- (-1.5,0);
				\draw[thick] (-1.5,0) -- (0,0);
				\draw[thick] (3,0) to[out=225,in=0] (1.5,-1) (-2.5,-1) to[out=180,in=315] (-4,0);
				\draw[thick,->] (1.5,-1) -- (0,-1);
				\draw[thick,->] (0,-1) -- (-0.5,-1);
				\draw[thick,->] (-0.5,-1) -- (-1,-1);
				\draw[thick] (-1,-1) -- (-2.5,-1);
				\draw[fill=white] (0,0) circle (0.2);
				\draw[fill=gray] (3,0) circle (0.2);
				\node[above] at (0,0.2) {$1$};
				\node[above] at (3,0.2) {$2$};
				\begin{scope}[shift={(-4,0)}]
					\draw[fill=red] (-0.15,-0.15) -- (-0.15,0.15) -- (0.15,0.15) -- (0.15,-0.15) -- cycle;
				\end{scope}
				\node[above] at (1.5,0.401924) {$a_1$, $a_2$};
				\node[above] at (1.5,-0.401924) {$b_1$, $b_2$};
				\node[above] at (-2,0) {$r_1,r_2,r_3,r_4$};
				\node[above] at (-0.5,-1) {$s_1,s_2,s_3$};
			\end{tikzpicture}
		\end{array},\\ 
\sW&=\Tr\Big[b_2a_2b_1a_1-b_2a_1b_1a_2\\
&\qquad+s_1\left(a_2r_1-a_1b_2a_1r_2\right)\\
&\qquad+s_2\left(a_2r_2-a_1b_2a_1r_3\right)\\
&\qquad+s_3\left(a_2b_2a_2b_1a_2r_3-a_1r_4\right)\Big]\,,
	\end{aligned}
	\ee
where the corresponding masses of fields are:
\begin{equation}
    \begin{array}{c}
         \mu(r_1)=-2 \mathsf{h}_1+2 \mathsf{h}_2,\quad\mu(r_2)=\mathsf{h}_1+\mathsf{h}_2,\quad \mu(r_3)=4 \mathsf{h}_1,\quad \mu(r_4)=8 \mathsf{h}_1,\\
         \mu(s_1)=\mathsf{h}_1-2\mathsf{h}_2,\quad \mu(s_2)=-2\mathsf{h}_1-\mathsf{h}_2,\quad \mu(s_3)=-7\mathsf{h}_1,\\
         \mu(a_1)=-\mathsf{h}_1,\quad \mu(a_2)=\mathsf{h}_1,\quad \mu(b_1)=-\mathsf{h}_2,\quad \mu(b_2)=\mathsf{h}_2\;.
    \end{array}
\end{equation}
The masses $\mu(r_i)$ and $\mu(s_i)$ correspond to the coordinate function of the starters and the pausers of the subcrystal ${}^{\sharp}\CC$ shown in Figure~\ref{fig:conifold_open}, therefore the ground state charge functions are:
\begin{equation}\label{eq:GSCF_conifold_Open}
    {}^{\sharp}\psi_0^{(1)}=\prod\lm_{i=1}^4\frac{1}{\left(z-\mu(r_i)\right)}\quad \textrm{and}\quad {}^{\sharp}\psi_0^{(2)}=\prod\lm_{i=1}^3\left(z+\mu(s_i)\right)
    \,.
\end{equation}
Counting the number of poles and zeros in \eqref{eq:GSCF_conifold_Open}, we see that the subcrystal in 
Figure~\ref{fig:conifold_open} gives rise to a representation of the shifted affine Yangian of $\mathfrak{gl}_{1|1}$, with shift
\begin{equation}
    \mys=(4,-3)\,,
\end{equation}
with net shift $1$. This is the same as the case when the asymptotics $\vec{Y}$ is trivial (see \eqref{shift_inf}), as expected.
	
\section{More general representations}
\label{sec:genera_rep}

In all the examples considered in Section~\ref{sec:wallcrossing} and Section~\ref{sec:Open}, the subcrystals themselves and their corresponding framed quivers either have already appeared in the literature, or can arise naturally if we perform some conventional deformations to our field theory setup considered also in the literature.
What we have achieved in Section~\ref{sec:wallcrossing} and Section~\ref{sec:Open} is to construct their associated representations of the relevant shifted quiver Yangians. 
All these representations are irreducible for generic choices of the equivariant parameters. 

However, the procedure of constructing a representation from a given subcrystal, and moreover of relating (1) the shape of the subcrystal, (2) the ground state charge function, and (3) frame quiver and superpotential, are rather general. 
Namely, any \emph{arbitrary} subcrystal can give rise to a representation of the shifted quiver Yangian and the associated framed quiver.
For simplicity here we consider those subcrystals that correspond to irreducible representations (recall Section \ref{subsec.convex}).

In this section, we will consider those subcrystals (subject to the irreducibility constraints) that do not arise as a result of some physical manipulation on the underlying theory, like wall-crossing or BPS states supporting open invariants, at least not as far as we are aware. 	
We will call these representations \emph{\funny} representations.

\subsection{\texorpdfstring{Example: finite representations for $\IC^3$}{Example: Finite Representations for C(3)}}
\label{ssec:funny_C3}
	
The first example we would like to consider in this  section is a family of finite representations for the case of $\IC^3$. 
As we have mentioned, any convex subcrystal subject to the irreducibility constraint would give a new representation. 
For simplicity, let us consider a subcrystal of the shape of a cuboid, with three lengths being of $m$, $n$ and $k$ atoms, namely, we  would like the crystal growth to terminate after steps $m$, $n$ and $k$ along the $x_{1,2,3}$ direction, respectively, shown below:	\be\label{fig:cuboid}
	\begin{array}{c}
	\begin{tikzpicture}[scale=0.5]
		\foreach \a in {0,...,4}
		\foreach \b in {0,...,2}
		\foreach \c in {3,...,3}
		{
			\draw[thick, fill=gray] (-0.866025*\a+0.866025*\b,-0.5*\a-0.5*\b+\c) -- (-0.866025*\a+0.866025*\b,-0.5*\a-0.5*\b+\c - 1) --(-0.866025*\a+0.866025*\b -0.866025,-0.5*\a-0.5*\b+\c -0.5) -- (-0.866025*\a+0.866025*\b -0.866025,-0.5*\a-0.5*\b+\c +0.5) -- cycle
			(-0.866025*\a+0.866025*\b,-0.5*\a-0.5*\b+\c) -- (-0.866025*\a+0.866025*\b,-0.5*\a-0.5*\b+\c - 1) --(-0.866025*\a+0.866025*\b +0.866025,-0.5*\a-0.5*\b+\c -0.5) -- (-0.866025*\a+0.866025*\b +0.866025,-0.5*\a-0.5*\b+\c +0.5) -- cycle
			(-0.866025*\a+0.866025*\b,-0.5*\a-0.5*\b+\c) -- (-0.866025*\a+0.866025*\b -0.866025,-0.5*\a-0.5*\b+\c +0.5) -- (-0.866025*\a+0.866025*\b,-0.5*\a-0.5*\b+\c+1) -- (-0.866025*\a+0.866025*\b +0.866025,-0.5*\a-0.5*\b+\c +0.5) -- cycle;
		}
	\foreach \a in {0,...,4}
	\foreach \b in {2,...,2}
	\foreach \c in {0,...,3}
	{
		\draw[thick, fill=gray] (-0.866025*\a+0.866025*\b,-0.5*\a-0.5*\b+\c) -- (-0.866025*\a+0.866025*\b,-0.5*\a-0.5*\b+\c - 1) --(-0.866025*\a+0.866025*\b -0.866025,-0.5*\a-0.5*\b+\c -0.5) -- (-0.866025*\a+0.866025*\b -0.866025,-0.5*\a-0.5*\b+\c +0.5) -- cycle
		(-0.866025*\a+0.866025*\b,-0.5*\a-0.5*\b+\c) -- (-0.866025*\a+0.866025*\b,-0.5*\a-0.5*\b+\c - 1) --(-0.866025*\a+0.866025*\b +0.866025,-0.5*\a-0.5*\b+\c -0.5) -- (-0.866025*\a+0.866025*\b +0.866025,-0.5*\a-0.5*\b+\c +0.5) -- cycle
		(-0.866025*\a+0.866025*\b,-0.5*\a-0.5*\b+\c) -- (-0.866025*\a+0.866025*\b -0.866025,-0.5*\a-0.5*\b+\c +0.5) -- (-0.866025*\a+0.866025*\b,-0.5*\a-0.5*\b+\c+1) -- (-0.866025*\a+0.866025*\b +0.866025,-0.5*\a-0.5*\b+\c +0.5) -- cycle;
	}
\foreach \a in {4,...,4}
\foreach \b in {0,...,2}
\foreach \c in {0,...,3}
{
	\draw[thick, fill=gray] (-0.866025*\a+0.866025*\b,-0.5*\a-0.5*\b+\c) -- (-0.866025*\a+0.866025*\b,-0.5*\a-0.5*\b+\c - 1) --(-0.866025*\a+0.866025*\b -0.866025,-0.5*\a-0.5*\b+\c -0.5) -- (-0.866025*\a+0.866025*\b -0.866025,-0.5*\a-0.5*\b+\c +0.5) -- cycle
	(-0.866025*\a+0.866025*\b,-0.5*\a-0.5*\b+\c) -- (-0.866025*\a+0.866025*\b,-0.5*\a-0.5*\b+\c - 1) --(-0.866025*\a+0.866025*\b +0.866025,-0.5*\a-0.5*\b+\c -0.5) -- (-0.866025*\a+0.866025*\b +0.866025,-0.5*\a-0.5*\b+\c +0.5) -- cycle
	(-0.866025*\a+0.866025*\b,-0.5*\a-0.5*\b+\c) -- (-0.866025*\a+0.866025*\b -0.866025,-0.5*\a-0.5*\b+\c +0.5) -- (-0.866025*\a+0.866025*\b,-0.5*\a-0.5*\b+\c+1) -- (-0.866025*\a+0.866025*\b +0.866025,-0.5*\a-0.5*\b+\c +0.5) -- cycle;
}
\foreach \a in {5,...,5}
\foreach \b in {0,...,0}
\foreach \c in {0,...,0}
{
	\draw[thick, fill=red] (-0.866025*\a+0.866025*\b,-0.5*\a-0.5*\b+\c) -- (-0.866025*\a+0.866025*\b,-0.5*\a-0.5*\b+\c - 1) --(-0.866025*\a+0.866025*\b -0.866025,-0.5*\a-0.5*\b+\c -0.5) -- (-0.866025*\a+0.866025*\b -0.866025,-0.5*\a-0.5*\b+\c +0.5) -- cycle
	(-0.866025*\a+0.866025*\b,-0.5*\a-0.5*\b+\c) -- (-0.866025*\a+0.866025*\b,-0.5*\a-0.5*\b+\c - 1) --(-0.866025*\a+0.866025*\b +0.866025,-0.5*\a-0.5*\b+\c -0.5) -- (-0.866025*\a+0.866025*\b +0.866025,-0.5*\a-0.5*\b+\c +0.5) -- cycle
	(-0.866025*\a+0.866025*\b,-0.5*\a-0.5*\b+\c) -- (-0.866025*\a+0.866025*\b -0.866025,-0.5*\a-0.5*\b+\c +0.5) -- (-0.866025*\a+0.866025*\b,-0.5*\a-0.5*\b+\c+1) -- (-0.866025*\a+0.866025*\b +0.866025,-0.5*\a-0.5*\b+\c +0.5) -- cycle;
}
\foreach \a in {0,...,0}
\foreach \b in {3,...,3}
\foreach \c in {0,...,0}
{
	\draw[thick, fill=red] (-0.866025*\a+0.866025*\b,-0.5*\a-0.5*\b+\c) -- (-0.866025*\a+0.866025*\b,-0.5*\a-0.5*\b+\c - 1) --(-0.866025*\a+0.866025*\b -0.866025,-0.5*\a-0.5*\b+\c -0.5) -- (-0.866025*\a+0.866025*\b -0.866025,-0.5*\a-0.5*\b+\c +0.5) -- cycle
	(-0.866025*\a+0.866025*\b,-0.5*\a-0.5*\b+\c) -- (-0.866025*\a+0.866025*\b,-0.5*\a-0.5*\b+\c - 1) --(-0.866025*\a+0.866025*\b +0.866025,-0.5*\a-0.5*\b+\c -0.5) -- (-0.866025*\a+0.866025*\b +0.866025,-0.5*\a-0.5*\b+\c +0.5) -- cycle
	(-0.866025*\a+0.866025*\b,-0.5*\a-0.5*\b+\c) -- (-0.866025*\a+0.866025*\b -0.866025,-0.5*\a-0.5*\b+\c +0.5) -- (-0.866025*\a+0.866025*\b,-0.5*\a-0.5*\b+\c+1) -- (-0.866025*\a+0.866025*\b +0.866025,-0.5*\a-0.5*\b+\c +0.5) -- cycle;
}
\foreach \a in {0,...,0}
\foreach \b in {0,...,0}
\foreach \c in {4,...,4}
{
	\draw[thick, fill=red] (-0.866025*\a+0.866025*\b,-0.5*\a-0.5*\b+\c) -- (-0.866025*\a+0.866025*\b,-0.5*\a-0.5*\b+\c - 1) --(-0.866025*\a+0.866025*\b -0.866025,-0.5*\a-0.5*\b+\c -0.5) -- (-0.866025*\a+0.866025*\b -0.866025,-0.5*\a-0.5*\b+\c +0.5) -- cycle
	(-0.866025*\a+0.866025*\b,-0.5*\a-0.5*\b+\c) -- (-0.866025*\a+0.866025*\b,-0.5*\a-0.5*\b+\c - 1) --(-0.866025*\a+0.866025*\b +0.866025,-0.5*\a-0.5*\b+\c -0.5) -- (-0.866025*\a+0.866025*\b +0.866025,-0.5*\a-0.5*\b+\c +0.5) -- cycle
	(-0.866025*\a+0.866025*\b,-0.5*\a-0.5*\b+\c) -- (-0.866025*\a+0.866025*\b -0.866025,-0.5*\a-0.5*\b+\c +0.5) -- (-0.866025*\a+0.866025*\b,-0.5*\a-0.5*\b+\c+1) -- (-0.866025*\a+0.866025*\b +0.866025,-0.5*\a-0.5*\b+\c +0.5) -- cycle;
}
\node[left] at (-4.33013, -0.5) {$k$ atoms};
\node[above left] at (-2.16506, 2.75) {$m$ atoms};
\node[above right] at (1.29904, 3.25) {$n$ atoms};
	\end{tikzpicture}
	\end{array}
	\ee

In terms of the positive/negative crystals picture, the subcrystal (\ref{fig:cuboid}) has one starter at 	
\begin{equation}
(x_1,x_2,x_3)=(0,0,0) \;,
\end{equation}
and three stoppers at
\begin{equation}\label{eq:stopper_funny_C3}
(x_1,x_2,x_3)=(m,0,0)\,, \quad (0,n,0)\,, \quad (0,0,k)\,.
\end{equation}
Plugging these into ~\eqref{eq:psi0_summary}, we obtain the ground state charge function:
\begin{equation}\label{eq:psi_0_fin_C3}
	    {}^{\sharp}\psi_0(z)=\frac{(z-m\mathsf{h}_1)(z-n\mathsf{h}_2)(z-k\mathsf{h}_3)}{z}\,,
\end{equation}
where $\mathsf{h}_1+\mathsf{h}_2+\mathsf{h}_3=0$.
The three zeros in \eqref{eq:psi_0_fin_C3} cancel the creating poles at $m\mathsf{h}_1$, $n\mathsf{h}_2$, and $k\mathsf{h}_3$, respectively.
Therefore,  applying the creation operator $e(z)$ repeatedly on the ground state would give a finite representation of the shifted affine Yangian of $\mathfrak{gl}_1$. 
The basis of the corresponding Verma module is finite and represented by molten crystals embedded in the cuboid (\ref{fig:cuboid}). 

\bigskip

Compared to the canonically framed quiver and superpotential \eqref{eq:QW0C3}, the presence of the three stoppers \eqref{eq:stopper_funny_C3} imposes three 
new relations in the quiver path algebra:
	\be
	B_1^{m-1}\cdot R=0\,,\quad B_2^{n-1}\cdot R=0\,,\quad B_3^{k-1}\cdot R=0\,,
	\ee
which are	implemented by three new Lagrange constraint fields $S_{1,2,3}$, represented by three arrows going from the gauge node to the framing node.
The resulting framed quiver and superpotential are:
	\be
	\begin{aligned}
	&Q_{m,n,k}^{(\mbox{\tiny\funny})}=\begin{array}{c}
		\begin{tikzpicture}
		\begin{scope}[rotate=-90]
			\draw[thick,->] ([shift=(120:0.75)]0,0.75) arc (120:420:0.75);
			\draw[thick,->] ([shift=(60:0.75)]0,0.75) arc (60:90:0.75);
			\draw[thick,->] ([shift=(90:0.75)]0,0.75) arc (90:120:0.75);
			\draw[fill=white] (0,0) circle (0.2);
			\draw[thick,->] (0.1,-1.35) -- (0.1,-0.75);
			\draw[thick] (0.1,-0.75) -- (0.1,-0.173205);
			\draw[thick] (-0.1,-1.35) -- (-0.1,-0.85);
			\draw[thick,<-] (-0.1,-0.85) -- (-0.1,-0.75);
			\draw[thick,<-] (-0.1,-0.75) -- (-0.1,-0.65);
			\draw[thick,<-] (-0.1,-0.65) -- (-0.1,-0.173205);
			\begin{scope}[shift={(0,-1.5)}]
				\draw[thick,fill=red] (-0.15,-0.15) -- (-0.15,0.15) -- (0.15,0.15) -- (0.15,-0.15) -- cycle;
			\end{scope}
			\node[right] at (0,1.5) {$B_{1,2,3}$};
			\node[below] at (0.15,-0.75) {$R$};
			\node[above] at (-0.15,-0.75) {$S_{1,2,3}$};
		\end{scope}
		\end{tikzpicture} 
	\end{array} \;,
	\\
&\begin{array}{l} W_{m,n,k}^{(\mbox{\tiny\funny})}=\Tr\big[B_1[B_2,B_3]+S_1B_1^{m-1}R +S_2B_2^{n-1}R+S_3B_3^{k-1}R\big].
\end{array}
\end{aligned}
	\ee

\subsection{Example: ``half-infinite" representation for conifold}
\label{ssec:funny_conifold}	

Another interesting example emerges when we revisit the case of wall-crossing in the conifold. 

Recall that the finite chamber $\CC^{(f)}_m$ differs from the infinite chamber $\CC^{(i)}_m$ by the presence of the two stoppers at the end of the ridge of the pyramid (i.e.\  the initial chain of atoms), see Figure~\ref{fig:fin_cry_conif}(b)  vs.\ Figure~\ref{fig:inf_quiv_cry}(b).
These two stoppers impose two termination constraints and cut the infinite subcrystal $\CC^{(i)}_m$ into a finite one $\CC^{(f)}_m$. 

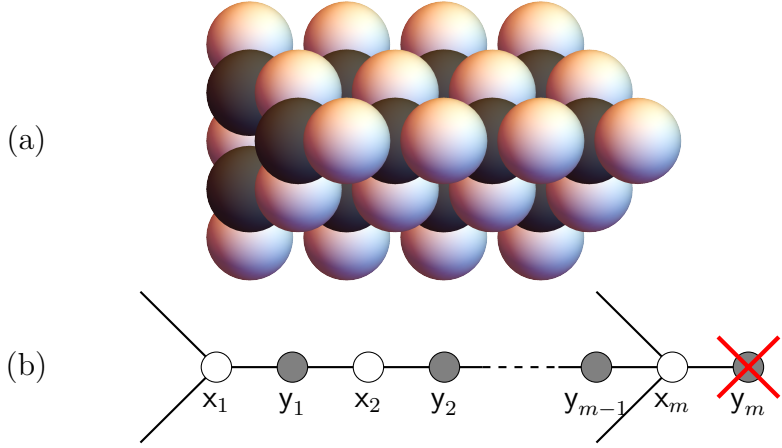
\begin{figure}[ht!]
		\begin{center}
		\begin{tikzpicture}
				\node at (0,0) {\includegraphics[scale=0.27]{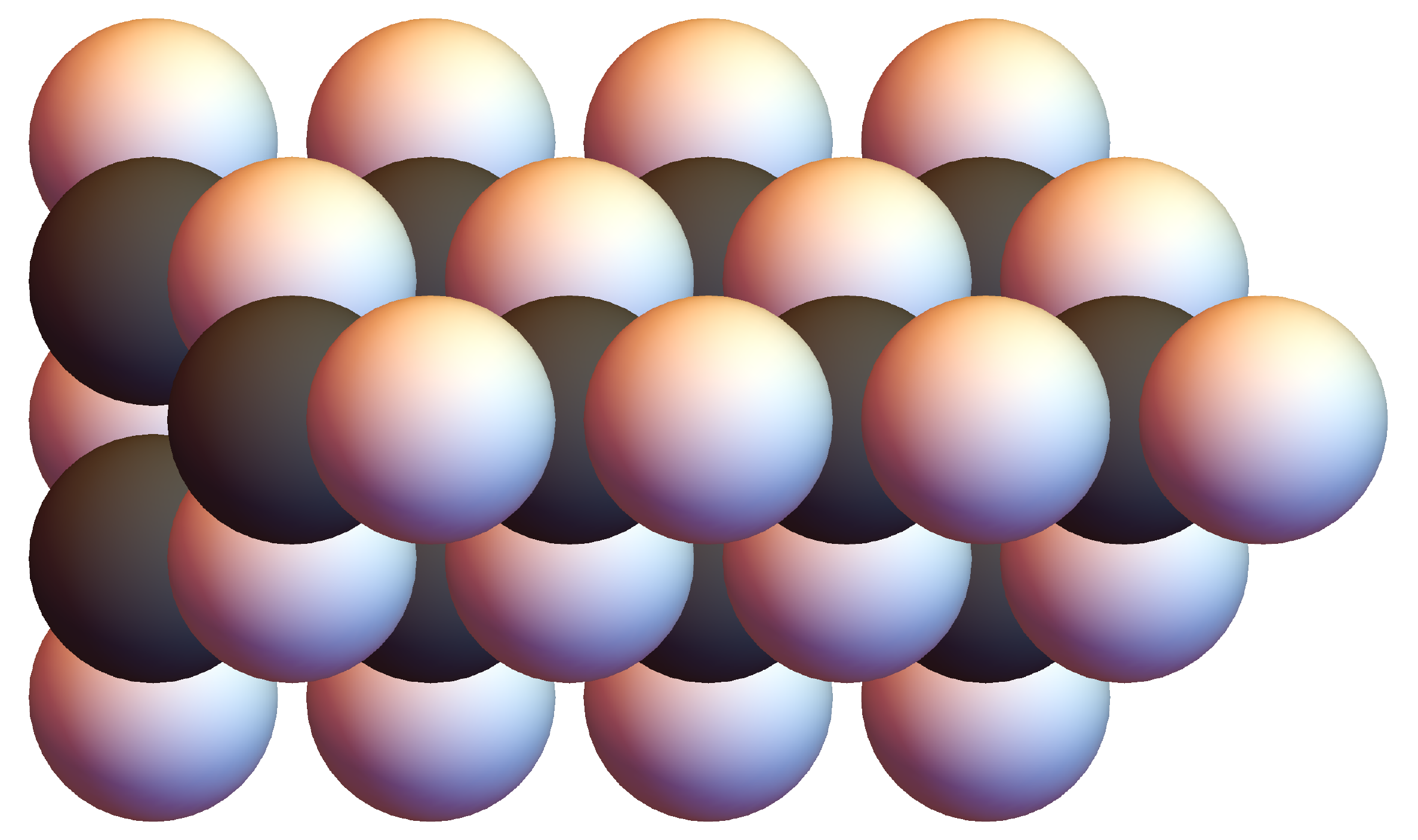}};
				\node at (-5.5,0) {(a)};
				\node at (-5.5,-3) {(b)};
				\begin{scope}[shift={(-3,-3)}]
					\draw[thick] (0,0) -- (3.5,0) (4.5,0) -- (7,0) (-1,1) -- (0,0) -- (-1,-1) (5,1) -- (6,0) -- (5,-1);
		\draw[thick,dashed] (3.5,0) -- (4.5,0);
		\draw[fill=white] (0,0) circle (0.2);
		\draw[fill=gray] (1,0) circle (0.2);
		\draw[fill=white] (2,0) circle (0.2);
		\draw[fill=gray] (3,0) circle (0.2);
		\draw[fill=gray] (5,0) circle (0.2);
		\draw[fill=white] (6,0) circle (0.2);
		\draw[fill=gray] (7,0) circle (0.2);
		\begin{scope}[shift={(7,0)}]
			\draw[ultra thick, red] (-0.4,-0.4) -- (0.4,0.4) (0.4,-0.4) -- (-0.4,0.4);
		\end{scope}
		\node[below] at (0,-0.2) {$\textsf{x}_1$};
		\node[below] at (1,-0.2) {$\textsf{y}_1$};
		\node[below] at (2,-0.2) {$\textsf{x}_2$};
		\node[below] at (3,-0.2) {$\textsf{y}_2$};
		\node[below] at (5,-0.2) {$\textsf{y}_{m-1}$};
		\node[below] at (6,-0.2) {$\textsf{x}_m$};
		\node[below] at (7,-0.2) {$\textsf{y}_{m}$};
				\end{scope}
			\end{tikzpicture}
	\caption{(a) The crystal for the ``half-infinite" chamber of the resolved conifold, i.e.\ a half-infinite pyramid, 
			with $m=4$.
			(b) The locations of the starters $\textsf{x}_k$ with $k=1,2,\dots,m$, the pausers $\textsf{y}_k$ with $k=1,2,\dots,m-1$, and a single stopper  $\textsf{y}_{m}$.}\label{fig:half_fin_cry_conif}
	\end{center}
	\end{figure}

Now let us place only one stopper, at the right end of the ridge of $\CC^{(i)}_m$ in Figure~\ref{fig:inf_quiv_cry}. 
	The resulting crystal $\CC^{h.i.}_m$ has the shape of a half-infinite prism, with $m$ atoms in its ridge (see Figure \ref{fig:half_fin_cry_conif}.)

To obtain the framed quiver and superpotential for the half-infinite chamber $\CC^{(h.i)}_m$, we simply eliminate the arrow that corresponds to the left stopper from the framed quiver \eqref{QW_conif_fin} for the finite chamber $\CC^{(f)}_m$ and the corresponding term in the superpotential in \eqref{QW_conif_fin}, 
and the resulting quiver and superpotential read:
	\be
	\begin{aligned}
		Q_m^{(\mbox{\tiny\funny})}&=\begin{array}{c}
			\begin{tikzpicture}
				\draw[thick,->] ([shift=(120:3)]1.5,-2.59808) arc (120:80:3);
				\draw[thick,->] ([shift=(80:3)]1.5,-2.59808) arc (80:70:3);
				\draw[thick] ([shift=(70:3)]1.5,-2.59808) arc (70:60:3);
				\draw[thick,->] ([shift=(300:3)]1.5,2.59808) arc (300:260:3);
				\draw[thick,->] ([shift=(260:3)]1.5,2.59808) arc (260:250:3);
				\draw[thick] ([shift=(250:3)]1.5,2.59808) arc (250:240:3);
				\draw[thick,->] (-4,0) -- (-2.5,0);
				\draw[thick,->] (-2.5,0) -- (-2,0);
				\draw[thick,->] (-2,0) -- (-1.5,0);
				\draw[thick] (-1.5,0) -- (0,0);
				\draw[thick] (3,0) to[out=225,in=0] (1.5,-1) (-2.5,-1) to[out=180,in=315] (-4,0);
				\draw[thick,->] (1.5,-1) -- (0,-1);
				\draw[thick,->] (0,-1) -- (-0.5,-1);
				\draw[thick,->] (-0.5,-1) -- (-1,-1);
				\draw[thick] (-1,-1) -- (-2.5,-1);
				\draw[fill=white] (0,0) circle (0.2);
				\draw[fill=gray] (3,0) circle (0.2);
				\node[above] at (0,0.2) {$v_0$};
				\node[above] at (3,0.2) {$v_1$};
				\begin{scope}[shift={(-4,0)}]
					\draw[fill=red] (-0.15,-0.15) -- (-0.15,0.15) -- (0.15,0.15) -- (0.15,-0.15) -- cycle;
				\end{scope}
				\node[above] at (1.5,0.401924) {$a_1$, $a_2$};
				\node[above] at (1.5,-0.401924) {$b_1$, $b_2$};
				\node[above] at (-2,0) {$r_1,\ldots, r_m$};
				\node[above] at (-0.5,-1) {$s_1,\ldots,s_{m}$};
			\end{tikzpicture}
		\end{array},\\ 
		W_m^{(\mbox{\tiny\funny})}&=\Tr\left[b_2a_2b_1a_1-b_2a_1b_1a_2+\sum\lm_{i=1}^{m-1}s_{i}(a_2r_i-a_1r_{i+1})+s_{m}a_2r_m\right].
	\end{aligned}
	\ee
\bigskip
	
The equivariant weights of the starters, pausers, and the single stopper for the crystal $\CC^{h.i.}_m$ are 
\be
\begin{aligned}
\textrm{starts of color $1$:}\qquad 	&\chi_k=\mu(r_k)=2(k-1)\mathsf{h}_1\,,\quad k=1,\dots,m\,;\\
\textrm{pausers of color $2$:}\qquad 	&\upsilon_k=-\mu(s_k)=(2k-1)\mathsf{h}_1\,,\quad k=1,\dots,m-1\,;\\
\textrm{stoppers of color $2$:}\qquad 	&\upsilon_m=-\mu(s_m)=(2m-1)\mathsf{h}_1\,,	\end{aligned}
	\ee	
from which we can immediately write down the ground state charge function using \eqref{eq:psi0_summary}:
	\be\label{eq:half_infinite_chamber_GSCF}
	\begin{aligned}
		{}^{\sharp}\psi^{(1)}_0=\prod\lm_{k=1}^m\frac{1}{z-\chi_k}\quad \textrm{and} \quad
		{}^{\sharp}\psi^{(2)}_0=\prod\lm_{k=1}^{m}\left(z-\upsilon_k\right)\,.
	\end{aligned}
	\ee
We see that the half-infinite representation from the crystal $\CC_m^{(h.i)}$ is a representation of the shifted quiver Yangian of $\mathfrak{gl}_{1|1}$, with shifts given by
\begin{equation}
\mys = (m, -m)\,.
\end{equation}	
We have checked that all the algebraic relations \eqref{eq-OPE-toric} and \eqref{eq.efpsi-action}  are applicable to this family of examples.

\section{Summary and open problems}\label{sec:summary}

The (unshifted) quiver Yangian $\mathsf{Y}(Q,W)$ is the BPS algebra for type IIA string theory compactified on an arbitrary non-compact toric Calabi-Yau three-fold, in the so-called non-commutative DT chamber and without the open BPS states involved \cite{Li:2020rij,Galakhov:2020vyb}.
The BPS Hilbert space corresponds to the set of molten crystals from a canonical crystal $\CC_0$, which spans the \emph{vacuum} representation of $\mathsf{Y}(Q,W)$.
The quiver and superpotential pair $(Q,W)$ is the canonically framed one. 
\bigskip

In this paper, we generalize this story to include all the known BPS counting problems, including other chambers in wall-crossing and open BPS states, in this framework; and the end result actually produces a much broader class of BPS counting problems, most of them new.
\bigskip

The central player is the subcrystal ${}^{\sharp}\CC$ of the canonical crystal $\CC_0$. 
We have shown that an \emph{arbitrary} simply-connected convex subcrystal ${}^{\sharp}\CC$ gives rise to a set of molten crystal configurations that furnish a (generically non-vacuum) representation ${}^{\sharp}\textrm{Rep}$ of a shifted quiver Yangian $\mathsf{Y}(Q,W, \mys)$.

We have shown that a subcrystal ${}^{\sharp}\CC$ can be characterized by a unique decomposition of positive and negative canonical crystals $\CC_0$.
Namely, the information of the representation ${}^{\sharp}\textrm{Rep}$ from ${}^{\sharp}\CC$ can be packaged into a set of leading atoms  --  called starters, pausers, and stoppers  --  that specify the locations of the corresponding positive and negative $\CC_0$ building blocks. 
The information of these leading atoms can then be easily translated to the ground state charge function $\{{}^{\sharp}\psi^{(a)}(z)\}$ of the representation ${}^{\sharp}\textrm{Rep}$, which determines the entire  representation.\footnote{This method of determining $\{{}^{\sharp}\psi^{(a)}_0(z)\}$ from the shape of the subcrystal is  far superior to an alternative method that was used in the $\mathbb{C}^3$ case and is generalized to arbitrary toric Calabi-Yau threefolds in Appendix~\ref{sec:app_manual_charge_f}.}
\bigskip

Similar to the fact that the unshifted quiver Yangian $\mathsf{Y}(Q,W)$ can be bootstrapped from its vacuum representation $\textrm{Rep}_0$ (which is spanned by the molten crystals from $\CC_0$) \cite{Li:2020rij}, the shifted quiver Yangian $\mathsf{Y}(Q,W, {}^{\sharp}\psi)$, or equivalently $\mathsf{Y}(Q,W, \mys)$, can also be bootstrapped from these non-vacuum representations ${}^{\sharp}\textrm{Rep}$, spanned by the molten crystals from the subcrystal ${}^{\sharp}\CC$.
\bigskip

The shape of the subcrystal can then be easily translated into the framing $\sharp$ of the quiver superpotential $(Q,W)$, giving rise to the framed quiver superpotential $(\sQ,\sW)$.
We have checked that the BPS algebra of the $\mathcal{N}=4$ supersymmetric quiver quantum mechanics system defined by $(\sQ,\sW)$ matches with the shifted quiver Yangian $\mathsf{Y}(Q,W,{}^{\sharp}\psi)$ that is directly bootstrapped from the subcrystal representation ${}^{\sharp}\textrm{Rep}$.  
\bigskip

Although our original motivation was to generalize the unshifted quiver Yangian in order to cover all the known BPS counting problems, the result goes far beyond this. 
It would be interesting to explore these new BPS counting problems further, both within the framework of quiver Yangians and beyond.

 \bigskip

In closing, let us comment on some further open questions for future research:

\begin{itemize}

\item In Section~\ref{ssec:truncations} we discussed situations where the subcrystal representations become reducible for non-generic equivariant parameters, leading to truncations of the algebra (so that the algebra acts irreducibly). It would be interesting to work out this truncation of the shifted quiver Yangians in detail, and also identify the truncations with the geometry of non-compact D-branes. Such an analysis will 
strengthen the analysis of the unshifted case in \cite{Li:2020rij}.

\item As we discussed in Section~\ref{sec:BPSalgebras}, the representations will no longer be described by 
statistical model of crystal melting when the choice of stability parameters are not chosen as those from the cyclic chamber.
In these non-cyclic chambers, we still expect that the fixed points of the moduli space are given by combinatorics associated with quivers,
and that we obtain representations from them. The problem is to study these representations in detail, 
and generalize our discussion to arbitrary chambers.

\item Our discussion of this paper relies crucially on the equivariant localization with respect to the torus action originating from the toric condition 
on the Calabi-Yau three-fold. One natural question is whether we can extend our discussion to non-toric Calabi-Yau three-folds,
especially to compact Calabi-Yau three-folds.

\end{itemize}

\section*{Acknowledgements}

We would like to thank Hiraku Nakajima for helpful discussions.
WL is grateful for support from NSFC No.\ 11875064 and 11947302, CAS Grant No.\ XDPB15, and the Max-Planck Partnergruppen fund. 
The work of MY and DG was supported in part by WPI Research Center Initiative, MEXT, Japan. MY was also supported by the JSPS Grant-in-Aid for Scientific Research (17KK0087, 19K03820, 19H00689, 20H05850, 20H05860). The research of DG was supported in part by the National Science Foundation under Grant No. NSF PHY-1748958.

\appendix

\section{Conifold mutations}\label{sec:App_conifold}
In this appendix we give the details of a single quiver mutation process 
-- or Seiberg duality transformation -- on a conifold quiver~\eqref{eq:quiv_inf_conif}. 
We will show that the duality describes a transition between two members of the family:
\begin{equation}
	\begin{array}{c}
		\begin{tikzpicture}
		\node (A) at (0,0) {$\left(Q_m^{(i)},W_m^{(i)}\right)$};
		\node (B) at (6,0) {$\left(Q_{m+1}^{(i)},W_{m+1}^{(i)}\right)$};
		\path (A) edge[<->] node[above] {\footnotesize Seiberg dual} (B);
	\end{tikzpicture}
	\end{array}.
\end{equation}
In particular, we will start with the pair $\left(Q_m^{(i)},W_m^{(i)}\right)$ in \eqref{eq:quiv_inf_conif}, which we reproduce here
\be\label{eq:quiv_inf_conif_app}
	\begin{aligned}
Q_m^{(i)}&=\begin{array}{c}
			\begin{tikzpicture}
				\draw[thick,->] ([shift=(120:3)]1.5,-2.59808) arc (120:80:3);
				\draw[thick,->] ([shift=(80:3)]1.5,-2.59808) arc (80:70:3);
				\draw[thick] ([shift=(70:3)]1.5,-2.59808) arc (70:60:3);
				\draw[thick,->] ([shift=(300:3)]1.5,2.59808) arc (300:260:3);
				\draw[thick,->] ([shift=(260:3)]1.5,2.59808) arc (260:250:3);
				\draw[thick] ([shift=(250:3)]1.5,2.59808) arc (250:240:3);
				\draw[thick,->] (-4,0) -- (-2.5,0);
				\draw[thick,->] (-2.5,0) -- (-2,0);
				\draw[thick,->] (-2,0) -- (-1.5,0);
				\draw[thick] (-1.5,0) -- (0,0);
				\draw[thick] (3,0) to[out=225,in=0] (1.5,-1) (-2.5,-1) to[out=180,in=315] (-4,0);
				\draw[thick,->] (1.5,-1) -- (0,-1);
				\draw[thick,->] (0,-1) -- (-0.5,-1);
				\draw[thick,->] (-0.5,-1) -- (-1,-1);
				\draw[thick] (-1,-1) -- (-2.5,-1);
				\draw[fill=white] (0,0) circle (0.2);
				\draw[fill=gray] (3,0) circle (0.2);
				\node[above] at (0,0.2) {$1$};
				\node[above] at (3,0.2) {$2$};
				\begin{scope}[shift={(-4,0)}]
					\draw[fill=red] (-0.15,-0.15) -- (-0.15,0.15) -- (0.15,0.15) -- (0.15,-0.15) -- cycle;
				\end{scope}
				\node[above] at (1.5,0.401924) {$a_1$, $a_2$};
				\node[above] at (1.5,-0.401924) {$b_1$, $b_2$};
				\node[above] at (-2,0) {$r_1,\ldots, r_m$};
				\node[above] at (-0.5,-1) {$s_1,\ldots,s_{m-1}$};
			\end{tikzpicture}
		\end{array},\\ 
W_m^{(i)}&=\Tr\left[b_2a_2b_1a_1-b_2a_1b_1a_2+\sum\lm_{i=1}^{m-1}s_i\left(a_2r_i-a_1r_{i+1}\right)\right]\,,
	\end{aligned}
	\ee
apply one quiver mutation process, and show that the result is given precisely by~$\left(Q_{m+1}^{(i)},W_{m+1}^{(i)}\right)$.
\bigskip

We apply the standard procedure of quiver mutation \cite{Benini:2014mia} on the quiver in~\eqref{eq:quiv_inf_conif_app} with respect to the node $1$.
First, each pair of chiral fields $p$ and $q$ forming a path going through the mutation node, which is node $1$ here, is represented in the dual theory (i.e.\ the mutated quiver) by a meson operator $M_{pq}=q\cdot p$. 
In our case, for the mutated quiver we derive the following meson operators: 
	\be
	\begin{aligned}
		&S_{ij}=a_ib_j \in \{ 2\to 2\},\;i,j=1,2\,,\\
		&M_{k,i}=a_i r_k\in \{\infty\to 2\},\;k=1,\ldots,m\,,
	\end{aligned}
	\ee
where $\infty$ labels the framing node.
Secondly, in addition to the meson field for each pair $p$ and $q$, in the mutated quiver one needs to add the corresponding dual chiral fields $\check p$ and $\check q$, going in the opposite direction of $p$ and $q$, respectively. 
Finally, the superpotential $W$ needs to be modified accordingly. 
Since $W$ is a gauge invariant quantity, it should depend only on the meson combination $q\cdot p$. 
Therefore the modification $W$ should be performed according to the following rule for each pair of $q$ and $p$:
\begin{equation}
	W\ni \Tr\, f(qp)\quad \longrightarrow\quad  \check W\ni \Tr\left[f(M_{pq})+M_{pq}\check p\check q\right].
\end{equation}
After these three steps, we obtain the quiver and superpotential for the Seiberg dual theory that results from the quiver mutation w.r.t. node $1$ of the theory given by~\eqref{eq:quiv_inf_conif_app}:
	\be\label{conif_Seib_dual}
	\begin{aligned}
		\check Q_m&=\begin{array}{c}
			\begin{tikzpicture}
				\draw[thick] ([shift=(120:3)]1.5,-2.59808) arc (120:110:3);
				\draw[thick,<-] ([shift=(110:3)]1.5,-2.59808) arc (110:100:3);
				\draw[thick,<-] ([shift=(100:3)]1.5,-2.59808) arc (100:60:3);
				\draw[thick] ([shift=(300:3)]1.5,2.59808) arc (300:290:3);
				\draw[thick,<-] ([shift=(290:3)]1.5,2.59808) arc (290:280:3);
				\draw[thick,<-] ([shift=(280:3)]1.5,2.59808) arc (280:240:3);
				\draw[thick] (-4,0) -- (-2.5,0);
				\draw[thick,<-] (-2.5,0) -- (-2,0);
				\draw[thick,<-] (-2,0) -- (-1.5,0);
				\draw[thick,<-] (-1.5,0) -- (0,0);
				\draw[thick] (3,0) to[out=225,in=0] (1.5,-1) (-2.5,-1) to[out=180,in=315] (-4,0);
				\draw[thick,->] (1.5,-1) -- (0,-1);
				\draw[thick,->] (0,-1) -- (-0.5,-1);
				\draw[thick,->] (-0.5,-1) -- (-1,-1);
				\draw[thick] (-1,-1) -- (-2.5,-1);
				\draw[thick,dashed] (3,0) to[out=240,in=0] (1.5,-1.5) (-2.5,-1.5) to[out=180,in=300] (-4,0); 
				\draw[thick,dashed,->] (-2.5,-1.5) -- (-0.5,-1.5);
				\draw[thick,dashed] (-0.5,-1.5) -- (1.5,-1.5);
				\draw[dashed,thick,->] (3,0) to[out=330,in=180] (3.5,-0.3) to[out=0,in=270] (3.8,0); 
				\draw[dashed,thick] (3.8,0) to[out=90,in=0] (3.5,0.3) to[out=180,in=30] (3,0);
				\draw[fill=white] (0,0) circle (0.2);
				\draw[fill=gray] (3,0) circle (0.2);
				\node[above] at (0,0.2) {$\check 1$};
				\node[above] at (3,0.2) {$\check 2$};
				\begin{scope}[shift={(-4,0)}]
					\draw[fill=red] (-0.15,-0.15) -- (-0.15,0.15) -- (0.15,0.15) -- (0.15,-0.15) -- cycle;
				\end{scope}
				\node[above] at (1.5,0.401924) {$\check a_1$, $\check a_2$};
				\node[above] at (1.5,-0.401924) {$\check b_1$, $\check b_2$};
				\node[above] at (-2,0) {$\check r_1,\ldots, \check r_m$};
				\node[above] at (-0.5,-1) {$s_1,\ldots,s_{m-1}$};
				\node[below] at (-0.5,-1.5) {$M_{k,i}$};
				\node[right] at (3.8,0) {$S_{ij}$};
			\end{tikzpicture}
		\end{array},\\ 
		\check W_m&=\Tr\left[S_{22}S_{11}-S_{21}S_{12}+\sum\lm_{k=1}^{m-1}s_k\left(M_{k,2}-M_{k+1,1}\right)\right. \\
		&\qquad \qquad  \qquad\qquad \qquad \left. +\sum\lm_{i,j=1}^2S_{ij}\, \check b_j\, \check a_i+\sum\lm_{k=1}^m\sum\lm_{i=1}^2M_{k,i}\, \check r_k\, \check a_i\right].
	\end{aligned}
	\ee
\bigskip

To complete the description of the dual quiver theory, it suffices to calculate the quiver dimensions and the stability parameters in the dual quiver~\eqref{conif_Seib_dual}. 
For the quiver dimensions and the FI parameters we have:
\be\label{dzetacheck}
\begin{aligned}
&(\check d_{1},\check d_{2})=(2d_{2}-d_{1},d_{2})\;,\\
&(\check \zeta_{1},\check \zeta_{2})=(-\zeta_{1},\zeta_{2}+2\zeta_{1}) \;.
\end{aligned}
\ee
Applying the relations \eqref{eq:conif_chamb_FI} and \eqref{eq:conif_chamb_dim} on~\eqref{dzetacheck}, we find that under this map these vectors are transformed in the following way:
\be
\left(\vec d_m^{(i)},\;\vec \zeta_m^{(i)}\right)\longrightarrow \left(P\vec d_{m+1}^{(i)},\;P\vec \zeta_{m+1}^{(i)}\right),
\ee
where the operator $P$ permutes the components of a two-dimensional vector.

Finally, to see that the dual quiver~\eqref{conif_Seib_dual} is in the same family as 	\eqref{eq:quiv_inf_conif_app}, with $m\rightarrow m+1$, we simplify the dual quiver ~\eqref{conif_Seib_dual} further.
First, one can see that the quiver and the superpotential defined in  \eqref{conif_Seib_dual} are redundant: some of the relations in the quiver path algebra have a single solution and the corresponding fields can be integrated out in the effective field theory.
Integrating over fields $S_{ij}$, $s_k$  --  using their equations of motion that follow from the superpotential constraint  --  we derive:
	\be
	\begin{aligned}
		S_{22}=-\check b_1\check a_1\;,\; S_{11}=-\check b_2\check a_2\;,\; S_{12}=\check b_1\check a_2\;,\; S_{21}=\check b_2\check a_1\;,\\
		M_{k,2}=M_{k+1,1}\;,\; k=1,\ldots, m-1\;.
	\end{aligned}
	\ee
Then it is useful to define a new chiral field $\check s_k$ in the following way:
	\be
	\begin{aligned}
		\check s_1&:=M_{1,1}\;,\; \\
		\check s_k&:=M_{k,1}=M_{k-1,2}\;,\; \quad \textrm{for}\quad k=2,\ldots, m\;,\;\\
		\check s_{m+1}&:=M_{m,2}\;.
	\end{aligned}
	\ee
Substituting these solutions (to the equations of motion) back into the superpotential, we derive the following reduced quiver and superpotential:
	\be\label{eq:red_mutated_QW}
	\begin{aligned}
		\check Q_m^{\rm red} &=\begin{array}{c}
			\begin{tikzpicture}
				\draw[thick] ([shift=(120:3)]1.5,-2.59808) arc (120:110:3);
				\draw[thick,<-] ([shift=(110:3)]1.5,-2.59808) arc (110:100:3);
				\draw[thick,<-] ([shift=(100:3)]1.5,-2.59808) arc (100:60:3);
				\draw[thick] ([shift=(300:3)]1.5,2.59808) arc (300:290:3);
				\draw[thick,<-] ([shift=(290:3)]1.5,2.59808) arc (290:280:3);
				\draw[thick,<-] ([shift=(280:3)]1.5,2.59808) arc (280:240:3);
				\draw[thick] (-4,0) -- (-2.5,0);
				\draw[thick,<-] (-2.5,0) -- (-2,0);
				\draw[thick,<-] (-2,0) -- (-1.5,0);
				\draw[thick,<-] (-1.5,0) -- (0,0);
				\draw[thick] (3,0) to[out=225,in=0] (1.5,-1) (-2.5,-1) to[out=180,in=315] (-4,0);
				\draw[thick] (1.5,-1) -- (0,-1);
				\draw[thick,<-] (0,-1) -- (-0.5,-1);
				\draw[thick,<-] (-0.5,-1) -- (-1,-1);
				\draw[thick,<-] (-1,-1) -- (-2.5,-1);
				\draw[fill=white] (0,0) circle (0.2);
				\draw[fill=gray] (3,0) circle (0.2);
				\node[above] at (0,0.2) {$\check 1$};
				\node[above] at (3,0.2) {$\check 2$};
				\begin{scope}[shift={(-4,0)}]
					\draw[fill=red] (-0.15,-0.15) -- (-0.15,0.15) -- (0.15,0.15) -- (0.15,-0.15) -- cycle;
				\end{scope}
				\node[above] at (1.5,0.401924) {$\check a_1$, $\check a_2$};
				\node[above] at (1.5,-0.401924) {$\check b_1$, $\check b_2$};
				\node[above] at (-2,0) {$\check r_1,\ldots, \check r_m$};
				\node[above] at (-0.5,-1) {$\check s_1,\ldots,\check s_{m+1}$};
			\end{tikzpicture}
		\end{array},\\ 
		\check W_m^{\rm red} &=-\left(\check a_1\check b_1\check a_2\check b_2-\check a_1\check b_2\check a_2\check b_1\right)+\sum\lm_{k=1}^m\check r_k\left(\check a_1\check s_k+\check a_2\check s_{k+1}\right)\,,
	\end{aligned}
	\ee
which is equivalent to the mutated quiver and superpotential pair~\eqref{conif_Seib_dual}.
The chiral masses of the dual fields are:
\begin{equation}
\begin{split}
	\mu(\check a_1)&=\mathsf{h}_1\;,\quad \mu(\check a_2)=-\mathsf{h}_1\;,\\
	\mu(\check b_1)&=\mathsf{h}_2\;,\quad \mu(\check b_2)=-\mathsf{h}_2\;,\\
	\mu(\check r_k)&=-2(k-1)\mathsf{h}_1\;,\; \quad \textrm{for}\quad k=1,\ldots, m\;,\\
	\mu(\check s_k)&=(2k-3)\mathsf{h}_1\;,\; \quad \textrm{for} \quad k=1,\ldots,m+1\;.
\end{split}
\end{equation}

One can immediately see that the reduced version of the mutated pair of a quiver and a superpotential $({\check Q}_m^{\rm red},{\check W}_m^{\rm red})$  in~\eqref{eq:red_mutated_QW}  is identical to the pair $(Q_{m+1},W_{m+1})$, i.e.  \eqref{eq:quiv_inf_conif_app} with $m\rightarrow m+1$, after renaming the quiver gauge nodes: $\check 1\to 2$ and $\check 2\to 1$.
		
\section{New representations via redefinition of ground states}\label{sec:app_manual_charge_f}

In this Appendix we explain in more detail an alternative method of deriving the representation associated with a subcrystal, mentioned at the end of Section~\ref{ssec:rep_subcry}.
This was the method used to study non-trivial representations of the affine Yangian of $\mathfrak{gl}_1$ for the $\mathbb{C}^3$ geometry \cite{Prochazka:2015deb, Datta:2016cmw, Gaberdiel:2017dbk,Gaberdiel:2018nbs}; and here we generalize it to the quiver Yangian for arbitrary toric Calabi-Yau threefolds.
The end result agrees with the positive/negative crystal method employed throughout the main text. 
But the method in this appendix serves to confirm the results and to highlight the advantage of the new method.
\bigskip

The basic idea is simple. 
In the crystal melting we usually consider removing a finite number of atoms from the canonical crystal $\CC_0$. 
Let us instead consider removing an \emph{infinite} number of atoms from the crystal $\CC_0$, to obtain a state $|\Kappa\rangle$, which consists of these infinite number of atoms ``melted away" from the crystal $\CC_0$.
We do this by choosing a set of atoms at asymptotic infinity, and 
subsequently remove all the atoms minimally needed to 
satisfy the melting rule.

We can now regard the state $|\Kappa\rangle$ as a new vacuum: 
\begin{equation}
|\varnothing \rangle_{\rm new} \equiv |\Kappa\rangle\,,
\end{equation}
and start growing the crystal from there. 
Indeed, since $\Kappa$ contains all the atoms minimally needed to 
consistently remove the atoms of $\CC_0$ at infinity, it is not possible to further remove finitely-many atoms from $\Kappa$;
the only possibility is to add atoms to $\Kappa$, and this ensures that 
the state $|\Kappa\rangle$ can be regarded as a new ground state.\footnote{When we choose a certain excited state as a new ``vacuum'', one usually encounters holes in addition to particles. In our situation $\Kappa$ is chosen such that creating a hole consistently requires removing an infinite number of atoms, and hence costs ``infinite energy''. This ensures that there are no ``hole-like'' excitations.}
In other words, this has the effect of permanently removing some atoms from the canonical crystal $\mathcal{C}_0$,
to obtain a new subcrystal ${}^{\sharp} \mathcal{C}$.

The vacuum charge function ${}^{\sharp}\psi$ associated with the 
representation can be derived from the charge function for the state $|\Kappa\rangle$ as
\begin{align}
{}^{\sharp}\psi^{(a)}_0(z) \equiv \lim_{\Kappa'\to \Kappa}   \Psi^{(a)}_{\Kappa'}(z) \;,
\end{align}
where $\Kappa'$ has only finitely-many atoms removed. 
As we will see, this limit in general requires a regularization procedure.

\subsection{Example for open BPS states: \texorpdfstring{$(\mathbb{C}^2/\mathbb{Z}_2)\times \mathbb{C}$}{C3/Z2}}

Let us reexamine the open BPS states of \texorpdfstring{$(\mathbb{C}^2/\mathbb{Z}_2)\times \mathbb{C}$}{C3/Z2}, studied in Section~\ref{ssec:Open_Z2}, using the regularization procedure.
Recall that a general subcrystal corresponding to the open BPS states of \texorpdfstring{$(\mathbb{C}^2/\mathbb{Z}_2)\times \mathbb{C}$}{C3/Z2} is given by the three Young diagrams asymptotics $\vec{Y}=(Y_1,Y_2,Y_3)$.
Here, to illustrate the procedure, we will first obtain the three fundamental representations, with only one of $Y_i=\square$ and the other two trivial.
Then we will explain how to construct representations with arbitrary $\vec{Y}$ by combining these fundamental representations.

\bigskip

The canonical crystal, the canonically framed quiver and superpotential, and the 
 corresponding unshifted affine Yangian of $\mathfrak{gl}_2$ for the \texorpdfstring{$(\mathbb{C}^2/\mathbb{Z}_2)\times \mathbb{C}$}{C3/Z2} geometry were summarized in Section~\ref{sssec:canonical_gl2}.
Now, to consider open BPS states corresponding to a simplest non-trivial Young diagram, $\Box$, along the $x_1$ direction, we start from the canonical crystal and remove all atoms located at 
\begin{align}
\blacksquare_1 : \quad (x_1, x_2, x_3)=(m\in \mathbb{N}_0, 0, 0)\,,
\end{align} 
along the $x_1$ direction.
By the definition \eqref{eq.efpsi-action-2} and the coloring scheme \eqref{eq:color_def_Zn}, we can compute the charge functions as
\begin{equation}
\label{eq-Psi-minimal1-gl2}
\Psi^{(a)}_{\blacksquare_1}(u)=(\psi_0(u))^{\delta_{a,1}} \prod^{\infty}_{n=0} \varphi^{1\Rightarrow a}(u- 2n\mathsf{h}_1) \varphi^{2\Rightarrow a}(u- (2n+1)\mathsf{h}_1)\,,
\end{equation}
where $n$ accounts for the $n^{\textrm{th}}$ atom sequence $1\rightarrow 2$ in the long row of atoms from the origin to infinity along the $x_1$ direction.
Plugging in the bond factors
\begin{equation}\label{eq-bond-gl2}
\begin{aligned}
&\varphi^{a\Rightarrow a}(u)=\frac{u+\mathsf{h}_3}{u-\mathsf{h}_3} \qquad \textrm{and}\qquad\varphi^{a+1\Rightarrow a}(u)=\frac{(u+\mathsf{h}_1)(u+\mathsf{h}_2)}{(u-\mathsf{h}_1)(u-\mathsf{h}_2)} \;,
\end{aligned}
\end{equation}
we have 
\begin{equation}\label{eq-Psi1-minimal1-gl2-2}
\begin{aligned}
\Psi^{(1)}_{\blacksquare_1}(u)
&=\frac{1}{u}\cdot\frac{u}{u-\mathsf{h}_3}\left(\lim_{k \rightarrow \infty}\frac{u-2k\,\mathsf{h}_1-\mathsf{h}_3}{u-2k\,\mathsf{h}_1}\right)
\rightarrow\frac{1}{u-\mathsf{h}_3} =: {}^{\sharp}\psi^{(1)}_{\blacksquare_1}(u)\,,\\
\Psi^{(2)}_{\blacksquare_1}(u)&=\frac{u+\mathsf{h}_1}{u-\mathsf{h}_2}\left(\lim_{k \rightarrow \infty}\frac{u-(2k-1)\,\mathsf{h}_1+\mathsf{h}_3}{u-(2k-1)\,\mathsf{h}_1}\right)
\rightarrow \frac{u+\mathsf{h}_1}{u-\mathsf{h}_2}=: {}^{\sharp}\psi^{(1)}_{\blacksquare_1}(u)\,.
\end{aligned}
\end{equation}
In deriving \eqref{eq-Psi1-minimal1-gl2-2} from \eqref{eq-Psi-minimal1-gl2}, we first see that although the naive definition \eqref{eq-Psi-minimal1-gl2} of the charge function of the non-trivial representation $\blacksquare_1$ involves infinitely many factors (from the infinitely many atoms), there are actually a lot of cancellation between factors from the neighboring atoms.
After the cancellation, there is a limiting function in the bracket (for each $a=1,2$), which describes the end of the long row with the position of the end going to the infinity.
Secondly, for this particular representation, the limiting function evaluates to a finite number (which is $1$ here). 
In the end, the charge functions \eqref{eq-Psi1-minimal1-gl2-2} of the state $\blacksquare_1$ are rather simple rational functions, which become the ground state charge functions of the representation $\blacksquare_1$.

We can verify that this is precisely the ground state charge function expected for the minimal representation $\blacksquare_1$.
The functions \eqref{eq-Psi1-minimal1-gl2-2} have poles at $z=\mathsf{h}_3$ and $\mathsf{h}_2$, with color $1$ and $2$, representing two positive crystals whose leading atoms are color $1$ and $2$, respectively; there is also a color $2$ zero 
at $z=-\mathsf{h}_1=\mathsf{h}_2+\mathsf{h}_3$, representing a negative crystal (whose leading atom is color $2$), which is needed to cancel the 
overlap of the two positive crystals.

\bigskip

Since the crystal has a symmetry exchanging the $x_1$ and $x_2$ directions, the case of the $x_2$-direction is parallel to that of the $x_1$-direction, with the result
\begin{equation}\label{eq-Psi1-minimal2-gl2}
\begin{aligned}
\Psi^{(1)}_{\blacksquare_2}(u)
&=\frac{1}{u}\cdot\frac{u}{u-\mathsf{h}_3}\left(\lim_{k \rightarrow \infty}\frac{u-2k\,\mathsf{h}_2-\mathsf{h}_3}{u-2k\,\mathsf{h}_2}\right)
\rightarrow\frac{1}{u-\mathsf{h}_3} =: {}^{\sharp}\psi^{(1)}_{\blacksquare_2}(u)\,,\\
\Psi^{(2)}_{\blacksquare_2}(u)
&=\frac{u+\mathsf{h}_2}{u-\mathsf{h}_1}\left(\lim_{k \rightarrow \infty}\frac{u-(2k-1)\,\mathsf{h}_2+\mathsf{h}_3}{u-(2k-1)\,\mathsf{h}_2}\right)
\rightarrow \frac{u+\mathsf{h}_2}{u-\mathsf{h}_1}=: {}^{\sharp}\psi^{(1)}_{\blacksquare_2}(u)\,.
\end{aligned}
\end{equation}

What about the analogous representation along the $x_3$ direction?
First of all, from the crystal structure, the charge function of $\blacksquare_3$ is 
\begin{equation}
\label{eq-Psi-minimal3-gl2}
\Psi^{(a)}_{\blacksquare_3}(u)=(\psi_0(u))^{\delta_{a,1}} \prod^{\infty}_{n=0} \varphi^{1\Rightarrow a}(u-n\mathsf{h}_3)\,.
\end{equation}
Plugging in the bond factor (\ref{eq-bond-gl2}), we get
\begin{equation}\label{eq-Psi12-minimal3-gl2}
\begin{aligned}
\Psi^{(1)}_{\blacksquare_3}(u)
&=\frac{(u+\mathsf{h}_3)}{\lim_{n\rightarrow \infty} (u-n \mathsf{h}_3)(u-(n+1) \mathsf{h}_3)} \rightarrow (u+\mathsf{h}_3) =:{}^{\sharp}\psi^{(1)}_{\blacksquare_3}(u)\;, \\
\Psi^{(2)}_{\blacksquare_3}(u)&
=\frac{\lim_{n\rightarrow \infty} (u+\mathsf{h}_1-n \mathsf{h}_3)(u+\mathsf{h}_2-n\mathsf{h}_3) 
}{(u-\mathsf{h}_1)(u-\mathsf{h}_2)} \rightarrow \frac{1}{(u-\mathsf{h}_1)(u-\mathsf{h}_2)} =:{}^{\sharp}\psi^{(2)}_{\blacksquare_3}(u)\;.
\end{aligned}
\end{equation}
Although there are still a lot of cancellations between the factors from the neighboring atoms, 
we see that the limit function doesn't evaluate to a finite number, unlike the cases for $\blacksquare_{1,2}$.
We nevertheless proceeded to regularize by throwing away factors corresponding to infinities in the $u$-plane.
The end results are regarded as the ground state charge functions of the representation $\blacksquare_3$.
This is again expected for the minimal representation for $\blacksquare_{3}$: the decomposition of the subcrystal gives two positive crystals at $u=\mathsf{h}_1, \mathsf{h}_2$ 
and one negative crystal for the overlap $u=-\mathsf{h}_3=\mathsf{h}_1+\mathsf{h}_2$.
\bigskip

With the ground state charge functions of the three fundamental representations \eqref{eq-Psi1-minimal1-gl2-2}, \eqref{eq-Psi1-minimal2-gl2}, and \eqref{eq-Psi12-minimal3-gl2} as the building blocks, one can construct the ground state charge function for the representation associated with a subcrystal with arbitrary asymptotics $\vec{Y}$.
For example, for $\vec{Y}=(\mbox{\tiny $\yng(2)$}, \emptyset,\emptyset )$, we need two $\blacksquare_1$, from \eqref{eq-Psi1-minimal1-gl2-2}, and for the second $\blacksquare$ factor, we need to apply $\Psi^{(a)}_{\blacksquare}(u)\rightarrow \Psi^{(a+1)}_{\blacksquare}(u-\mathsf{h}_2)$. Combining them together we get the ground state charge functions as\footnote{Note that the vacuum factor $\frac{1}{u}$ should only appear once and is never shifted.}
\begin{equation}
\begin{aligned}
\Psi^{(1)}_{\vec{Y}}(u)
&=\frac{1}{u}\cdot\frac{u}{u-\mathsf{h}_3} \cdot \frac{u+\mathsf{h}_1-\mathsf{h}_2}{u-2\mathsf{h}_2} =\frac{1}{u-\mathsf{h}_3} \cdot \frac{u+\mathsf{h}_1-\mathsf{h}_2}{u-2\mathsf{h}_2} =: {}^{\sharp}\psi^{(1)}_{\vec{Y}}(u)\,,\\
\Psi^{(2)}_{\vec{Y}}(u)
&=\frac{u+\mathsf{h}_1}{u-\mathsf{h}_2}\cdot  \frac{u-\mathsf{h}_2}{u-\mathsf{h}_3-\mathsf{h}_2}=1=: {}^{\sharp}\psi^{(2)}_{\vec{Y}}(u)\,.
\end{aligned}
\end{equation}
Again, one can check that all the poles and zeros are expected from the shape of the subcrystal.
Take another example $\vec{Y}=({\tiny \yng(1)},\emptyset, {\tiny \yng(1)})$, we need a  $\blacksquare_1$ from \eqref{eq-Psi1-minimal1-gl2-2}, and a $\blacksquare_3$ from
\eqref{eq-Psi12-minimal3-gl2}, and remove the double counting from the contribution of the first atom (which is shared between $\blacksquare_1$ and $\blacksquare_3$). 
Altogether, we have
\begin{equation}
\begin{aligned}
\Psi^{(1)}_{\vec{Y}}(u)
&=\frac{1}{u}\cdot\frac{u}{u-\mathsf{h}_3} \cdot u(u+\mathsf{h}_3)\cdot \frac{u-\mathsf{h}_3}{u+\mathsf{h}_3}\\
&=u=: {}^{\sharp}\psi^{(1)}_{\vec{Y}}(u)\,,\\
\Psi^{(2)}_{\vec{Y}}(u)
&=\frac{u+\mathsf{h}_1}{u-\mathsf{h}_2}\cdot  \frac{1}{(u-\mathsf{h}_1)(u-\mathsf{h}_2)}\cdot \frac{(u-\mathsf{h}_1)(u-\mathsf{h}_2)}{(u+\mathsf{h}_1)(u+\mathsf{h}_2)} \\
&=  \frac{1}{(u-\mathsf{h}_2)(u+\mathsf{h}_2)} 
=: {}^{\sharp}\psi^{(2)}_{\vec{Y}}(u)\,,
\end{aligned}
\end{equation}
where the positions of the poles and zeros are expected from the shape of the subcrystal. 
Although repeating this procedure can give the ground state charge function of any subcrystal, it is clear that the positive/negative crystal method used in the main text is much simpler.

\subsection{Example for wall crossing: conifold}
\label{sec:app:WC_con}

In the previous subsection, we showed how to generalize the method of studying non-trivial representations of the affine Yangian of $\mathfrak{gl}_1$ for the $\mathbb{C}^3$ geometry \cite{Prochazka:2015deb, Datta:2016cmw, Gaberdiel:2017dbk,Gaberdiel:2018nbs} to the case of open BPS states for other geometries, with  \texorpdfstring{$(\mathbb{C}^2/\mathbb{Z}_2)\times \mathbb{C}$}{C3/Z2} as an example.
In this subsection, we will show that this method can also be used to construct representations corresponding to other chambers of wall-crossing. 
We will consider the example of the infinite chamber $\CC^{(i)}_m$ for the conifold geometry, already studied using the positive/negative crystal method in Section~\ref{ssec:infinite_chamber_conifold}.

\bigskip

\begin{figure}[ht!]\label{fig:conifold_cut}
		\begin{center}
			\begin{tikzpicture}
			\node(A) at (0,0) {\includegraphics[scale=0.25]{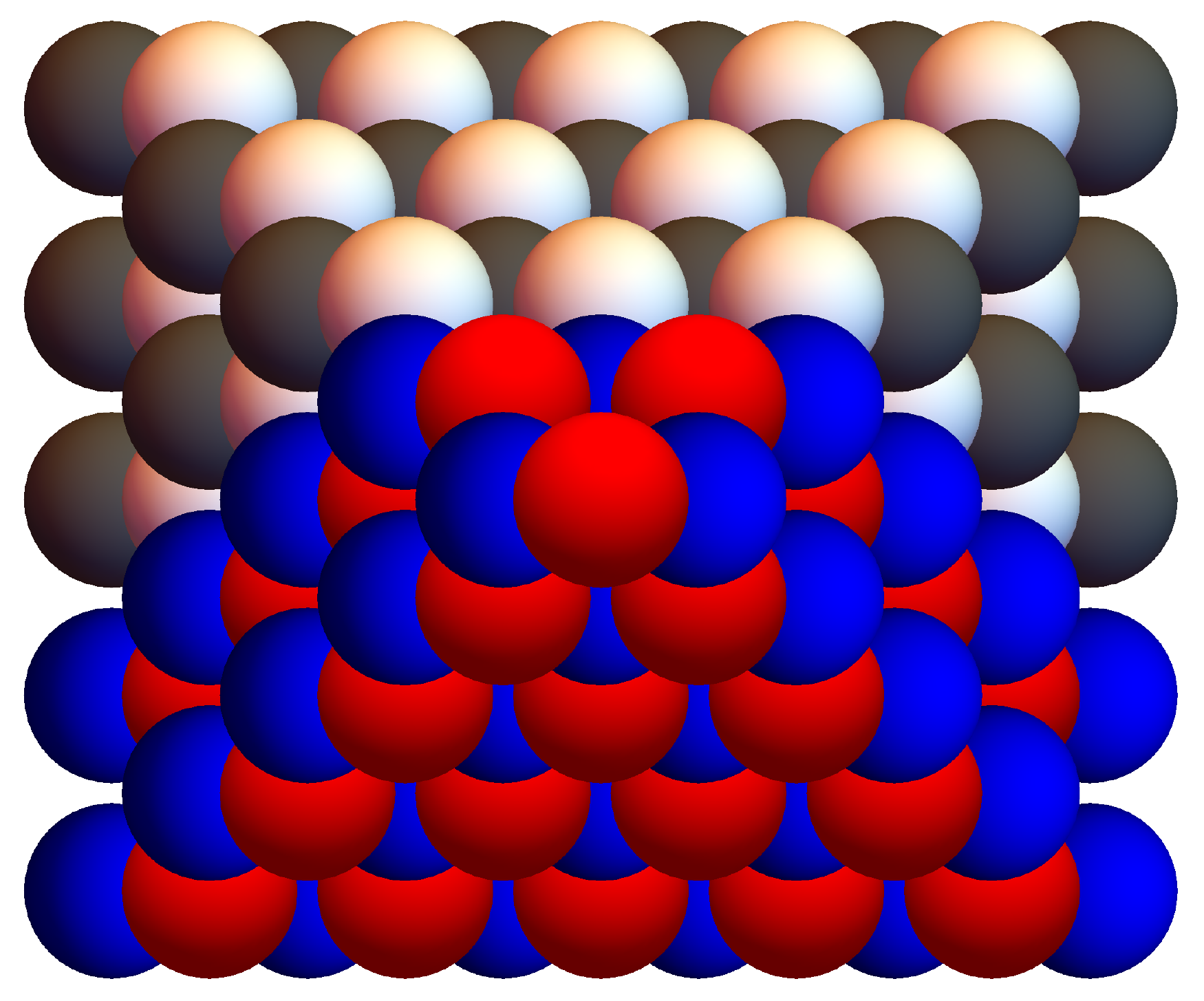}};
			\draw[ultra thick,->] (-2.5,0) -- (2.5,0);
			\draw[ultra thick,->] (0,-2) -- (0,2);
			\node[left] at (3.15,0) {$x_1$};
			\node[above left] at (0,2) {$x_2$};
			\begin{scope}[shift={(-3,-1.5)}]
			    \draw (0,0) circle (0.2);
			    \draw[ultra thick] (-0.141421,-0.141421) -- (0.141421,0.141421) (-0.141421,0.141421) -- (0.141421,-0.141421);
			    \node[below] at (0,-0.2) {$x_3$};
			\end{scope}
			\draw[black!10!orange,ultra thick,->] (-2.5,2) to[out=270,in=90] (-0.7,0.9);
			\draw[black!10!orange,ultra thick,->] (-2.5,2) to[out=270,in=90] (0,0.9);
			\draw[black!10!orange,ultra thick,->] (-2.5,2) to[out=270,in=90] (0.7,0.9);
			\draw[black!10!orange,ultra thick] (-2.5,2) to[out=90,in=180] (-2,2.7);
			\node[right] at (-2,2.7) {$m$ (white) starters at positions $(x_1,x_2,x_3) = (m-1-2n,m-1,m-1)$};
			\draw[black!30!green,ultra thick,->] (-1.5,-2) to[out=90,in=270] (-0.35,0.9);
			\draw[black!30!green,ultra thick,->] (-1.5,-2) to[out=90,in=270] (0.35,0.9);
			\draw[black!30!green,ultra thick] (-1.5,-2) to[out=270,in=90] (-2.3,-2.3) to[out=270,in=180] (-2,-2.5);
			\node[right] at (-2,-2.5) {$m-1$ (black) pausers at positions $(x_1,x_2,x_3) = (m-2-2n,m-1,m-1)$};
			\begin{scope}[shift={(7,0)}]
			    \node(B) at (0,0) {\includegraphics[scale=0.2]{figures/Prism.pdf}};
			\end{scope}
			\path ([shift={(0.5,0)}]A.east) edge[->] (B);
			\end{tikzpicture}
			\caption{The subcrystal for the infinite chamber $\CC^{(i)}_m$ can be obtained from the canonical crystal $\CC_0=\CC^{(i)}_1$ by removing $m-1$ double layers of atoms. Left: canonical crystal $\CC_0$ with two double slices of atoms marked (red for color $1$ and blue for color $2$). Right: removing the red and blue atoms gives the subcrystal for the infinite chamber $\CC^{(i)}_m$ with $m=3$.
			 }\label{fig:inf_quiv_cry_cut}
		\end{center}
	\end{figure}

To apply this method, we first compare the subcrystal $\CC^{(i)}_m$ with the canonical crystal $\CC_0=\CC^{(i)}_{m=1}$, and determine which atoms to remove in order to obtain $\CC^{(i)}_m$ from $\CC^{(i)}_{1}$.
In Figure~\ref{fig:inf_quiv_cry_cut}, we reproduce the crystals $\CC^{(i)}_1$ from Figure~\ref{fig:pyramid} and $\CC^{(i)}_m$ from Figure~\ref{fig:inf_quiv_cry} (for $m=3$).
Comparing the two, we see that in order to obtain the latter from the former, we need to remove $m-1$ (which equals $2$) double slices of atoms along one of the surfaces, which are colored red for node $1$ and blue for node $2$.

In order to compute the charge function, we need to specify the coordinates of the atoms in the canonical crystal $\CC_0=\CC^{(i)}_1$.
Recall that each box in the plane partition can be described by the 3D coordinates $(x_1,x_2,x_3)$ with $x_i\in \mathbb{N}_0$.
Similarly, we label the atoms in the crystal $\CC^{(i)}_1$ by the coordinates $(x_1,x_2,x_3=\ell)$. 
First of all, the crystal has a layered structure, labeled by $x_3=\ell$ with $\ell=0,1,2,\cdots,\infty$. 
Each level has two sublevels: first a level of atoms with color $1$ followed by a level of atoms with color $2$.
Within each level, the atom is further distinguished by their $(x_1,x_2)$ coordinate. 
At level $\ell$, first there is a sublevel that contains $(\ell+1)^2$ atoms of color $1$, with positions
\begin{equation}
\begin{aligned}
x_1&=-\ell\,, -\ell+2\,,\cdots, \ell-2\,, \ell\,;\quad\\
 x_2&=-\ell\,, -\ell+2\,,\cdots, \ell-2\,, \ell\,;\\
 x_3&=\ell\,.
\end{aligned}
\end{equation}
These are followed by the sublevel with $(\ell+1)(\ell+2)$ atoms of color $2$, with positions\footnote{Note that in this subsection, we are using a slightly different coordinate system for $x_3$ for the conifold geometry from the one used in the main text, given by \eqref{eq-basisvector-conifold}. The main difference is that here we are using a double-layer coordinate system in which the white atoms and the black atoms in the same double-layer have the same $x_3$ coordinate, whereas in the main text we have the single-layer system in which atoms from different layers have different $x_3$ coordinate. However, since the coordinate function \eqref{eq:cfconifoldm} is independent of the $x_3$ coordinate of the atom, this difference is immaterial to our discussion.}
\begin{equation}
\begin{aligned}
x_1&=-\ell-1\,, -\ell+1\,,\cdots, \ell-1\,, \ell+1\,;\quad \\
x_2&=-\ell\,, -\ell+2\,,\cdots, \ell-2\,, \ell\,;\\
x_3&=\ell\,.
\end{aligned}
\end{equation}
The coordinate function of the atom at $(x_1,x_2,x_3=\ell)$ is
\begin{equation}
\begin{aligned}
h(\square)=\mathsf{h}_1\, x_1(\square)+\mathsf{h}_2\, x_2(\square)\,.
\end{aligned}
\end{equation}

As shown in Figure~\ref{fig:inf_quiv_cry_cut}, the subcrystal that corresponds to the chamber $\CC^{(i)}_m$ can be obtained from the canonical crystal $\CC_0=\CC^{(i)}_1$ by removing $m-1$ double layers of atoms along the surface.
The atoms to be removed are: the atoms of color $1$ at positions
\begin{equation}
\begin{aligned}
x_1&=-\ell\,, -\ell+2\,,\cdots, \ell-2\,, \ell\,;\quad \\
 x_2&=-\ell\,, -\ell+2\,,\cdots, -\ell+2(m-2)\,;\\
x_3&=\ell\,,
\end{aligned}
\end{equation}
together with atoms of color $2$ at positions
\begin{equation}
\begin{aligned}
x_1&=-\ell-1\,, -\ell+1\,,\cdots, \ell-1\,, \ell+1\,;\\
x_2&=-\ell\,, -\ell+2\,,\cdots, -\ell+2(m-2)\,;\\
x_3&=\ell\,,
\end{aligned}
\end{equation}
for $\ell=0,1,\dots,\infty$.

Adding these atoms to the vacuum, we obtain the state to be considered as the ground state of the representation $\CC^{(i)}_m$.
The charge function can be computed by definition \eqref{eq.efpsi-action-2} as
\begin{equation}
\begin{aligned}
\Psi^{(a)}(z)=&(\psi_0(z))^{\delta_{a,1}}\\
\times\prod^{\infty}_{\ell=0}
\prod^{\textrm{min}(m-2,\ell)}_{n_2=0}
&\Big(\prod^{\ell}_{n_1=0} \varphi^{1\Rightarrow a} \left(z-\left((\ell-2\,n_1)\,\mathsf{h}_1-(\ell-2\,n_2)\, \mathsf{h}_2\right)\right) \\
&\times\prod^{\ell+1}_{n_1=0} \varphi^{2\Rightarrow a} \left(z-\left((\ell+1-2\,n_1)\,\mathsf{h}_1-(\ell-2n_2)\, \mathsf{h}_2\right)\right)\Big) \;.
\end{aligned}
\end{equation}
Plugging in the bond factor \eqref{eq-charge-function-conifold} of affine Yangian of $\mathfrak{gl}_{1|1}$, we have
\begin{equation}
\begin{aligned}
\Psi^{(1)}(z)
=&\frac{1}{\prod^{m-1}_{n=0} (z-((m-1-2n)\mathsf{h}_1+(m-1)\,\mathsf{h}_2))}\\
&\cdot \lim_{L\rightarrow \infty}
 \prod^{m-1}_{n_2=0} 
 \frac{\prod^{L+2}_{n_1=0}\left(z-\left((L+2-2n_1)\,\mathsf{h}_1-(L-2n_2)\,\mathsf{h}_2\right)\right)}
{\prod^{L+1}_{n_1=0}\left(z-\left((L+1-2n_1)\,\mathsf{h}_1-(L+1-2n_2)\,\mathsf{h}_2\right)\right)} \;,\\
\Psi^{(2)}(z)
=&\prod^{m-2}_{n=0}\left(z-((m-2-2n)\,\mathsf{h}_1+(m-1)\,\mathsf{h}_2)\right)\\
&\cdot 
\lim_{L \rightarrow 0}
\prod^{m-2}_{n_2=0}
\frac{\prod^{L}_{n_1=0} \left(z-\left((L-2n_1)\,\mathsf{h}_1-(L+1-2n_2)\,\mathsf{h}_2\right)\right)}
{\prod^{L+1}_{n_1=0}\left(z-\left((L+1-2n_1)\,\mathsf{h}_1-(L-2n_2)\,\mathsf{h}_2\right)\right)} \;.
\end{aligned}
\end{equation}

While there are a lot of cancellations between factors of neighboring atoms, the cancellation is not complete: the factors at the final level $L$ (before taking $L$ to infinity) remain. As a regularization we propose to drop these factors in the limit $L\to \infty$, to obtain
\begin{equation}
\label{eq-Psi12-gl11-m2}
\begin{aligned}
\Psi^{(1)}(z)&=\frac{1}{\prod^{m-1}_{n=0} (z-((m-1-2n)\mathsf{h}_1+(m-1)\,\mathsf{h}_2))} \;,\\
\Psi^{(2)}(z)&=\prod^{m-2}_{n=0}\left(z-((m-2-2n)\,\mathsf{h}_1+(m-1)\,\mathsf{h}_2)\right) \;.
\end{aligned}
\end{equation}
We see that this regularization scheme reproduces the correct vacuum charge function, as expected from the general discussions in Section~\ref{ssec:infinite_chamber_conifold}. 

The charge function $\Psi^{(1)}$ for chamber $\CC^{(i)}_m$ has $m$ poles, corresponding to the $m$ positions where one can add the first atom of color $1$:
\begin{equation}\label{eq:chamberm_pole_conifold}
\begin{aligned}
(x_1, x_2, x_3)=\left(m-1-2n, m-1, m-1\right)\,,&\\
\textrm{with}\quad & n=0, 1, \dots,m-1\,.
\end{aligned}
\end{equation}
On the other hand, the charge function $\Psi^{(2)}$ has no pole, corresponding to the fact that there is no place to add an atom of color $2$ immediately starting from the ground state of this representation. 
Instead, it has $m-1$ zeros at
\begin{equation}\label{eq:chamberm_zero_conifold}
\begin{aligned}
(x_1, x_2, x_3)=\left(m-2-2n, m-1, m-1\right)\,,&\\
\textrm{with}\quad & n=0, 1, \dots,m-2 \;.
\end{aligned}
\end{equation}
These poles and zeros agree with the positions of starters and pausers of the subcrystal $\CC^{(i)}_m$, derived using the positive/negative method in Section~\ref{ssec:infinite_chamber_conifold}.\footnote{Note that the computation in Section~\ref{ssec:infinite_chamber_conifold} and the one in the current section use slightly different coordinate systems, in particular, the atom at the origin in Section~\ref{ssec:infinite_chamber_conifold} has the coordinates $(-(m-1), m-1, m-1)$ in the coordinate system of the current section.  
Therefore when comparing the positions of the starter \eqref{eq:chamberm_pole_conifold} and the pausers \eqref{eq:chamberm_zero_conifold}  with \eqref{eq:inf_chamber_s} and \eqref{eq:inf_chamber_p}, respectively, we need to apply a coordinate shift $(x_1,x_2,x_3)\rightarrow (x_1+(m-1),x_2-(m-1),x_3 -(m-1))$ on the coordinates of the current section, which is immaterial to the final result since it only leads to an overall shift of the spectral parameter.}

\newpage
\bibliographystyle{utphys}
\bibliography{biblio}

\end{document}